\documentclass[preprint,aps,12pt,showpacs,nofootinbib,tightenlines,amsmath,amssymb]{revtex4}
\usepackage{amsmath}
\usepackage{graphicx}
\usepackage{amssymb}
\usepackage{color}

\newcommand{\be}{\begin{eqnarray}}
\newcommand{\ee}{\end{eqnarray}}

\newcommand{\p}{\partial}

\newcommand{\nn}{\nonumber}

\newcommand{\diag}{\mathop{\rm diag}}

\newcommand{\cQ}{{\mathcal Q}}

\newcommand{\hU}{\hat{U}}
\newcommand{\hD}{\hat{D}}
\newcommand{\hq}{\hat{q}}
\newcommand{\hu}{\hat{u}}

\newcommand{\cM}{{\mathcal M}}

\newcommand{\cD}{{\mathcal D}}

\newcommand{\cW}{{\mathcal W}}
\newcommand{\cS}{{\mathcal S}}

\newcommand{\fx}{{\mathbf x}}

\newcommand{\adjoin}{\,\widetilde{\Join}\, }

\newcommand{\1}{\mspace{1mu}}

\newcommand{\tga}{\tilde{\gamma}}

\newcommand{\vGa}{\varGamma}

\newcommand{\tvGa}{\tilde{\varGamma}}

\newcommand{\tvSi}{\tilde{\varSigma}}
\newcommand{\vSi}{\varSigma}
\newcommand{\tSi}{\tilde{\Sigma}}

\newcommand{\tla}{\tilde{\lambda}}

\newcommand{\mH}{\mathsf H }

\newcommand{\mL}{\mathsf L}
\newcommand{\tR}{\tilde{R}}
\newcommand{\mT}{\mathsf T}

\newcommand{\mR}{\mathsf R}

\newcommand{\mS}{\mathsf S}

\newcommand{\mM}{\mathsf M}
\newcommand{\mN}{\mathsf N}

\newcommand{\mQ}{\mathsf Q}
\newcommand{\mU}{\mathsf U}

\newcommand{\bmA}{\bar{\mathsf A}}

\newcommand{\tmA}{\tilde{\mathsf A}}

\newcommand{\tA}{\tilde{A}}
\newcommand{\tB}{\tilde{B}}

\newcommand{\bA}{\bar{A}}
\newcommand{\bB}{\bar{B}}

\newcommand{\mA}{\mathsf A}
\newcommand{\mB}{\mathsf B}
\newcommand{\mC}{\mathsf C}

\newcommand{\mE}{\mathsf E}

\newcommand{\fM}{\mathbf M}

\newcommand{\fQ}{\mathbf Q}

\newcommand{\fA}{\mathbf A}
\newcommand{\fB}{\mathbf B}

\newcommand{\fD}{\mathbf D}

\newcommand{\mW}{\mathsf W}

\newcommand{\bc}{\bar{c}}

\newcommand{\bM}{\bar{M}}

\newcommand{\ha}{\hat{a}}
\newcommand{\hb}{\hat{b}}
\newcommand{\hc}{\hat{c}}
\newcommand{\hd}{\hat{d}}

\newcommand{\tc}{\tilde{c}}

\newcommand{\tU}{\tilde{U}}
\newcommand{\tD}{\tilde{D}}
\newcommand{\tQ}{\tilde{Q}}
\newcommand{\tM}{\tilde{M}}

\newcommand{\ckU}{\check{U}}
\newcommand{\ckD}{\check{D}}
\newcommand{\cku}{\check{u}}
\newcommand{\ckd}{\check{d}}

\newcommand{\tpsi}{\tilde{\psi}}

\newcommand{\tGa}{\tilde{\Gamma}}

\newcommand{\Ups}{\Upsilon}
\newcommand{\tUps}{\tilde{\Upsilon}}
\newcommand{\vtUps}{\tilde{\varUpsilon}}

\newcommand{\vUps}{\varUpsilon}
\newcommand{\hUps}{\hat{\Upsilon}}

\newcommand{\fPsi}{\mathbf \Psi}

\newcommand{\hm}{\hat{\mu}}
\newcommand{\hn}{\hat{\nu}}

\newcommand{\tC}{\tilde{C}}

\newcommand{\mG}{\mathsf{G}}

\newcommand{\cC}{\mathcal{C}}

\newcommand{\CQc}{\mathcal{C}_{\cQ_c}}

\newcommand{\CMc}{\mathcal{C}_{\cM_c}}

\newcommand{\QS}{\mQ_{\mS}}
\newcommand{\QC}{\mQ_{\mC}}
\newcommand{\QCp}{\mQ_{\mC+}}
\newcommand{\QCn}{\mQ_{\mC-}}

\newcommand{\QT}{\mQ_{\mT}}

\newcommand{\hQS}{\hat{\mQ}_{\mS}}

\newcommand{\fQS}{\fQ_{\mS}}
\newcommand{\fQC}{\fQ_{\mC}}

\newcommand{\tfQS}{\tilde{\fQ}_{\mS}}

\newcommand{\hQC}{\hat{\mQ}_{\mC}}
\newcommand{\hQCp}{\hat{\mQ}_{\mC+}}
\newcommand{\hQCn}{\hat{\mQ}_{\mC-}}

\newcommand{\QWpn}{\mQ_{\mW\pm}}
\newcommand{\QWp}{\mQ_{\mW+}}
\newcommand{\QWn}{\mQ_{\mW-}}

\newcommand{\fQW}{\fQ_{\mW}}
\newcommand{\fQWp}{\fQ_{\mW+}}
\newcommand{\fQWn}{\fQ_{\mW-}}
\newcommand{\fQWpn}{\fQ_{\mW\pm}}

\newcommand{\fQWpj}{\fQ_{\mW_j^+}^6}
\newcommand{\fQWni}{\fQ_{\mW_i^-}^6}

\newcommand{\fQWpnj}{\fQ_{\mW_j^{\pm}}^6}
\newcommand{\fQWpni}{\fQ_{\mW_i^{\pm}}^6}

\newcommand{\fQG}{\fQ_{\mG}}

\newcommand{\fQGf}{\fQ_{\mG_f}}

\newcommand{\fQGni}{\fQ_{\mG_i}^-}

\newcommand{\fQGpj}{\fQ_{\mG_j}^+}

\newcommand{\fQGpni}{\fQ_{\mG_i}^{\pm}}
\newcommand{\fQGpnj}{\fQ_{\mG_j}^{\pm}}

\newcommand{\fQGpf}{\fQ_{\mG_1}^+}
\newcommand{\fQGps}{\fQ_{\mG_2}^+}
\newcommand{\fQGpt}{\fQ_{\mG_3}^+}
\newcommand{\fQGpft}{\fQ_{\mG_4}^+}

\newcommand{\fQGnf}{\fQ_{\mG_1}^-}
\newcommand{\fQGns}{\fQ_{\mG_2}^-}
\newcommand{\fQGnt}{\fQ_{\mG_3}^-}
\newcommand{\fQGnft}{\fQ_{\mG_4}^-}

\newcommand{\fQGst}{\fQ_{\mG_1}}
\newcommand{\fQGnd}{\fQ_{\mG_2}}
\newcommand{\fQGth}{\fQ_{\mG_3}}
\newcommand{\fQGft}{\fQ_{\mG_4}}

\newcommand{\QU}{\mQ_{\mU}}
\newcommand{\fQU}{\fQ_{\mU}}
\newcommand{\QUn}{\mQ_{\mU}^-}
\newcommand{\fQUn}{\fQ_{\mU}^-}
\newcommand{\QUp}{\mQ_{\mU}^+}
\newcommand{\fQUp}{\fQ_{\mU}^+}

\newcommand{\QE}{\mQ_{\mE}}
\newcommand{\fQE}{\fQ_{\mE}}
\newcommand{\QEp}{\mQ_{\mE+}}
\newcommand{\QEn}{\mQ_{\mE-}}
\newcommand{\QEpn}{\mQ_{\mE\pm}}

\newcommand{\fQEpni}{\fQ_{\mE_i^{\pm}}^7}
\newcommand{\fQEpnj}{\fQ_{\mE_j^{\pm}}^7}

\newcommand{\fQEni}{\fQ_{\mE_i^-}^7}
\newcommand{\fQEpj}{\fQ_{\mE_j^+}^7}

\newcommand{\fQEnf}{\fQ_{\mE_1^-}^7}
\newcommand{\fQEns}{\fQ_{\mE_2^-}^7}
\newcommand{\fQEpt}{\fQ_{\mE_3^+}^7}
\newcommand{\fQEpft}{\fQ_{\mE_4^+}^7}

\newcommand{\fQEpf}{\fQ_{\mE_1^+}^7}
\newcommand{\fQEps}{\fQ_{\mE_2^+}^7}
\newcommand{\fQEnt}{\fQ_{\mE_3^-}^7}
\newcommand{\fQEnft}{\fQ_{\mE_4^-}^7}

\newcommand{\QH}{\mQ_{\mH}}
\newcommand{\fQH}{\fQ_{\mH}}
\newcommand{\QHp}{\mQ_{\mH}^+}
\newcommand{\fQHp}{\fQ_{\mH}^+}
\newcommand{\QHn}{\mQ_{\mH}^-}
\newcommand{\fQHn}{\fQ_{\mH}^-}
\newcommand{\QHz}{\mQ_{\mH}^0}
\newcommand{\fQHz}{\fQ_{\mH}^0}
\newcommand{\QL}{\mQ_{\mL}}
\newcommand{\QR}{\mQ_{\mR}}

\newcommand{\hPsi}{\hat{\Psi}}
\newcommand{\tPsi}{\tilde{\Psi}}

\begin{document}
\def\intdk{\int\frac{d^4k}{(2\pi)^4}}
\def\sla{\hspace{-0.22cm}\slash}
\hfill


\title{ The foundation of the hyperunified field theory I \\ - fundamental building block and symmetry}

\author{Yue-Liang Wu}\email{ylwu@itp.ac.cn}
\affiliation{$^1$Institute of Theoretical Physics, Chinese Academy of Sciences, 
Beijing 100190, China\\
$^2$International Centre for Theoretical Physics Asia-Pacific (ICTP-AP), (Beijing/Hangzhou), UCAS, Beijing 100190, China \\
$^3$ Taiji Laboratory for Gravitational Wave Universe (Beijing/Hangzhou), University of Chinese Academy of Sciences (UCAS), Beijing 100049, China \\
$^4$ School of Fundamental Physics and Mathematical Sciences, Hangzhou Institute for Advanced Study, UCAS, Hangzhou 310024, China }


\begin{abstract}
Starting from the motional property of functional field based on the action principle of path integral formulation while proposing maximum coherence motion principle and maximum locally entangled-qubits motion principle as guiding principles, we show that such a functional field as fundamental building block appears naturally as an entangled qubit-spinor field expressed by a locally entangled state of qubits. Its motion brings about the appearance of Minkowski spacetime with dimension determined by the motion-correlation $\cM_c$-spin charge and the emergence of $\cM_c$-spin/hyperspin symmetry as fundamental symmetry. Intrinsic $\cQ_c$-spin charge displays a periodic feature as the mod 4 qubit number, which enables us to classify all entangled qubit-spinor fields and spacetime dimensions into four categories with respective to four $\cQ_c$-spin charges $\cC_{\cQ_c}=0,1,2,3$. An entangled decaqubit-spinor field in 19-dimensional hyper-spacetime is found to be a hyperunified qubit-spinor field which unifies all discovered leptons and quarks and brings on the existence of mirror lepton-quark states. The inhomogeneous hyperspin symmetry WS(1,18) as hyperunified symmetry in association with inhomogeneous Lorentz-type symmetry PO(1,18) and global scaling symmetry provides a unified fundamental symmetry. The maximum locally entangled-qubits motion principle is shown to lay the foundation of hyperunified field theory, which enables us to comprehend longstanding questions raised in particle physics and quantum field theory. 
\end{abstract}


\maketitle

\begin{widetext}
\tableofcontents
\end{widetext}

\section{Introduction}

Recently, the hyperunified field theory (HUFT) \cite{HUFT,HUFTSB,HUFTTK} has been built in the framework of gravitational quantum field theory (GQFT)\cite{GQFT,GQFTTK} to provide an alternative approach for exploring unification theory. In the GQFT, the basic gravitational field is regarded as the gauge-type bicovariant vector field instead of the spacetime metric field in Einstein general theory of relativity(GR)\cite{GR,FGR}. So that the gravitational interaction as a gauge type interaction is treated on the same footing as the electroweak and strong interactions, which enables us to make the unified description on four basic forces within the framework of quantum field theory. In the construction of GQFT, the concept of biframe spacetime plays an essential role. One spacetime frame is globally flat Minkowski spacetime of coordinates, which acts as an inertial reference frame for describing the free motion of basic fields and allows us to derive the well-defined conservation laws and make the physically meaningful definition for space and time, thus the differences of the spatial coordinates or time coordinate can directly be measured by the standard ways proposed in Einstein special theory of relativity\cite{SR}. The other spacetime frame is the so-called locally flat non-coordinate spacetime described by the gauge-type bicovariant vector field, which is viewed as a dynamically emerging spacetime characterized by the non-commutative geometry and serves as an internal spacetime frame for describing the dynamics of basic fields. In HUFT, all intrinsic quantum numbers of leptons and quarks as elementary particles are regarded as spin-like charges, which are referred to as the hyperspin charge and treated equally on the same footing. The spinor structure of elementary particles is found to correlate with the geometric property of hyper-spacetime. All known leptons and quarks as elementary particles can be merged into a single hyperunified spinor field in $D_h=19$ dimensional hyper-spacetime and all basic forces can uniformly be described by the gauge interaction governed by the hyperspin gauge symmetry SP(1,$D_h$-1) based on the gauge invariance principle with the introduction of gauge field\cite{YM}. The biframe hyper-spacetime is composed of globally flat Minkowski hyper-spacetime and locally flat non-coordinate spacetime, the former is regarded as the base spacetime and the latter is viewed as the hyper-fiber that appears as a dynamically emerging hyper-spacetime characterized by a non-commutative geometry, which forms the fiber bundle structure in mathematics. 

The HUFT has taken the first step towards unifying all the elementary particles and basic forces. Nevertheless, in order to clarify more profound problems raised in particle physics and quantum field theory, we should think of more fundamental concepts and carry out a more systematic analysis along with the line of HUFT. The well-known long-standing open questions include: what is made to be the fundamental building block of nature? What is acted as the fundamental interaction of nature? What brings about the fundamental symmetry of nature? Which symmetry governs the gauge-type gravitational interaction? What is the basic structure of spacetime? How many dimensions does spacetime have? What makes time difference from space? Why is there only one temporal dimension? Why do we live in a universe with only four dimensional spacetime? Why are there leptons and quarks more than one family? Why are the existed leptons and quarks the chiral fermions with maximum parity violation?

To get a better understanding and make a reasonable explanation on those puzzles come to the main purpose of our present considerations for a further study on HUFT, which motivates us to seek the foundation of HUFT. In this paper as part I of the foundation of the HUFT, we will present a systematic analysis and detailed investigation to show how the fundamental building block and fundamental symmetry in HUFT can uniquely be achieved by proposing more profound guiding principles based on the action principle with path integral formulation and also how the above mentioned long-standing questions can be comprehended through exploring the foundation of the HUFT. To have a global overview on the content of this paper, let us first provide a brief outline for the main ideas and concepts in the introduction. 

The action principle has successfully be exploited to derive the laws of nature for both classical and quantum theories. In classical mechanics, the action principle is thought to be a mathematical functional that takes a unique classical trajectory called a path or history of the system. In general, the action of mathematical functional takes different values for different paths. The principle of least action is used to derive equations of motion of a system by finding the path that has the least value, it has been widely applied to obtain Newtonian, Lagrangian and Hamiltonian equations of motion as well as Einstein equation of GR. The path integral formulation has been developed to provide a consistent description on quantum theory by generalizing the classical notion of a unique trajectory to make a summation or functional integral over an infinity of possible trajectories, which allows us to compute in principle the quantum amplitude. The advantages of such a formalism can be attributed to the manifest Lorentz covariance and explicit symmetry property. The action principle can characterize both the kinematics and dynamics of a physical system, which is governed by the symmetry principle. In general, the action principle enables us to establish a reliable Lagrangian of theory based on the symmetry principle, and apply the path integral formulation to carry out the computation within the framework of quantum field theory. One of the most successful applications for the action principle with path integral formulation has been realized in describing the standard model (SM) of electroweak and strong interactions\cite{SM1,SM2a,SM2b,SM3a,SM3b,SM3c,SM4, SM5,SM6,SM7,SM8} within the framework of renormalizable quantum field theory\cite{QFT1a,QFT1b,QFT2}.

To seek for the foundation of the HUFT in light of the advantages of the action principle with path integral formulation, we will begin with emphasizing on the motional properties of matter field instead of imposing on presupposed symmetries. For that, we are going to propose firstly a maximum coherence motion principle as guiding principle, which combines the simplest motion postulate and maximum correlation motion postulate together with the quadratic free motion postulate. Following along such a guiding principle, we are able to show that any field as a continuous and differentiable function turns out to be a basic spinor field, which brings naturally on the appearance of canonical anticommutative relation and Pauli exclusion principle. The basic spacetime describing the inertial motion of spinor field is shown to emerge as globally flat Minkowski spacetime with only one dimension appearing as temporal dimension. The symmetry either for basic spinor field or basic spacetime is found to emerge as a natural consequence resulting from the maximum coherence motion principle as guiding principle.

To find out a general relation between the dimension of Hilbert space for spinor field as basic constituent of nature and the dimension of Minkowski spacetime for coordinates, we are led to propose a local coherent-qubits motion postulate. As it is known that a qubit (quantum bit) is usually regarded as the single unit of quantum information, which is represented by a linear superposition of two states denoted either by Dirac ``bra-ket" notation or spin-like ``column-row" notation with the usual convention for two basis states, i.e., ket ``0 $\&$ 1" denoted as $|0\rangle$  and $|1\rangle$  or spin ``up $\&$ down" denoted as $\binom{1}{0}$ and $\binom{0}{1}$. In general, qubits are viewed as the basic building blocks for quantum information processes, which are described by the normalized probability amplitudes of qubit basis states. Instead of the probability amplitudes used in quantum information, we will introduce local distribution amplitudes in terms of the spin ``up $\&$ down" spinor representation based on the local coherent-qubits motion postulate. Such a postulate indicates that a local coherent state of qubits as a column vector field of local distribution amplitudes obeying the maximum coherence motion principle brings on a basic spinor field which is referred to as qubit-spinor field. The independent degrees of freedom $\cD_H$ for qubit-spinor field as basic spinor field form a $\cD_H$-dimensional Hilbert space, which is determined by the qubit number $Q_N$ through the simple relation $\cD_H=2^{Q_N}$. The dimension $D_h$ of coordinates $x^{\fM}$ in Minkowski spacetime is determined by $\cM_c$-spin charge $\CMc=D_h$, which is identified from maximally correlated motion of qubit-spinor field $\Psi(x)$ following along the maximum coherence motion principle as guiding principle and leads to the emergence of $\cM_c$-spin or hyperspin symmetry SP(1, $D_h$-1). The intrinsic property of qubit-spinor field $\Psi(x)$ is characterized by the $\cQ_c$-spin charge $\CQc=q_c$ and corresponding $\cQ_c$-spin symmetry SP($q_c$). It is intriguing to found that there exist four $\cQ_c$-spin charges $\CQc=q_c=0,1,2,3$ characterized as the mod 4 qubit number with a periodic feature. Namely, a categoric qubit number is generally given by the identity $Q_N^{(q_c,k)} = \CQc + 4k $, where $\cQ_c$-spin charge is found to obey a periodic relation $\CQc^{(q_c,k)} = \CQc^{(q_c)} \equiv \CQc = q_c$ with the integer $k$ $(k=0,1,\cdots)$ referred to as $k$-th period. Such a periodic $\cQ_c$-spin charge plays an essential role as qubit classification number, which allows us to categorize all qubit-spinor fields and spacetime dimensions into four categories with respective to four $\cQ_c$-spin charges, $\CQc = q_c =0, 1, 2, 3$. 

Based on the fact that leptons and quarks in the standard model (SM) are chiral fermions in the electroweak interaction, we are motivated to make a locally entangled-qubits motion postulate upon the local coherent-qubits motion postulate. In fact, it is known from quantum information that the quantum entanglement of multiple qubits is viewed as the distinguishing feature between qubit and bit. In general, if a state in the tensor product Hilbert space becomes inseparable, it is referred to as an entangled state, which is valid for both global and local states. The locally entangled-qubits motion postulate indicates that a locally entangled state of qubits as a column vector field of local distribution amplitudes obeying the maximum coherence motion principle is proposed to be an entangled qubit-spinor field which constitutes fundamental building block of nature. Such an entangled qubit-spinor field is generally represented as a self-conjugated chiral qubit-spinor field in a chiral qubit-spinor representation, which allows us to extend the hyperspin symmetry SP(1, $D_h$-1) to inhomogeneous hyperspin symmetry WS(1, $D_h$-1)=SP(1, $D_h$-1)$\rtimes$W$^{1,D_h-1}$. In particular, to make inhomogeneous hyperspin symmetry WS(1, $D_h$-1) as a maximal hyperunified symmetry of entangled hyperqubit-spinor field, we are motivated to propose a least $\cQ_c$-spin postulate for obtaining a hyperunified qubit-spinor field. So that we are led to propose a maximum locally entangled-qubits motion principle as guiding principle and achieve a unified fundamental symmetry given by inhomogeneous hyperspin symmetry WS(1, $D_h$-1) in association with inhomogeneous Lorentz-type/Poincar\'e-type group symmetry PO(1, $D_h$-1). Consequently, a minimal entangled qubit-spinor field that unifies all known lepton-quark states into a single hyperunified qubit-spinor field is obtained uniquely to be a locally entangled state of ten qubits ($Q_N=10$) with $\cQ_c$-spin charge $\CQc=1$, which is referred to as entangled decaqubit-spinor field in 19-dimensional hyper-spacetime ($D_h=19$).

We will demonstrate in a systematic way that the foundation of the HUFT proposed in this paper as the part I is laid by two guiding principles: one is the maximum coherence motion principle that combines the simplest motion postulate and maximum correlation motion postulate together with the quadratic free motion postulate, and the other is the maximum locally entangled-qubits motion principle that combines the local coherent-qubits motion postulate and locally entangled-qubits motion postulate together with the least $\cQ_c$-spin postulate. Following along such two guiding principles, we are able to provide reasonable comprehensions on some longstanding open questions in particle physics and quantum field theory mentioned above. The detailed discussions and analyses will be carried out in this paper as the part I of the foundation of the HUFT.

The paper is organized as follows: after the brief introduction in Sect.1, we will start from thinking of motional nature of matter field in Sect.2, which is unlike to the usual consideration by imposing on presupposed symmetry as a starting point. We will show how a real column vector field in Hilbert space exists as a spinor field from considering the motional properties of real column vector field through proposing a maximum coherence motion principle as guiding principle. Meanwhile, we demonstrate explicitly the emergence of canonical anticommutation relation in association with Pauli exclusion principle and the appearance of Minkowski spacetime with only one temporal dimension. As a consequence, the action of spinor field with scalar coupling associated with intrinsic $\cQ_c$-matrices is demonstrated to bring about the motion-correlation $\cM_c$-spin symmetry and intrinsic $\cQ_c$-spin symmetry in Hilbert space in association with Poincar\'e-type group symmetry and global scaling symmetry in Minkowski spacetime. In Sect. 3,  we examine and analyze a simple two-component entity spinor field, which is conceptually considered as a local coherent state of qubit and referred to as real/self-conjugated uniqubit-spinor field in two-dimensional real/self-conjugated Hilbert space. The maximum coherence motion principle leads such a uniqubit-spinor field to possess the $\cM_c$-spin charge $\cC_{\cM_c}=3$ and $\cQ_c$-spin charge $\CQc=1$, which brings the uniqubit-spinor field to have free motion in three-dimensional Minkowski spacetime with single scalar coupling. The symmetry property and intrinsic feature of the uniqubit-spinor field as the so-called Taijion spinor field are discussed in detail to show the correlation between Hilbert space of qubit-spinor field and Minkowski spacetime of coordinates.

To demonstrate explicitly how the independent degrees of freedom of spinor field determines the dimension of Minkowski spacetime, we are motivated, in Sect.4, to make a local coherent-qubits motion postulate. As simple examples, Majorana and Dirac fermions as basic constituents of matter are shown to be characterized by local coherent states of 2-qubit ($Q_N$=2) and 3-qubit ($Q_N=$3) with $\cQ_c$-spin charges $\CQc=q_c=Q_N=2$ and 3, which are correspondingly referred to as real/self-conjugated biqubit-spinor and triqubit-spinor fields in real/self-conjugate Hilbert space. Meanwhile, both of them can equivalently be viewed as complex uniqubit-spinor field and complex biqubit-spinor field in complex Hilbert space. Follow along the maximum coherence motion principle as guiding principle, the maximally correlated motions of biqubit-spinor field and triqubit-spinor field bring about four-dimensional and six-dimensional Minkowski spacetimes, respectively. A detailed discussion on symmetry property and intrinsic feature is presented for both qubit-spinor fields, which displays coincidental transformations in Hilbert space and Minkowski spacetime. They should be attributed to the isomorphic property of special group symmetries, SL(2,$\mathbb{C}$)$\cong$SP(1,3)$\cong$SO(1,3) and SL(2,$\mathbb{Q}$)$\cong$SP(1,5)$\cong$SO(1,5), with respective to 4-dimensional and 6-dimensional Minkowski spacetimes. In general, Dirac-type and Majorana-type as well as Weyl-type fermions with equal $\cM_c$-spin and $\cQ_c$-spin charges can equivalently be described by qubit-spinor fields with respective to complex and self-conjugated as well as chiral representations. In Sect. 5, we will analyze in detail a local coherent state of four qubits as tetraqubit-spinor field based on the local coherent-qubits motion postulate, which brings on the appearance of 10-dimensional Minkowski hyper-spacetime with hyperspin symmetry SP(1,9). Such a tetraqubit-spinor field as hyperqubit-spinor field is found to have zero $\cQ_c$-spin charge, $\CQc=q_c =Q_N-4 = 0$, which displays a periodic feature of $\cQ_c$-spin charge. We will demonstrate that qubit-spinor fields in Hilbert space with dimensions $\cD_H$=1,2,4,8,16 corresponding to qubit numbers $Q_N=0,1,2,3,4$ exhibit a periodic $\cQ_c$-spin charge with $\CQc=q_c$=0,1,2,3,0 and bring on Minkowski spacetime with corresponding dimensions $D_h$=2,3,4,6,10.

In Sect. 6, we will further demonstrate the periodic feature of $\cQ_c$-spin charge $\CQc$ by analyzing high product basis states of qubits based on the local coherent-qubits motion postulate. Four $\cQ_c$-spin charges $\CQc=q_c=0,1,2,3$ are shown to be characterized by the mod 4 qubit number with a simple relation $Q_N^{(q_c,k)}= \CQc + 4k$ ($k=0,1,\cdots$) and have a periodic behavior $\CQc^{(q_c,k)}= \CQc^{(q_c)} \equiv \CQc=q_c $ with the integer $k$ referred to as $k$-th period. Such a periodic property enables us to categorize all hyperqubit-spinor fields in Hilbert space and all hyper-spacetime dimensions in Minkowski spacetime into four categories designated as category-$q_c$ with $q_c=0,1,2,3$. By introducing a self-conjugated chiral-like qubit-spinor structure in Sect. 7, a string-like qubit-spinor field can be constructed from any qubit-spinor field to be as a self-conjugated chiral-like qubit-spinor field. Such a string-like qubit-spinor field is verified to have a free motion only in two-dimensional spacetime with intrinsic spin symmetry SP($D_h$-2)$\times$SP($q_c$). On the other hand, we will prove that any qubit-spinor field with categoric qubit number $Q_N^{(q_c,k)}$ in category-$q_c$ can be described by a self-conjugated chiral qubit-spinor field in the same category-$q_c$ but in an entended Hilbert space with dimension determined by $Q_N^{(q_c,k)}+1$ qubit number. As a consequence, all motion-correlation $\cM_c$-matrices and motion-irrelevance $\cQ_c$-matrices become anti-commuting and the group generators of $\cM_c$-spin symmetry and $\cQ_c$-spin symmetry are directly given by the commutators of $\Gamma$-matrices. As a specific example, a massless Dirac fermion\cite{DF} as self-conjugated triqubit-spinor field in 6-dimensional spacetime\cite{GGFT6D} will explicitly be constructed as a self-conjugated chiral tetraqubit-spinor field in category-$3$, so that all $\cM_c$-matrices and $\cQ_c$-matrices become anti-commuting. By analyzing the correspondence between entangled state in quantum theory and self-conjugated chiral qubit-spinor structure, we are led to propose, in Sect. 8, a locally entangled-qubits motion postulate. As a consequence, any motion-correlation hyperspin symmetry SP(1, $D_h$-1) of entangled hyperqubit-spinor field in $D_h$-dimensional hyper-spacetime can be extended to be inhomogeneous hyperspin symmetry WS(1, $D_h$-1)= SP(1, $D_h$-1)$\times W^{1,D_h-1}$. Such a semi-direct product Abelian group symmetry $W^{1,D_h-1}$ reflects an entanglement-correlated translation-like $\cW_e$-spin symmetry. Based on the locally entangled-qubits motion postulate, we examine three types of entangled hyperqubit-spinor fields corresponding to three categories with respective to $\cQ_c$-spin charges $q_c=0,1,2$, which are constructed explicitly from chiral spinor structures of self-conjugated hyperqubit-spinor fields in correspondence to qubit numbers $Q_N=4,5,6$ and referred to as entangled pentaqubit-, hexaqubit- and heptaqubit-spinor fields, respectively. We further show how leptons and quarks\cite{QK1,QK2} in SM are identified and merged into entangled hyperqubit-spinor fields in three categories with respective to hyper-spacetime dimensions $D_h=10,11,12$. 

In Sect. 9,  we present a detailed analysis on a locally entangled state of eight qubits which has a free motion in 14-dimensional Minkowski hyper-spacetime. Such an entangled hyperqubit-spinor field is referred to as entangled octoqubit-spinor field in category-3, which possesses inhomogeneous hyperspin symmetry WS(1,13) and $\cQ_c$-spin symmetry SP(3). The hyperspin symmetry SP(1,13)$\cong$SO(1,13) as proposed in a unification model\cite{UM} is shown to realize SO(10) grand unified theory\cite{GUT1,GUT2} for each family of leptons and quarks in SM and reproduce the basic symmetry of SM, i.e., SO(1,3)$\times$SU$_C$(3)$\times$ SU$_L$(2) $\times$ U$_Y$(1), as a subgroup symmetry of SP(1,13). The $\cQ_c$-spin symmetry SP(3) as a direct product group symmetry to SP(1,13) is found to characterize the family-spin symmetry SU$_F$(2)$\cong$SP(3) of two families of leptons and quarks in SM with including right-handed neutrinos. We will demonstrate that each family of leptons and quarks in SM alone can have a free motion only in four-dimensional Minkowski spacetime. All the features which arise mainly from the local coherent-qubits motion postulate and locally entangled-qubits motion postulate enable us to provide reasonable explanations on the essential issues why there exist leptons and quarks beyond one family in SM and why our observed universe becomes only four-dimensional spacetime. 

To achieve the hyperunified field theory with inhomogeneous hyperspin symmetry as fundamental symmetry of nature, we are motivated, in Sect.10, to propose a least $\cQ_c$-spin postulate. The combination of the least $\cQ_c$-spin postulate together with the local coherent-qubits motion postulate and locally entangled-qubits motion postulate brings on a maximum locally entangled-qubits motion principle as guiding principle. Following along such a guiding principle, we will provide a detailed investigation on locally entangled states of nine qubits ($Q_N=9$) and ten qubits ($Q_N=9$), which are referred to as entangled enneaqubit-spinor field and entangled decaqubit-spinor field in correspondence to $\cQ_c$-spin charges $q_c=0$ and $q_c=1$. We will also show that the hyperspin symmetry SP(1,13) and family-spin symmetry SP(3) as direct product group symmetry SP(1,13)$\times$SP(3) for entangled octoqubit-spinor field can be transmuted into a motion-correlation hyperspin symmetry SP(1,17) of entangled enneaqubit-spinor field, so that the entangled enneaqubit-spinor field is regarded as a minimum hyperunified qubit-spinor field and its inhomogeneous hyperspin symmetry WS(1,17) is viewed as a minimal hyperunified symmetry in 18-dimensional hyper-spacetime. Eventually, the entangled decaqubit-spinor field is verified to be a hyperunified qubit-spinor field in 19-dimensional hyper-spacetime, which merges all discovered leptons and quarks into a single entangled hyperqubit-spinor field with the prediction on the existence of mirror lepton-quark states. Such a hyperunified qubit-spinor field is regarded as a unified fundamental building block of nature, which brings on the inhomogeneous hyperspin symmetry WS(1,18) as hyperunified fundamental symmetry. Our summaries and conclusions with certain remarks are presented in the final section.


\section{ Appearance of spinor field with canonical anticommutation relation and the emergence of Minkowski spacetime and basic symmetry with proposing maximum coherence motion principle as guiding principle }

To explore the foundation of the HUFT,  we are going to start with focusing on the motion nature of matter field rather than the presupposed symmetry property in the usual consideration. For that, we will begin with thinking of an intuitive notion that the universe is composed of the fundamental building block which always keeps motion continuously in a simple and correlation way. In SM, leptons and quarks as fermionic fields are regarded as the basic constituents of matter and described by the quantum field theory. As the path integral approach based on the action principle is proved to be equivalent to the operator formalism of quantum field theory with canonical quantization, we will apply for the action principle with path integral formulation in the present consideration. In the path integral formalism with a functional integral, the basic field is treated as a function. So that we will start with the hypothesis that the fundamental building block of nature is characterized by a column vector field $\Psi(x)$ which is presented by a set of real component fields, i.e., $\Psi^{T}(x) = (\psi_1(x), \psi_2(x), \cdots, \psi_{\cD_H}(x) )$ with $\psi_{i}(x) = \psi_{i}^{\ast}(x)$ ($i=1,\cdots,\cD_H$), and meanwhile such a column vector field $\Psi(x)$ is considered to be a real continuous and differentiable function of real variables $\{x\} \equiv \{ x^{\fM}; \fM=0,1,\cdots, D_h-1\}$. To make the above intuitive notion about basic constituent and its motion to become physically meaningful concept, we are led to propose a maximum coherence motion principle as guiding principle, which brings the column vector field $\Psi(x)$ to be a basic spinor field with natural emergence of canonical anticommutation relation and presence of high dimensional Minkowski spacetime with only one temporal dimension. In particular, the resulting action is shown to bring about the appearance of motion-correlation $\cM_c$-spin symmetry and motion-irrelevance intrinsic $\cQ_c$-spin symmetry in association with Poincar\'e-type group symmetry and global scaling symmetry in Minkowski spacetime.


\subsection{ Anti-commuting field operator as the spinor field based on the simplest motion postulate}

For an illustration, let us first consider a simple case with a single real field $\psi(x)= \psi^{\ast}(x)$ and treat it as a differentiable function of variables $x^{\mu}$ with the superscript ($\mu=0,1,\dots$) denoting the numbers of the variables.  The real field $\psi(x)$ is proposed to keep a free motion, which is defined as a change of $\psi(x)$ with respect to the variation of $x^{\mu}$. Mathematically, such a change is characterized by the differentiation of real field $\psi(x)$ with respect to the variables $x^{\mu}$, i.e., 
\be
\frac{\partial}{\partial x^{\mu}}  \psi(x) \equiv \p_{\mu} \psi(x) . \nn
\ee  
When thinking of the real field $\psi(x)$ as matter field that keeps a free and continuous motion in the simplest way, we are led to make the following simplest motion postulate.

{\it Simplest motion postulate}:  the action of real field $\psi(x)$ within path integral formulation is proposed to involve merely the first order derivative with respect to variables $x^{\mu}$ and concern only a bilinear form of $\psi(x)$ with a requirement of hermiticity. 

Based on such a simplest motion postulate, the hermitian action of real field $\psi(x)$ is simply expressed as the following form:
 \be \label{actionB0}
 {\cS}_{\psi} & = &  \int [dx]\,  \frac{1}{4} \{ \lambda_C \psi(x) \alpha^{\mu} \p_{\mu}\psi(x) + H.c. \} , \nn \\
 & = & \int [dx]\,  \frac{1}{4} \{ \lambda_R \alpha^{\mu} \p_{\mu} [ \psi(x) \psi(x) ] + \lambda_I [ \psi(x)i \alpha^{\mu} \p_{\mu} \psi(x) - i \alpha^{\mu} \p_{\mu}\psi(x) \, \psi(x) ]  \}, 
\ee
where $\lambda_C$ is a complex constant $\lambda_C = \lambda_R + i \lambda_I$ with $\lambda_R$ denoting the real part and $\lambda_I$ the imaginary part. The notation $ \p_{\mu}\equiv \frac{\partial}{\partial x^{\mu}}$ is a derivative operator vector and $\alpha^{\mu}$ is a constant real vector with vector index $\mu$ ($\mu=0,1,\dots$). The product of two vectors forms a scalar $\alpha^{\mu} \p_{\mu}$ with  two indices indicating a summation, i.e., $\alpha^{\mu} \p_{\mu} = \sum_{\mu} \alpha^{\mu} \p_{\mu}$. In general, the magnitude of each constant $\alpha^{\mu}$ can be made to be a unit by the redefinitions of variables $x^{\mu}$ and real field $\psi(x)$, so that we can always choose $|\alpha^{\mu}|=1$.

In the second equality of Eq.(\ref{actionB0}), the first term relating to the real coupling constant $\lambda_R$ is given by a total derivative, which becomes a trivial term. Only the second term concerning the imaginary coupling constant $\lambda_I$ brings about a nontrivial hermitian action that describes a free motion of the real field $\psi(x)$. 

Therefore, to obtain a physically meaningful hermitian action, the real field $\psi(x)$ must satisfy the following relation:
\be \label{ACR}
 \alpha^{\mu} \p_{\mu} \psi(x) \, \psi(x) = - \psi(x)  \alpha^{\mu} \p_{\mu} \psi(x)  , \nn
\ee 
which can easily be verified by using reduction to absurdity. This is because if the real field $\psi(x)$ fits to the commutative relation, i.e., $\psi(x)  \alpha^{\mu} \p_{\mu} \psi(x) =  \alpha^{\mu} \p_{\mu} \psi(x)\, \psi(x)$, the action relating to the imaginary coupling constant $\lambda_I$ becomes vanishing.

The above relation can be expressed as follows: 
\be
\psi(x)  \alpha^{\mu} \p_{\mu} \psi(x) +   \alpha^{\mu} \p_{\mu}\psi(x) \, \psi(x) & \equiv & \lim_{x'\to x}  [ \psi(x')  \alpha^{\mu} \p_{\mu} \psi(x) +   \alpha^{\mu} \p_{\mu}\psi(x) \, \psi(x') ] \nn \\ 
 & = & \lim_{x'\to x}   \alpha^{\mu} \p_{\mu} [ \psi(x') \psi(x) +  \psi(x) \psi(x') ] = 0,  \nn 
\ee
which possesses a permutation symmetry $x^{\mu}\leftrightarrow x^{'\mu}$ in the bracket and becomes translational invariance under the transformation ($x^{\mu}, x^{' \mu}) \to (x^{\mu}, x^{' \mu}) + a^{\mu}$ with $a^{\mu}$ a constant vector. The above relation holds when the real field $\psi(x)$ fulfills the following general relation:
\be \label{ACR}
& & \psi(x) \psi(x') +  \psi(x') \psi(x) \equiv \{  \psi(x),  \psi(x') \} = \Delta(x-x') ,
\ee
where the function $\Delta(x-x')$ must satisfy the following conditions:
\be
& & \Delta(x-x') =  \Delta(x'-x), \nn \\
& & \alpha^{\mu} \p_{\mu} \Delta(x-x') =  -  \alpha^{\mu} \p_{\mu} \Delta(x'-x), \quad   \p_{\mu} \Delta(0) = 0 . \nn
\ee

Such a general relation is the so-called {\it anticommutation relation}. The real field satisfying such an anticommutation relation is referred to as {\it spinor field}.

By making a redefinition on such a spinor field, $\psi\to \psi/\sqrt{\lambda_I}$, the hermitian action can be rewritten into the following simple form:
\be
 {\cS}_{\psi} & = & \int [dx]\,  \frac{1}{4} \{ \psi(x) \alpha^{\mu} i\p_{\mu}\psi(x) - \alpha^{\mu} i\p_{\mu}\psi(x) \, \psi(x) \}  \nn \\
 & = & \int [dx]\,  \frac{1}{2} \psi(x) \alpha^{\mu} i\p_{\mu}\psi(x) , 
\ee
where $i \partial_{\mu}$ defines a {\it linear self-adjoint operator},
\be
i \partial_{\mu} \equiv i \frac{\partial}{\partial x^{\mu}} , 
\ee
which characterizes a free motion of real spinor field.

Therefore, when making the simplest motion postulate to construct an action for describing a free motion of single real field, we are allowed to have only the first order derivative operator acting on the real field and get solely the bilinear form of real field with hermiticity requirement. So that we are able to obtain a nontrivial action iff the coupling constant is a pure imaginary one and the real field satisfies anticommutation relation. Such a real field is verified to be a spinor field $\psi(x)$. The first order derivative operator associated with the pure imaginary factor provides a linear self-adjoint operator $i \partial_{\mu}$. It is interesting to notice that the pure imaginary factor appears coincidentally with the anti-commuting feature of real field when the action obeys the simplest motion postulate.

\subsection{ Appearance of temporal and spatial dimensions with quadratic free motion postulate and the free motion of single real spinor field in two-dimensional spacetime }

To describe the free motion of real spinor field $\psi(x)$, let us write down its equation of motion. From the least action principle, we can easily obtain the free motion of single real spinor field $\psi(x)$,
\be \label{EM}
 \alpha^{\mu} i\p_{\mu}\psi(x) = 0, \quad \mbox{or} \quad  i\p_{0}\psi(x) = - \alpha^{k} i\p_{k}\psi(x),  \nn
 \ee
with $i\p_0 \equiv i\frac{\p}{\p x^0}$ and $ i\p_k \equiv i\frac{\p}{\p x^k}$ ($k=1,\cdots $) being self-adjoint derivative operators.  The second-order derivative equation of motion reads, 
\be
 \p_0^2 \psi(x) =  \p_1^2 \psi(x)  + 2 \alpha^{1} \alpha^{k} \p_1  \p_k  \psi(x)  +  \alpha^{k} \alpha^{l} \p_k  \p_l  \psi(x) , \quad k, l =2, 3, \cdots , \nn
\ee 
where we have used the convention $\alpha^0=1$ and $(\alpha^{1})^2=1$. It indicates that as long as there are more than two variables $x^{\mu}$, the equation of motion always involves the cross terms of the derivatives, i.e.,  $\p_k  \p_l  \psi(x)$ ($k\neq l = 2, 3,\cdots $). To prevent the cross terms from the derivative operators, we are motivated to make the following postulate.

{\it Quadratic free motion postulate}: In the partial differential equation of motion, the second-order derivative with respect to the variable $x^0$ is postulated to be proportional to the second-order derivative with respect to the variables $x^k$ without involving any cross terms of derivative operators.  

Following such a quadratic free motion postulate, we arrive at the wave equation of motion for the single real spinor field with the following standard form:
\be
 (\p_0^2 - \p_1^2) \psi(x) \equiv  \eta^{\mu\nu} \p_{\mu}\p_{\nu}  \psi(x) = 0 ,
\ee   
where $\eta^{\mu\nu}$ is a unit metric matrix, 
\be \label{sign2}
\eta^{\mu\nu} = \diag. (1, -1) , \nn
\ee
which indicates that the single real spinor field is a differential function which depends on two variables $x^{\mu}$ ($\mu=0,1$) when imposing the quadratic free motion postulate.

It is noticed from the sign of unit metric matrix $\eta^{\mu\nu}$ that the derivatives $i\p_0 \equiv i\frac{\p}{\p x^0}$  and $i\p_1 \equiv i\frac{\p}{\p x^1}$ define two distinguishable differential operators. For convention, $i\p_0$ is regarded as a temporal derivative operator and $i\p_1$ as a spatial derivative operator. Correspondingly, the variable $x^0$ defines a temporal coordinate and the variable $x^1$ represents a spatial Cartesian coordinate. Such two coordinates $x^{\mu} $ ($\mu = 0, 1$) span two-dimensional {\it Minkowski spacetime } with one temporal dimension and one spatial dimension. In such a case, it is in general not possible to designate obviously the temporal and spatial dimensions as they appear equally each other, we just make a definition for convention. Nevertheless, it gets meaningful for high dimensional spacetime which will be discussed later on. 

When taking the simplest motion postulate together with the quadratic free motion postulate, the hermitian action of the real field as anti-commuting field operator is found to have the following unique form:
  \be \label{BSFA}
 {\cS}_{\psi\pm} & = & \int d^2x\,  \frac{1}{4} \{ \psi(x) i(\p_0\pm \p_1) \psi(x) - i(\p_0\pm \p_1)\psi(x) \, \psi(x) \} \nn \\
 & = & \int d^2x\,  \frac{1}{2} \psi(x) i(\p_0\pm \p_1) \psi(x) ,
 \ee
which describes the free motion of single real field in two-dimensional Minkowski spacetime.

Therefore, a single real field $ \psi(x)$ obeying the simplest motion postulate and quadratic free motion postulate turns out to be a spinor field satisfying the anticommutation relation and have a free wave motion in two-dimensional Minkowski spacetime.


\subsection{ Existence of real column vector field as the spinor field in Hilbert space through introduction of motion-correlation $\cM_c$-matrices and motion-irrelevance $\cQ_c$-matrices with proposing maximum coherence motion principle as guiding principle }

After an instructive analysis made above for a single real field, we are going to present a detailed investigation on the foundation of the HUFT. For that, let us begin with an intuitive notion that the universe is composed of elementary constituents which always keep a free and continuous motion in the simplest and correlation way. To turn such an intuitive notion into a practically realizable and physically meaningful concept, we start from considering a set of real fields which are represented to be a real column vector field as follows:
\be
& & \Psi^{T}(x) = (\psi_1(x), \psi_2(x), \cdots, \psi_{\cD_H}(x) ) , \nn \\
& &  \psi_{i}(x) = \psi_{i}^{\ast}(x), \quad i =1,2, \cdots, \cD_H, \nn
\ee
with the superscript $``T"$ denoting the transpose of the real column vector field $\Psi(x)$. 

Regarding the real column vector field $\Psi(x)$ as a basic matter field and treating it as a $\cD_H$-component entity differentiable function of variables $x^{\mu}$ in the path integral formulation, we are going to construct a hermitian action of real column vector field $\Psi(x)$ by proposing the following simplest motion postulate. 

{\it Simplest motion postulate}: the action of real column vector field $\Psi(x)$ as basic matter field in path integral formulation is built to bring a free and continuous motion of  $\Psi(x)$ with a postulate that the action concerns only the first order derivative on $\Psi(x)$ with respect to variables $x^{\mu}$ and involves solely the bilinear form of $\Psi(x)$ and meanwhile satisfies the hermiticity requirement.

Following along such a simplest motion postulate, we are able to construct explicitly the following action for the real column vector field $\Psi(x)$:
\be \label{actionQ}
 \cS_{\Psi} & = &  \int [dx] \,  \frac{1}{4} \{  \lambda_C  \delta_a^{\;\mu}  \Psi^{\dagger}(x) \Ups^{a} \partial_{\mu} \Psi(x) + \tla_C  \delta_p^{\;\alpha}  \Psi^{\dagger}(x) \Ups^{p} \partial_{\alpha} \Psi(x)  + H.c. \}  \nn 
 \ee
 with  $\delta_a^{\; \mu}$ and $\delta_p^{\; \alpha}$ the Kronecker sympols. Where $\lambda_C$ and $\tla_C$ are two complex coupling constants. The real variables are decomposed into two row vectors $ x \equiv (x^{\mu}, x^{\alpha} )$ with their derivative operators in association with two types of $\cD_H\times \cD_H$ real $\Ups$-matrices $(\Ups^{a}, \Ups^p)$. Such $\Ups$-matrices act on the real column vector field $\Psi(x)$ in {\it $\cD_H$-dimensional Hilbert space} which is spanned by $\cD_H$ number of real fields $\psi_i(x)$ ($i =1,2, \cdots, \cD_H$). 

Such two types of real $\Ups$-matrices are categorized by real symmetric matrices $\Ups^a$ and real antisymmetric matrices $\Ups^p$. The operators $\partial_{\mu} \equiv \frac{\p}{\p x^{\mu}}$ and $ \partial_{\alpha} \equiv \frac{\p}{\p x^{\alpha}}$ are two derivatives with respect to the real variables $x^{\mu}$ and $x^{\alpha}$. The Greek alphabet ($\mu, \nu = 0, 1,\cdots $;  $\alpha, \beta = 0,1, \cdots$) and Latin alphabet ($a, b = 0, 1, \cdots$; $p, q = 0,1, \cdots$) are used to distinguish the indices for the vectors of real variables $(x^{\mu}, x^{\alpha})$ in parameter space and the vectors of real matrices $(\Ups^{a}, \Ups^p)$ in Hilbert space of real fields. Two types of real $\Ups$-matrices $\Ups^a$ and $\Ups^p$ satisfy the following conditions:  
\be
& & \Ups^a=(\Ups^a)^{\ast} = (\Ups^a)^T , \nn \\
& & \Ups^p= (\Ups^p)^{\ast} = -(\Ups^p)^T , \quad \Ups^0 \equiv 0 \;\; \mbox{for}\;\; p=0 , \nn
\ee
which will be shown to characterize correspondingly the motion-correlation property and motion-irrelevant intrinsic feature of real column vector field $\Psi(x)$. Note that the matrix $\Ups^p$ with $p=0$ is defined to be a zero matrix by convention, which implies that there is no coupling to the real column vector field $\Psi(x)$ through the antisymmetric matrix. 

By expressing the complex coupling constants $\lambda_C$ and $\tla_C$ into the real and imaginary parts,
\be
\lambda_C = \lambda_R + i \lambda_I , \quad \tla_C =\tla_R + i \tla_I, \nn
\ee
we can rewritten the above hermitian action as follows:
\be \label{actionQ}
 \cS_{\Psi}& = &  \int [dx] \,  \frac{1}{4} \{ \, \lambda_I \delta_a^{\;\mu}  \Ups^a_{ij} [ \psi_i(x)  i\p_{\mu} \psi_j(x) -  i\p_{\mu}\psi_i(x)   \psi_j(x)]  \nn \\ 
 & - & \tla_R \delta_p^{\;\alpha} i \Ups^p_{ij} [  \psi_i(x) i\p_{\alpha}   \psi_j(x)  - i\p_{\alpha} \psi_i(x)   \psi_j(x) ] 
 \nn \\
 & + &  \lambda_R  \delta_a^{\;\mu} \Ups^a_{ij} \p_{\mu} [ \psi_i(x)  \psi_j(x) ]  + \tla_I  \delta_p^{\;\alpha} i\Ups^p_{ij} \p_{\alpha} [ \psi_i(x)  \psi_j(x) ]\, \} , \nn
  \ee
where the last two terms concerning the couplings $\lambda_R$ and $\tla_I$ are given by total derivatives, which indicates that these two terms cannot characterize a free motion of real fields.  

To arrive at a hermitian action for describing the free motion of real fields, we are going to analyze in detail the first two terms involving the couplings $\lambda_I$ and $\tla_R$ in the above action. They are required to satisfy the following conditions:
\be
& & \psi_i(x) i \p_{\mu} \psi_j(x) = - i\p_{\mu}\psi_i(x)   \psi_j(x), \nn \\
& &   \psi_i(x) i\p_{\alpha} \psi_j(x) = - i \p_{\alpha}\psi_i(x)   \psi_j(x) , \nn 
\ee
from the hermiticity requirement, and
\be 
& & \psi_i(x) i \p_{\mu} \psi_j(x) = \psi_j(x) i\p_{\mu} \psi_i(x) , \nn \\
& & \psi_i(x) i\p_{\alpha} \psi_j(x) = - \psi_j(x) i\p_{\alpha} \psi_i(x) , \nn
\ee
from the symmetric and antisymmetric properties of real matrices $\Ups^a$ and $\Ups^p$, respectively. Combining the above relations, we obtain two identities:
\be \label{Identity}
& & \p_{\mu}\psi_i(x)   \psi_j(x) + \psi_j(x) \p_{\mu} \psi_i(x) \equiv 0, 
\ee
and
\be
& &  \p_{\alpha}\psi_i(x)   \psi_j(x) - \psi_j(x) \p_{\alpha} \psi_i(x) \equiv 0 . \nn
\ee 

The sign difference appearing in two identities results from the properties of symmetric and antisymmetric matrices $\Ups^a$ and $\Ups^p$, respectively. In general, two identities cannot be satisfied simultaneously. This is because the first identity concerns an anticommutative relation of the real fields and the second one involves a commutative relation of the real fields. To pick up appropriately one of two identities for the real fields, we are led to make the following postulate. 

{\it Maximum correlation motion postulate}: the free motion of real column vector field as $\cD_H$-component entity differential function is supposed to be correlated maximally among $\cD_H$ independent degrees of freedom in Hilbert space. 

To realize a maximally correlated free motion, it is appropriate to choose the anti-commuting identity presented in Eq.(\ref{Identity}), which associates with the real symmetric matrices $\Ups^a$ and brings on anticommutative relations for the real fields $\psi_i(x)$. Such a choice allows not only each component of real column vector field to have a free motion but also all components of real column vector field to get maximum correlation motion. This actually follows the simple fact that there are more symmetric matrices than antisymmetric ones for a given real square matrix. 

By taking the anticommutative relations given in Eq.(\ref{Identity}), we arrive at the following hermitian action for the anti-commuting field operators $\psi_i(x)$: 
\be \label{actionQ}
 \cS_{\Psi}& = &  \int [dx] \,  \frac{1}{4} \{ \lambda_I  \delta_a^{\;\mu}  \Ups^a_{ij} [ \psi_i(x)  i\p_{\mu} \psi_j(x) -  i\p_{\mu}\psi_i(x)   \psi_j(x)]   
 + \tla_I  \delta_p^{\;\alpha} i\Ups^p_{ij} \p_{\alpha} [ \psi_i(x)  \psi_j(x) ]  \} \nn \\
 & \equiv &  \int [dx] \,   \{  \frac{1}{2} \Psi^{\dagger}(x)  \delta_a^{\;\mu}  \Ups^a i\p_{\mu} \Psi(x) 
 + \frac {1}{4}\lambda_Y \delta_p^{\;\alpha}  \p_{\alpha} [ \Psi^{\dagger}(x) \tUps^p \Psi(x) ]  \}
  \ee
In the second equality, we have made a normalization for the real column vector field $\Psi(x)\to \Psi(x)/\sqrt{\lambda_I}$ and also redefinition for the coupling constant $\lambda_Y = \tla_I/\lambda_I$. Where the second term in association with the real antisymmetric matrices $\tUps^{p}$ is presented as a total derivative with respect to variables $x^{\alpha}$, which indicates that there exists no free motion in the parameter space of $x^{\alpha}$. 

As the real symmetric $\Ups$-matrices $\Ups^a$ characterize the free motion of real column vector field $\Psi(x)$, they are regarded as {\it motion-correlation matrices}. For convenience of mention, we may refer to $\Ups$-matrices $\Ups^a$ as {\it $\cM_c$-matrices}. 

We have also made a redefinition for the real antisymmetric $\Ups$-matrices $\Ups^{p}$ to be pure imaginary antisymmetric matrices, 
\be
\tUps^p \equiv  i\Ups^p = - (\tUps^p)^T = (\tUps^p)^{\dagger},
\ee
which become hermitian $\tUps$-matrices. Such pure imaginary antisymmetric matrices reflect the motion-irrelevant intrinsic feature of real column vector field $\Psi(x)$. 

We will show later on that there exist only finite numbers of $\tUps$-matrices in light of a {\it local coherent-qubits motion postulate}, which plays a significant role as {\it characteristic charge of qubit-spinor field}. For convenience and meaningfulness, we should mention such antisymmetric matrices $\tUps^{p}$ as {\it $\cQ_c$-matrices}. 

For the real symmetric $\cM_c$-matrices $\Ups^a \equiv (\Ups^0, \Ups^k)$, one can always choose one of them as an identity matrix without losing generality. As a convention, we choose $\Ups^0$ as an identity matrix, i.e.:
\be \label{IM}
\Ups^0 = I \quad \mbox{or} \quad \Ups^0_{ij} = \delta_{ij}  . \nn
\ee

To find out the set of all real symmetric $\cM_c$-matrices $\Ups^a$ for describing a maximally correlated free motion and determine the number of variables $x^{\mu}$, let us consider the equation of motion for the real column vector field $\Psi(x)$. From the principle of least action, it is easy to read from the action in Eq.(\ref{actionQ}) the following equation of motion for the freely moving real field $\Psi(x)$: 
\be \label{DEM}
& & \delta_a^{\;\mu} \Ups^a i\partial_{\mu} \Psi(x)  = 0, \nn \\
& &  i\p_0 \Psi(x) = - \Ups^k  i\p_k \Psi(x) , \; \; k =1,2, \cdots , 
\ee
which indicates that there is no contribution to the equation of motion from the parameter space of $x^{\alpha}$ due to the antisymmetric property of $\cQ_c$-matrices. Where the derivatives $i\p_0 \equiv i\frac{\p}{\p x^0}$ and $ i\p_k \equiv i\frac{\p}{\p x^k}$ are linear self-adjoint operators. In the second formalism, we have used the convention $\Ups^0=1$.

Its second-order differential equation of motion can be expressed as follows: 
\be
\p_0^2 \Psi(x) & = & (\Ups^k  \p_k )(\Ups^l  \p_l ) \Psi(x) =  \frac{1}{2}\{\Ups^k, \Ups^l \} \p_k \p_l  \Psi(x) , \nn 
\ee 
which implies that as long as there exist more than two variables $x^{\mu}$ in association with arbitrary real symmetric $\cM_c$-matrices, the quadratic equation of motion will in general concern the cross terms of derivatives, i.e., $\p_k  \p_l  \Psi(x)$ ($k\neq l$), and the linear correlations among different components of real column vector field, i.e., $\{\Ups^k, \Ups^l \} \Psi(x)$. To prevent the cross terms from derivatives and eliminate the mixing terms among various components of real column vector field, we are led to make the following postulate.

{\it Quadratic free motion postulate}:  in the partial differential equation of motion, the second-order derivative with respect to variable $x^0$ is postulated to be proportional to the second-order derivatives with respect to variables $x^k$ without involving any cross term of derivative operators, and each component $\psi_i(x)$ of real column vector field is supposed to obey an independent quadratic free motion without concerning any correlation among different components.
 
To meet such a quadratic free motion postulate, the motion-correlation $\cM_c$-matrices $\Ups^k$ must satisfy the following anti-commuting relations with diagonal conditions:
\be \label{AC0}
& & \{\Ups^k, \Ups^l \} = \Ups^k \Ups^l + \Ups^l \Ups^k = 2\hat{\Delta}^k\delta^{kl},  \quad k, l =1,2, \cdots , \nn
 \ee
where all $\hat{\Delta}^k$ have to be real diagonal matrices. As the real symmetric $\cM_c$-matrices $\Ups^k$ can always be diagonalized by orthogonal matrices $O^k$,  
\be
(O^k)^T \Ups^k O^k = \Delta^k, \quad (O^k)^T O^k = O^k (O^k)^T= I, \quad \quad k =1,2, \cdots , \nn
\ee
where all $\Delta^k$ are real diagonal matrices and $I$ is the unit matrix. From the above relations, we arrive at the following identities:
\be
O^k (\Delta^k)^2 (O^k)^T = \hat{\Delta}^k,  \quad k =1,2, \cdots , \nn
\ee
which can be held for general orthogonal matrices $O^k$ only when the square of real diagonal matrices $(\Delta^k)^2$ is proportional to the real diagonal unit matrix. We then obtain the following relations:
\be
(\Delta^k)^2 = (c^k)^2 I, \quad \hat{\Delta}^k = (c^k)^2 I,  \quad k =1,2, \cdots , \nn
\ee  
with $c^k$ being constants. The anti-commuting relations of $\cM_c$-matrices $\Ups^k$ get the following forms: 
\be \label{AC1}
& & \{\Ups^k, \Ups^l \} = \Ups^k \Ups^l + \Ups^l \Ups^k = 2(c^k)^2 I\, \delta^{kl},  \quad k, l =1,2, \cdots . \nn
 \ee

As the matrices $\Ups^k$ can always be normalized as basis matrices with $(\Ups^k)^2=1$ by the redefinitions of variables $x^{\mu}$ and column vector field $\Psi(x)$, $\cM_c$-matrices $\Ups^k$ satisfy the following normalized anti-commuting relations: 
\be \label{CA0}
& & \{\Ups^k, \Ups^l \} = \Ups^k \Ups^l + \Ups^l \Ups^k = 2\delta^{kl},  \quad k, l =1,2, \cdots , \nn \\
 & &  \{\Ups^k, \Ups^0 \} = \Ups^k \Ups^0  + \Ups^0 \Ups^k  = \Ups^k ,
\ee
which can generally be expressed as follows:
\be \label{CA}
 & & \Ups^a \hUps^b + \Ups^b \hUps^a = 2\eta^{ab}; \quad \hUps^{a,b}=(\Ups^0, -\Ups^{k,l}), \nn \\
 & &  \eta^{ab} = \diag. (1, -1, \cdots, -1), \quad a,b = 0, 1, \cdots ,
\ee
where $\eta^{ab}$ defines the unit metric matrix. 

The above general relations of $\cM_c$-matrices $\Ups^a$ form a {\it Clifford algebra}, which is resulted from the quadratic equation of motion of real column vector field $\Psi(x)$ when following along the quadratic free motion postulate. Consequently, the real column vector field $\Psi(x)$ satisfies the standard wave equation of motion, 
\be \label{WEM}
\square \Psi(x)  \equiv (\p_0^2 - \p_k^2) \Psi(x) \equiv \eta^{\mu\nu} \p_{\mu}\p_{\nu}  \Psi(x) = 0 ,
\ee   
with $\square$ denoting the d'Alembert operator. Where $\eta^{\mu\nu}$ is a  diagonal matrix, 
\be \label{sign}
\eta^{\mu\nu} = \diag. (1, -1, \cdots, -1), \quad \mu, \nu = 0, 1, \cdots ,
\ee
which defines the unit metric matrix in correspondence to the one given in Eq.(\ref{CA}). 

It is the quadratic free motion postulate that leads each component $\psi_i(x)$ of column vector field $\Psi(x)$ to obey independently a standard wave equation of motion. The equation of motion presented in Eq. (\ref{DEM}) for the column vector field $\Psi(x)$ brings about a generalized Dirac equation. The real column vector field $\Psi(x)$ as $\cD_H$-component entity differential function of $x^{\mu}$ spans a $\cD_H$-dimensional real Hilbert space. In addition to the simplest motion postulate, it is the correlation motion and quadratic free motion postulates that bring about the real column vector field $\Psi(x)$ as {\it spinor field} satisfying the anticommutative relation given in Eq.(\ref{Identity}), which will become manifest later on when we demonstrate the natural emergence of canonical anticommutative relation for the column vector field $\Psi(x)$.

We are now in the position to summarize the above descriptions and analyses. Instead of the usual start point with presupposed symmetry, we begin with the motional consideration by taking the real column vector field $\Psi(x)$ as motional matter field. Its motion is proposed to obey the simplest motion postulate on which the hermitian action concerns only the first order derivative of variables $x^{\mu}$ and keeps solely the bilinear form of real column vector field $\Psi(x)$. By further proposing the maximum correlation motion postulate, it is verified that the maximally correlated motion among independent degrees of freedom of real column vector field $\Psi(x)$ is characterized by real symmetric $\cM_c$-matrices. As a consequence, such a real column vector field $\Psi(x)$ turns out to be a spinor field satisfying the anticommutation relation. Furthermore, each component $\psi_i(x)$ of $\Psi(x)$ is supposed to obey the quadratic free motion postulate for its equation of motion in the second-order derivative with respect to the variables $x^{\mu}$. 

It is appropriate to combine the simplest motion postulate and maximum correlation motion postulate together with the quadratic free motion postulate to be as guiding principle in constructing the action for describing the free motion of real column vector field $\Psi(x)$ as basic matter field. For convenience of mention, we may refer to such a guiding principle as {\it maximum coherence motion principle}.


\subsection{ Presence of high dimensional Minkowski spacetime with single temporal dimension and the action of spinor field with scalar coupling associated to intrinsic $\cQ_c$-matrices }

In the unit metric matrix $\eta^{\mu\nu}$ presented in Eq.(\ref{sign}), there exists a sign flip for one component which is in connection with the unit $\cM_c$-matrix $\Ups^0$. As a convention, the derivative $i\p_0 \equiv i\frac{\p}{\p x^0}$ associating with the unit $\cM_c$-matrix $\Ups^0$ is identified to be temporal derivative operator, and the other derivatives $ i\p_k \equiv i\frac{\p}{\p x^k}$ associating with anti-commuting real symmetric $\cM_c$-matrices $\Ups^k$ define spatial derivative operators. Correspondingly, the variables $x^0$ and $x^k$ characterize the temporal coordinate and spatial Cartesian coordinates of globally flat spacetime, respectively. Analogously, the sign of unit metric matrix $\eta^{ab}$ presented in Eq.(\ref{CA}) describes a similar feature for a motion-correlation spacetime in the vector representation of $\cM_c$-matrices. Obviously, it is the identity $\cM_c$-matrix $\Ups^0$ that associates with the temporal derivative operator and reflects the temporal property of spacetime, while the real symmetric normalized basis $\cM_c$-matrices $\Ups^k$ which satisfy anti-commuting relations characterize the spatial property of spacetime and associate with the spatial derivative operators of Cartesian coordinates.

Let us consider a finite set of $\cM_c$-matrices $\Ups^a$, 
\be
& & \Ups^a=(\Ups^a)^{\ast} = (\Ups^a)^T , \quad a = 0, 1, \cdots, D_h-1, \nn \\
& & i\partial_{\mu} = i \frac{\partial}{\partial x^{\mu}}, \quad \mu =  0, 1, \cdots, D_h-1, \nn
\ee
where the number $D_h$ of $\cM_c$-matrices determines $D_h$-dimensional vector spacetime in spinor representation. Correspondingly, the coordinates $x^{\mu} $ span $D_h$-dimensional {\it Minkowski spacetime} with one temporal dimension and $D_h$-1 spatial dimensions as indicated from the sign of the unit metric matrix in Eq.(\ref{sign}). 

The above analyses provide us a natural comprehension on the longstanding open question {\it why there exists only one temporal dimension in high dimensional Minkowski spacetime and why the temporal dimension is not geometrically visible though it is physically measurable}. The answer is just deduced from the motional properties of matter field when following along the maximum coherence motion principle as guiding principle. The simplest motion postulate leads to the introductions of motion-correlation $\cM_c$-matrices and motion-irrelevant $\cQ_c$-matrices. The maximal maximum correlation motion postulate brings on the choice of real symmetric matrices as $\cM_c$-matrices in which a unit $\cM_c$-matrix can always be chosen to commute with all other $\cM_c$-matrices without losing generality. The quadratic free motion postulate results all $\cM_c$-matrices except the unit $\cM_c$-matrix to be anti-commuting. It is such a commuting unit $\cM_c$-matrix that associates to the isotropic temporal derivative operator and brings about invisible temporal dimension geometrically.

We now turn to discuss the motion-irrelevant parameter space spanned by $x^{\alpha}$ in the hermitian action, which associates with the hermitian antisymmetric $\cQ_c$-matrices $\tUps^p$. As the action presented in motion-irrelevant parameter space involves the total derivative with respect to the variables $x^{\alpha}$, its non-zero effect relies on the boundary condition in the parameter space. In general, the parameter space of $x^{\alpha}$ can be a finite space, so that the boundary conditions for real spinor field operators $\psi_i(x)$ are not necessary to be zero in the parameter space of $x^{\alpha}$. 

As the parameter space of variables $x^{\alpha}$ becomes motion irrelevant, to illustrate its possible effect, let us simply apply the method of separation of variables to express the spinor field as follows:
\be
\Psi(x^{\mu}, x^{\alpha}) \sim  \Psi(x^{\mu}) \xi(x^{\mu}, x^{\alpha} ), \nn
\ee
with $\xi(x^{\mu}, x^{\alpha} )$ being a real scaling function. In such a simple case, the action in Eq.(\ref{actionQ}) can be rewritten as follows:
\be \label{action1}
 \cS_{\Psi}& = & \int [dx] \,   \{  \frac{1}{2} \Psi^{\dagger}(x)  \delta_a^{\;\mu}  \Ups^a i\p_{\mu} \Psi(x) 
 + \frac{1}{4} \lambda_Y  \delta_p^{\;\alpha}  \p_{\alpha} [ \Psi^{\dagger}(x) i \Ups^p \Psi(x) ]  \} , \nn \\
 & = & \int [dx^{\mu}]  \, \{ \bar{\xi}^2(x) \frac{1}{2} \Psi^{\dagger}(x)  \delta_a^{\;\mu}  \Ups^a i\p_{\mu} \Psi(x)  + \frac{1}{2} \lambda_Y \delta^{\alpha}_p \hat{\xi}_{\alpha}^2 (x) \Psi^{\dagger}(x) \tUps^p \Psi(x)  \} , \nn
 \ee
where we have introduced the following real functions resulted from integrating over the variables $x^{\alpha}$:
 \be
  \bar{\xi}^2(x^{\mu}) = \int [dx^{\alpha}]  \xi^2(x) \, ; \quad  \hat{\xi}_{\alpha}^2(x^{\mu}) = \frac{1}{2}\int [dx^{\alpha}] \p_{\alpha} \xi^2(x)  .\nn
 \ee
The existence of real function $\hat{\xi}_{\alpha}^2(x^{\mu})$ requires nonzero boundary conditions for the scaling function $\xi^2(x^{\mu}, x^{\alpha})$ in the motion-irrelevance parameter space of variables $x^{\alpha}$.

The above demonstration indicates that the action of spinor field $\Psi(x)$ in motion-correlation Minkowski spacetime of coordinates $x^{\mu}$ can be written into the following general form:
\be \label{actionM}
 \cS_{\Psi}& = &  \int [dx] \,  \{ \frac{1}{2} \Psi^{\dagger}(x)  \delta_a^{\;\mu}  \Ups^a i\p_{\mu} \Psi(x) - \frac{1}{2}\phi_p(x) \Psi^{\dagger}(x) \tUps^p \Psi(x) \} ,
  \ee
where $\phi_p(x)$ are regarded as real scalar fields associated with the normalized basis $\cQ_c$-matrices $\tUps^p$, 
\be
& &  (\tUps^p)^2 =1, \quad  p=1, \cdots, q_c , \nn \\
& & \tUps^p \equiv  0,\; \;  p=0,
\ee
where $q_c$ denotes the number of $\cQ_c$-matrices $\tUps^p$. We will show below that the number $q_c$ appears to be periodic and the scalar field coupling depends on the spinor structure. Note that $\tUps^0\equiv 0$ is presented by a convention in order to characterize the intrinsic feature of spinor field $\Psi(x)$ with $q_c=0$.


\subsection{Appearance of canonical anticommutation relation and Pauli exclusion principle with maximum coherence motion principle as guiding principle }

It is shown that when proposing the maximum coherence motion principle as guiding principle, the real column vector field $\Psi(x)$ turns out to be a spinor field with $\cD_H$ components and satisfy the anticommutative relations. Meanwhile, each component of spinor field moves freely and independently as a wave field. We now focus on the anticommutative relations presented in Eq.(\ref{Identity}) and discuss in detail their properties. 

The anticommutative relations in Eq.(\ref{Identity}) can be rewritten into the following forms:
\be
 \p_{\mu}\psi_i(x)\,   \psi_j(x) + \psi_j(x) \p_{\mu} \psi_i(x)  & = & \lim_{x' \to x }  [ \p_{\mu}\psi_i(x)\,   \psi_j(x') + \psi_j(x') \p_{\mu} \psi_i(x)  ]   \nn \\
 &  \equiv & \lim_{x'\to x}  \p_{\mu} [ \psi_i(x)   \psi_j(x')  + \psi_j(x') \psi_i(x) ]  =0, \nn 
\ee
which are invariant under both translational operation $(x^{\mu}, x^{' \mu})  \to ( x^{\mu}, x^{' \mu}) + a^{\mu}$ and permutation operations $x^{\mu}\leftrightarrow x'^{\mu}$, $i \leftrightarrow j$ in the bracket. For a given point $x^0 = x'^0$, it can be verified that the above identities hold once the real spinor fields $\psi_i(x)$ fulfill the following anticommutation relations:
\be
& & \{\psi_i(x^0,\cdots x^{k} \cdots) , \;  \psi_j(x^0, \cdots x'^{k} \cdots) \}  \nn \\
& = & \psi_i(x^0, \cdots x^{k} \cdots)   \psi_j(x^0, \cdots x'^{k} \cdots)  +  \psi_j(x^0, \cdots x'^{k} \cdots) \psi_i(x^0, \cdots x^{k}\cdots) \nn \\
& = & \delta_{ij} \Delta(x^{k}-x'^{k}), \quad k = 1, \cdots ,D_h-1  , \nn 
\ee
where $\Delta(x^{k}-x'^{k})$ is a symmetric function which satisfies the following general conditions:  
\be
& & \Delta(x^{k}-x'^{k}) =  \Delta(x'^{k}-x^{k}), \nn \\
& &  \Delta'(x^{k}-x'^{k}) =  -\Delta'(x'^{k}-x^{k}), \nn \\
& & \Delta'(x^{k}-x'^{k})|_{x^{k}=x'^{k}} = 0 , \quad k = 1, \cdots ,D_h-1 , \nn
\ee
with $\Delta'(x) \equiv \frac{\p}{\p x^{k}} \Delta(x) = - \Delta'(-x)$ being an odd function.

To validate all above anticommutation relations concerning the spatial coordinates $x^{k}$ with $k=1,2,\cdots, D_h-1$, a simple solution is achieved by taking the function $\Delta(x^{k}-x'^{k})$ to be the $\delta$-function, i.e.:
\be
\Delta(x^k-x'^k) = \lambda_d \delta(x^k-x'^k),  \nn
\ee
which enables us to express all anticommutation relations at the given point $x^0=x^{'0}$ via the following general solution:
\be \label{CACR}
& & \{ \psi_i(x^0, \fx) ,  \psi_j(x^0, \fx') \} =  \lambda_d \delta_{ij} \delta(\fx-\fx'),
\ee
where the parameter $\lambda_d$ is determined from the normalization of spinor field $\Psi(x)$.  Such a solution provides the well-known {\it canonical anticommutation relation} for spinor fields $\psi_i(x)$ at an equal temporal coordinate $x^0=x^{'0}$ in quantum field theory. 

On the other hand, when applying the equations of motion presented in Eqs.(\ref{DEM}) and (\ref{WEM}) for the freely moving spinor field, we are able to obtain the following {\it general anticommutation relation}:
 \be
& & \{ \psi_i(x) ,  \psi_j(x') \} =  \lambda_d \delta_{ij} \Delta(x-x'), \nn \\
& & \Delta(x-x') = \int \frac{d^{D_h}p}{(2\pi)^{D_h-1}}\, \delta(\, p_{0}^2 - p^2\, )( p_0 + p_k \Ups^k )\, e^{ -i p\cdot(x - x') } ,
\ee
which recovers the canonical anticommutation relation given in Eq.(\ref{CACR}) at the equal point $x^0=x^{'0}$. 

Such a general anticommutation relation can be traced back again to the motional properties of matter field by following along the maximum coherence motion principle as guiding principle.

The appearance of canonical anticommutation relation implies Pauli exclusion principle which states that the spinor field cannot occupy the same quantum state. This can be understood from the least action principle and path integral formulation together with the canonical anticommutation relation. To be explicit, let us illustrate the locally coupled term $\int [dx]  \psi_i(x) \psi_j(x)$ in the calculation of path integral, such a term represents the distribution functional corresponding to various paths in the action and gives the probability amplitudes of various paths when $i=j$. It is easy to verify that the locally coupled term $\int [dx] \psi_i(x) \psi_j(x)$ brings on an infinity distribution functional for $i=j$ due to the canonical anticommutation relation shown as follows:
\be
& & \int [dx] \psi_i(x) \psi_j(x) = \int [dx] \lim_{\fx\to \fx' } \{\psi_i(x^0, \fx), \psi_j(x^0, \fx')\}/2 \nn \\
& & \quad = \lambda_d \delta_{ij} \int [dx] \lim_{\fx\to \fx' } \delta(\fx-\fx' ) = \lambda_d \delta_{ij} \int [dx] \delta(0) \to \infty, \quad \mbox{for}\quad i=j, 
\ee
which results in a zero probability amplitude for $i=j$ in the path integral formulation. This means that the spinor field occupying the same quantum state will get a zero probability amplitude, which explains directly the statement of Pauli exclusion principle.  

In other words, by proposing maximum coherence motion principle as guiding principle, we are able to deduce directly the canonical anticommutation relation and understand naturally Pauli exclusion principle for spinor field.


\subsection{ Emergence of $\cM_c$-spin and $\cQ_c$-spin symmetries in Hilbert space in association with Poincar\'e-type symmetry and scaling symmetry in Minkowski spacetime }

Unlike the usual consideration with starting from a presupposed symmetry, we begin with the motion nature of basic constituents by proposing the maximum coherence motion principle as guiding principle, which brings on real symmetric normalized basis matrices as motion-correlation $\cM_c$-matrices to form Clifford algebra presented in Eq.(\ref{CA}). From such $\cM_c$-matrices $\Ups^k$, we can always define a new set of matrices as follows:
\be
\Sigma^{kl} = \frac{1}{4i} [\Ups^k, \Ups^l], \quad k, l = 1,2, \cdots, D_h-1 , \nn
\ee 
which satisfy the following commutation relations:
\be
& & [\Sigma^{kl} , \Sigma^{k'l'} ]  = - i (\Sigma^{kl'}\delta^{lk'} -\Sigma^{ll'}  \delta^{kk'} - \Sigma^{kk'} \delta^{ll'} + \Sigma^{lk'} \delta^{kl'}) . \nn
\ee 
Such algebra relations can be verified to form the group generators of {\it spinning group} denoted as SP($D_h$-1) which is isomorphic to the orthogonal rotational group SO($D_h$-1), i.e., SP($D_h$-1)$\cong$ SO($D_h$-1). 

It can be checked from Clifford algebra in Eq.(\ref{CA}) that such matrices $\Sigma^{kl}$ satisfy the following group algebras via commutation relations:
\be
& & [\Sigma^{kl} , \frac{i}{2}  \Ups^{k'}] = - i ( \delta^{lk'} \frac{i}{2} \Ups^k - \delta^{kk'} \frac{i}{2} \Ups^l  ) , \nn
\ee
which indicates that when adding matrices $\frac{i}{2}  \Ups^{k}$ to be the group generators, we obtain the following extended commutation relations:
\be
& & [\Sigma^{ab}, \Sigma^{cd}] = i (\Sigma^{ad}\eta^{bc} -\Sigma^{bd}  \eta^{ac} - \Sigma^{ac} \eta^{bd} + \Sigma^{bc} \eta^{ad}) , \nn \\
& & \Sigma^{ab} \equiv (\Sigma^{0k}, \Sigma^{kl}), \quad \Sigma^{0k} = -\Sigma^{k0} = \frac{i}{2} \Ups^k ,  
\ee
with $\eta^{ab}$ the unit metric matrix defined in Eq.(\ref{CA}). 

The matrices $\Sigma^{ab}$ provide the group generators of an enlarged spinning group denoted as SP(1, $D_h$-1). As $\cM_c$-matrices reflect maximally correlated motion of spinor field, we may refer to such an enlarged spinning group as {\it $\cM_c$-spin group} SP(1, $D_h$-1) with a spinor represented in $\cD_H$-dimensional Hilbert space. The number $D_h$ of $\cM_c$-matrices is regarded as {\it $\cM_c$-spin charge} notated as $\CMc$, i.e.: 
\be
\CMc=D_h, \nn
\ee
which determines the dimension of vector spacetime in the spinor representation. 

It can be checked that the {\it $\cM_c$-spin group} SP(1,$D_h$-1) has the following transformation property:
\be
& & S^T(L) \Ups^a S(L) = L^{a}_{\; b}\,  \Ups^b , \nn \\
& & S(L) = e^{i \varpi_{ab}\Sigma^{ab} } \in SP(1, D_h-1), \quad  L^{a}_{\; b} \in SO(1, D_h-1) ,
\ee
where $S(L)$ represents the group element of $\cM_c$-spin group SP(1, $D_h$-1) and $L^{a}_{\; b} $ belong to the group elements of orthogonal rotational group SO(1, $D_h$-1) in the vector representation. With such a feature of transformations, it can be verified that when the spinor field and Minkowski spacetime transform simultaneously as follows:
\be  \label{SPT}
& & \Psi(x) \to \Psi'(x') = S(L) \Psi(x), \quad x^{\mu} \to x^{'\mu} = L^{\mu}_{\; \nu}\,  x^{\nu},  \nn \\
& &  L^{\mu}_{\; \nu} = L^{a}_{\; b} \in SO(1, D_h-1), 
\ee 
the motional kinetic term of the action presented in Eq.(\ref{actionM}) becomes invariant. Namely, the transformation $L^{a}_{\; b}$ under the $\cM_c$-spin group SP(1, $D_h$-1) must be coincidental with the transformation $L^{\mu}_{\; \nu}$ under the orthogonal Lorentz-type group SO(1, $D_h$-1) in order to preserve the invariance of the action.

When the imaginary antisymmetric normalized basis $\cQ_c$-matrices $\tUps^p$ are anti-commuting with the real symmetric normalized basis $\cM_c$-matrices $\Ups^k$, namely:
\be \label{QCM0}
\{\Ups^k, \tUps^p \} = \Ups^k \tUps^p + \tUps^p \Ups^k = 0 , 
\ee
the scalar coupling term of the action given in Eq.(\ref{actionM}) turns out to be invariant under the transformation of $\cM_c$-spin group SP(1, $D_h$-1) as given in Eq.(\ref{SPT}), which can be checked from the following identity:
\be
S^T(L) \tUps^p = \tUps^p S^{-1}(L) . \nn
\ee

Let us suppose that the scalar fields get constant values, i.e., $\phi_p = v_p$, we arrive at from the action presented in Eq.(\ref{actionM}) the following equation of motion for the spinor field:
\be \label{DEMM}
& & \delta_a^{\;\mu} \Ups^a i\partial_{\mu} \Psi(x) -  v_p \tUps^p \Psi(x) = 0,  \nn \\
& &  (\p_0^2 - \p_k^2) \Psi(x) \equiv v_pv_q \frac{1}{2} \{ \tUps^p, \tUps^q\} \Psi(x) . \nn
\ee
Once the hermitian $\cQ_c$-matrices $\tUps^p$ become anti-commuting and satisfy Clifford algebra,
\be \label{QCM}
\{\tUps^p, \tUps^q \} = \tUps^p \tUps^q + \tUps^q \tUps^p = 2\delta^{pq} ,  \quad p, q = 1, \cdots, q_c, 
\ee
the constant scalar coupling term turns out to be a mass-like term. Consequently, the equation of motion is simplified to be:
\be
  \eta^{\mu\nu} \p_{\mu}\p_{\nu} \Psi(x) \equiv m^2 \Psi(x), \quad m^2 = \sum_{p=1}^{q_c} v_p^2, 
\ee
where $q_c$ denotes the number of $\cQ_c$-matrices. 

For such a case, the action in Eq.(\ref{actionM}) becomes invariant under the following transformations:
\be
& & \Psi(x) \to \Psi'(x) = S(V) \Psi(x), \quad \phi_p(x)\to \phi'_p(x) = V_p^{\; q} \phi_q(x),  \nn \\
& & S^{T}(V) \tUps^p S(V) = V^p_{\; q} \tUps^q , \quad S(V) = e^{i  \varpi_{pq} \tSi^{pq} } \in SP(q_c), 
\ee
where $V^p_{\; q} \in SO(q_c)$ for $q_c > 1$ and $V^p_{\; q} \in O(q_c)$ for $q_c =1$ are the group elements of rotating group. The matrices $\tSi^{pq}$ are regarded as the group generators of {\it motion-irrelevance spinning group} SP($q_c$), which satisfy the following commutation relations:
\be
& & [\tSi^{pq}, \tSi^{rs}] = i (\tSi^{ps}\delta^{qr} -\tSi^{qs}  \delta^{pr} - \tSi^{pr} \delta^{qs} + \tSi^{qr} \delta^{ps}) , \nn \\
& & \tSi^{pq} = \frac{i}{4} [\tUps^p, \tUps^q ] , \quad [\tSi^{pq} , \tUps^{r}] =  i ( \delta^{qr} \tUps^p - \delta^{pr} \tUps^q  ).
\ee
Where the real scalar fields $\phi_p(x)$ transform as a vector field of orthogonal rotational group SO($q_c$). 

For convenience, we may refer to such a motion-irrelevance spinning group as {\it $\cQ_c$-spin group}. The number $q_c$ is regarded as {\it $\cQ_c$-spin charge} of spinor field denoted as follows:
\be
\cC_{\cQ_c}= q_c . \nn
\ee 

Note that once the anti-commuting relation in Eq.(\ref{QCM0}) becomes invalid for a typical $\cM_c$-matrix, namely, if the following relations occur for a typical matrix $\Ups^{k_1}$, 
\be 
\{\Ups^{k_1}, \tUps^p \} \ne 0 , \quad  [\Ups^{k_1}, \tUps^p ] = 0 \nn
\ee
the scalar coupling term of spinor field is prohibited from the maximal $\cM_c$-spin group symmetry. Nevertheless, the action remains to have $\cQ_c$-spin group symmetry SP($q_c$). We will show later on how the situation happens. 

In general, as long as the real symmetric normalized basis $\cM_c$-matrices $\Ups^a$ ($a=0,1, \cdots D_h-1$) and the imaginary antisymmetric normalized basis $\cQ_c$-matrices $\tUps^p$ ($p=1, \cdots q_c$) satisfy Clifford algebra, the action presented in Eq.(\ref{actionM}) possesses the following maximal {\it $\cM_c$-spin symmetry} and {\it $\cQ_c$-spin symmetry}:
\be
& & G_S = SO(1, D_h-1)\adjoin SP(1, D_h-1) \times SP(q_c).
\ee
where the symbol ``$\adjoin$" is adopted to indicate that the transformation of $\cM_c$-spin symmetry SP(1,$D_h$-1) in Hilbert space must be coincidental to that of Lorentz-type group symmetry SO(1,$D_h$-1) in globally flat Minkowski spacetime, such a transformation may be called as an {\it associated transformation}. For convenience, we may refer to the symmetry SO(1, $D_h$-1)$\adjoin$SP(1, $D_h$-1) as an {\it associated symmetry}.

It is noted that the groups SO(1, $D_h$-1) and SP(1,$D_h$-1) do not form a direct-product group symmetry. While the $\cM_c$-spin symmetry group  SP(1,$D_h$-1) in the spinor representation is isomorphic to the Lorentz-type group SO(1, $D_h$-1) in the vector representation of Minkowski spacetime of coordinates. Similarly, the $\cQ_c$-spin symmetry group SP($q_c$) in the spinor representation is isomorphic to the orthogonal rotational symmetry group SO($q_c$) in the vector representation of scalar-type fields. Mathematically, it is expressed as follows:
\be
& & SP(1, D_h-1) \cong SO(1, D_h-1) , \nn \\
& & SP(q_c) \cong
   \begin{cases}
 SO(q_c) , &  q_c > 1 , \\
 O(q_c) ,  & q_c \leq 1 ,  \\
  \end{cases}
\ee
where we keep in notation the spin symmetry SP(0) for the zero $\cQ_c$-spin charge $\CQc = q_c=0$ as it will be shown later on that such a $\cQ_c$-spin charge is applied to make the categorization for basic spinor fields. A zero $\cQ_c$-spin charge indicates no scalar coupling to spinor field as it is defined to have $\tUps^0\equiv 0$ by a convention.
 
It is easy to verify that the action given in Eq.(\ref{actionM}) has a translational group symmetry $P^{1,D_h-1}$ in Minkowski spacetime, 
\be
x^{\mu} \to x^{\mu} + a^{\mu}, \quad \Psi(x) \to \Psi(x + a), \quad \phi_p(x) \to \phi_p(x + a) , \quad a^{\mu} \in  P^{1, D_h-1} .   
\ee

In addition, the action presented in Eq.(\ref{actionM}) possesses global scaling invariance under scaling transformations for the basic fields in Hilbert space and coordinates in Minkowski spacetime, i.e.:
\be \label{GSS}
& & \Psi(x) \to \Psi'(x') = e^{(D_h-1)\varpi/2} \Psi(x), \quad \phi_p(x) \to \phi_p'(x') = e^{\varpi}  \phi_p(x), \nn \\
& &  x^{\mu} \to x^{'\mu} = e^{-\varpi} x^{\mu}, \quad e^{-\varpi} \in SC(1), \;\;  e^{(D_h-1)\varpi/2} , e^{\varpi} \in SG(1),
\ee 
with $\varpi$ the constant scaling parameter. Such an invariance is referred to as {\it scaling symmetry} notated by SC(1) and SG(1) for the scaling symmetry of coordinates and fields, respectively.

In general, an associated symmetry emerges in the action shown in Eq.(\ref{actionM}), which is expressed as follows:
\be
G & = &  SC(1)\ltimes P^{1, D_h-1} \ltimes SO(1, D_h-1) \adjoin SP(1, D_h-1) \rtimes SG(1)\times SP(q_c)  \nn \\
& \equiv & SC(1)\ltimes PO(1, D_h-1) \adjoin SP(1, D_h-1) \rtimes SG(1) \times SP(q_c)  , 
\ee
with the definition,
\be
PO(1, D_h-1) \equiv P^{1, D_h-1} \ltimes SO(1, D_h-1),
\ee
which represents Poincar\'e-type group symmetry (also called as inhomogeneous Lorentz-type group symmetry). The symbol ``$\ltimes$" means that the Poincar\'e-type group symmetry is a semidirect product group symmetry.  SC(1)$\adjoin$SG(1) represents an {\it associated symmetry} with coincidental global scaling transformations of coordinates and fields. 

Before ending this section, we would like to address that it becomes essential to recognize that the emergent symmetry group SP(1, $D_h$-1) of spinor field in Hilbert space and the emergent symmetry group SO(1, $D_h$-1) of coordinates in Minkowski spacetime should be distinguishable as they operate on different spacetimes. In particular, the $\cM_c$-spin symmetry of spinor field in Hilbert space will be taken as a gauge symmetry to describe the fundamental interaction of nature, which is going to be discussed separately in the part II of the foundation of the hyperunified field theory. 

The number $D_h$ of $\cM_c$-matrices brings on $\cM_c$-spin charge denoted as $\CMc=D_h$, which determines the dimension of spacetime. In four-dimensional spacetime, which corresponds to $\cM_c$-spin charge $\CMc=D_h $= 4. The $\cM_c$-spin symmetry SP(1,3) characterizes the rotational spin symmetry SU(2) and boost spin symmetry SU$^\ast$(2) for Majorana and massive Dirac fermions defined in four-dimensional real/self-conjugated and complex Hilbert spaces, respectively. The rotational symmetry SO(1,3) is the well-known Lorentz group symmetry in four-dimensional Minkowski spacetime of coordinates. In the hyperunified field theory\cite{HUFT}, it concerns 19-dimensional {\it hyper-spacetime} in correspondence to $\cM_c$-spin charge $\CMc=D_h$ = 19. The $\cM_c$-spin symmetry SP(1,18) is referred to as {\it hyperspin symmetry}, and the rotational symmetry SO(1, 18) is a generalized Lorentz-type group symmetry. Where the high dimensional spacetime of coordinates is mentioned to be Minkowski {\it hyper-spacetime}. The spinor field defined in hyper-spacetime is referred to as {\it hyper-spinor field} in Hilbert space with dimension $\cD_H=2^{(D_h-1)/2}$ determined by the independent degrees of freedom of self-conjugated hyper-spinor field. 

For the $\cQ_c$-spin symmetry SP($q_c$), we will demonstrate below that when a basic spinor field is considered to be a {\it local coherent state of qubits}, the number $q_c$ of $\cQ_c$-matrices is found to characterize the periodic behavior of local coherent states with arbitrary {\it qubit number} $Q_N$. So that the $\cQ_c$-spin charge $q_c$ enables us to categorize any local coherent state of qubits following along a {\it local coherent-qubits motion postulate}. As the $\cQ_c$-spin charge $q_c$ reflects motion-irrelevant intrinsic spinning symmetry SP($q_c$) of spinor field, it is regarded as {\it categoric spin charge} for a local coherent state of qubits as basic spinor field. For convenience, such a spin charge is referred to as {\it $\cQ_c$-spin charge} denoted as $\CQc=q_c$, and the intrinsic spinning symmetry SP($q_c$) is mentioned as {\it $\cQ_c$-spin symmetry} for short.


\section{ Spinor field in 2D Hilbert space as local coherent state of qubit and basic properties of uniqubit-spinor field with appearance of 3D Minkowski spacetime  }

By proposing the maximum coherence motion principle as guiding principle, we have shown that a real column vector field turns out to be a spinor field satisfying canonical anticommutation relation and the real variables of spinor field span a high dimensional Minkowski spacetime of coordinates with only one temporal dimension. For a spinor field with single real component, it is verified to have a free motion only in two-dimensional Minkowski spacetime. To demonstrate explicitly how the independent degrees of freedom of spinor field determine the dimension of Minkowski spacetime, we are going to examine as a simple case the two-component entity spinor field, which is regarded conceptually as a local coherent state of qubit and referred to as {\it uniqubit-spinor field}. Such a uniqubit-spinor field with two independent degrees of freedom is shown to have $\cM_c$-spin charge $\cC_{\cM_c}=3$ and $\cQ_c$-spin charge $\CQc=1$ when following along the maximum coherence motion principle as guiding principle, which enables us to build the action of uniqubit-spinor field with scalar coupling in three-dimensional Minkowski spacetime. We will discuss in detail the basic properties of the action.

\subsection{ Spinor field in 2D Hilbert space as local coherent state of qubit and the appearance of 3D Minkowski spacetime for uniqubit-spinor field }

To study intrinsic properties of spinor field, let us start with considering a simple two-component column vector field,
\be
\psi(x) = \binom{\psi_{+}(x)}{\psi_{-}(x)} , \nn
\ee 
which turns out to be a spinor field when applying for the maximum coherence motion principle proposed in the previous section. In general, it can be expressed as a linear superposition of two orthonormal basis states,
\be
\psi(x) =  \sum_{s=\pm} \psi_s(x)\, \varsigma_s ,
\ee
where $\varsigma_s$ ($s=\pm$) denote two orthonormal basis states,
\be \label{QB}
\varsigma_{+} = \binom{1}{0}, \quad  \varsigma_{-} =\binom{0}{1}, 
\ee
which form a {\it real basic qubit-basis} $\{\varsigma_s\}$. 

The two components $\psi_{+}(x)$ and $\psi_{-}(x)$ are regarded as local distribution amplitudes with respective to the basis states $\varsigma_{+} $ and $\varsigma_{-} $. They are considered to be differentiable functions of coordinates $x^{\mu}$ and their freely moving dimension is determined by the maximally correlated motion of spinor field $\psi(x)$. In general, $\psi_{+}(x)$ and $\psi_{-}(x)$ can be either real or complex field operators. In such a representation, the spinor field $\psi(x)$ is viewed as a local linear superposition state of two basis states. 

It is known that a qubit (quantum bit) as the basic unit of quantum information is in a linear superposition of two states regarded as basis states, which are often represented by Dirac ``bra-ket" notation or spin-like ``column-row" notation just like any sort of quantum states. The two basis states can conventionally be written as $|0\rangle$  and $|1\rangle$  (called as ``ket 0" and ``ket 1") or $\binom{1}{0}$ and $\binom{0}{1}$ (called as ``spin-up" and ``spin-down"). A pure qubit state can be represented as a linear combination of two orthonormal basis states $|0\rangle$  and $|1\rangle$ or $\binom{1}{0}$ and $\binom{0}{1}$,
\be
|\psi \rangle =\alpha |0\rangle +\beta |1\rangle , \quad \mbox{or}\quad \psi = \alpha \binom{1}{0} + \beta \binom{0}{1} = \binom{\alpha}{\beta}, \nn
\ee
where $\alpha$ and $\beta$ are constant numbers that represent the probability amplitudes and can in general be complex numbers. The qubit state $|\psi \rangle$ or $\psi$ as two-component entity is regarded as a global coherent state. When measuring such a qubit in the standard basis, the probability of outcome $|0\rangle$ or $\binom{1}{0}$ is $|\alpha |^{2}$ and the probability of outcome $|1\rangle$ or $\binom{0}{1}$ is $|\beta |^{2}$. Meanwhile, $\alpha$  and $\beta$ must be constrained by the normalization condition $|\alpha |^{2}+|\beta |^{2}=1$ for the total probability.

Conceptually, a two-component entity spinor field can be regarded as a {\it local coherent state of qubit} and the field operators $\psi_{+}(x)$ and $\psi_{-}(x)$ are viewed as {\it evolving local distribution amplitudes}. As two orthonormal basis states span two-dimensional Hilbert space, the local coherent state of qubit forms a vector field with its evolving local distribution amplitudes $\psi_{+}(x)$ and $\psi_{-}(x)$ as two components in two-dimensional Hilbert space. 

Let us first consider the simplest case that the evolving local distribution amplitudes $\psi_{+}(x)$ and $\psi_{-}(x)$ represent two independent degrees of freedom, which occurs when $\psi_{+}(x)$ and $\psi_{-}(x)$ are two real fields or a pair of complex conjugated fields. In this case, we can express such a spinor field to be a {\it local coherent state of qubit} in a linear superposition of two orthonormal basis states with two local distribution amplitudes,
\be \label{UQSF}
\psi_{\mQ^1}(x) & \equiv & \sum_{s=\pm} \psi_s(x) \varsigma_s \equiv \binom{\psi_{\mQ^0_{+}}(x)}{\psi_{\mQ^0_{-}}(x)} \equiv \binom{\psi_{0}(x)}{\psi_{1}(x)} , \nn \\
\psi_{\QS^1}(x) & \equiv & \sum_{s=\pm} \psi_s(x) \zeta_s \equiv \binom{\psi_{\QC^0}(x)}{\psi_{\QC^0}^{\ast}(x)}  \equiv \sum_{s=\pm} \tilde{\psi}_s(x) \varsigma_s =  \binom{\tilde{\psi}_{+}(x)}{\tilde{\psi}_{-}(x)} ,
\ee
where the superscript ``1" in $\mQ^1$ and $\QS^1$ labels the qubit number $Q_N=1$.  We have used the following definitions:
\be
& & \psi_{\mQ^1}(x): \quad  \psi_{+}(x) \equiv \psi_{\mQ^0_{+}}(x) \equiv \psi_{0}(x)=\psi_{0}^{\ast}(x), \quad \psi_{-}(x)\equiv \psi_{\mQ^0_{-}} \equiv \psi_{1}(x)=\psi_{1}^{\ast}(x), \nn \\
& &  \psi_{\QS^1}(x): \quad   \psi_{\QC^0}(x) \equiv \tilde{\psi}_{+}(x) =  \frac{1}{\sqrt{2}}(\psi_{\mQ^0_{+}}(x) + i \psi_{\mQ^0_{-}}(x) )  = \frac{1}{\sqrt{2}}(\psi_0(x) + i\psi_1(x) )  , \nn \\ 
& & \qquad \qquad \quad \psi_{\QC^0}^{\ast}(x) \equiv \tilde{\psi}_{-}(x) = \frac{1}{\sqrt{2}}(\psi_{\mQ^0_{+}}(x) - i \psi_{\mQ^0_{-}}(x) )  = \frac{1}{\sqrt{2}}(\psi_0(x) -i\psi_1(x) ) , \nn
\ee
where two real local distribution amplitudes $\psi_{\mQ^0_{+}}(x)$ and $\psi_{\mQ^0_{-}}(x)$ are regarded as two {\it real local states of zeroqubit} as hinted in the superscript ``0" in $\mQ^0$ with zero qubit number $Q_N=0$. We have introduced alternatively in Eq.(\ref{UQSF}) an equivalent orthonormal basis states $\zeta_s$,
\be \label{CQB}
\zeta_+ = \frac{1}{\sqrt{2}}\binom{1}{1}, \quad  \zeta_- = \frac{1}{\sqrt{2}}\binom{i}{-i} ,
\ee
where $\zeta_s$ ($s=\pm$) form an orthonormal {\it complex-conjugated basic qubit-basis} $\{\zeta_s\}$, which spans a complex-conjugated two-dimensional Hilbert space. 

The two kinds of local coherent states of qubit, $\psi_{\mQ^1}(x)$ and $\psi_{\QS^1}(x)$, satisfy the following complex-conjugated conditions:
\be
& & \psi_{\mQ^1}^{\ast}(x) = \psi_{\mQ^1}(x), \nn \\
& & \psi_{\QS^1}^{\ast}(x) = \sigma_1 \psi_{\QS^1}(x),  \quad  \sigma_1 = \begin{pmatrix}
0 & 1 \\
1 & 0
 \end{pmatrix} , 
\ee
where $\psi_{\mQ^1}(x)$ is considered as a {\it real local coherent state of qubit} and $\psi_{\QS^1}(x)$ as a {\it self-conjugated local coherent state of qubit} indicated from the subscript ``$\mS$" in $\QS^1$.

When two local distribution amplitudes $\psi_{+}(x)$ and $\psi_{-}(x)$ are all complex fields, $\psi(x)$ is regarded as a {\it complex local coherent state of qubit},
\be
\psi_{\QC^1}(x) & \equiv & \sum_{s=\pm} \psi_s(x) \varsigma_s \equiv \binom{\psi_{\QCp^0}(x)}{\psi_{\QCn^0}(x)} ,
\ee
where the subscript ``$\mC$" in $\QC$ reflects the complex feature for local coherent state $\psi_{\QC^1}(x)$ and zeroqubit local states $\psi_{\QCp^0}(x)$ and $\psi_{\QCn^0}(x)$. Note that $\psi_{\QC^1}(x)$ as a complex local coherent state of qubit involves in general four real independent degrees of freedom. 

The two basic qubit-bases enable us to define conceptually the real two-component entity spinor field $\psi_{\mQ^1}(x)$ and self-conjugated two-component entity spinor field $\psi_{\QS^1}(x)$ as well as complex two-component entity spinor field $\psi_{\QC^1}(x)$ to be as local coherent states of qubit. We would like to address that, unlike the ordinary global coherent state of qubit in which the global amplitudes are explained as the probability amplitudes of basis states for measuring the qubits in a standard basis based on the Born's rule of quantum mechanics, while the evolving local distribution amplitudes of spinor field are considered as {\it dynamic field operators}.

To reflect above conceptual definitions on two-component spinor field and also for compact terminology, we may refer to $\psi_{\mQ^1}(x)$ as {\it real uniqubit-spinor field}, $\psi_{\QS^1}(x)$ as {\it self-conjugated uniqubit-spinor field} and $\psi_{\QC^1}(x)$ as {\it complex uniqubit-spinor field}.  In the following discussions, {\it uniqubit-spinor field} will mean {\it real uniqubit-spinor field} unless otherwise specified. 

In order to investigate the kinematical evolution of uniqubit-spinor field, we are going to apply the maximum coherence motion principle as guiding principle to build the action of uniqubit-spinor field. Let us first consider the action of freely moving uniqubit-spinor field, which has the following general form:
\be \label{actionQ}
 \cS_{\mQ^1}  & = &  \int [dx] \,  \{ \frac{1}{2}  \psi_{\mQ^1}^{\dagger}(x)  \delta_a^{\;\mu}  \Ups^a i\p_{\mu} \psi_{\mQ^1}(x) 
 -  \, \frac{1}{2}\lambda_1 \phi_{p}(x) \psi_{\mQ^1}^{\dagger}(x)  \tUps^{p} \psi_{\mQ^1}(x)   \} ,
  \ee
with $\lambda_1$ the real coupling constant and $\phi_{p}(x)$ real scalar fields. Following along the maximum coherence motion principle as guiding principle, we are able to figure out explicitly all real symmetric normalized basis $\cM_c$-matrices from all $2\times 2$ matrices as follows: 
\be \label{UpsQ1}
\Ups^0 = \sigma_0 =  \begin{pmatrix}
1 & 0 \\
0 & 1
 \end{pmatrix} ;  \quad
 \Ups^1 = \sigma_3 =  \begin{pmatrix}
1 & 0 \\
0 & -1
 \end{pmatrix}; \quad  \Ups^2 = \sigma_1 = \begin{pmatrix}
0 & 1 \\
1 & 0
 \end{pmatrix} ,  
\ee
which satisfy Clifford algebra relations given in Eq.(\ref{CA}) with $a=0, 1, 2$. Obviously, there exists only a single pure imaginary antisymmetric normalized basis $\cQ_c$-matrix,  
\be \label{tUpsQ1}
\tUps = \sigma_2 =  \begin{pmatrix}
0 & -i \\
i & 0
 \end{pmatrix} , 
 \ee
which meets to Clifford algebra relations in Eqs.(\ref{QCM0}) and (\ref{QCM}) with $q_c=1$. Where $\sigma_i$ ($i=1,2,3$) are Pauli matrices. 
 
From the action of uniqubit-spinor field given in Eq.(\ref{actionQ}), the maximally correlated motion is characterized by three real symmetric normalized basis $\cM_c$-matrices, and the motion-irrelevance scalar coupling is described by the single $\cQ_c$-matrix, which brings $\cM_c$-spin charge and $\cQ_c$-spin charge as follows:
\be
\CMc=D_h=3, \quad \CQc=q_c = Q_N =1.
 \ee
Where the $\cM_c$-spin charge determines three-dimensional Minkowski spacetime with one temporal dimension and two spatial dimensions. The $\cQ_c$-spin charge brings on the scalar coupling with one real scalar field $\phi_{1}(x)$ in associating with the single $\cQ_c$-matrix. It is noticed that the $\cQ_c$-spin charge $\CQc$ is directly associated to the qubit number $Q_N$.


\subsection{ Properties of  uniqubit-spinor field and three-dimensional Minkowski spacetime under intrinsic discrete symmetries}

To study the intrinsic properties of uniqubit-spinor field and scalar field as well as Minkowski spacetime based on the action in Eq.(\ref{actionQ}), we should firstly discuss basic discrete symmetries under the operations: charge-conjugation ($\mathcal{C}$), parity-inversion ($\mathcal{P}$) and time-reversal ($\mathcal{T}$). 

Let us begin with rewritting the action in Eq.(\ref{actionQ}) into the conventional one by ultilizing the so-called $\gamma$-matrices. Both uniqubit-spinor field and self-conjugated uniqubit-spinor field may be referred to as {\it Taijion } or {\it Taijion spinor field} denoted as $\psi_{\QT}(x)$ ($\QT =\mQ^1, \QS^1$), which has the conventional action as follows:
\be \label{actionB}
 \cS_{\QT} & = & \int d^3x \, \{  \frac{1}{2} \bar{\psi}_{\QT}(x) \delta_a^{\;\mu} \gamma^a i\partial_{\mu} \psi_{\QT}(x)   -  \frac{1}{2} \lambda_{1} \phi_{1}(x) \bar{\psi}_{\QT}(x) \tga \psi_{\QT}(x)  \}, 
\ee
where we have used the following definitions for $\bar{\psi}_{\QT}(x)$ and $\gamma$-matrices,
\be \label{GM1}
&& \bar{\psi}_{\mQ^1}(x)  = \psi^{\dagger}_{\mQ^1} \gamma^0,  \quad \gamma^0 = \sigma_2 , \quad \gamma^1  = i \sigma_1 , \quad   \gamma^2 = - i \sigma_3 , \quad \tga = \sigma_0, \nn \\
& &  \bar{\psi}_{\QS^1}(x)  = \psi^{\dagger}_{\QS^1} \gamma^0,  \quad \gamma^0 =  \sigma_3 , \quad \gamma^1 = i \sigma_2 , \quad   \gamma^2 = i \sigma_1 , \quad \tga = \sigma_0, 
\ee
with $ \gamma^a= \tUps \Ups^a$ ($a=0,1,2$) for $\psi_{\mQ^1}(x)$, and $ \gamma^0= \tUps \Ups^{2}, \gamma^1= \tUps \Ups^{0}, \gamma^2= \tUps \Ups^{1}$ for $\psi_{\QS^1}(x)$. It is noticed that in the real spinor presentation $\psi_{\mQ^1}(x)=\psi_{\mQ^1}^{\ast}(x)$, all $\gamma$-matrices are imaginary ones. The above defined $\gamma$-matrices satisfy the following anti-commuting relations: 
\be
& & \{ \gamma^a, \gamma^b\} = 2 \eta^{ab} , \quad \eta^{ab} = \diag(1, -1, -1).
\ee

Two actions for the uniqubit-spinor field and self-conjugated uniqubit-spinor field can be related via the following unitary transformation between two qubit-spinor fields: 
\be
\psi_{\QS^1}(x) = U \psi_{\mQ^1}(x), \quad U^{\dagger} U = 1, \quad U =  \frac{1}{\sqrt{2}} \begin{pmatrix}
1 & i \\
1 & -i
 \end{pmatrix} .  \nn
\ee
It is easy to verify that $\psi_{\mQ^1}(x)$  and $\psi_{\QS^1}(x)$ satisfy the following self-conjugate conditions: 
\be
& & \psi_{\mQ^1}^{c}(x) = C_2\bar{\psi}_{\mQ^1}^T (x) = C_2 \gamma^0 \psi_{\mQ^1}^{\ast}(x) =\psi_{\mQ^1}(x), \nn \\
& &  C_2 \equiv C_{\mQ^1}= \gamma^0 = \sigma_2 , \nn \\
& & \psi_{\QS^1}^{c}(x) = C_{2}\bar{\psi}_{\QS^1}^T (x) = C_{2}\gamma^0 \psi_{\QS^1}^{\ast}(x) =\psi_{\QS^1}(x), \nn \\
& &  C_{2} \equiv C_{\QS^1} = i\gamma^0\gamma^2 = -i \sigma_2 , \nn
\ee
where $\psi^{c}_{\mQ^1}(x)$  and $\psi^{c}_{\QS^1}(x)$ define the charge-conjugated ones of $\psi_{\mQ^1}(x)$  and $\psi_{\QS^1}(x)$. 

Let us now demonstrate how the action in Eq.(\ref{actionB}) preserves the invariance under discrete symmetries: charge-conjugation ($\mathcal{C}$), parity-inversion ($\mathcal{P}$), time-reversal ($\mathcal{T}$) and $W$-parity operation ($\mathcal{W}$). Explicitly,  the Taijion spinor field $\psi_{\QT}(x)$ and scalar field $\phi_1(x)$ have the following transformation properties:
\be
& & \mathcal{C} \psi_{\QT}(x) \mathcal{C}^{-1} \equiv \psi_{\QT}^{c}(x) = C_2 \bar{\psi}_{\QT}^{T} ( x ) ,\nonumber \\
& & C_2^{-1} \gamma^a C_2 = - \gamma^{a\, T}, \quad  C_2^{T} = -C_2 , \quad C_2^{\dagger} = C_2^{-1} , \nn \\
& & C_2 = \begin{cases} 
 \gamma^0= \sigma_2 , & \psi_{\mQ^1} , \\
  i\gamma^0\gamma^2 = -i\sigma_2 , &  \psi_{\QS^1} ,
 \end{cases} \nn \\ 
 & & \mathcal{C} \phi_{1}(x) \mathcal{C}^{-1}  =   \phi_{1}(x), \nn
\ee 
for the charge conjugation, and 
\be
& & \mathcal{P} \psi_{\QT}(x) \mathcal{P}^{-1} \equiv \psi_{\QT}^{p}(x) = P_2 \psi_{\QT}( x^0, -x^k ) , \nonumber \\
& &  P_2^{-1} \gamma^a P_2 = \gamma^{a\, \dagger} ,  \quad a = 0, 1, 2,   \nn \\
& & P_2 = \begin{cases} 
 -i\gamma^0= -i\sigma_2 , & \psi_{\mQ^1} , \\
 \gamma^0 = \sigma_3 , &  \psi_{\QS^1} ,
 \end{cases}  \nn \\
& & \mathcal{P} \phi_{1}(x) \mathcal{P}^{-1}  =   \phi_{1}(x^0, - x^k), \quad k =1, 2, \nn
\ee
for the parity-inversion, and 
\be
& & \mathcal{T} \psi_{\QT}(x) \mathcal{T}^{-1} \equiv \psi_{\QT}^{t}(x) = T_2 \psi_{\QT}( -x^0, x^1, -x^2 ) , \nonumber \\
& &  T_2^{-1} \gamma^a T_2 = \gamma^{a\, T}, \; (a=0,1), \quad  T_2^{-1} \gamma^2 T_2 = - \gamma^{2\, T},   \nn \\
& &  T_2 = \begin{cases} 
 \gamma^0\gamma^2 = \sigma_1 , & \psi_{\mQ^1} , \\
 i\gamma^1\gamma^2 = -\sigma_3 , &  \psi_{\QS^1} ,
 \end{cases} \nn \\
 & & \mathcal{T} \phi_{1}(x) \mathcal{T}^{-1}  =   \phi_{1}(- x^0, x^1, - x^2), \nn
\ee
for the time-reversal with the normal definition. 

It is noticed that the spatial coordinate $x^2$ gets an unusual transformation property under the time-reversal operation as it undergoes a flip in sign so as to ensure the invariance of the action. That means there is an unusual $\mathcal{CPT}$ operation for the Taijion spinor field in three-dimensional Minkowski spacetime. To obtain a usual transformation property similar to the ordinary $\mathcal{CPT}$ operator acting on a massive Dirac fermion in four-dimensional Minkowski spacetime, let us introduce the so-called W-parity operation\cite{GGFT6D} as follows: 
\be
& & \mathcal{W} \psi_{\QT}(x) \mathcal{W}^{-1} \equiv \psi_{\QT}^{w}(x) = W_2 \psi_{\QT}( x^{\mu} ,-x^2 ) , \quad \mu=0,1, \nn \\
& &  W_2^{-1} \gamma^a W_2 = -\gamma^a, \; (a=0, 1), \quad W_2^{-1} \gamma^{2} W_2 = \gamma^{2} ,  \nn \\
& & W_2 = \begin{cases} 
 i\gamma^2= \sigma_3 , & \psi_{\mQ^1} , \\
 i\gamma^2 = -\sigma_1 , &  \psi_{\QS^1} ,
 \end{cases} \nn \\
& & \mathcal{W} \phi_{1}(x) \mathcal{W}^{-1}  =  - \phi_{1}(x^{\mu}, - x^2), \quad \mu=0,1 ,  \nn
\ee
which is viewed as a kind of chirality reflection.

Such a W-parity operation motivates us to define the combined operator $\tilde{\mathcal{T}} =  \mathcal{T}\mathcal{W} $ as a generalized time-reversal operator, it can bring on the following proper transformation:
\be
& & \tilde{\mathcal{T}} \psi_{\QT}(x) \tilde{\mathcal{T}}^{-1} \equiv \psi_{\QT}^{\tilde{t}}(x) = \tilde{T}_2 \psi_{\QT}( -x^0, x^k ) , \nn \\
& &  \tilde{T}_2^{-1} \gamma^a \hat{T}_2 = - \gamma^{a\, T}, \; \; a=0,1, 2,   \nn \\
& &  \tilde{T}_2 = \begin{cases} 
 i\gamma^0 = i\sigma_2 , & \psi_{\mQ^1} , \\
 -\gamma^1  = -i\sigma_2 , &  \psi_{\QS^1} ,
 \end{cases} \nn \\
 & & \tilde{\mathcal{T}} \phi_{1}(x) \tilde{\mathcal{T}}^{-1}  =  - \phi_{1}(-x^0,  x^k), \quad k=1,2, , \nn
\ee
which recovers the usual time-reversal operation.

It is clear that we should adopt the combined joint operator $\varTheta \equiv \mathcal{CPTW} = \mathcal{ C P\tilde{T}}$ for the Taijion spinor field $\psi_{\QT}(x)$ in three-dimensional Minkowski spacetime, so that we are able to arrive at the following conventional transformation property:
\be
& & \varTheta \psi_{\QT}(x) \varTheta^{-1} = \Theta \bar{\psi}_{\QT}^{T} ( - x) , \quad  \Theta^{-1} \gamma^{a} \Theta =  \gamma^{a \, \dagger}, \nn \\  
& &  \Theta = WTPC=  \begin{cases} 
\gamma^0 = \sigma_2 , & \psi_{\mQ^1} , \\
\gamma^0 = \sigma_3 , &  \psi_{\QS^1} ,
 \end{cases} \nn \\
& & \varTheta \phi_{1}(x) \varTheta^{-1} =  -\phi_{1}(- x) , \nn
\ee
which indicates that the combined joint operator $\varTheta $, in comparison to the ordinary $\mathcal{CPT}$ operator for a massive Dirac fermion in four-dimensional Minkowski spacetime, becomes an essential one for the Taijion spinor field in three-dimensional Minkowski spacetime.


\subsection{ Presence of spatial dimensions in correspondence to rotational $\cM_c$-spin group symmetry}

The maximally correlated motion of uniqubit-spinor field as local coherent state of qubit determines the dimension of spacetime. Let us explicitly illustrate how the rotational $\cM_c$-spin group symmetry of uniqubit-spinor field correlates to spatial dimensions of Minkowski spacetime.  

In general, we are not able to distinguish two local distribution amplitudes $\psi_0(x)$ and $\psi_1(x)$ of uniqubit-spinor field $\psi_{\mQ^1}(x)$, the action should be invariant under a rotational transformation between two amplitudes $\psi_{0} $ and $\psi_{1}$. Explicitly, such a rotational transformation can be expressed as follows: 
\be
& & \binom{\psi_{0} }{\psi_{1} } \to \binom{\psi'_{0} }{\psi'_{1}} \equiv e^{ i\varpi \Sigma } \binom{\psi_{0} }{\psi_{1} }= \begin{pmatrix}
\cos \frac{\varpi}{2} & \sin \frac{\varpi}{2} \\
-\sin \frac{\varpi}{2} & \cos \frac{\varpi}{2}
 \end{pmatrix}  \binom{\psi_{0} }{\psi_{1} }    , \nn
\ee
with $\varpi$ a constant parameter. $\Sigma$ is the group generator of $\cM_c$-spin symmetry group SP(2), which is given by the commutator of $\gamma$-matrices,
\be
& & \Sigma = \frac{1}{2} \sigma_2 =  \frac{i}{4} [\gamma^1, \gamma^2] ,  \quad e^{ i\varpi \Sigma }  \in SP(2). \nn
\ee
Under such a SP(2) group transformation, the $\gamma$-matrices $\gamma^a$ behave as a vector in the spinor representation and have the following transformation property:
\be
& & e^{ -i\varpi \Sigma } \gamma^a e^{ i\varpi \Sigma } =  L^{a}_{\; b}  \gamma^{b}, \; \; a,b =1,2, \nn \\
& & L^{a}_{\; b} = 
\begin{pmatrix}
\cos \varpi & \sin \varpi \\
-\sin \varpi & \cos \varpi
 \end{pmatrix} \in SO(2),  \nn
\ee
where the matrix $L^{a}_{\; b}$ belongs to the element of orthogonal rotation group SO(2). 

To ensure the invariance of the action under transformation of rotational $\cM_c$-spin symmetry group SP(2), it must bring the spatial Cartesian coordinates $x^{\mu}$ ($\mu=1,2$) to have a coincidental transformation under orthogonal rotation group SO(2),
\be
x'^{\mu} = L^{\mu}_{\; \nu} \, x^{\nu}, \quad L^{\mu}_{\; \nu}  = \begin{pmatrix}
\cos \varpi & \sin \varpi \\
-\sin \varpi & \cos \varpi
 \end{pmatrix} \in SO(2) .  \nn
\ee
So that the action becomes invariant only under simultaneous SP(2) and SO(2) transformations for uniqubit-spinor field $\psi_{\mQ^1}(x)$ and spatial coordinates $x^{\mu}$, i.e.:
\be 
& & \psi_{\mQ^1}(x) \to \psi_{\mQ^1}'(x') = e^{ i\varpi \Sigma} \psi_{\mQ^1}(x), \quad x'^{\mu} = L^{\mu}_{\; \nu} x^{\nu},  \nn \\
& & e^{ i\varpi \Sigma }  \in SP(2), \quad  L^{\mu}_{\; \nu}  = L^{a}_{\; b} \in SO(2), \quad (a, b, \mu, \nu = 1, 2).
\ee
In terms of the self-conjugated uniqubit-spinor field $ \psi_{\QS^1}(x)$, such a rotational $\cM_c$-spin transformation is equivalent to an U(1) spin charge transformation:
\be
& &  \psi_{\QS^1}(x) \to \psi'_{\QS^1}(x) = e^{ i\varpi \Sigma} \psi_{\QS^1}(x) = \binom{ e^{ - \frac{1}{2} i\varpi} \psi(x)}{e^{ \frac{1}{2} i\varpi} \psi^{\ast}(x)} , \nn \\
& & \Sigma = -\frac{1}{2} \sigma_3 =  \frac{i}{4} [\gamma^1, \gamma^2] , \quad  e^{ i\varpi \Sigma}  \in U(1).
\ee

Therefore, the presence of spatial dimensions is in correspondence to the rotational $\cM_c$-spin symmetry. This is because the invariance of the action under the transformation of rotational $\cM_c$-spin symmetry group SP(2) (or U(1) ) for freely moving uniqubit-spinor field naturally brings on the transformation of orthogonal rotation group SO(2) for Cartesian coordinates $x^{\mu}$ ($\mu =1,2$) so as to match coincidentally to the transformation of rotational $\cM_c$-spin symmetry group. Mathematically, it is indicated from the isomorphic property of symmetry groups, 
\be
SP(2)\cong U(1) \cong SO(2) ,
\ee
where the rotational $\cM_c$-spin group SP(2) in two-dimensional Hilbert space doubly covers the orthogonal rotation group SO(2) in Euclidean space of coordinates. This is because each rotation can be obtained in two inequivalent ways as the endpoint of a path.

\subsection{ Presence of temporal dimension in correspondence to non-homogeneous scaling symmetry}

The action in Eq.(\ref{actionB}) is naturally invariant under homogeneous scaling transformation as shown in Eq.(\ref{GSS}). It is interesting to consider a non-homogeneous scaling transformation between two local distribution amplitudes of uniqubit-spinor field $\psi_{\mQ^1}$,
\be
& & \binom{\psi_{0} }{\psi_{1} } \to \binom{\psi'_{0} }{\psi'_{1}} \equiv e^{ i\varpi \Sigma } \binom{\psi_{0} }{\psi_{1} } 
=  \binom{ e^{- \frac{1}{2}\varpi} \psi_{0}}{ e^{ \frac{1}{2}\varpi}\psi_{1}} , \nn
\ee
with $\varpi$ a constant parameter. $\Sigma$ is viewed as the group generator of $U^{\ast}(1)$ for non-homogeneous scaling transformation, which can be given by the commutator of $\gamma$-matrices,
\be
& & \Sigma  = \frac{i}{4} [\gamma^0, \gamma^1] = \frac{1}{2} i \sigma_3 , \qquad e^{ i\varpi \Sigma} \in U^{\ast}(1) . \nn
\ee
Such a non-homogeneous scaling transformation is regarded as a boost $\cM_c$-spin transformation. The $\gamma$-matrices $\gamma^a$ ($a=0,1$) transform as a vector in Hilbert space with the following transformation property:
\be
& & e^{ - i\varpi \Sigma } \gamma^a e^{ i\varpi \Sigma } =  L^{a}_{\; b}  \gamma^{b}, \; \; a, b = 0,1, \nn \\
& & L^{a}_{\; b}  = \begin{pmatrix}
\cosh \varpi & -\sinh \varpi \\
-\sinh \varpi & \cosh \varpi
 \end{pmatrix}  \in SO(1,1),  \nn
\ee
where the matrix $L^{a}_{\; b}$ belongs to the element of Lorentz-type group SO(1,1).

To obtain the invariant action under non-homogeneous scaling transformation, it requires that both temporal and spatial coordinates $x^{\mu}$ ($\mu=0,1$) must simultaneously transform as follows:
\be
& & x'^{\mu} = L^{\mu}_{\; \nu} x^{\nu}, \quad \mu, \nu = 0,1, \nn \\
& & L^{\mu}_{\; \nu}  =  L^{a}_{\; b}  = \begin{pmatrix}
\cosh \varpi & -\sinh \varpi \\
-\sinh \varpi & \cosh \varpi
 \end{pmatrix} \in SO(1,1) .  \nn
\ee
Namely, the action presented in Eq.(\ref{actionB}) gets invariance only when the uniqubit-spinor field $\psi_{\mQ^1}(x)$ together with the temporal and spatial coordinates $x^{\mu}$ ($\mu = 0,1$) transform explicitly as follows:
\be
& & \psi_{\mQ^1}(x) \to \psi_{\mQ^1}'(x') = e^{- i\varpi \Sigma} \psi_{\mQ^1}(x) = \binom{ e^{ -\frac{1}{2}\varpi} \psi_{0}(x)}{ e^{ \frac{1}{2}\varpi}\psi_{1}(x)} , \nn \\
& &  x^{\mu} \to x'^{\mu} = L^{\mu}_{\; \nu}\, x^{\nu} \equiv \delta^{\mu}_{\; a} \delta_{\nu}^{\; b} L^{a}_{\; b} \, x^{\nu},  \quad (\mu, \nu = 0,1). 
\ee

There is a similar feature for the self-conjugated uniqubit-spinor field $ \psi_{\QS^1}(x)$, the action in Eq.(\ref{actionB}) becomes invariant only when $ \psi_{\QS^1}(x)$ and $x^{\mu}$ ($\mu = 0,1$) transform simultaneously as follows:  
\be
& &  \psi_{\QS^1}(x) \to \psi'_{\QS^1}(x') = e^{ i\varpi \Sigma} \psi_{\QS^1}(x) = \begin{pmatrix}
\cosh \frac{\varpi}{2} & -\sinh \frac{\varpi}{2} \\
-\sinh \frac{\varpi}{2} & \cosh \frac{\varpi}{2}
 \end{pmatrix}  \binom{\psi(x) }{\psi^{\ast}(x) }  , \nn \\
 & & x^{\mu} \to x'^{\mu} = L^{\mu}_{\; \nu} x^{\nu} \equiv \delta^{\mu}_{\; a} \delta_{\nu}^{\; b} L^{a}_{\; b} x^{\nu},  \quad L^{\mu}_{\; \nu} \in SO(1,1), \quad  (\mu, \nu = 0,1), \nn \\
& & \Sigma = \frac{i}{4} [\gamma^0, \gamma^1] = i\frac{1}{2} \sigma_1, \quad  e^{ i\varpi \Sigma}  \in SP(1,1), 
\ee
where $\Sigma$ is the group generator of boost $\cM_c$-spin symmetry group SP(1,1).

The invariance of the action under the transformation of non-homogeneous scaling group $U^{\ast}(1)$ (or boost $\cM_c$-spin group SP(1,1) ) for uniqubit-spinor field demands simultaneously the coincidental transformation of Lorentz-type group SO(1,1) for temporal and spatial coordinates, which can be characterized from the isomorphic property of symmetry groups, 
\be
U^{\ast}(1)\cong SP(1,1) \cong SO(1,1).  
\ee

The variables $x^{\mu}$ ($\mu =0,1$) and $\gamma$-matrices $\gamma^a$ ($a=0,1$) transform as two vectors of symmetry group SO(1,1)$\cong$ SP(1,1) with respective to both coordinate and non-coordinate two-dimensional spacetimes. In general, we can define SO(1,1) invariant scalar product of coordinates $x^{\mu}$ and $\gamma$-matrices $\gamma^a$ as follows:
\be
& & x^{\mu}x_{\mu} = \eta_{\mu\nu} x^{\mu}x^{\nu} = (x^0)^2 - (x^1)^2 , \nn \\
& & \gamma^a \gamma_a = \eta_{ab} \gamma^a \gamma^b = (\gamma^0)^2 - (\gamma^1)^2 ,  \nn
\ee
where two components $x^0$ ($\gamma^0$) and $x^1$ ($\gamma^0$) exhibit a different feature as indicated with a flip in sign. When the coordinate $x^1$ is regarded as a Cartesian coordinate of space, the coordinate $x^0$ is designated as a temporal coordinate. 

Therefore, the presence of the temporal dimension is in correspondence to the non-homogeneous scaling symmetry, which is attributed to the requirement on the invariance of the action under non-homogeneous scaling transformation (or boost $\cM_c$-spin transformation) for uniqubit-spinor field. In another word, the existence of one temporal dimension implies that the local distribution amplitudes of uniqubit-spinor field are varying as the evolution of temporal coordinate.


\subsection{ Maximal coherent symmetry of uniqubit-spinor field and the equation of motion with dynamic $\cQ_c$-spin scalar field in three-dimensional Minkowski spacetime}

The existence of basic scalar field $\phi_{1}(x)$ and its coupling to uniqubit-spinor field bring on a global scaling symmetry for the action presented in Eq.(\ref{actionB}),
\be \label{CSS}
& & \psi_{\QT}(x)\to \psi_{\QT}'(x') = e^{\varpi} \psi_{\QT}(x), \quad \phi_{1}(x) \to \phi_{1}'(x') = e^{\varpi} \phi_{1}(x), \nn \\
& & x^{\mu} \to x^{'\mu} = e^{-\varpi} x^{\mu}, \quad e^{-\varpi} \in SC(1), \;\; e^{\varpi} \in SG(1), 
\ee
with $\varpi$ a constant parameter.

Let us consider the scalar field $\phi_{1}(x)$ to be a dynamical field. A global scaling invariant action in Eq.(\ref{actionB}) for uniqubit-spinor field with dynamic scalar field is generalized to be as follows:
\be \label{Baction}
 \cS_{\QT} & = & \int d^3x \, \{ \frac{1}{2} \bar{\psi}_{\QT}(x) \delta_a^{\;\mu} \gamma^a i\partial_{\mu} \psi_{\QT}(x)  -  \frac{1}{2} \lambda_{1} \tilde{\phi}^2(x) \bar{\psi}_{\QT}(x) \psi_{\QT}(x) \nn \\
 & + & \frac{1}{2}\eta^{\mu\nu}\p_{\mu}\tilde{\phi}(x)\p_{\nu}\tilde{\phi}(x) - \frac{1}{6} \lambda_Q \tilde{\phi}^6(x) \} ,
\ee
with $\QT=(\mQ^1, \QS^1)$ and $\lambda_Q$ a real coupling constant for scalar interaction. Where we have made a redefinition,
\be
\phi_{1}(x) \equiv \tilde{\phi}^2(x) , \nn
\ee
which is referred to as {\it $\cQ_c$-spin scalar field} for convenience of mention.

When making simultaneously the rotational $\cM_c$-spin and non-homogeneous scaling (or boost $\cM_c$-spin) as well as global scaling transformations together with translational and $\cQ_c$-spin transformations, we arrive at an associated symmetry for the action presented in Eq.(\ref{actionB}) in three-dimensional spacetime. Explicitly, the action becomes invariant under the following transformations for uniqubit-spinor field and $\cQ_c$-spin scalar field together with coordinates:
\be
& & \psi_{\QT}(x) \to \psi_{\QT}'(x') = S(L) \psi_{\QT}(x); \quad \tilde{\phi}(x) \to \tilde{\phi}'(x') = \tilde{\phi}(x), \nn \\
& & x^{\mu} \to x'^{\mu} = L^{\mu}_{\; \nu} x^{\nu}, \quad S^{-1}(L) \gamma^a S(L) = L^{a}_{\; b} \gamma^b,  \nn \\
& & S(L) = e^{i\Sigma^{ab}\varpi_{ab} }   \in SP(1,2),  \quad L^{\mu}_{\; \nu}  = L^{a}_{\; b} \in SO(1,2),  \nn \\
& & \psi_{\QT}(x) \to \psi_{\QT}'(x) = S_{\pm}\psi_{\QT}(x) ;\quad \tilde{\phi}(x) \to \tilde{\phi}'(x) = S_{\pm} \tilde{\phi}(x), \nn \\
& &  S_{\pm} \in SP(1) \cong O(1),
\ee
with $a, b, \mu, \nu = 0,1, 2$. Where $S(L)$ and $L^{a}_{\; b}$ are the group elements given explicitly by the following matrices:
\be
& & S(L) = e^{i\Sigma^{ab}\varpi_{ab} } = \cos \frac{\varpi}{2} + i \gamma^a n_{a} \sin \frac{\varpi}{2}, \nn \\
& & L^{a}_{\; b} =  (\eta^{a}_{\; b} - n^a n_b) \cos \varpi +  n^a n_b +   \epsilon^a_{\;\; bc} n^c \sin \varpi , 
\ee
with the definitions,
\be
& & \varpi = \frac{1}{2\sqrt{2}} \sqrt{  \varpi_{ab} \varpi^{ab} } = \frac{1}{2} \sqrt{\varpi_{12}^2 - \varpi_{01}^2 - \varpi_{02}^2}, \nn \\
& &  n^{a} = \frac{1}{4\varpi} \epsilon^{abc} \varpi_{bc} , \quad \epsilon^{012}=1, \quad n_a n^a = 1 . \nn
\ee

Therefore, the action in Eq.(\ref{Baction}) possesses $\cM_c$-spin symmetry SP(1,2) and global scaling symmetry SG(1) together with $\cQ_c$-spin symmetry SP(1) in Hilbert space, which is simultaneously associated with Poincar\'e-type (inhomogeneous Lorentz-type) group symmetry PO(1,2) and conformal scaling symmetry SC(1) in Minkowski spacetime. Such a general symmetry can be expressed as follows:
\be
G_S & = & SC(1)\ltimes P^{1,2}\ltimes SO(1,2) \adjoin SP(1,2)\rtimes SG(1) \times SP(1) \nn \\
 & = & SC(1)\ltimes PO(1,2) \adjoin SP(1,2)  \rtimes SG(1) \times SP(1) ,
\ee 
where Poincar\'e-type group symmetry is a semidirect product group symmetry of Lorentz-type group symmetry SO(1,2) and translational group symmetry P$^{1,2}$, 
\be
PO(1,2) = P^{1,2}\ltimes SO(1,2) . 
\ee
The $\cM_c$-spin symmetry group SP(1,2) in Hilbert space of uniqubit-spinor field is isomorphic to the orthogonal Lorentz-type symmetry group SO(1,2) in Minkowski spacetime of coordinates, and $\cQ_c$-spin symmetry SP(1) for uniqubit-spinor field is isomorphic to the group O(1) for $\cQ_c$-spin scalar field. Note that we have adopted the symbol ``$\adjoin$" to indicate an {\it associated symmetry} in which the transformations of $\cM_c$-spin symmetry SP(1,2) and global scaling symmetry SG(1) in Hilbert space must be coincidental to those of Lorentz-type group symmetry SO(1,2) and scaling symmetry SC(1) in globally flat Minkowski spacetime.

As all $\gamma$-matrices are imaginary in the real representation of uniqubit-spinor field $\psi_{\mQ^1}(x)$, the $\cM_c$-spin symmetry group SP(1,2) is isomorphic to special linear symmetry group SL(2,R). For the self-conjugated uniqubit-spinor field $\psi_{\QS^1}(x)$, its $\cM_c$-spin symmetry group is isomorphic to special symmetry group $SU(1,1)$. In general, we can present the symmetry property of the action in Eq.(\ref{Baction}) as follows:
\be
& & G_S= \begin{cases}  SP(1,2) \cong SL(2,R) \cong SO(1,2)  , & \psi_{\mQ^1}(x) \\
 SP(1,2) \cong SU(1,1) \cong SO(1,2) , &  \psi_{\QS^1}(x) .
 \end{cases}
\ee

From the principle of least action, we obtain the following equations of motion for the uniqubit-spinor field and $\cQ_c$-spin scalar field:
 \be
& & \delta_a^{\;\mu} \gamma^a i\partial_{\mu} \psi_{\QT}(x) - \lambda_{1} \tilde{\phi}^2(x) \psi_{\QT}(x) = 0 ,  \nn \\
& &  \eta^{\mu\nu} \p_{\mu}\p_{\nu} \tilde{\phi}(x) +  \lambda_Q \tilde{\phi}^5(x) +  \lambda_{1}\tilde{\phi}(x) \bar{\psi}_{\QT}(x) \psi_{\QT}(x) = 0 ,
\ee
which are globally scaling invariant.


\section{ Local coherent-qubits motion postulate and the nature of biqubit-spinor field and triqubit-spinor field as Majorana and Dirac fermions with appearance of 4D and 6D Minkowski spacetimes }

Following along the maximum coherence motion principle as guiding principle, we have demonstrated that the local coherent state of qubit with real distribution amplitudes of two orthonormal qubit-basis states presents a uniqubit-spinor field with $\cM_c$-spin charge $\cC_{\cM_c}$=3 and $\cQ_c$-spin charge $\CQc=Q_N$=1. Such a Taijion spinor field spans two-dimensional Hilbert space with maximally correlated motion in three-dimensional Minkowski spacetime. For a generalized consideration, we are led to make a {\it local coherent-qubits motion postulate} upon which the qubit-spinor field is supposed to be the basic constituent of matter. As an explicit illustration, we will demonstrate in this section that neutrinos and charged leptons/quarks as basic constituents of matter are viewed as local coherent states of 2-qubit ($Q_N$=2) and 3-qubit ($Q_N=$3) with respective to $\cQ_c$-spin charges $\CQc=q_c=Q_N=2$ and $\CQc=q_c=Q_N=3$. It will be shown that neutrinos as Majorana fermions and charged leptons/quarks as Dirac fermions can be characterized correspondingly by {\it biqubit-spinor field} and {\it triqubit-spinor field} or equivalently by {\it complex uniqubit-spinor field} and {\it complex biqubit-spinor field}. They span four-dimensional and eight-dimensional real Hilbert spaces or two-dimensional and four-dimensional complex Hilbert spaces, respectively. We will prove that the maximally correlated motions of biqubit-spinor field and triqubit-spinor field bring on four-dimensional and six-dimensional Minkowski spacetimes, respectively.


\subsection{ Local coherent state of qubits as basic spinor field and local coherent-qubits motion postulate }

In general, the qubit basis states can be combined to form high product basis states. In terms of the {\it real basic qubit-basis} $\{\varsigma_s\}$ in Eq.(\ref{QB}) or {\it complex-conjugated basic qubit-basis} $\{\zeta_s\}$ in Eq.(\ref{CQB}),  we are able to define high product basis states as follows:
\be
& & \varsigma_{s_1\cdots s_{n}} \equiv  \varsigma_{s_1}\otimes \cdots \otimes \varsigma_{s_{n}}, \quad  \varsigma_{+} =\binom{1}{0}, \quad  \varsigma_{-} =\binom{0}{1} ; 
\ee
and 
\be
& & \zeta_{s_1\cdots s_{n}} \equiv  \zeta_{s_1}\otimes \cdots \otimes \zeta_{s_{n}}  , \quad
\zeta_+ = \frac{1}{\sqrt{2}}\binom{1}{1}, \quad  \zeta_- = \frac{1}{\sqrt{2}}\binom{i}{-i} ,
\ee
with $s_1, \cdots, s_{n} = \pm$ denoting the spin ``up $\&$ down" states and ``n" labeling the qubit number $Q_N=n$. Where $\{\varsigma_{s_i}\}$  and $\{\zeta_{s_i}\}$ are the real and complex-conjugated basic qubit-bases, respectively. The new bases $\{ \varsigma_{s_1\cdots s_{n}} \}$ and $\{ \zeta_{s_1\cdots s_{n}} \} $ are referred to as real and complex-conjugated {\it $n$-product qubit-basis}, respectively. In general, there exist $2^{n}$ {\it product qubit-basis states}, which form $2^{n}$-dimensional linear vector space as Hilbert space spanned by $n$-product qubit-basis $\{ \varsigma_{s_1\cdots s_{n}} \}$ or $\{ \zeta_{s_1\cdots s_{n}} \} $. 

Let us apply both the real $n$-product qubit-basis $\{ \varsigma_{s_1\cdots s_{n}} \}$ and complex-conjugated $n$-product qubit-basis $\{ \zeta_{s_1\cdots s_{n}} \} $ to define {\it real local coherent state of qubits} $\Psi_{\mQ^n}(x)$ and {\it self-conjugated local coherent state of qubits} $\Psi_{\QS^n}(x)$, respectively, as follows:
\be \label{QSF}
& & \Psi_{\mQ^n}(x) =  \sum_{s_1, \cdots, s_{n}}  \psi_{s_1\cdots s_{n}}(x)  \varsigma_{s_1\cdots s_{n}} \equiv \sum_{s_1, \cdots, s_{n}}  \psi_{s_1\cdots s_{n}}(x)\,  \varsigma_{s_1}\otimes \cdots \otimes \varsigma_{s_{n}} ,  \nn \\
& & \Psi_{\QS^n}(x) =  \sum_{s_1, \cdots, s_{n}}  \psi_{s_1\cdots s_{n}}(x)  \zeta_{s_1\cdots s_{n}} \equiv \sum_{s_1, \cdots, s_{n}}  \psi_{s_1\cdots s_{n}}(x)\,  \zeta_{s_1}\otimes \cdots \otimes \zeta_{s_{n}} ,  
\ee
where the real field operators $\psi_{s_1\cdots s_{n}}(x) $ are regarded as evolving local distribution amplitudes. As the orthonormal product qubit-basis states span $2^{n}$-dimensional real Hilbert space, $\Psi_{\mQ^n}(x)$ and $\Psi_{\QS^n}(x)$ are viewed as vector fields with $2^{n}$ local distribution amplitudes $\psi_{s_1\cdots s_{n}}(x) $ as their components in $2^{n}$-dimensional Hilbert space. 

For the simplest case with single qubit number $Q_N=1$, we have demonstrated following along the maximum coherence motion principle that the real and complex-conjugated basic qubit-bases bring on correspondingly the real uniqubit-spinor field $\psi_{\mQ^1}(x)$ and self-conjugated uniqubit-spinor field $\psi_{\QS^1}(x)$ in three-dimensional Minkowski spacetime, which are shown to be presented as a local coherent state of qubit with two independent degrees of freedom. When extending such a concept to a general case with arbitrary qubit number $Q_N=n$, we are motivated to make the following postulate.

{\it Local coherent-qubits motion postulate}: a local coherent state of qubits as a column vector field of local distribution amplitudes obeying the maximum coherence motion principle brings on a basic spinor field which is postulated to be as the basic constituent of matter and referred to as {\it qubit-spinor field}.  

Based on such a local coherent-qubits motion postulate, we may refer to local coherent states of qubits $\Psi_{\mQ^n}(x)$ and $\Psi_{\QS^n}(x)$ as {\it real qubit-spinor field} and {\it self-conjugated qubit-spinor field}, respectively. For a convention, {\it qubit-spinor field} will mean {\it real qubit-spinor field} unless otherwise specified. 

We would like to address that local distribution amplitudes of qubit-spinor field are dynamical field operators and their kinematical and dynamical evolutions lead to the laws of nature, which are proposed to be governed by the maximum coherence motion principle and symmetry principle along with the least action principle. It is noted that the local coherent state of qubits as qubit-spinor field is unlike the usual coherent state of qubits in quantum mechanics, where global amplitudes are explained as probability amplitudes of basis states when measuring the qubits according to the Born's rule.


\subsection{ Local coherent states of 2-qubit and 3-qubit in 4D and 8D Hilbert spaces and the appearance of 4D and 6D Minkowski spacetimes for biqubit-spinor and triqubit-spinor fields}

It is well known that quarks as basic constituents of matter are regarded as Dirac fermions in the strong interaction of SM. In the electromagnetic interaction, both charged leptons and quarks as basic constituents of matter are viewed as Dirac fermions with eight independent degrees of freedom. Neutrinos with four independent degrees of freedom may appear as Majorana fermions in their free motion mass basis. Nevertheless, in the weak interaction, all leptons and quarks behave as Weyl fermions with four independent degrees of freedom. Following along the local coherent-qubits motion postulate, we are going to demonstrate how the leptons and quarks as basic constituents of matter can be characterized by qubit-spinor fields. 

Let us begin with examining the simple cases for the product qubit-basis states formed from two qubits $Q_N$=2 and three qubits $Q_N$=3, which are represented in 4-dimensional and 8-dimensional Hilbert spaces spanned by {\it 2-product qubit-basis states} and {\it 3-product qubit-basis states}, respectively.  

In terms of real basic qubit-basis $\{\varsigma_s\}$ given in Eq.(\ref{QB}), the real product qubit-basis states $\varsigma_{s_1s_2}$ and $\varsigma_{s_1s_2s_3}$ are constructed from the direct tensor product as follows:
\be \label{RPQB}
& & \varsigma_{s_1s_2} \equiv  \varsigma_{s_1}\otimes \varsigma_{s_2},  \quad  \varsigma_{s_1s_2s_3} \equiv  \varsigma_{s_1s_2}\otimes \varsigma_{s_3} \equiv \varsigma_{s_1}\otimes \varsigma_{s_2s_3} =  \varsigma_{s_1}\otimes \varsigma_{s_2} \otimes \varsigma_{s_3}.
\ee
To be manifest, we write down 2-product qubit-basis states with the following explicit forms:
\be
& & \varsigma_{++} = \begin{pmatrix} 
1\\
0\\
0 \\
0
 \end{pmatrix} , \quad 
 \varsigma_{+-} = \begin{pmatrix} 
0\\
1\\
0 \\
0
 \end{pmatrix}, \quad 
 \varsigma_{-+} = \begin{pmatrix} 
0\\
0\\
1 \\
0
 \end{pmatrix} , \quad 
 \varsigma_{--} = \begin{pmatrix} 
0\\
0\\
0 \\
1
 \end{pmatrix} ;  \nn 
 \ee
Similarly, one can write down the explicit forms for the {\it 3-product qubit-basis states}.

From the general definition of qubit-spinor field shown in Eq.(\ref{QSF}), the local coherent states of two qubits and three qubits with real distribution amplitudes can directly be constructed in terms of 2-product qubit-basis states and 3-product qubit-basis states, respectively, as follows:
\be
& & \psi_{\mQ^2}(x) = \sum_{s_1,s_2 } \psi_{s_1s_2} (x) \, \varsigma_{s_1s_2}  \equiv  \binom{  \psi_{Q_+^1}(x) }{ \psi_{Q_-^1}(x) } \equiv   \begin{pmatrix} 
\psi_{++}(x) \\
\psi_{+-}(x) \\
\psi_{-+}(x) \\
\psi_{--}(x)
 \end{pmatrix} , \nn \\
 & & \psi_{\mQ^3}(x) = \sum_{s_1,s_2 s_3} \psi_{s_1s_2s_3} (x) \, \varsigma_{s_1s_2s_3} \equiv  \binom{  \psi_{Q_+^2}(x) }{ \psi_{Q_-^2}(x) }  \equiv \begin{pmatrix} 
 \psi_{Q_{++}^1}(x)  \\
 \psi_{Q_{+-}^1}(x)  \\
 \psi_{Q_{-+}^1}(x)   \\
 \psi_{Q_{--}^1}(x) 
 \end{pmatrix} \equiv  \begin{pmatrix} 
\psi_{+++}(x) \\
\psi_{++-}(x) \\
\psi_{+-+}(x) \\
\psi_{+--}(x) \\
\psi_{-++}(x) \\
\psi_{-+-}(x) \\
\psi_{--+}(x) \\
\psi_{---}(x)
 \end{pmatrix} ,
\ee
where $\psi_{\mQ^2}(x)$ and $\psi_{\mQ^3}(x)$ turn out to be qubit-spinor fields with respective to the vectors in {\it four-dimensional and eight-dimensional Hilbert spaces} spanned correspondingly by 2-product qubit-basis $\{\varsigma_{s_1s_2} \}$ and 3-product qubit-basis $\{\varsigma_{s_1s_2 s_3} \}$. Where $\psi_{s_1s_2}(x)$ and $\psi_{s_1s_2s_3}(x)$ ($s_1, s_2, s_3 = \pm$) correspond to four and eight real local distribution amplitudes, $\psi_{s_1s_2}(x) = \psi_{s_1s_2}^{\ast}(x)$ and $\psi_{s_1s_2s_3}(x) = \psi_{s_1s_2s_3}^{\ast}(x)$. 

For convenience, the local coherent states of two qubits and three qubits $\psi_{\mQ^2}(x)$ and $\psi_{\mQ^3}(x)$ are referred to as {\it biqubit-spinor field} and {\it triqubit-spinor field}, respectively, they all have real local distribution amplitudes.

Based on the local coherent-qubits motion postulate with following along the maximum coherence motion principle as guiding principle, we are able to construct the hermitian action of biqubit-spinor field $\psi_{\mQ^2}(x)$ as follows:
\be \label{actionQ2}
 \cS_{\mQ^2}  & = &  \int d^4x \,   \{  \frac{1}{2} \psi_{\mQ^2}^{\dagger}(x)  \delta_a^{\;\mu}  \Ups^a i\p_{\mu} \psi_{\mQ^2}(x) 
 -  \, \frac{1}{2} \lambda_{2} \phi_p(x) \psi_{\mQ^2}^{\dagger}(x)  \tUps^p \psi_{\mQ^2}(x) \},
  \ee
with $a, \mu = 0, 1, 2, 3$ and $p=1,2$. Where the real symmetric normalized $\cM_c$-matrices read,
\be \label{UpsQ2}
& & \Ups^0 = \sigma_0 \otimes \sigma_0, \quad \Ups^1 = \sigma_0\otimes \sigma_3, \nn \\
& & \Ups^2 = \sigma_0\otimes \sigma_1, \quad \Ups^3 = \sigma_2\otimes \sigma_2,  
\ee
which satisfy Clifford algebra relations shown in Eq.(\ref{CA}) and characterize the maximally correlated motion of biqubit-spinor field $\psi_{\mQ^2}(x)$ in four-dimensional Minkowski spacetime. 

The pure imaginary antisymmetric normalized $\cQ_c$-matrices are found to be,
\be \label{tUpsQ2}
& &  \tUps^1 = \sigma_3\otimes \sigma_2, \quad  \tUps^2 = \sigma_1\otimes \sigma_2 ,
\ee
which also satisfy Clifford algebra relations. The number of $\cQ_c$-matrices determines the $\cQ_c$-spin charge of biqubit-spinor field $\psi_{\mQ^2}(x)$, 
\be
\CQc=q_c=Q_N=2,  
\ee
which is the same as qubit number $Q_N$.

By applying the maximum coherence motion principle to triqubit-spinor field $\psi_{\mQ^3}(x)$, we obtain the following hermitian action:
 \be \label{actionQ3}
 \cS_{\mQ^3}^{\lambda} & = &  \int d^6x \,  \{ \frac{1}{2}  \psi_{\mQ^3}^{\dagger}(x)  \delta_a^{\;\mu}  \Ups^a i\p_{\mu} \psi_{\mQ^3}(x)  -  \, \frac{1}{2} \lambda_{3} \phi_p(x) \psi_{\mQ^3}^{\dagger}(x)  \tUps^p \psi_{\mQ^3}(x) \}, 
\ee
with $a, \mu = 0, 1, \cdots, 5$ and $p=1, 2, 3$. There are six real symmetric normalized $\cM_c$-matrices, 
\be \label{CAQ3}
& & \Ups^0 = \sigma_0 \otimes \sigma_0 \otimes \sigma_0, \quad \Ups^1 = \sigma_3\otimes  \sigma_0\otimes \sigma_3, \nn \\
& &  \Ups^2 = \sigma_3\otimes  \sigma_0\otimes \sigma_1, \quad \Ups^3 = \sigma_3\otimes  \sigma_2\otimes \sigma_2, \nn \\
& &  \Ups^4 = \sigma_1\otimes  \sigma_0\otimes \sigma_0, \quad \Ups^5 = \sigma_2\otimes  \sigma_2\otimes \sigma_0 ,
\ee
which satisfy the Clifford algebra relations shown in Eq.(\ref{CA}) and characterize the maximally correlated motion of triqubit-spinor field $\psi_{\mQ^3}(x)$ in six-dimensional Minkowski spacetime. 

There exist in general three pure imaginary antisymmetric normalized $\cQ_c$-matrices,
\be \label{ACAQ3}
\tUps^1 = \sigma_3\otimes \sigma_1\otimes \sigma_2, \quad  \tUps^2 = \sigma_3\otimes \sigma_3\otimes \sigma_2 , \quad  \tUps^3 = \sigma_2\otimes \sigma_0\otimes \sigma_0 ,
\ee
which satisfy Clifford algebra. Nevertheless, it is easy to check that they only anticommute with four of the six real symmetric normalized $\cM_c$-matrices as they all commute with the $\Ups$-matrix $\Ups^5$.

It can be verified that the actions presented in Eqs.(\ref{actionQ2}) and (\ref{actionQ3}) for biqubit-spinor field $\psi_{\mQ^2}(x)$ and triqubit-spinor field $\psi_{\mQ^3}(x)$ with zero coupling constant $\lambda_{3} =0$ possess the following maximal $\cM_c$-spin and $\cQ_c$-spin symmetries:
\be
G_S= \begin{cases}
SP(1,3)\times SP(2) \cong SO(1,3)\times SO(2), & \;  $for$ \quad \psi_{\mQ^2}(x), \\
SP(1,5)\times SP(3)\cong SO(1,5)\times SO(3), & \;  $for$ \quad   \psi_{\mQ^3}(x) . 
\end{cases}
\ee
The $\cQ_c$-spin charge of triqubit-spinor field $\psi_{\mQ^3}(x)$ is the same as the qubit number $Q_N$, 
\be
\CQc=q_c=Q_N=3. 
\ee

It is noted that for the case with non-zero coupling constant $\lambda_{3} \neq 0$, the action given in Eq. (\ref{actionQ3}) for triqubit-spinor field $\psi_{\mQ^3}(x)$ has only a reduced symmetry, 
\be
G_S= SP(1,4)\times SP(3)\cong SO(1,4)\times SO(3) , \; \; \text{for} \;\;  \lambda_3\ne 0 . \nn
\ee
This is because the $\cM_c$-matrix $\Ups^5$ presented in Eq.(\ref{CAQ3}) commutes with all three antisymmetric normalized $\cQ_c$-matrices given in Eq.(\ref{ACAQ3}), which do not meet to Clifford algebra.

To obey the maximum coherence motion principle and preserve the maximal $\cM_c$-spin symmetry SP(1,5), the action of triqubit-spinor field $\psi_{\mQ^3}(x)$ is simplified to be as follows:
 \be \label{actionQ30}
 \cS_{\mQ^3}  & = &  \int d^6x \,  \frac{1}{2} \psi_{\mQ^3}^{\dagger}(x)  \delta_a^{\;\mu}  \Ups^a i\p_{\mu} \psi_{\mQ^3}(x),
 \ee
with $a, \mu = 0, 1, \cdots, 5$ and $\lambda_{3}=0$. 

Therefore, the local coherent states of 2-qubit and 3-qubit as biqubit-spinor field $\psi_{\mQ^2}(x)$ and triqubit-spinor field $\psi_{\mQ^3}(x)$ can reach maximally correlated motions in four-dimensional and six-dimensional Minkowski spacetimes, respectively. It is shown explicitly how the genesis of spacetime dimension is correlated to the local coherent state of qubits by following along the maximum coherence motion principle as guiding principle.


\subsection{  Dynamics of $\cQ_c$-spin bi-scalar field and intrinsic discrete symmetries of biqubit-spinor and triqubit-spinor fields }

To show symmetry properties of the action presented in Eq.(\ref{actionQ2}), it is useful to rewrite it into the conventional form in terms of $\gamma$-matrices and consider scalar fields as dynamical fields. In general, we arrive at the following action for the biqubit-spinor field with scalar fields as dynamical fields in four-dimensional spacetime: 
\be \label{actionQ2B}
 \cS_{\mQ^2}  & = &  \int d^4x \,  \{  \frac{1}{2}  \bar{\psi}_{\mQ^2}(x)  \delta_a^{\;\mu}  \gamma^a i\p_{\mu} \psi_{\mQ^2}(x)
 -  \, \frac{1}{2}  \lambda_{2}  \phi_p(x) \bar{\psi}_{\mQ^2}(x)  \tga^p \psi_{\mQ^2}(x)    \nn \\
 & + & \frac{1}{2}\eta^{\mu\nu} \p_{\mu}\phi_p(x)\p_{\nu}\phi^p(x) - \frac{1}{4} \lambda_Q ( \phi_p(x)\phi^p(x))^2 \} ,
  \ee
with $a, \mu = 0,1, 2, 3$ and $p=1,2$. Where we have adopted the conventional definition $\bar{\psi}_{\mQ^2}(x) =\psi_{\mQ^2}^{\dagger}(x)\gamma^0$ and introduced $\gamma$-matrices with the following explicit structures:  
\be \label{GQ2}
& &  \gamma^0= \sigma_1 \otimes \sigma_2,  \quad  \gamma^1 = i \sigma_1\otimes \sigma_1, \nn \\
& & \gamma^2 = -i\sigma_1\otimes \sigma_3, \quad \gamma^3 = i\sigma_3\otimes \sigma_0,  \nn \\
& &  \tga^1 =  -i\sigma_2\otimes \sigma_0 ,\quad  \tga^2 = \sigma_0\otimes \sigma_0, 
\ee
where the $\gamma$-matrices $\gamma^a \equiv\tUps^2\Ups^a$ ($a=0,1, 2, 3$) are all imaginary ones and the $\tga$-matrices $\tga^p\equiv \tUps^2\tUps^p$ ($p=1, 2$) are all real ones. It is noted that $\tga^2$ commutes with all other $\gamma$-matrices.  

The presence of two scalar fields $\phi_{p}(x)$ (p=1,2) is associated with two $\cQ_c$-matrices characterized by $\cQ_c$-spin charge $\cC_{Q_c}=2$. For convenience, such two scalar fields $\phi_{p}(x)$ (p=1,2) are referred to as {\it $\cQ_c$-spin bi-scalar field}.

In analogous, the action in Eq.(\ref{actionQ30}) can be written into the following conventional form:
\be \label{actionQ3B}
 \cS_{\mQ^3} & = &  \int d^6x \, \frac{1}{2} \bar{\psi}_{\mQ^3}(x)  \delta_{\ha}^{\;\hm}  \Gamma^{\ha} i\p_{\hm} \psi_{\mQ^3}(x)   
 \ee
with $\ha= (a, 5, 6)$, $\hm = (\mu, 5, 6)$  ($a, \mu = 0, 1, 2, 3$) and $\bar{\psi}_{\mQ^3}(x) =\psi_{\mQ^3}^{\dagger}(x)\Gamma^0$. The explicit forms of $\Gamma$-matrices are found to be,
\be \label{GQ3}
& & \Gamma^0 = \sigma_2 \otimes \sigma_0 \otimes \sigma_0,  \quad  \Gamma^1 =  -i\sigma_1\otimes  \sigma_0\otimes \sigma_3, \nn \\
& & \Gamma^2 = -i\sigma_1\otimes  \sigma_0\otimes \sigma_1, \quad  \Gamma^3 = -i\sigma_1\otimes  \sigma_2\otimes \sigma_2, \nn \\
& & \Gamma^5 = i\sigma_3\otimes  \sigma_0\otimes \sigma_0, \quad \Gamma^6 = \sigma_0\otimes  \sigma_2\otimes \sigma_0, 
\ee
with $ \Gamma^a = \tUps^3 \Ups^a$ ($a=0, 1, 2, 3 $ ) and $ \Gamma^a = \tUps^3 \Ups^{(a-1)}$ ($a= 5, 6$). 
Where all $\gamma$-matrices $\Gamma^{\ha}$ are imaginary ones. Unlike the usual case, the matrix $\Gamma^6$ becomes commuting with all other $\Gamma$-matrices.  

Let us now discuss intrinsic discrete symmetries of the actions under the operations: charge-conjugation ($\mathcal{C}$), parity-inversion ($\mathcal{P}$), time-reversal ($\mathcal{T}$). To preserve the invariance of the action in Eq.(\ref{actionQ2B}) under $\mathcal{C}$, $\mathcal{P}$ and $\mathcal{T}$ operations, the biqubit-spinor field $\psi_{\mQ^2}(x)$ and $\cQ_c$-spin bi-scalar field $\phi_p(x)$ ($p=1,2$) should transform as follows:  
\be
& & \mathcal{C} \psi_{\mQ^2}(x) \mathcal{C}^{-1} \equiv \psi_{\mQ^2}^{c}(x) = C_4 \bar{\psi}_{\mQ^2}^{T} ( x ) ,  \nonumber \\
& & C_4^{-1} \gamma^a C_4 = - \gamma^{a\, T}, \quad  C_4 = \gamma^0 = \sigma_1 \otimes \sigma_2,  \nn \\
& & C_4^{-1} \tga^p C_4 =  \tga^{p\, T}, \quad \mathcal{C} \phi_p(x) \mathcal{C}^{-1} =  \phi_p(x), \nn
\ee 
for the charge conjugation, and 
\be
& & \mathcal{P} \psi_{\mQ^2}(x) \mathcal{P}^{-1} \equiv \psi_{\mQ^2}^{p}(x) = P_4 \psi_{\mQ^2}( x^0, -x^k) , \nonumber \\
& &  P_4^{-1} \gamma^a P_4 = \gamma^{a\, \dagger}, \quad P_4= -i \gamma^0 = -i \sigma_1 \otimes \sigma_2,  \nn \\
& & P_4^{-1} \tga^p P_4 =  \tga^{p\, \dagger}, \quad \mathcal{P} \phi_p(x) \mathcal{P}^{-1} =  \begin{cases}
-\phi_p(x^0, -x^k), \; p=1,   \\
 + \phi_p(x^0, -x^k), \; p= 2,   \nn
\end{cases}
\ee
for the parity-inversion, and 
\be
& & \mathcal{T} \psi_{ \mQ^2}(x) \mathcal{T}^{-1} \equiv \psi_{\mQ^2}^{t}(x) = T_4 \psi_{\mQ^2}( -x^0, x^k ) , \nonumber \\
& &  T_4^{-1} \gamma^a T_4 = \gamma^{a\, T}, \quad  T_4 = i \gamma^1\gamma^2\gamma^3 = i \sigma_3\otimes \sigma_2, \nn \\
& & T_4^{-1} \tga^p T_4 =  \tga^{p\, T}, \quad  \mathcal{T} \phi_p(x) \mathcal{T}^{-1} =  \begin{cases}
-\phi_p(-x^0, x^k), \; p=1,   \\
 + \phi_p(-x^0, x^k), \; p= 2,   \nn
\end{cases}
\ee
for the time-reversal.  

It is seen that the biqubit-spinor field $\psi_{\mQ^2}(x)$ obeys conventional transformation properties under the well-defined discrete symmetries of charge-conjugation ($\mathcal{C}$), parity-inversion ($\mathcal{P}$) and time-reversal ($\mathcal{T}$). For the $\cQ_c$-spin bi-scalar field $\phi_p(x)$ ($p=1,2$), $\phi_1(x)$ transforms as pseudoscalar field and $\phi_2(x)$ as scalar field. The combined joint operation $\varTheta \equiv \mathcal{CPT} $ on biqubit-spinor field $\psi_{\mQ^2}(x)$ and $\cQ_c$-spin bi-scalar field in 4-dimensional spacetime leads to the following conventional transformations:
\be
& & \varTheta \psi_{\mQ^2}(x) \varTheta^{-1} = \Theta \bar{\psi}_{\mQ^2}^{T} ( - x) , \quad  \Theta^{-1} \gamma^{a} \Theta = - \gamma^{a \, \dagger} , \quad \Theta \gamma^0 = - \gamma^0 \Theta , \nn \\
& &   \Theta^{-1} \tga^{p} \Theta = \tga^{p \, \dagger} , \quad  \varTheta \phi_p(x) \varTheta^{-1} = \phi_p(-x). \nn
\ee

To ensure the invariance of the action in Eq.(\ref{actionQ3B}) under $\mathcal{C}$, $\mathcal{P}$ and $\mathcal{T}$ operations, the triqubit-spinor field $\psi_{\mQ^3}(x)$ is found to transform as follows:
\be
& & \mathcal{C} \psi_{\mQ^3}(x) \mathcal{C}^{-1} \equiv \psi_{\mQ^3}^{c}(x) = C_6 \bar{\psi}_{\mQ^3}^{T} ( x ) ,\nonumber \\
& & C_6^{-1} \Gamma^{\ha} C_6 = - \Gamma^{\ha\, T}, \quad  C_6 = \Gamma^0 = \sigma_2 \otimes \sigma_0 \otimes \sigma_0,  \nn 
\ee 
for the charge conjugation, and 
\be
& & \mathcal{P} \psi_{\mQ^3}(x) \mathcal{P}^{-1} \equiv \psi_{\mQ^3}^{p}(x) = P_6 \psi_{\mQ^3}( x^0, -x^k, x^6) , \nonumber \\
& &  P_6^{-1} \Gamma^{\ha} P_6 = \Gamma^{\ha\, \dagger}, \quad P_6= -i \Gamma^0 = -i \sigma_2 \otimes \sigma_0 \otimes \sigma_0,   \nn 
\ee
for the parity-inversion, and 
\be
& & \mathcal{T} \psi_{ \mQ^3}(x) \mathcal{T}^{-1} \equiv \psi_{\mQ^3}^{t_6}(x) = T_6 \psi_{\mQ^3}( -x^0, x^k, -x^6 ) , \nonumber \\
& &  T_6^{-1} \Gamma^{\ha} T_6 = \Gamma^{\ha\, T}, \quad  T_6 = i \sigma_3 \otimes \sigma_1\otimes \sigma_2,  \nn 
\ee
for the time-reversal.  

It is noticed that the spatial coordinate $x^6$ has a distinguished transformation feature under both parity-inversion and time-reversal operations as it has to undergo a flip in sign in order to keep the invariance of the action. For that, it is useful to consider the combined operator $\vartheta \equiv \mathcal{PT} $, which leads to a proper transformation as the usual one,
\be
& & \vartheta \psi_{\mQ^3}(x) \vartheta^{-1} = \theta \psi_{\mQ^3}( - x) , \quad  \theta^{-1} \Gamma^{\ha} \theta =  \Gamma^{\ha \, \ast} . \nn 
\ee
For the combined joint operation $\varTheta \equiv \mathcal{CPT} $ on the triqubit-spinor field $\psi_{\mQ^3}(x)$ in 6-dimensional spacetime, we arrive at the conventional transformation property as follows:
\be
& & \varTheta \psi_{\mQ^3}(x) \varTheta^{-1} = \Theta \bar{\psi}_{\mQ^3}^{T} ( - x) , \quad  \Theta^{-1} \Gamma^{\ha} \Theta =  -\Gamma^{\ha \, \dagger}, \quad \Theta \gamma^0 = - \gamma^0 \Theta , \nn 
\ee
which indicates that both the combined operator $\vartheta $ and combined joint operator $\varTheta $ become essential ones for the triqubit-spinor field $\psi_{\mQ^3}(x)$ in 6-dimensional Minkowski spacetime. 

The real property of {\it qubit-spinor fields} $\psi_{\mQ^2}(x)$ and $\psi_{\mQ^3}(x)$ can be characterized by their self-conjugate conditions, 
\be
& & \psi_{\mQ^2}^{c}(x) = C_4\gamma^0 \psi_{\mQ^2}^{\ast} (x) = \psi_{\mQ^2}(x), \quad C_4 =  \gamma^0 = \sigma_1 \otimes \sigma_2, \nn \\
& & \psi_{\mQ^3}^{c}(x) = C_6\Gamma^0 \psi_{\mQ^3}^{\ast} (x) = \psi_{\mQ^3}(x), \quad C_6 = \Gamma^0 = \sigma_2 \otimes \sigma_0 \otimes \sigma_0 ,
\ee
with $\psi_{\mQ^2}^{c}(x)$ and $\psi_{\mQ^3}^{c}(x)$ defined as charge conjugations of biqubit-spinor field $\psi_{\mQ^2}(x)$ and triqubit-spinor field $\psi_{\mQ^3}(x)$ in 4-dimensional and 6-dimensional spacetimes, respectively.


\subsection{Maximal coherent symmetries of biqubit-spinor and triqubit-spinor fields}

Let us further analyze invariant properties of the actions presented in Eqs.(\ref{actionQ2B}) and (\ref{actionQ3B}) under transformations of maximal coherent symmetries.

It can be verified that the action in Eq.(\ref{actionQ2B}) keeps invariant under the following transformations for biqubit-spinor field $\psi_{\mQ^2}(x)$ and $\cQ_c$-spin bi-scalar field $\phi_p(x)$ as well as coordinates of Minkowski spacetime:
\be
& & \psi_{\mQ^2}(x) \to \psi_{\mQ^2}'(x') = S(L) \psi_{\mQ^2}(x), \quad \phi_p(x) \to \phi'_p(x') = \phi_p(x),  \nn \\
& & S^{-1}(L) \gamma^a S(L) = L^{a}_{\; b} \gamma^b,  \quad  x^{\mu} \to x'^{\mu} = L^{\mu}_{\; \nu} x^{\nu},  \quad (a, b, \mu, \nu = 0,1, 2, 3), \nn \\
& & S(L) = e^{i\Sigma^{ab}\varpi_{ab} }   \in SP(1,3),  \quad L^{\mu}_{\; \nu}  = L^{a}_{\; b} \in SO(1,3), \nn 
\ee
for $\cM_c$-spin symmetry, and 
\be
& & \psi_{\mQ^2}(x) \to \psi_{\mQ^2}'(x) = e^{i\tSi^{pq} \, \varpi_{pq}} \psi_{\mQ^2}(x), \quad  \phi_p(x)  \to \phi'_p(x) = L_p^{\; \, q}\phi_q(x), \nn \\
& & e^{i\tSi^{pq} \, \varpi_{pq}} \in SP(2),   \quad  L_p^{\; \, q} \in SO(2) , \quad (p,q = 1,2),  \nn 
\ee
for $\cQ_c$-spin symmetry,  and
\be
& & \psi_{\mQ^2}(x) \to \psi_{\mQ^2}'(x') = e^{\frac{3}{2} \varpi} \psi_{\mQ^2}(x), \quad \phi_p(x) \to \phi'_p(x') = e^{\varpi} \phi_p(x),  \nn \\
& & x^{\mu} \to x'^{\mu} = e^{-\varpi} x^{\mu} , \quad e^{\frac{3}{2}\varpi}, e^{\pm\varpi} \in SC(1)\cong SG(1), \nn
\ee
for global scaling symmetry. Where $\Sigma^{ab}$ are the group generators of SP(1,3) and $\tSi^{pq}$ the group generator of SP(2), which can be defined from the $\gamma$-matrices given in Eq.(\ref{GQ2}) and $\tUps$-matrices defined in Eq.(\ref{tUpsQ2}), 
\be
\Sigma^{ab} = \frac{i}{4} [ \gamma^a, \gamma^b],  \quad \tSi^{pq} = \frac{i}{4} [ \tUps^p, \tUps^q] = \frac{i}{2} \tUps^1\tUps^2 = -\frac{1}{2} \sigma_2 \otimes \sigma_0 . \nn
\ee 
The $\cQ_c$-spin bi-scalar field $\phi_p(x)$ ($p=1,2$) can be regarded as a vector field under SO(2) transformation. 

After considering translation group symmetry $P^{1,3}$ in 4-dimensional Minkowski spacetime, we arrive at a maximal associated symmetry of the action given in Eq.(\ref{actionQ2B}), 
\be
G_S & = &  SC(1)\ltimes P^{1,3} \ltimes SO(1,3) \adjoin SP(1,3) \times SG(1)\times SP(2)  \nn \\
& \equiv & SC(1)\ltimes PO(1,3) \adjoin SP(1,3) \times SG(1) \times SP(2) ,
\ee
with PO(1,3)$\equiv$P$^{1,3}\ltimes$SO(1,3) denoting Poincar\'e group symmetry in 4-dimensional Minkowski spacetime. Note that the symbol ``$\adjoin$" is adopted to indicate the associated symmetry in which the transformations of $\cM_c$-spin symmetry SP(1,3) and global scaling symmetry SG(1) in Hilbert space have to be coincidental to those of Lorentz group symmetry SO(1,3) and conformal scaling symmetry SC(1) in globally flat Minkowski spacetime.

Similarly, it can be checked that the action in Eq.(\ref{actionQ3B}) becomes invariant under the following transformations for triqubit-spinor field $\psi_{\mQ^3}(x)$ and coordinates in six-dimensional Minkowski spacetime:
\be
& & \psi_{\mQ^3}(x) \to \psi_{\mQ^3}'(x') = S(L) \psi_{\mQ^3}(x), \quad  x^{\hm} \to x'^{\hm} = L^{\mu}_{\; \hn} x^{\hn},   \nn \\
& & S^{-1}(L) \Gamma^{\ha} S(L) = L^{\ha}_{\; \hb} \Gamma^{\hb},  \quad \ha, \hb, \hm, \hn = 0,1, 2, 3, 5, 6, \nn \\
& &   S(L) = e^{i\Sigma^{\ha\hb}\varpi_{\ha\hb} }   \in SP(1,5), \quad L^{\hm}_{\; \hn}  = L^{\ha}_{\; \hb} \in SO(1,5), \nn 
\ee
for maximal $\cM_c$-spin symmetry, and 
\be
& & \psi_{\mQ^3}(x) \to \psi_{\mQ^3}'(x) = e^{i\tSi^{pq} \, \varpi_{pq}} \psi_{\mQ^2}(x), \nn \\
& & e^{i\tSi^{pq} \, \varpi_{pq}}  \in SP(3)\cong SO(3), \quad p,q =1,2,3, \nn 
\ee
for $\cQ_c$-spin symmetry, and 
\be
& & \psi_{\mQ^3}(x) \to \psi_{\mQ^3}'(x') = e^{\frac{5}{2} \varpi} \psi_{\mQ^3}(x), \nn \\
& &  x^{\mu} \to x'^{\mu} = e^{-\varpi} x^{\mu} , \quad e^{\frac{5}{2} \varpi} , e^{-\varpi} \in SC(1)\cong SG(1), \nn
\ee
for global scaling symmetry. Where $\Sigma^{\ha\hb}$ are the group generators of SP(1,5) and $\tSi^{pq}$ the group generators of SP(3), they are defined correspondingly from $\gamma$-matrices presented in Eq.(\ref{GQ3}) and $\cQ_c$-matrices in Eq.(\ref{ACAQ3}), 
\be
& & \Sigma^{ab} = \frac{i}{4} [ \Gamma^a, \Gamma^b],  \quad  \Sigma^{a6} = -  \Sigma^{6a} =  \frac{i}{2} \Gamma^a \Gamma^6,  \quad (a,b = 0, 1, 2, 3, 5), \nn \\
& & \tSi^{pq} = \frac{i}{4} [ \tUps^p, \tUps^q] , \quad p, q = 1, 2, 3 . \nn
\ee 
It is noted that the $\Gamma$-matrix $\Gamma^6$ is not anti-commuting with all other $\Gamma$-matrices.

By including the translation group symmetry $P^{1,5}$ and global scaling symmetry in 6-dimensional Minkowski spacetime, the action in Eq.(\ref{actionQ3B}) possesses the following maximal associated symmetry: 
\be
G_S & = &  SC(1)\ltimes P^{1,5} \ltimes SO(1,5) \adjoin SP(1,5) \rtimes SG(1)\times SP(3)  \nn \\
& \equiv & SC(1)\ltimes PO(1,5) \adjoin SP(1,5) \rtimes SG(1) \times SP(3)  , 
\ee
where PO(1,5)$\equiv$P$^{1,5}\ltimes$SO(1,5) represents Poincar\'e-type group symmetry in 6-dimensional Minkowski spacetime.

It is noticed that the maximal $\cM_c$-spin symmetry SP(1,5) does not allow to have scalar coupling term between triqubit-spinor field $\psi_{\mQ^3}(x)$ and $\cQ_c$-spin scalar field $\phi_p(x)$ ($p=1,2,3$), i.e., $\lambda_3=0$, as such a scalar coupling term will spoil maximal $\cM_c$-spin symmetry SP(1,5). The reason is that the three antisymmetric $\cQ_c$-matrices given in Eq.(\ref{ACAQ3}) are not anti-commuting with all real symmetric $\cM_c$-matrices $\Ups^a$ ($ a =0,1\cdots, 5$) defined in Eq.(\ref{CAQ3}). Specifically, it is the real symmetric $\cM_c$-matrix $\Ups^5$ that commutes with all three antisymmetric $\cQ_c$-matrices, so that the introduction of scalar coupling term ($\lambda_3\neq 0$) will bring the maximal $\cM_c$-spin symmetry SP(1,5) broken down to the reduced $\cM_c$-spin symmetry SP(1,4).


\subsection{ Local coherent states of 2-qubit and 3-qubit as basic constituents of matter and Majorana and Dirac fermions as biqubit-spinor and triqubit-spinor fields }

The neutrino oscillation indicates that the freely moving neutrinos may appear as Majorana fermions with small masses and large mixings. In the electromagnetic interaction, the charged leptons and quarks behave as Dirac fermions with eight independent degrees of freedom, which are described by the complex four-component entity spinor fields. We will show that Majorana and Dirac fermions as basic constituents of matter can be viewed as local coherent states of qubits in correspondence to biqubit-spinor field $\psi_{\mQ^2}(x)$ and triqubit-spinor field $\psi_{\mQ^3}(x)$. In particular, we will demonstrate explicitly the relationship between the complex and self-conjugated spinor representations of qubit-spinor field.

To express the biqubit-spinor and triqubit-spinor fields $\psi_{\mQ^2}(x)$ and $\psi_{\mQ^3}(x)$ into the usual Majorana and Dirac fermions composed of complex components in the spinor representation, it is useful to apply the complex-conjugated basic qubit-basis $\{\zeta_s\}$ defined in Eq.(\ref{CQB}) to form the complex-conjugated {\it product qubit-basis states} $\zeta_{s_1s_2}$ and $\zeta_{s_1s_2s_3}$ with the following forms:
\be \label{CPQB}
& & \zeta_{s_1s_2} =  \zeta_{s_1}\otimes \zeta_{s_2},  \quad \zeta_{s_1s_2s_3} = \zeta_{s_1s_2} \otimes \zeta_{s_3} =  \zeta_{s_1} \otimes \zeta_{s_2s_3} = \zeta_{s_1}\otimes \zeta_{s_2} \otimes \zeta_{s_3}. 
\ee
For an illustration, let us write down the explicit forms for the complex-conjugated {\it 2-product qubit-basis states} $\zeta_{s_1s_2}$ as follows:
\be
& & \zeta_{++} = \frac{1}{2} \begin{pmatrix} 
1\\
1\\
1 \\
1
 \end{pmatrix} , \quad 
 \zeta_{+-} = \frac{1}{2} \begin{pmatrix} 
i\\
-i\\
i \\
-i
 \end{pmatrix}, \quad 
 \zeta_{-+} = \frac{1}{2} \begin{pmatrix} 
i\\
i\\
-i \\
-i
 \end{pmatrix} , \quad 
 \zeta_{--} = \frac{1}{2} \begin{pmatrix} 
 -1 \\
1 \\
1 \\
-1
 \end{pmatrix} . 
  \nn 
  \ee
In a similar way, one can directly write down the complex-conjugated {\it 3-product qubit-basis states} $\zeta_{s_1s_2s_3}$.
 
In light of the complex-conjugated {\it 2-product qubit-basis } $\{\zeta_{s_1s_2}\}$ and {\it 3-product qubit-basis } $\{\zeta_{s_1s_2s_3}\}$, the local coherent states of qubits for biqubit-spinor and triqubit-spinor fields $\psi_{\mQ^2}(x)$ and $\psi_{\mQ^3}(x)$ can equivalently be represented by the corresponding {\it self-cojugated biqubit-spinor and triqubit-spinor fields} $\psi_{\QS^2}(x)$ and $\psi_{\QS^3}(x)$, respectively. Explicitly, they can be expressed as the following forms:
\be
& & \psi_{\QS^2}(x) = \sum_{s_1,s_2 } \psi_{s_1s_2} (x) \, \zeta_{s_1s_2}   , \nn \\
& & \psi_{\QS^3}(x) = \sum_{s_1,s_2 ,s_3} \psi_{s_1s_2s_3} (x) \, \zeta_{s_1s_2s_3} , 
\ee
where $\psi_{s_1s_2} (x)$ and $\psi_{s_1s_2s_3} (x)$ are real local distribution amplitudes. 

In light of self-conjugated biqubit-spinor field $\psi_{\QS^2}(x)$ and self-conjugated triqubit-spinor field $\psi_{\QS^3}(x)$, we are able to express the actions given in Eqs.(\ref{actionQ2B}) and (\ref{actionQ3B}) into the following equivalent ones:
\be \label{actionM2P}
 \cS_{\QS^2}  & = &  \int d^4x \,   \{   \frac{1}{2} \bar{\psi}_{\QS^2}(x)  \delta_a^{\;\mu}  \gamma^a i\p_{\mu} \psi_{\QS^2}(x) 
 -   \frac{1}{2}\lambda_2  \phi_p(x) \bar{\psi}_{\QS^2}(x) \tga^p \psi_{\QS^2}(x)   \nn \\
 & + & \frac{1}{2}\eta^{\mu\nu} \p_{\mu}\phi_p(x)\p_{\nu}\phi^p(x) -  \frac{1}{4}\lambda_Q (\phi_p(x)\phi^p(x))^2 \} ,
  \ee
for the action of $\psi_{\QS^2}(x)$ and $\phi_p(x)$ with the definition $\bar{\psi}_{\QS^2}(x) =\psi_{\QS^2}^{\dagger}(x)\gamma^0$. Where the $\gamma$-matrices $\gamma^a$ ($a=0,1,2,3$) including $\tga^p$ ($p=1,2$) have the following structures:
\be \label{GM2}
& &  \gamma^0 = \sigma_2 \otimes \sigma_3,   \quad \gamma^1 = -i\sigma_2\otimes \sigma_2, \nn \\
& & \gamma^2 =  i \sigma_2\otimes \sigma_1 ,  \quad \gamma^3 = -i \sigma_1\otimes \sigma_0 , \nn \\
& & \tga^1 =  i \sigma_3 \otimes \sigma_0 = i \gamma_5, \quad \tga^2 =  \sigma_0 \otimes \sigma_0 .
\ee
And similarly we have
\be \label{actionM3P}
\cS_{\QS^3}  & = &  \int d^6x \frac{1}{2} \bar{\psi}_{\QS^3}(x)  \delta_{\ha}^{\;\hm}  \Gamma^{\ha} i\p_{\hm} \psi_{\QS^3}(x), 
\ee
for the action of $\psi_{\QS^3}(x)$ with the definition $\bar{\psi}_{\QS^3}(x) =\psi_{\QS^3}^{\dagger}(x)\Gamma^0$. The $\Gamma$-matrices $\Gamma^{\ha}$ ($\ha=0,1,2,3,5,6$) have the following explicit forms:
\be \label{GM3}
& & \Gamma^0 = \sigma_0 \otimes \sigma_2 \otimes \sigma_3,   \quad \Gamma^1 =  -i \sigma_0 \otimes \sigma_2\otimes \sigma_2, \nn \\
& & \Gamma^2 = i \sigma_0 \otimes \sigma_2\otimes \sigma_1, \quad  \Gamma^3 = -i \sigma_0 \otimes \sigma_1\otimes \sigma_0, \nn \\
& & \Gamma^5 = \sigma_3 \otimes \sigma_0\otimes \sigma_0 , \quad \Gamma^6 = i \sigma_3\otimes  \sigma_3\otimes \sigma_0 ,   \ee
and three $\tGa$-matrices characterizing $\cQ_c$-spin charge are given by,
\be
& &  \tGa^1 = i\sigma_1 \otimes \sigma_3\otimes \sigma_0, \quad \tGa^2 =  i \sigma_2 \otimes \sigma_3\otimes \sigma_0 , \quad \tGa^3 =  \sigma_0 \otimes \sigma_0\otimes \sigma_0 .
\ee

It is noticed that not all $\gamma$-matrices $\gamma^a$ and $\tga$-matrices $\tga^p$ become anti-commuting since $\tga^2$ is actually commuting with all others. Analogously, the $\Gamma$-matrices $\Gamma^{\ha}$ and $\tGa$-matrices $\tGa^p$ are not all anti-commuting as the matrix $\Gamma^5$ commutes with other $\Gamma$-matrices $\Gamma^{\ha}$ and the matrix $\tGa^3$ is commuting with all others. 

Both $\psi_{\QS^2}(x)$ and $\psi_{\QS^3}(x)$ retain the same independent degrees of freedom as biqubit-spinor and triqubit-spinor fields $\psi_{\mQ^2}(x)$ and $\psi_{\mQ^3}(x)$, respectively, they all satisfy the following self-conjugate conditions:
\be
 & & \psi_{\QS^2}^{c}(x) = C_{\QS^2}\gamma^0 \psi_{\QS^2}^{\ast} (x) = \psi_{\QS^2}(x), \nn \\
 & & \psi_{\QS^3}^{c}(x) = C_{\QS^3}\Gamma^0 \psi_{\QS^3}^{\ast} (x) = \psi_{\QS^3}(x), \nn \\
 & & C_{\QS^2} = - \gamma^1\gamma^3= -i\sigma_3\otimes \sigma_2 \equiv \sigma_3\otimes C_{\QS^1} , \nn \\
 & &  C_{\QS^3}  = -i \sigma_1 \otimes \sigma_3\otimes \sigma_2 \equiv \sigma_1 \otimes C_{\QS^2}, 
\ee
where $\psi_{\QS^2}^{c}(x)$ and $\psi_{\QS^3}^{c}(x) $ are defined as the charge-conjugations in 4-dimensional and 6-dimensional Minkowski spacetimes, respectively. 

In fact, the self-conjugated biqubit-spinor field $\psi_{\QS^2}(x)$ defines the well-known Majorana fermion $\psi_M(x)$, i.e.:
\be
\psi_{\QS^2}(x) \equiv \psi_M(x), \nn
\ee
which was introduced initially in 4-dimensional Minkowski spacetime. 

In such self-conjugated qubit-spinor representations, the $\cQ_c$-spin symmetries for local coherent states of qubits as self-conjugated quibt-spinor fields can explicitly be characterized as follows: 
\be
& &  \psi_{\QS^2}(x) \to  \psi'_{\QS^2}(x) = e^{i\varpi_{12} \tilde{\Sigma}^{12}}  \psi_{\QS^2}(x), \quad e^{i\varpi_{12} \tilde{\Sigma}^{12}} \in SP(2) \cong U(1), \nn \\
 & & \psi_{\QS^3}(x) \to  \psi'_{\QS^3}(x) = e^{i \varpi_{ij} \tilde{\varSigma}^{ij} }  \psi_{\QS^3}(x), \quad e^{i \varpi_{ij} \tilde{\varSigma}^{ij} }  \in SP(3) \cong SU(2), 
\ee
with $\tilde{\Sigma}^{12}$ the U(1) group generator and $\tilde{\varSigma}^{ij}$ the SU(2) group generators, 
\be \label{gamma_5_7}
& & \tilde{\Sigma}^{12} = -\frac{1}{2} \sigma_3\otimes \sigma_0 \equiv -\frac{1}{2} \gamma_5, \nn \\
& & \tilde{\varSigma}^{12}  = \frac{1}{2} \sigma_3 \times \sigma_0\times \sigma_0  \equiv \frac{1}{2}\gamma_7, \nn \\
& & \tilde{\varSigma}^{23} = \frac{1}{2} \sigma_1\times \gamma_5, \quad \tilde{\varSigma}^{31} = \frac{1}{2} \sigma_2 \times \gamma_5 ,
\ee
which indicates that their $\cQ_c$-spin charges are the same as qubit numbers, i.e., $\CQc=q_c =Q_N$.

When rewriting self-conjugated biqubit-spinor field $\psi_{\QS^2}(x)$ and triqubit-spinor field $\psi_{\QS^3}(x)$ by utilizing the real {\it 2-product qubit-basis } $\{\varsigma_{s_1s_2}\}$ and {\it 3-product qubit-basis } $\{\varsigma_{s_1s_2s_3}\}$ defined in Eq.(\ref{RPQB}), we obtain the explicit expressions for such two self-conjugated qubit-spinor fields with complex local distribution amplitudes as follows:
\be \label{MD23}
& & \psi_{\QS^2}(x)   \equiv \sum_{s_1,s_2 } \tilde{\psi}_{s_1s_2} (x) \, \varsigma_{s_1s_2}   =  \begin{pmatrix} 
\tpsi_{++}(x) \\
\tpsi_{+-}(x) \\
\tpsi_{-+}(x) \\
\tpsi_{--}(x)
 \end{pmatrix}  \equiv \binom{\psi_{\QC^1}(x)}{\psi_{\hQC^1}(x)} , \nn \\
& & \psi_{\QS^3}(x) \equiv \sum_{s_1,s_2 ,s_3} \tilde{\psi}_{s_1s_2s_3} (x) \, \varsigma_{s_1s_2s_3}  = 
  \begin{pmatrix} 
\tpsi_{+++}(x) \\
\tpsi_{++-}(x) \\
\tpsi_{+-+}(x) \\
\tpsi_{+--}(x) \\
\tpsi_{-++}(x) \\
\tpsi_{-+-}(x) \\
\tpsi_{--+}(x) \\
\tpsi_{---}(x)
 \end{pmatrix}  \equiv \binom{\psi_{\QC^2}(x)}{\psi_{\hQC^2}(x)} ,
\ee
where $\tilde{\psi}_{s_1s_2} (x)$ and $\tilde{\psi}_{s_1s_2s_3} (x)$ are complex local distribution amplitudes. 

In the above expressions, we have introduced both {\it complex uniqubit-spinor field} $\psi_{\QC^1}(x)$ and {\it complex biqubit-spinor field} $\psi_{\QC^2}(x)$ and defined correspondingly their {\it complex charge-conjugated fields} $\psi_{\hQC^1}(x)$ and $\psi_{\hQC^2}(x)$, which can generally be expressed by adopting real product qubit-basis states with complex local distribution amplitudes as follows: 
\be \label{DF1}
& & \psi_{\QC^1}(x) \equiv \sum_{s_1=\pm} \hat{\psi}_{s_1}(x) \varsigma_{s_1} = \binom{\hat{\psi}_{+} (x)}{\hat{\psi}_{-} (x)} \equiv  \binom{\psi_{\QCp^0} (x)}{\psi_{\QCn^0} (x)} , \nn \\
& &  \psi_{\hQC^1}(x) \equiv  \sigma_1 \psi_{\QC^1}^{\ast}(x)  = \binom{\hat{\psi}_{-}^{\ast} (x)}{\hat{\psi}_{+}^{\ast} (x)}
\equiv  \binom{\psi_{\QCn^0}^{\ast} (x)}{\psi_{\QCp^0}^{\ast} (x)} ,
\ee
for the complex uniqubit-spinor field $\psi_{\QC^1}(x) $ and its complex charge-conjugated one $\psi_{\hQC^1}(x)$. There are following relations between complex and real distribution amplitudes:  
\be
\hat{\psi}_{+} (x) & \equiv & \tilde{\psi}_{++} (x) = \frac{1}{2} [ \psi_{++}(x) + i \psi_{+-}(x)+ i \psi_{-+}(x) -\psi_{--}(x) ] , \nn \\
 & \equiv & \frac{1}{\sqrt{2}} [ \psi_{+0} + i \psi_{+1} ]  \equiv \psi_{\QCp^0}(x), \nn \\
 \hat{\psi}_{-} (x) & \equiv &  \tilde{\psi}_{+-} (x) = \frac{1}{2} [ \psi_{++}(x) - i \psi_{+-}(x)+ i \psi_{-+}(x) +\psi_{--}(x) ] , \nn \\
 & \equiv & \frac{1}{\sqrt{2}} [ \psi_{-0} + i \psi_{-1} ]  \equiv  \psi_{\QCn^0}(x) , \nn \\
  \hat{\psi}_{-} ^{\ast} (x) & \equiv &  \tilde{\psi}_{-+} (x) =   \tilde{\psi}_{+-}^{\ast} (x) \equiv \psi_{\QCn^0}^{\ast} (x) , \nn \\
 \hat{\psi}_{+} ^{\ast} (x) & \equiv & \tilde{\psi}_{--} (x)  =   \tilde{\psi}_{++}^{\ast} (x) \equiv \psi_{\QCp^0}^{\ast} (x) . \nn
\ee
Similarly, the complex biqubit-spinor field $\psi_{\QC^2}(x) $ and its complex charge-conjugated one $\psi_{\hQC^2}(x)$ can be expressed as follows:
\be \label{DF2}
& & \psi_{\QC^2}(x) \equiv \sum_{s_1s_2} \hat{\psi}_{s_1s_2}(x) \varsigma_{s_1s_2}    =    \begin{pmatrix} 
\hat{\psi}_{++}(x) \\
\hat{\psi}_{+-}(x) \\
\hat{\psi}_{-+} (x) \\
\hat{\psi}_{--} (x)
 \end{pmatrix} \equiv \begin{pmatrix} 
\psi_{\mQ^0_{\mC++}}(x) \\
\psi_{\mQ^0_{\mC+-}}(x) \\
\psi_{\mQ^0_{\mC-+}} (x) \\
\psi_{\mQ^0_{\mC--}} (x)
 \end{pmatrix}  \equiv \binom{\psi_{\QCp^1} (x)}{\psi_{\QCn^1} (x)}  , \nn \\
 & &  \psi_{\hQC^2}(x) \equiv  \sigma_1\otimes \sigma_1 \psi_{\QC^2}^{\ast}(x) =  \begin{pmatrix} 
\hat{\psi}_{--}^{\ast}(x) \\
\hat{\psi}_{-+}^{\ast}(x) \\
\hat{\psi}_{+-}^{\ast} (x) \\
\hat{\psi}_{++}^{\ast} (x)
 \end{pmatrix}  \equiv  \begin{pmatrix} 
\psi_{\mQ^0_{\mC--}}^{\ast}(x) \\
\psi_{\mQ^0_{\mC-+}}^{\ast}(x) \\
\psi_{\mQ^0_{\mC+-}}^{\ast} (x) \\
\psi_{\mQ^0_{\mC++}}^{\ast} (x)
 \end{pmatrix}   \equiv \binom{\psi_{\hQCn^1} (x)}{\psi_{\hQCp^1} (x)}  ,
\ee
where the relations between complex and real local distribution amplitudes are given by,
\be
\hat{\psi}_{++} (x) & \equiv & \tilde{\psi}_{+++} (x) = \frac{1}{\sqrt{8}} [ \psi_{+++}(x) + i \psi_{++-}(x)+ i \psi_{+-+}(x) -\psi_{+--}(x) 
 \nn \\
 & + & i\psi_{-++}(x) -  \psi_{-+-}(x) - \psi_{--+}(x) - i\psi_{---}(x) ] \nn \\
 & \equiv &  \frac{1}{\sqrt{2}} [ \psi_{++0} + i \psi_{++1} ]  \equiv \psi_{\mQ^0_{\mC++}} (x), \nn \\
 \hat{\psi}_{+-} (x) & \equiv & \tilde{\psi}_{++-} (x) = \frac{1}{\sqrt{8}} [ \psi_{+++}(x) - i \psi_{++-}(x)+ i \psi_{+-+}(x) +\psi_{+--}(x)  \nn \\
 & + & i\psi_{-++}(x) + \psi_{-+-}(x) - \psi_{--+}(x) + i\psi_{---}(x) ]  \nn \\
 & \equiv &  \frac{1}{\sqrt{2}} [ \psi_{+-0} + i \psi_{+-1} ]  \equiv \psi_{\mQ^0_{\mC+-}} (x), \nn \\
 \hat{\psi}_{-+} (x) & \equiv & \tilde{\psi}_{+-+} (x) = \frac{1}{\sqrt{8}} [ \psi_{+++}(x) + i \psi_{++-}(x) - i \psi_{+-+}(x) +\psi_{+--}(x)  \nn \\
 & + & i\psi_{-++}(x) -  \psi_{-+-}(x) + \psi_{--+}(x) + i\psi_{---}(x) ] \nn \\
 & \equiv &  \frac{1}{\sqrt{2}} [ \psi_{-+0} + i \psi_{-+1} ]  \equiv \psi_{\mQ^0_{\mC-+}} (x), \nn \\
 \hat{\psi}_{--} (x) & \equiv & \tilde{\psi}_{+--} (x) = \frac{1}{\sqrt{8}} [ \psi_{+++}(x) - i \psi_{++-}(x)- i \psi_{+-+}(x) -\psi_{+--}(x) \nn \\
 & + & i\psi_{-++}(x) +  \psi_{-+-}(x) + \psi_{--+}(x) - i\psi_{---}(x) ] \nn \\
 & \equiv &  \frac{1}{\sqrt{2}} [ \psi_{--0} + i \psi_{--1} ]  \equiv \psi_{\mQ^0_{\mC--}} (x), \nn 
 \ee
 and
 \be
  \hat{\psi}_{--} ^{\ast} (x) & \equiv &  \tilde{\psi}_{-++} (x) =   \tilde{\psi}_{+--}^{\ast} (x) \equiv \psi_{\mQ^0_{\mC--}}^{\ast} (x), \nn \\
  \hat{\psi}_{-+} ^{\ast} (x) & \equiv & \tilde{\psi}_{-+-} (x)  =   \tilde{\psi}_{+-+}^{\ast} (x) \equiv \psi_{\mQ^0_{\mC-+}}^{\ast} (x), \nn \\
   \hat{\psi}_{+-} ^{\ast} (x) & \equiv &  \tilde{\psi}_{--+} (x) =   \tilde{\psi}_{++-}^{\ast} (x) \equiv \psi_{\mQ^0_{\mC+-}}^{\ast} (x), \nn \\
 \hat{\psi}_{++} ^{\ast} (x) & \equiv & \tilde{\psi}_{---} (x)  =   \tilde{\psi}_{+++}^{\ast} (x) \equiv \psi_{\mQ^0_{\mC++}}^{\ast} (x) . \nn
\ee

In general, there exist relations between the complex qubit-spinor field and its qubit-reduced complex qubit-spinor field. In particular, with the introduction of complex charge-conjugated qubit-spinor fields $\psi_{\hQC^1}(x)$ and $\psi_{\hQC^2}(x)$, the self-conjugated biqubit-spinor and triqubit-spinor fields $\psi_{\QS^2}(x)$ and $\psi_{\QS^3}(x)$ can be characterized in formal by the corresponding qubit-reduced complex qubit-spinor fields $\psi_{\QC^1}(x) $ and $\psi_{\QC^2}(x)$ defined via the {\it reduced product qubit-basis states} with complex local distribution amplitudes. We may refer to complex qubit-spinor fields $\psi_{\QC^1}(x) $ and $\psi_{\QC^2}(x)$ as complex uniqubit-spinor field and complex biqubit-spinor field, respectively. 

In fact, the complex biqubit-spinor field $\psi_{\QC^2}(x)$ with four complex components behaves as a Dirac fermion, 
\be
\psi_{\QC^2}(x) \to \psi_{D}(x) . \nn
\ee
Note that the ordinary Dirac fermion $\psi_{D}(x)$ was introduced initially by Dirac in 4-dimensional Minkowski spacetime as it was assumed to be a massive spinor field. Therefore, it is distinguished to massive Dirac fermion $\psi_{D}(x)$, a complex biqubit-spinor field $\psi_{\QC^2}(x)$ as massless Dirac fermion possesses generally a free motion in 6-dimensional Minkowski spacetime\cite{GGFT6D} based on the maximum coherence motion principle.

In terms of the complex qubit-spinor fields $\psi_{\QC^1}(x) $ and $\psi_{\QC^2}(x)$, the actions given in Eqs.(\ref{actionM2P}) and (\ref{actionM3P}) can be rewritten into equivalent formalisms. For the complex uniqubit-spinor field $\psi_{\QC^1}(x)$, we obtain the following hermitian action:
\be \label{actionD1P}
 \cS_{\QC^1}  \equiv \cS_{\QS^2} \equiv \cS_{M} & = &  \int d^4x \,   \{  [ \frac{1}{2} \bar{\psi}_{\QC^1}(x)  \delta_a^{\;\mu}  \gamma^a i\p_{\mu} \psi_{\QC^1}(x) 
 -  \frac{1}{2} \lambda_{1}  \phi_p(x) \bar{\psi}_{\QC^1}(x) \tga^p \psi_{\hQC^1}(x)  + H.c. ]  \nn \\
 & + & \frac{1}{2}\eta^{\mu\nu} \p_{\mu}\phi_p(x)\p_{\nu}\phi^p(x) -  \frac{1}{4} \lambda_Q (\phi_p(x)\phi^p(x))^2 \} ,
  \ee
with the definitions $\bar{\psi}_{\QC^1}(x) =\psi_{\QC^1}^{\dagger}(x)\gamma^0$ and $\psi_{\hQC^1}(x) = \sigma_1 \psi_{\QC^1}^{\ast}(x)$ as shown in Eq.(\ref{DF1}). The $\gamma$-matrices $\gamma^a$ ($a=0,1,2,3$) and $\tga^p$ ($p=1,2$) are explicitly given by,
\be
& & \gamma^0 = \sigma_3,   \quad \gamma^1 =  i \sigma_2, \quad \gamma^2 =  i\sigma_1 ,  \quad \gamma^3 = - \sigma_0 , \nn \\
& & \tga^1 = \sigma_0, \quad  \tga^2 =  -i\sigma_0 , 
\ee
which are defined in two-dimensional complex Hilbert space. 

For the massless complex biqubit-spinor field $\psi_{\QC^2}(x)$, we arrive at the following hermitian action in six-dimensional Minkowski spacetime:
\be \label{actionD2P}
 \cS_{\QC^2} \equiv \cS_{D} \equiv \cS_{\QS^3} =   \int d^6x \,  \frac{1}{2}[ \bar{\psi}_{\QC^2}(x)  \delta_{\ha}^{\;\hm}  \Gamma^{\ha} i\p_{\hm} \psi_{\QC^2}(x) + H.c. ],
\ee
with the definition $\bar{\psi}_{\QC^2}(x) =\psi_{\QC^2}^{\dagger}(x)\Gamma^0$. The $\Gamma$-matrices $\Gamma^{\ha}$ ($\ha=0,1,2,3,5,6$) are found to have the following explicit forms:
\be \label{D2GM}
& & \Gamma^0 = \sigma_2 \otimes \sigma_3, \quad \Gamma^1 =  -i \sigma_2\otimes \sigma_2 , \quad \Gamma^2 = i \sigma_2\otimes \sigma_1, \nn \\
& & \Gamma^3 = -i \sigma_1\otimes \sigma_0,  \quad \Gamma^5 = \sigma_0\otimes \sigma_0,  \quad \Gamma^6 = i\sigma_3\otimes \sigma_0 ,
\ee
where $\Gamma$-matrices and $\psi_{\QC^2}(x)$ are represented in four-dimensional complex Hilbert space.


\subsection{Intrinsic properties of complex uniqubit-spinor and complex biqubit-spinor fields with SL(2,$\mathbb{C}$) and SL(2,$\mathbb{Q}$) group symmetries in 4D and 6D Minkowski spacetimes }

In light of the complex uniqubit-spinor and biqubit-spinor fields, the $\cM_c$-spin symmetry of the action in Eq.(\ref{actionD1P}) is described by a special linear group of degree 2 over the complex number $\mathbb{C}$, i.e., SL(2,$\mathbb{C}$), which has the following transformation properties:
\be
& & \psi_{\QC^1}(x) \to \psi'_{\QC^1}(x')  = e^{i \hat{\varpi}_k \sigma^k } \psi_{\QC^1}(x), \; (k=1,2,3),  \nn \\
& & x^{\mu} \to x^{'\mu} = L^{\mu}_{\; \nu} x^{\nu} , \quad e^{-i\hat{\varpi}_k^{\ast} \sigma^k } \gamma^a e^{i\hat{\varpi}_k \sigma^k }  = L^{a}_{\; b} \gamma^{b}, \nn \\
& & e^{i\hat{\varpi}_k \sigma^k}\in SL(2, \mathbb{C}) \cong SP(1,3), \nn \\
& & L^{a}_{\; b} = L^{\mu}_{\; \nu} \in SO(1,3) \cong SL(2, \mathbb{C}) ,
\ee
where $\hat{\varpi}_k$ ($k=1,2,3$) are three complex parameters and $\sigma^k$ are Pauli matrices, which indicates the following isomorphism of symmetry groups:
\be
SL(2,\mathbb{C}) \cong SP(1,3)\cong SO(1,3) . 
\ee
The special linear group SL(2,$\mathbb{C}$) reflects the maximal $\cM_c$-spin symmetry of complex uniqubit-spinor field $\psi_{\QC^1}(x)$ with two complex local distribution amplitudes. Such an isomorphism of symmetry groups brings on the equivalence between complex uniqubit-spinor field $\psi_{\QC^1}(x)$ and Majorana fermion $\psi_M(x)$ which is regarded as a self-conjugated biqubit-spinor field with $\cM_c$-spin symmetry SP(1,3).

The $\cM_c$-spin symmetry of the action in Eq.(\ref{actionD2P}) is described by another special linear group of degree 2 over the {\it quaternion} $\mathbb{Q}$, i.e., SL(2,$\mathbb{Q}$), which is isomorphic to special unitary group $SU^{\ast}(4)$ under the following transformations:
\be
& & \psi_{\QC^2}(x) \to \psi'_{\QC^2}(x')  = S(L) \psi_{\QC^2}(x),  \nn \\
& & x^{\hm} \to x^{'\hm} = L^{\hm}_{\; \hn} x^{\hn} , \quad S^{-1}(L) \Gamma^{\ha} S(L)  = L^{\ha}_{\; \hb}\,  \Gamma^{\hb}, \nn \\
& & S(L) = e^{i \varpi_{\ha\hb} \Sigma^{\ha\hb} }\in SL(2, \mathbb{Q})\cong SU^{\ast}(4) \cong SP(1,5), 
\nn \\
& & L^{\ha}_{\; \hb} = L^{\hm}_{\; \hn} \in SO(1,5) \cong SP(1,5) , \nn
\ee
where the fifteen $4\times 4$ matrices $\Sigma^{\ha\hb}$ ($\ha, \hb = 0, \cdots, 3, 5, 6$) defined via the $\Gamma$-matrices in Eq.(\ref{D2GM}) provide the group generators of $SL(2, \mathbb{Q})\cong SU^{\ast}(4)$ symmetry in the spinor representation,
\be
& & \Sigma^{ab} = \frac{i}{4}[\Gamma^a, \Gamma^{b}] = \frac{i}{4}[\Gamma^a, \Gamma^{b}] ,  \nn \\
& & \Sigma^{a 5} = -\Sigma^{5 a} = \frac{i}{2} \Gamma^a , \quad  a, b = 0, 1, 2, 3, 6 . 
\ee  
We have the following isomorphic property of symmetry groups,
\be \label{Quaternion}
SL(2, \mathbb{Q})\cong SU^{\ast}(4) \cong SP(1,5)\cong SO(1,5) ,  
\ee
which indicates that the maximal special unitary symmetry of complex biqubit-spinor field $\psi_{\QC^2}(x)$ with four complex local distribution amplitudes is equivalent to the maximal $\cM_c$-spin symmetry SP(1,5) of self-conjugated triqubit-spinor field $\psi_{\QS^3}(x)$ with four pairs of complex-conjugated local distribution amplitudes.  

Therefore, a massless complex biqubit-spinor field $\psi_{\QC^2}(x)$ possesses the maximal $\cM_c$-spin symmetry $SL(2, \mathbb{Q})\cong SU^{\ast}(4) \cong SP(1,5)\cong SO(1,5)$ in six-dimensional spacetime\cite{GGFT6D} instead of the ordinary symmetry $SP(1,3)\cong SO(1,3)$ for massive Dirac fermion $\psi_D(x)$ in four-dimensional spacetime.  

We now turn to discuss the basic properties of the actions presented in Eqs.(\ref{actionD1P}) and (\ref{actionD2P}) under intrinsic discrete symmetries: parity-inversion ($\mathcal{P}$), time-reversal ($\mathcal{T}$), charge-conjugation ($\mathcal{C}$) and W-parity operation ($\mathcal{W}$). The complex qubit-spinor fields $\psi_{\QC^1}(x)$ and $\psi_{\QC^2}(x)$ in four- and six-dimensional spacetimes transform as follows:
\be
& & \mathcal{P} \psi_{\QC^1}(x) \mathcal{P}^{-1} \equiv \psi_{\QC^1}^{p}(x) = P_{\QC^1} \psi_{\QC^1}( x^0,-x^1, -x^2, x^3 )\, , \nonumber \\
& &  P_{\QC^1}^{-1} \gamma^{a} P_{\QC^1} = \gamma^{a\, \dagger}, \quad P_{\QC^1}=\gamma^0 =\sigma_3,  \quad a = 0, \cdots, 3,  \nn \\
& & \mathcal{P} \psi_{\QC^2}(x) \mathcal{P}^{-1} \equiv \psi_{\QC^2}^{p}(x) = P_{\QC^2} \psi_{\QC^2}( x^0,-x^i, x^5, -x^6 )\, , \quad i=1,2,3, \nonumber \\
& &  P_{\QC^2}^{-1} \Gamma^{\ha} P_{\QC^2} = \Gamma^{\ha\, \dagger}, \quad P_{\QC^2}=\Gamma^0 ,  \quad \ha = 0, \cdots, 3, 5, 6,  \nn
\ee
for the parity-inversion,  and
\be
& & \mathcal{C} \psi_{\QC^1}(x) \mathcal{C}^{-1} \equiv \psi_{\QC^1}^{c}(x) = C_{\QC^1} \bar{\psi}_{\QC^1}^{T} ( x^0, x^1, x^2, -x^3 )\, , \nonumber \\
& & C_{\QC^1}^{-1} \gamma^a C_{\QC^1} = - \gamma^{a\, T}, \;\; a = 0, 1, 2 ,  \quad  C_{\QC^1}^{-1} \gamma^{3 } C_{\QC^1} = \gamma^{2 \, T},   \nn \\
& & C_{\QC^1}= -i\gamma^0\gamma^2 = i\sigma_2,  \nn \\
& & \mathcal{C} \psi_{\QC^2}(x) \mathcal{C}^{-1} \equiv \psi_{\QC^2}^{c}(x) = C_{\QC^2} \bar{\psi}_{\QC^2}^{T} ( x^{\mu},  -x^5, -x^6 ),  \;\; \mu=0,1,2,3, \nn \\
& & C_{\QC^2}^{-1} \Gamma^a C_{\QC^2} = - \Gamma^{a\, T}, \quad  C_{\QC^2}^{-1} \Gamma^{k } C_{\QC^2} = \Gamma^{k \, T},   \;\; a = 0, 1, 2, 3 , \; k = 5,6, \nn \\
& & C_{\QC^2}= C_{\QS^2}= \Gamma^0\Gamma^2\Gamma^6 = - i\sigma_3\times \sigma_2, 
\ee 
for the charge conjugation, and 
\be
& & \mathcal{T} \psi_{\QC^1}(x) \mathcal{T}^{-1} \equiv \psi_{\QC^1}^{t}(x) = T_{\QC^1} \psi_{\QC^1}( -x^0, x^i ), \;\; i=1,2,3, \nn \\
& &  T_{\QC^1}^{-1} \gamma^{a} T_{\QC^1} = -\gamma^{a\, T},  \;\; a = 0, 1, 2 ,  \quad  T_{\QC^1}^{-1} \gamma^{3 } T_{\QC^1} = \gamma^{3 \, T},  \nn \\
& & T_{\QC^1}= \gamma^1\gamma^3 = -i\sigma_2  ,   \nn \\
& & \mathcal{T} \psi_{\QC^2}(x) \mathcal{T}^{-1} \equiv \psi_{\QC^2}^{t}(x) = T_{\QC^2} \psi_{\QC^2}( -x^0, x^i,  -x^5, x^6 ),  \;\;  i=1,2,3, \nonumber \\
& &  T_{\QC^2}^{-1} \Gamma^{\ha} T_{\QC^2} = \Gamma^{\ha\, T}, \quad T_{\QC^2}= \Gamma^1\Gamma^3 = i\sigma_3\times \sigma_2,  \quad \ha = 0, \cdots, 3, 5, 6,  \nn
\ee
for the time-reversal, and 
\be
& & \varTheta \psi_{\QC^1}(x) \varTheta^{-1} = \Theta \bar{\psi}_{\QC^1}^{T} ( - x)  , \nonumber \\
& & \Theta^{-1} \gamma^{a} \Theta =  \gamma^{a\, \dagger},   \quad \Theta = CPT=\gamma^0 , \nn \\
& & \varTheta \psi_{\QC^2}(x) \varTheta^{-1} = \Theta \bar{\psi}_{\QC^2}^{T} ( - x^{\mu}, x^5, x^6) , \mu = 0, \cdots, 3 , \nn \\
& & \Theta^{-1} \Gamma^{a} \Theta =  -\Gamma^{a\, \dagger}, \; \; a=0,\cdots, 3, \nn \\
& &  \Theta^{-1} \Gamma^{k} \Theta =  \Gamma^{k\, \dagger},  \;\; k = 5,6, \quad \Theta = CPT= \Gamma^0 ,
\ee
for the combined operation $\varTheta \equiv \mathcal{CPT} $. 

It is noticed that for the complex uniqubit-spinor field $\psi_{\QC^1}(x)$, there are opposite transformation properties for the third coordinate $x^3$ under parity-inversion $\mathcal{P}$ and charge-conjugation $\mathcal{C}$, but their combined operation $\mathcal{CP}$ becomes the same as ordinary combined one in four-dimensional spacetime. For the complex biqubit-spinor field $\psi_{\QC^2}(x)$, there appear different transformation properties for coordinate $x^5$ under parity-inversion $\mathcal{P}$ and time-reversal $\mathcal{T}$, while their combined operation $\mathcal{PT}$ comes to be the same as usual combined one in six-dimensional spacetime. Nevertheless, the combined operation  $\varTheta \equiv \mathcal{CPT} $ on the complex biqubit-spinor field $\psi_{\QC^2}(x)$ does not bring on a proper transformation for all six coordinates in six-dimensional Minkowski spacetime, which is caused from an unusual transformation feature of charge-conjugation operation under which two extra dimensions corresponding to coordinates $x^5$ and $x^6$ have to flip in sign in order to ensure the invariance of the action.

It is useful to define the so-called W-parity operation discussed in \cite{GGFT6D}, i.e.:
\be
& & \mathcal{W} \psi_{\QC^2}(x) \mathcal{W}^{-1} \equiv \psi_{\QC^2}^{w}(x) = W_{\QC^2} \psi_{\QC^2}( x^{\mu} -x^5, -x^6 ) , \nn \\
& &  W_{\QC^2}^{-1} \Gamma^a W_{\QC^2} = -\Gamma^a,  \;\;  a = 0, 1, 2, 3 , \;\; \mu=0,1,2,3 , \nn \\
& & W_{\QC^2}^{-1} \Gamma^{k } W_{\QC^2} = \Gamma^{k}, \; \; k = 5,6, \quad W_{\QC^2}= i\gamma_5 , 
\ee
which motivates us to introduce the combined operation $\mathcal{\tC} = {\cal W}{\cal C}$,  
\be \label{WCC}
& & \mathcal{\tC} \psi_{\QC^2} (x ) \mathcal{\tC}^{-1}  \equiv \psi_{\QC^2}^{\tc}(x) =  \tC_{\QC^2} \bar{\psi}_{\QC^2}^{T} ( x ) , \nn \\ 
& & \left(\psi_{\QC^2}^{\tc}(x)\right)^{\tc} = - \psi_{\QC^2}(x), \quad  \tC_{\QC^2}  = i\gamma_5  C_{\QC^2}= -\Gamma^0\Gamma^2  = \sigma_0\otimes \sigma_2, 
\ee
so that all six-dimensional coordinates become unchanged. Such a combined operation $\mathcal{\tC} = {\cal W}{\cal C}$ is referred to as {\it W-charge conjugation}. 

When applying the joint operator $\tilde{\varTheta} \equiv \mathcal{\tC PT} = \mathcal{WCPT}$ to massless complex biqubit-spinor field $\psi_{\QC^2}(x)$ in six-dimensional Minkowski spacetime, we obtain the following proper transformation: 
\be
& & \tilde{\varTheta} \psi_{\QC^2}(x) \tilde{\varTheta}^{-1} = \tilde{\Theta} \bar{\psi}_{\QC^2}^{T} ( - x) , \quad \tilde{\Theta}^{-1} \Gamma^{\ha} \tilde{\Theta} =  \Gamma^{\ha \, \dagger}, \nn \\
& &   \tilde{\Theta} = \tC PT = WCPT= i\gamma_5\Gamma^0 = \sigma_1\otimes \sigma_3 ,
\ee
which indicates that the joint operation $\tilde{\varTheta} $ instead of the usual joint operation $\varTheta \equiv \mathcal{CPT} $ becomes an essential one for complex biqubit-spinor field $\psi_{\QC^2}(x)$ as massless Dirac spinor field in six-dimensional Minkowski spacetime.


\subsection{ Massless Dirac fermion with W-charge conjugation in 6D Minkowski spacetime and self-conjugated triqubit-spinor field with intrinsic $\cQ_c$-spin symmetry SU(2) }

The hermitian action in Eq.(\ref{actionD2P}) with the maximal $\cM_c$-spin symmetry $SL(2, \mathbb{Q})\cong SU^{\ast}(4)\cong SP(1,5)$ can be rewritten into the following form:
\be  \label{actionD2PC}
\cS_{\QC^2}   & = & \int d^6x \,  \frac{1}{2}  [ \bar{\psi}_{\QC^2}(x)  \delta_{\ha}^{\;\hm}  \Gamma^{\ha} i\p_{\hm} \psi_{\QC^2}(x) + H.c. ] \nn \\
& \equiv & \int d^6x \,  \frac{1}{2}  [ \bar{\psi}_{\QC^2}(x)  \delta_{\ha}^{\;\hm}  \Gamma^{\ha} i\p_{\hm} \psi_{\QC^2}(x) 
 + \bar{\psi}_{\QC^2}^{\tilde{c}}(x)  \delta_{\ha}^{\;\hm}  \Gamma^{\ha} i\p_{\hm} \psi_{\QC^2}^{\tilde{c}}(x)   ] , \nn
\ee
where $\psi_{\QC^2}^{\tilde{c}}(x)$ is called {\it W-charge conjugated complex biqubit-spinor field} defined in Eq.(\ref{WCC}) with the introduction of W-parity operation.  

As the structure of $\Gamma$-matrices presented in Eq.(\ref{D2GM}) for the action of complex biqubit-spinor $\psi_{\QC^2}(x)$ in Eq.(\ref{actionD2P}) differs from that of widely used $\gamma$-matrices for the ordinary massive Dirac fermion $\psi_{D}(x)$, let us make the following unitary transformation:
\be
\psi_{\QC^2}(x) = U_D \psi_{D}(x), \quad U_D = \begin{pmatrix}
\sigma_0  & 0 \\
0 &  i\sigma_3  
\end{pmatrix} ,
\ee
which brings on the usual spinor representation for Dirac fermion. The hermitian action can be expressed as follows: 
\be  \label{actionDC}
\cS_{\fD}   \equiv \cS_{\QC^2} & = &  \int d^6x \,  \frac{1}{2}  [ \bar{\psi}_{D}(x)  \delta_{\ha}^{\;\hm}  \Gamma^{\ha} i\p_{\hm} \psi_{D}(x)  + H.c. ] \nn \\ 
& \equiv & \int d^6x \,  \frac{1}{2}  [ \bar{\psi}_{D}(x)  \delta_{\ha}^{\;\hm}  \Gamma^{\ha} i\p_{\hm} \psi_{D}(x) 
 + \bar{\psi}_{D}^{\tilde{c}}(x)  \delta_{\ha}^{\;\hm}  \Gamma^{\ha} i\p_{\hm} \psi_{D}^{\tilde{c}}(x)   ] , 
\ee
with the definition $\bar{\psi}_{D}(x) =\psi_{D}^{\dagger}(x)\Gamma^0$. Where $\Gamma$-matrices $ \Gamma^{\ha} \equiv ( \gamma^a, \Gamma^l )$ ($a=0,1,2,3$, $l=5,6$) get the widely used structures in the spinor representation,
\be \label{DGM}
& & \Gamma^0 \equiv \gamma^0 = \sigma_1 \otimes \sigma_0, \quad \Gamma^1  \equiv \gamma^1 = i \sigma_2\otimes \sigma_1 , \nn \\
& & \Gamma^2 \equiv \gamma^2  = i \sigma_2\otimes \sigma_2,  \quad \Gamma^3  \equiv \gamma^3 = i \sigma_2\otimes \sigma_3, \nn \\
& &  \Gamma^5  \equiv i\gamma_5 = i\sigma_3\otimes \sigma_0,  \quad \Gamma^6  \equiv I_4 = \sigma_0\otimes \sigma_0, 
\ee
where $\Gamma^a=\gamma^a$ ($a=0,1,2,3,5$) provide the usual Dirac $\gamma$-matrices in chiral representation. It is manifest that the unit matrix $\Gamma^6$ is not anti-commuting with others. 

Such a unitary transformed complex biqubit-spinor field leads to a spinor structure for an ordinary massless Dirac fermion $\psi_{D}(x)$. $\psi_{D}^{\tc}(x)$ defines a W-charge conjugated Dirac fermion, 
\be
 & & \psi_{D}^{\tc}(x) \equiv \mathcal{\tC} \psi_{D} (x ) \mathcal{\tC}^{-1}   =  \tC_{D} \bar{\psi}_{D}^{T} ( x ) , \nn \\
 & & \tC_{D}  =  \Gamma^2\Gamma^0 \Gamma^6 \gamma_5 = i\gamma_5 C_D = \sigma_0\otimes \sigma_2, \nn \\
 & & C_D \equiv C_{\QC^2} \equiv C_{\QS^2}= i\Gamma^0\Gamma^2 \Gamma^6 = - i\sigma_3\otimes \sigma_2 .
\ee

In general, a massless Dirac fermion possesses the maximal $\cM_c$-spin symmetry $SL(2, \mathbb{Q})$, which indicates that the quaternions should be applicable to describe a massless Dirac fermion in six-dimensional spacetime. 

In terms of the massless Dirac fermion $\psi_{D}(x)$ and its W-charge conjugated one $\psi_{D}^{\tc}(x)$, the action in Eq.(\ref{actionDC}) can be rewritten into the following simple form: 
\be  \label{actionDM}
\cS_{\tfQS^3}   & \equiv & \int d^6x \,  \frac{1}{2} \bar{\psi}_{\tfQS^3}(x)  \delta_{\ha}^{\;\hm}  \Gamma^{\ha} i\p_{\hm} \psi_{\tfQS^3}(x)  , 
\ee
where $\psi_{\tfQS^3}(x)$ is referred to as {\it self-conjugated triqubit-spinor field} defined as follows in six-dimensional spacetime:
\be \label{DQS3}
& & \psi_{\tfQS^3}(x) \equiv \binom{\psi_{D}(x) } { \psi_{D}^{\tilde{c}}(x) },  \nn \\
& & \psi_{\tfQS^3}^{c}(x) = C_{\tfQS^3} \bar{\psi}_{\tfQS^3}^T(x) =  \psi_{\tfQS^3}(x), \nn \\
& & \left(\psi_{D}^{\tc}(x)\right)^{\tc} = - \psi_{D}(x), \quad C_{\tfQS^3} = -i\sigma_2 \otimes  \tC_{D} ,
\ee
which is related to the previous defined self-conjugated triqubit-spinor field $\psi_{\QS^3}(x)$ via the following unitary transformation:
\be
\psi_{\tfQS^3}(x) = U_M \psi_{\QS^3}(x), \quad U_M = \begin{pmatrix}
U_D^{\dagger}  & 0 \\
0 & U_D 
\end{pmatrix} ,
 \ee
where we have used the tilted bold letter ``$\tfQS^3$" to distinguish different representations between two self-conjugated triqubit-spinor fields $\psi_{\tfQS^3}(x)$ and $\psi_{\QS^3}(x)$. While the $\Gamma$-matrices $\Gamma^{\ha}$ appearing in Eq.(\ref{actionDM}) should have the same structure as those in Eq.(\ref{DGM}) but with a representation in eight-dimensional Hilbert space.

The action in Eq.(\ref{actionDM}) possesses explicitly the $\cQ_c$-spin symmetry SP(3)$\cong$SU(2) with the following group generators:
\be
& &  \tilde{\Sigma}^{ij} = \frac{i}{4}[ \tGa^i , \tGa^j ] = -\frac{i}{2}\epsilon^{ijk} \tGa^k  = \epsilon^{ijk} \frac{1}{2} \sigma_k\otimes \sigma_0\otimes \sigma_0 , \nn
\ee 
which reflects a charge-spin symmetry between Dirac fermion and its W-charge conjugated one. In such a self-conjugated representation, the $\cQ_c$-spin charge becomes manifest and equals to the qubit number, i.e., $\CQc=q_c=Q_N=3$.

Let us introduce further an alternative structure for {\it self-conjugated triqubit-spinor field} by adopting massless Dirac fermion $\psi_{D}(x)$ together with its complex charge conjugated one $\psi_{D}^{\bc}(x)$ instead of W-charge conjugated one. Explicitly, we define such an alternative self-conjugated triqubit-spinor field as follows: 
\be \label{fQS3}
& & \psi_{\fQS^3}(x) \equiv \binom{\psi_{D}(x) } { \psi_{D}^{\bc}(x) }, \quad 
\psi_{D}^{\bc}(x) \equiv C_D \bar{\psi}^T_{D}(x) , \nn \\
& & \psi_{\fQS^3}^{c}(x) = C_{\fQS^3} \bar{\psi}_{\fQS^3}^T(x)= \psi_{\fQS^3}(x) , \quad C_{\fQS^3} = \sigma_1 \otimes  C_{D} .
\ee
The corresponding action is found to be,  
\be  \label{actionfQS3}
\cS_{\fQS^3}   & \equiv & \int d^6x \,  \frac{1}{2} \bar{\psi}_{\fQS^3}(x)  \delta_{\ha}^{\;\hm}  \Gamma^{\ha} i\p_{\hm} \psi_{\fQS^3}(x)  , 
\ee
where $\Gamma$-matrices $ \Gamma^{\ha} \equiv ( \gamma^a, \Gamma^l )$ ($a=0,1,2,3$, $l=5,6$) get a different structure. Their explicit forms are given as follows:
\be \label{GMfQS3}
& & \Gamma^a = \sigma_1 \otimes \gamma^a , \quad a=0,1,2,3, \nn \\
& &  \Gamma^5  = i\sigma_3\otimes \gamma_5 , \quad \Gamma^6 = \sigma_3\otimes I_4 ,
\ee
where $\gamma^a$ are usual Dirac $\gamma$-matrices defined in Eq.(\ref{DGM}). The $\cQ_c$-spin matrices $\tGa^p$ ($p=1,2,3$) are found to be,
\be \label{tGMfQS3}
\tGa^1  = i\sigma_1\otimes \gamma_5 , \quad \tGa^2 = i\sigma_2\otimes \gamma_5 , \quad \tGa^3 = \sigma_0\otimes I_4 . 
\ee

Actually, in any representation of self-conjugated qubit-spinor field, all $\Gamma$-matrices should be constructed to meet the following general conditions under charge-conjugation operation:
\be
& & C_{\fQS^3}^{-1} \Gamma^{\ha} C_{\fQS^3} = -  \Gamma^{\ha T} , \quad  C_{\fQS^3}^{-1} \tGa^{p} C_{\fQS^3} = \tGa^{p T} .
\ee
It can be verified that $\Gamma^{\ha}$ matrices presented in Eq.(\ref{GMfQS3}) and $\tGa^p$ matrices given in Eq.(\ref{tGMfQS3}) do satisfy such conditions. It is noticed that either $\Gamma^{\ha}$ or $\tGa^p$ are not all anti-commuting.


\subsection{ Equivalent description of Dirac-$\And$Majorana-$\And$Weyl-type fermions as qubit-spinor fields with the same $\cM_c$-spin and $\cQ_c$-spin charges in corresponding to complex$\And$self-conjugated$\And$chiral representations }

It is shown that a self-conjugated qubit-spinor field can be described by a complex qubit-spinor field with a reduced qubit state. Similarly, a self-conjugated qubit-spinor field should be characterized equivalently by a chiral qubit-spinor field. In general, self-conjugated qubit-spinor field and complex qubit-spinor field can be regarded as Majorana-type fermion and Dirac-type fermion, respectively, and chiral qubit-spinor field is viewed as Weyl-type fermion.

Let us rewrite self-conjugated biqubit-spinor and triqubit-spinor fields $\psi_{\QS^2}(x)$ and $\psi_{\QS^3}(x)$ into coherent states of two parts, 
\be \label{MC}
& & \psi_{\QS^2}(x) \equiv \psi_{\QWp^2}(x) + \psi_{\QWn^2}(x), \quad \psi_{\QWpn^2}(x) \equiv \gamma_{\pm} \psi_{\QS^2}(x), \quad , \nn \\ 
& & \psi_{\QS^3}(x) \equiv \psi_{\QWp^3}(x) + \psi_{\QWn^3}(x), \quad \psi_{\QWpn^3 }(x) = \Gamma_{\pm} \psi_{\QS^3}(x) , \nn \\
& & \gamma_{\pm} = \frac{1}{2} ( 1 \pm \gamma_5 ) , \quad \Gamma_{\pm} = \frac{1}{2} ( 1 \pm \gamma_7 ) , 
\ee
with $\gamma_5$ and $\gamma_7$ given in Eq.(\ref{gamma_5_7}). Where $\psi_{\QWpn^2}(x)$ and $\psi_{\QWpn^3}(x)$ are defined as Weyl-type chiral spinor fermions in four- and six-dimensional spacetimes, which may be referred to as {\it chiral biqubit-spinor field} and {\it chiral triqubit-spinor field}, respectively. Actually, the two parts, $\psi_{\QWp^2}(x)$ and $\psi_{\QWn^2}(x)$, or $\psi_{\QWp^3}(x)$ and $\psi_{\QWn^3}(x)$, should not be independent ones due to self-conjugate conditions. They are correlated via the following charge conjugation operation: 
\be
& & \psi_{\QWp^2}(x) = \psi_{\QWn^2}^{c}(x) , \quad  \psi_{\QS^2 }(x) = \psi_{\QWn^2}(x) + \psi_{\QWn^2}^{c}(x) ,  \nn \\
& &  \psi_{\QWp^3}(x) = \psi_{\QWn^3}^{c}(x), \quad \psi_{\QS^3 }(x) = \psi_{\QWn^3}(x) + \psi_{\QWn^3}^{c}(x).
\ee

From Eqs.(\ref{MD23}) and (\ref{MC}), we arrive at the following general relations for chiral and self-conjugated as well as complex qubit-spinor fields:
\be \label{MWD}
& & \psi_{\QWn^2}(x) = \gamma_-\psi_{\QS^2 }(x) = \binom{0}{\psi_{\QC^1}(x)}  = \gamma_-\psi_{\QC^2 }(x) 
\equiv  \psi_{\QC^2-}(x), \nn \\
& & \psi_{\QWn^3}(x)  = \Gamma_-\psi_{\QS^3 }(x)  = \binom{0}{\psi_{\QC^2}(x)}, 
\ee
which provide explicit relations among Weyl-type and Majorana-type as well as Dirac-type fermions. In fact, $\psi_{\QWn^2}(x)$ defines the usual Weyl fermion in four dimensional spacetime, 
\be
\psi_{\QWn^2}(x) \equiv \psi_{W}(x). \nn
\ee

In terms of the chiral qubit-spinor fields $\psi_{\QWn^2}(x)$ and $\psi_{\QWn^3}(x)$, the actions given in Eqs.(\ref{actionM2P}) and (\ref{actionM3P}) for self-conjugated biqubit-spinor and triqubit-spinor fields can equivalently be expressed as follows:
\be \label{actionW2P}
   \cS_{\QWn^2}  \equiv \cS_{W} & = & \int d^4x \,   \{  [ \frac{1}{2} \bar{\psi}_{\QWn^2}(x)  \delta_a^{\;\mu}  \gamma^a i\p_{\mu} \psi_{\QWn^2}(x) 
 -   \frac{1}{2}\lambda_2  \phi_p(x) \bar{\psi}_{\QWn^2}(x) \tga^p \psi_{\QWn^2}^{c}(x)  + H.c. ]  \nn \\
 & + & \frac{1}{2}\eta^{\mu\nu} \p_{\mu}\phi_p(x)\p_{\nu}\phi^p(x) -  \frac{1}{4}\lambda_Q (\phi_p(x)\phi^p(x))^2 \} ,
 \ee
where the chiral biqubit-spinor field $\psi_{\QWn^2}(x)$ with its charge-conjugated one $\psi_{\QWn^2}^{c}(x)$ are defined as:
 \be
\psi_{\QWn^2}^{c}(x) = C_{\QS^2}\bar{\psi}_{\QWn^2}^T(x) = C_{\QS^2}\gamma^0 \psi_{\QWn^2}^{\ast} (x) , \nn
\ee
and
\be \label{actionW3P}
 \cS_{\QWn^3}  & = & \cS_{\QS^3} =  \int d^6x   \frac{1}{2} [ \bar{\psi}_{\QWn^3}(x)  \delta_{\ha}^{\;\hm}  \Gamma^{\ha} i\p_{\mu} \psi_{\QWn^3}(x) + H.c.  ] , 
\ee
for the chiral triqubit-spinor field $\psi_{\QWn^3}(x)$. 

From the above analyses, we arrive at the following equivalences for various formalisms of the actions: 
\be
& & \cS_{\QS^2} \equiv  \cS_{\QWn^2} \equiv  \cS_{\QC^1} \to \cS_{M} \equiv \cS_{W}  , \nn \\
& & \cS_{\QS^3} \equiv   \cS_{\QWn^3}  \equiv \cS_{\QC^2} \to \cS_{D}.
\ee
which indicates that all qubit-spinor fields with complex and self-conjugated as well as chiral spinor representations can bring on equivalent actions as long as they contain the same independent degrees of freedom with equal $\cM_c$-spin and $\cQ_c$-spin charges.

Therefore, when Dirac-type and Majorana-type as well as Weyl-type fermions as local coherent states of qubits have the same independent degrees of freedom in correspondence to complex and self-conjugated as well as chiral spinor representations with equal $\cM_c$-spin and $\cQ_c$-spin charges, their hermitian actions are actually equivalent in Minkowski spacetime.


\section{ Local coherent state of 4-qubit as hyperqubit-spinor field with hyperspin symmetry SP(1,9)$\cong$SL(2,$\mathbb{O}$) in 10D Minkowski hyper-spacetime and the periodic feature of $\cQ_c$-spin charge}

Majorana fermion and Dirac fermion are shown to be described by biqubit-spinor field and triqubit-spinor field with respective to $\cQ_c$-spin charges $\CQc=Q_N=2$ and $\CQc=Q_N=3$. In this section, we will discuss a local coherent state of four qubits ($Q_N$=4) based on the maximum coherence motion principle and local coherent-qubits motion postulate. Such a qubit-spinor field is referred to as {\it tetraqubit-spinor field} which spans 16-dimensional Hilbert space. Its maximally correlated motion leads to $\cM_c$-spin charge $\CMc=D_h=10$, which determines ten-dimensional Minkowski spacetime with $\cM_c$-spin symmetry SP(1,9). In general,  we are going to refer to high dimensional spacetime as {\it hyper-spacetime} and its $\cM_c$-spin symmetry as {\it hyperspin symmetry}. Also a qubit-spinor field formed with high qubit number ($Q_N\ge 4$) is referred to as {\it hyperqubit-spinor field}.  The $\cQ_c$-spin charge of tetraqubit-spinor field as hyperqubit-spinor field is found to be zero $\CQc=q_c=0$, which is no longer the same as qubit number $\CQc=q_c\neq Q_N$ and unlike the cases with qubit number $Q_N\le 3$. Instead, the $\cQ_c$-spin charge of tetraqubit-spinor field turns out to the same as that of single-component spinor field. Actually, the single-component spinor field can be regarded as self-conjugated chiral uniqubit-spinor field which may be referred to as {\it zeroqubit-spinor field} or {\it bit-spinor field}. Such an intriguing feature displays a periodic behavior of $\cQ_c$-spin charge for tetraqubit-spinor field and zeroqubit-spinor field, i.e., $\CQc = Q_N-4 = q_c =0$ for the case of four qubits and $\CQc=Q_N=q_c=0$ for the case of zero qubit.


\subsection{ Local coherent state of 4-qubit in 16D Hilbert space with zero $\cQ_c$-spin charge $\CQc$=0 as tetraqubit-spinor field in 10D Minkowski hyper-spacetime }

Let us now consider a local coherent state of four qubits, which is constructed with 4-product qubit-basis states as follows: 
\be
\Psi_{\mQ^4}(x) & = &  \sum_{s_1, \cdots, s_{4}}  \psi_{s_1\cdots s_{4}}(x)  \varsigma_{s_1\cdots s_{4}} \equiv \sum_{s_1, \cdots, s_{4}}  \psi_{s_1\cdots s_{4}}(x)\,  \varsigma_{s_1}\otimes \cdots \otimes \varsigma_{s_{4}}, \nn \\
 \Psi_{\QS^4}(x) & = &  \sum_{s_1, \cdots, s_{4}}  \psi_{s_1\cdots s_{4}}(x)  \zeta_{s_1\cdots s_{4}} \equiv \sum_{s_1, \cdots, s_{4}}  \psi_{s_1\cdots s_{4}}(x)\,  \zeta_{s_1}\otimes \cdots \otimes \zeta_{s_{4}}, \nn \\
 & \equiv & \sum_{s_1, \cdots, s_{4}}  \tilde{\psi}_{s_1\cdots s_{4}}(x)  \varsigma_{s_1\cdots s_{4}} \equiv \sum_{s_1, \cdots, s_{4}}  \tilde{\psi}_{s_1\cdots s_{4}}(x)\,  \varsigma_{s_1}\otimes \cdots \otimes \varsigma_{s_{4}},
\ee
where $\Psi_{\mQ^4}(x)$ and $\Psi_{\QS^4}(x) $ are referred to as {\it tetraqubit-spinor field} and {\it self-conjugated tetraqubit-spinor field}, respectively. $\psi_{s_1\cdots s_{4}}(x)$ present local distribution amplitudes with $16$ independent degrees of freedoms, which characterize the local coherent state of four qubits ($Q_N=4$). 

Based on the local coherent-qubits motion postulate with following along the maximum coherence motion principle, we arrive at the hermitian action of tetraqubit-spinor field $\Psi_{\mQ^4}(x)$ as follows:
 \be \label{actionQ4}
 \cS_{\mQ^4}  & = &  \int d^{10}x  \frac{1}{2}  \Psi_{\mQ^4}^{\dagger}(x)  \delta_{\mA}^{\; \mM}  \Ups^{\mA} i\p_{\mM} \Psi_{\mQ^4}(x) , 
\ee
where $i\p_{\mM}\equiv i\frac{\p}{\p x^{\mM}}$ is the linear self-adjoint derivative operator and $\delta_{\mA}^{\;\;\mM} $ denotes the Kronecker symbol, with the vector indices $\mA, \mM= 0, 1, \cdots, 9$. The $\cM_c$-matrices $\Ups^{\mA}$ form a vector in the qubit-spinor representation. The Latin alphabet $\mA, \mB, \ldots$ and the Latin alphabet starting from $\mM, \mN$ are adopted to distinguish vector indices in non-coordinate spacetime and coordinate spacetime, respectively.  All the Latin indices are raised and lowered by the constant metric matrices, i.e., $\eta^{\mA\mB} $ or $\eta_{\mA\mB} = diag. (1,-1,\ldots,-1)$, and $\eta^{\mM\mN} $ or $\eta_{\mM\mN} = diag.(1,-1,\ldots,-1)$.  

The maximally correlated motion leads to the following ten real symmetric normalized $\cM_c$-matrices $\Ups^{\mA}$ with $\mA = 0, 1, \cdots, 9$, 
\be \label{UpsM4}
& & \Ups^{\mA} = ( \Ups^0, \Ups^{\bA}, \Ups^9 ), \quad  \bA=1,\cdots, 8, \nn \\
& & \Ups^0 = \sigma_0 \otimes\sigma_0 \otimes \sigma_0 \otimes \sigma_0 , \nn \\
& & \Ups^1 = \sigma_0 \otimes\sigma_3\otimes  \sigma_0\otimes \sigma_1,  \nn \\
& & \Ups^2 =\sigma_0 \otimes \sigma_3\otimes  \sigma_2\otimes \sigma_2, \nn \\
& & \Ups^3 = \sigma_0 \otimes\sigma_1\otimes  \sigma_0\otimes \sigma_0, \nn \\
& & \Ups^4 = \sigma_1 \otimes\sigma_2\otimes  \sigma_2\otimes \sigma_0 , \nn \\
& & \Ups^5 = \sigma_2 \otimes  \sigma_3\otimes \sigma_1\otimes \sigma_2, \nn \\
& & \Ups^6 =  \sigma_2 \otimes \sigma_3\otimes \sigma_3\otimes \sigma_2 , \nn \\
& & \Ups^7 =  \sigma_2 \otimes \sigma_2\otimes \sigma_0\otimes \sigma_0 ,  \nn \\
& & \Ups^8 =  \sigma_3 \otimes \sigma_2\otimes \sigma_2\otimes \sigma_0 , \nn \\
& &  \Ups^9 =  \sigma_0 \otimes \sigma_3\otimes \sigma_0\otimes \sigma_3 , 
\ee
which satisfy Clifford algebra relations,
\be \label{CAQ4}
\{\Ups^{\mA}, \Ups^{\mB} \} = \delta^{\mA\mB}, \quad \{ \Ups^{0}, \Ups^{\mA} \} =  \Ups^{\mA}, \quad \mA, \mB = 1, \cdots , 9 ,
\ee
with $\Ups^0$ the $16\times 16$ unit matrix, i.e., $\Ups^0 \equiv I_{16}$.

The number of $\cM_c$-matrices $\Ups^{\mA}$ brings on $\cM_c$-spin charge $\CMc=D_h=10$ and determines {\it ten-dimensional Minkowski spacetime}.  As there exists no antisymmetric normalized $\tUps$-matrix that anticommutes with real symmetric normalized $\cM_c$-matrices, the $\cQ_c$-spin charge of tetraqubit-spinor field $\Psi_{\mQ^4}(x)$ is zero, which is no longer to be equal to the qubit number. Instead, it is given by:
\be
\CQc =q_c= Q_N-4 = 0 . \nn
\ee

In light of self-conjugated tetraqubit-spinor field $\Psi_{\QS^4}(x) $, the action can be written as follows:
 \be \label{actionM4}
 \cS_{\QS^4}  & = &  \int d^{10}x  \frac{1}{2}  \bar{\Psi}_{\QS^4}(x)  \delta_{\mA}^{\; \mM}  \Gamma^{\mA} i\p_{\mM} \Psi_{\QS^4}(x) , 
\ee
with $\mA, \mM= 0,1,2,3,5, \cdots, 10$. The $\Gamma$-matrices $\Gamma^{\mA}$ get the following forms:
\be \label{TGM}
& & \Gamma^{\mA}= (\Gamma^0, \Gamma^{\tA}, \Gamma^{10} ) , \quad \tA = 1,2,3,5, \cdots, 9 , \nn \\
& & \Gamma^0 =\; \sigma_3 \otimes\sigma_3 \otimes \sigma_3 \otimes \sigma_3, \nn \\
& & \Gamma^1 = i \sigma_3 \otimes \sigma_3\otimes \sigma_3\otimes \sigma_2 , \nn \\
& &  \Gamma^2 = i\sigma_3 \otimes\sigma_3\otimes  \sigma_3\otimes \sigma_1, \nn \\
& &  \Gamma^3 = i \sigma_0 \otimes\sigma_1\otimes  \sigma_3\otimes \sigma_0 , \nn \\
& &  \Gamma^5 = i \sigma_0 \otimes\sigma_2\otimes  \sigma_3\otimes \sigma_0, \nn \\
& & \Gamma^6 = i \sigma_3 \otimes\sigma_0\otimes  \sigma_1\otimes \sigma_0 , \nn \\
& & \Gamma^7 = i \sigma_3 \otimes \sigma_0\otimes \sigma_2\otimes \sigma_0 , \nn \\
& & \Gamma^8 = i \sigma_1 \otimes \sigma_3\otimes \sigma_0\otimes \sigma_0  , \nn \\
& & \Gamma^9 = i \sigma_2 \otimes  \sigma_3\otimes \sigma_0\otimes \sigma_0, \nn \\
& & \Gamma^{10} =  \sigma_0 \otimes \sigma_0\otimes  \sigma_0\otimes \sigma_0 \equiv I_{16} ,
\ee
with $\Gamma^{10}$ the $16\times 16$ unit matrix, i.e., $\Gamma^{10}\equiv I_{16}$. The self-conjugated tetraqubit-spinor field $\Psi_{\QS^4}(x) $ can be expressed as follows:
\be
& & \Psi_{\QS^4}(x) =\binom{\psi_{\QC^3}}{\psi_{\hQC^3}} =  \begin{pmatrix} 
\psi_{\QCp^2}(x) \\
\psi_{\QCn^2}(x) \\
\psi_{\hQCn^2} (x) \\
\psi_{\hQCp^2} (x)
 \end{pmatrix}  , \nn \\
& & \psi_{\hQC^3}  = \sigma_1\otimes \sigma_1\otimes \sigma_1\psi_{\QC^3}^{\ast} ,
\ee
which satisfies the following self-conjugate condition:
\be
& & \Psi^c_{\QS^4}(x) = C_{\QS^4} \bar{\Psi}^T_{\QS^4}(x) = \Psi_{\QS^4}(x) , \nn \\
& & C_{\QS^4} = \sigma_2 \otimes\sigma_2 \otimes \sigma_2 \otimes \sigma_2 .
\ee

In general, we are going to refer to high dimensional Minkowski spacetime ($D_h\ge10$) as {\it Minkowski hyper-spacetime}. Meanwhile, the tetraqubit-spinor field $\Psi_{\mQ^4}(x)$ or self-conjugated tetraqubit-spinor field $\Psi_{\QS^4}(x) $ in 16-dimensional Hilbert space is referred to as {\it hyperqubit-spinor field}.


\subsection{ Hyperspin symmetry SP(1,9)$\cong$SL(2,$\mathbb{O}$) and intrinsic discrete symmetries of hyperqubit-spinor field in 10D Minkowski hyper-spacetime }

It can be verified that the actions presented in Eq.(\ref{actionQ4}) and (\ref{actionM4}) possess the following associated symmetry:
\be
G_S & = & SC(1)\ltimes P^{1,9}\ltimes SO(1,9)\adjoin SP(1, 9)  \rtimes SG(1) \times SP(q_c=0) \nn \\
& = & SC(1)\ltimes  PO(1, 9) \adjoin SP(1,9) \rtimes SG(1), 
\ee
where SP(1,9) is $\cM_c$-spin symmetry which is referred to as {\it hyperspin symmetry}. SP($q_c$=0) denotes in formal $\cQ_c$-spin symmetry with zero $\cQ_c$-spin charge $q_c=0$, which indicates that there exists no antisymmetric $\cQ_c$-matrix and scalar coupling term. The semidirect product group PO(1,9) = P$^{1,9}\ltimes$ SO(1,9) is Poincar\'e-type group symmetry/inhomogeneous Lorentz-type group symmetry with $P^{1,9}$ denoting translation group symmetry and SO(1,9) representing Lorentz-type group symmetry in ten-dimensional Minkowski hyper-spacetime of coordinates. SC(1)$\adjoin$SG(1) labels an associated symmetry with coincidental global scaling transformations of coordinates and fields. In fact, the symbol ``$\adjoin$" is used to express an associated symmetry in which the transformations of hyperspin symmetry SP(1,9) and global scaling symmetry SG(1) in Hilbert space should be coincidental to those of Lorentz-type group symmetry SO(1,9) and conformal scaling symmetry SC(1) in Minkowski hyper-spacetime.

The hyperspin symmetry SP(1,9) of hyperqubit-spinor field and Lorentz-type group symmetry SO(1,9) of Minkowski hyper-spacetime as well as global scaling symmetry should obey the following transformation properties:
\be
& & \Psi(x)  \to \Psi'(x') =  S(\Lambda) \Psi(x), \quad S(\Lambda) \equiv e^{i\Sigma^{\mA\mB} \varpi_{\mA\mB} }, \nn \\
& & x^{\mM} \to x^{'\mM} = L^{\mM}_{\; \mN}\, x^{\mN},  \quad  S^{-1}(\Lambda) \Ups^{\mA} S(\Lambda)  = L^{\mA}_{\; \mB} \, \Ups^{\mB}, \nn \\
& &  S(\Lambda)  \in SP(1, 9), \quad L^{\mA}_{\; \mB} = L^{\mM}_{\; \mN} \in SO(1, 9) \cong SP(1,9), \nn \\
& & \Psi(x)\to \Psi'(x')=e^{9\varpi/2} \Psi(x), \quad x^{\mu} \to x^{'\mu} = e^{-\varpi} x^{\mu}, \nn \\
& & e^{-\varpi} \in SC(1), \quad e^{9\varpi/2} \in SG(1), 
\ee
with $\Sigma^{\mA\mB}$ the group generators of SP(1,9) defined as follows:  
\be
& & \Sigma^{\mA\mB} \equiv ( \Sigma^{\bA\bB}, \Sigma^{0\bA}, \Sigma^{0 9 }, \Sigma^{\bA 9} ),  \quad \Sigma^{\bA\bB} = \frac{i}{4}[\Ups^{\bA},  \Ups^{\bB} ], \nn \\
& &  \Sigma^{0\bA} = - \Sigma^{\bA 0}  =  \frac{i}{2} \Ups^{\bA}, \quad \Sigma^{09}  = - \Sigma^{9 0} = \frac{i}{2} \Ups^{9},  \nn \\
& & \Sigma^{\bA 9} = - \Sigma^{9 \bA }  = \frac{i}{2}\Ups^{\bA} \Ups^{9} , \quad \bA, \bB= 1, \cdots, 8 ,
\ee
which operates on the tetraqubit-spinor field $\Psi_{\mQ^4}(x)$ as hyperqubit-spinor field. Alternatively, the group generators $\Sigma^{\mA\mB}$ get the following forms:
\be \label{SP10G}
& &  \Sigma^{\mA\mB} = ( \Sigma^{\tA\tB}, \Sigma^{0\tA}, \Sigma^{0\, 10 }, \Sigma^{\tA\, 10} ) , \quad \Sigma^{\tA\tB} = \frac{i}{4}[\Gamma^{\tA},  \Gamma^{\tB} ], \nn \\
& & \Sigma^{0\tA} = \frac{i}{4}[\Gamma^{0},  \Gamma^{\tA} ] = \frac{i}{2}\Gamma^{0}\Gamma^{\tA}, \quad  \Sigma^{0\,10} = - \Sigma^{10\, 0 }  = \frac{i}{2} \Gamma^{0}, \nn \\
& & \Sigma^{\tA\,10} = - \Sigma^{10\, \tA }  = -\frac{i}{2} \Gamma^{\tA}, \quad \tA, \tB=1,2,3,5, \cdots, 9 , 
\ee
which acts on the self-conjugated tetraqubit-spinor field $\Psi_{\QS^4}(x) $ as hyperqubit-spinor field. 

As the $\Gamma$-matrix $\Gamma^{10}$ is a unit matrix which is commuting with all others, not all group generators can be expressed as the commutative relations of $\Gamma$-matrices. The group generators of hyperspin symmetry SP(1,9) are in analogous to those of conformal group of spin symmetry SP(8)$\cong$SO(8), which is formally isomorphic to a special linear group of degree 2 over the {\it octonion} $\mathbb{O}$, i.e.:
\be
SP(1,9)\cong SL(2, \mathbb{O}) \cong SO(1,9) .
\ee

To ensure the invariance of the action in Eq.(\ref{actionM4}) under operations of discrete symmetries: $\mathcal{C}$, $\mathcal{P}$ and $\mathcal{T}$, the self-conjugated tetraqubit-spinor field $\psi_{\QS^4}(x)$ should transform as follows: 
\be \label{CCQS4}
& & \mathcal{C} \Psi_{\QS^4}(x) \mathcal{C}^{-1} \equiv \Psi_{\QS^4}^{c}(x) = C_{\QS^4} \bar{\Psi}_{\QS^4}^{T} ( x ) = \Psi_{\QS^4}(x) ,\nn \\
& & C_{\QS^4}^{-1} \Gamma^{\mA} C_{\QS^4} = \Gamma^{\mA\, T}, \quad  C_{\QS^4} = i\Gamma^0 \Gamma^2  \Gamma^6 \Gamma^8 \Gamma^9 = \sigma_2 \otimes \sigma_2 \otimes \sigma_2 \otimes \sigma_2 ,  \nn 
\ee 
for the charge conjugation, and 
\be
& & \mathcal{P} \Psi_{\QS^4}(x) \mathcal{P}^{-1} \equiv \Psi_{\QS^4}^{p}(x) = P_{\QS^4} \Psi_{\QS^4}( x^0, -x^k, x^{10}) , \nonumber \\
& &  P_{\QS^4}^{-1} \Gamma^{\mA} P_{\QS^4} = \Gamma^{\mA\, \dagger}, \quad P_{\QS^4}= \Gamma^0 = \sigma_3 \otimes \sigma_3 \otimes \sigma_3 \otimes \sigma_3,   \nn 
\ee
for the parity-inversion, and 
\be
& & \mathcal{T} \Psi_{\QS^4}(x) \mathcal{T}^{-1} \equiv \Psi_{\QS^4}^{t}(x) = T_{\QS^4} \Psi_{\QS^4}( -x^0, x^k, -x^{10} ) , \nonumber \\
& &  T_{\QS^4}^{-1} \Gamma^{\mA} T_{\QS^4} = \Gamma^{\mA\, T}, \quad  
T_{\QS^4} =  \Gamma^1 \Gamma^3  \Gamma^5 \Gamma^7  = \sigma_2 \otimes \sigma_2 \otimes \sigma_2 \otimes \sigma_2 ,  \nn 
\ee
for the time-reversal.  

It is noticed that the spatial coordinate $x^{10}$ in the self-conjugated  tetraqubit-spinor representation gets an unusual transformation property under both the parity-inversion and time-reversal operations as it undergoes a flip in sign for keeping the action to be invariant. When considering the combined operator $\vartheta \equiv \mathcal{PT} $, we arrive at a proper and conventional transformation as follows:
\be
& & \vartheta \Psi_{\QS^4}(x) \vartheta^{-1} = \theta \Psi_{\QS^4}( - x) , \quad  \theta^{-1} \Gamma^{\mA} \theta =  \Gamma^{\mA \, \ast} , \nn 
\ee
which indicates that the combined operator $\vartheta $ should be more appropriate for the self-conjugated tetraqubit-spinor field $\Psi_{\QS^4}(x)$ in 10-dimensional Minkowski hyper-spacetime. 

For the combined joint operation $\varTheta \equiv \mathcal{CPT} $ acting on the self-conjugated tetraqubit-spinor field $\Psi_{\QS^4}(x)$ in 10-dimensional Hyper-spacetime, we obtain the following conventional transformation property:
\be
& & \varTheta \Psi_{\QS^4}(x) \varTheta^{-1} = \Theta \bar{\Psi}_{\QS^4}^{T} ( - x) ,\nn \\
& & \Theta^{-1} \Gamma^{\mA} \Theta =  \Gamma^{\mA \, \dagger}, \quad \Theta \Gamma^0 =  \Gamma^0 \Theta . \nn 
\ee


\subsection{ Self-conjugated chiral uniqubit-spinor field in 2D Minkowski spacetime as zeroqubit-spinor field or bit-spinor field with zero $\cQ_c$-spin charge $\CQc= Q_N=0$ }

It is manifest that for a real spinor field with only one-component entity, there exists no scalar coupling term due to anti-commuting property of spinor field, so that it has a zero $\cQ_c$-spin charge $\CQc=q_c=0$. Such a one-component entity real spinor field is regarded as {\it bit-spinor field} or {\it zeroqubit-spinor field} as it concerns zero qubit number $Q_N=0$,  which can also be resulted by imposing a qubit-reduction condition on uniqubit-spinor field. 

The uniqubit-spinor field with two independent degrees of freedom is shown to have a maximally correlated motion in three-dimensional spacetime. Let us define the {\it qubit-reduction operation} which reduces the independent degrees of freedom of qubit-spinor field into a half and results in a reduced qubit-spinor field. The {\it qubit-reduction condition} for uniqubit-spinor field is defined as follows: 
\be
& & \psi_{\QT \pm }(x) \equiv \gamma_{\pm} \psi_{\QT}(x) , \quad   \gamma_{\pm} = \frac{1}{2} (1\pm \hat{\gamma}_3 ) , \quad  \hat{\gamma}_3 = - i\gamma^2 .
\ee
Explicitly, we obtain the following reduced qubit-spinor field:
\be
& &  \psi_{\mQ^1\pm}(x) =  \frac{1}{2} ( 1\pm\sigma_3 ) \psi_{\mQ^1}(x) ; \quad \psi_{\QS^1\pm}(x) = \frac{1}{2} ( 1\pm\sigma_1 ) \psi_{\QS^1}(x) , \nn \\
& & \psi_{\mQ^1+}(x) = \binom{\psi_{\mQ^0_{+}}(x)}{0},   \quad \psi_{\mQ^1-}(x) =  \binom{ 0 }{ \psi_{\mQ^0_{-}}(x) } , \nn \\
& &  \psi_{\QS^1+}(x) = \frac{1}{\sqrt{2}} \binom{ \psi_{\mQ^0_{+}}(x) }{\psi_{\mQ^0_{+}}(x)} , \quad  \psi_{\QS^1-}(x) = \frac{i}{\sqrt{2}} \binom{\psi_{\mQ^0_{-}}(x)}{-\psi_{\mQ^0_{-}}(x)}, \nn
\ee  
which brings on {\it qubit-reduction spinor field}. Such a qubit-reduction spinor field has only one real component $\psi_{\mQ^0_{-}}(x)$ or $\psi_{\mQ^0_{+}}(x)$, which may be referred to as {\it zeroqubit-spinor field} or {\it bit-spinor field}.  Namely, each real component of uniqubit-spinor field is identified to be a zeroqubit-spinor field or a bit-spinor field in one-dimensional Hilbert space with zero qubit number $Q_N=0$.

The action for zeroqubit-spinor field can be deduced from Eq.(\ref{actionB}) or Eq.(\ref{actionQ}) with the replacement of $\psi_{\QT}(x)$ by $\psi_{\QT\pm }(x)$. Explicitly, we arrive at the following action of zeroqubit-spinor field: 
\be \label{actionQ0}
 \cS_{\mQ^0_{\pm}}  & = & \int d^2x \, \frac{1}{2}  \bar{\psi}_{\QT\pm}(x) \delta_a^{\;\mu} \gamma^a i\partial_{\mu} \psi_{\QT\pm}(x) \nn \\
 & = & \int d^2x \,  \frac{1}{2} \psi^{\dagger}_{\mQ^1\pm }(x) \delta_a^{\;\mu} \Ups^a i\partial_{\mu} \psi_{\mQ^1\pm}(x)  \nn \\
 & = & \int d^2x \,  \frac{1}{2} \psi_{\mQ^0_{\pm}}(x)  i(\p_0\pm \p_1) \psi_{\mQ^0_{\pm}}(x),
\ee
with $a, \mu = 0, 1$. The action in Eq.(\ref{actionQ0}) involves no scalar field coupling to the zeroqubit-spinor field, so that the $\cQ_c$-spin charge is just equal to the qubit number,
\be
\CQc = q_c=Q_N=0.
\ee

The action in Eq.(\ref{actionQ0}) possesses the following associated symmetry:
\be
G_S & =&  SC(1)\ltimes P^{1,1}\ltimes SO(1,1) \adjoin SP(1,1) \times SG(1) \times SP(q_c=0)  \nn \\
& = & SC(1)\ltimes PO(1,1) \adjoin SP(1,1) \times SG(1), 
\ee
where $SP(q_c)=SP(0)$ is in formal presented to indicate that there is no $\cQ_c$-spin symmetry and no scalar coupling for zeroqubit-spinor field, which is just applied to categorize both qubit-spinor field and spacetime dimension.

The qubit-reduction condition actually defines a chiral-like property of qubit-spinor field. It can be verified that the zeroqubit-spinor field as reduced uniqubit-spinor field satisfies the following {\it self-conjugated chiral condition} in two-dimensional Minkowski spacetime, 
\be
& & \mathcal{C} \psi_{\QT\pm}(x) \mathcal{C}^{-1}  = \psi_{\QT\pm}^{c}(x) = C_2\bar{\psi}_{\QT\pm}^T (x) = C_2 \gamma^0 \psi_{\QT\pm}^{\ast}(x) =\psi_{\QT\pm}(x), \nn \\
& & C_2^{-1} \gamma^a C_2 = - \gamma^{a\, T}, \; (a=0,1, 2) , \nn \\
& & C_2 = \begin{cases} 
  \gamma^0= \sigma_2 , & \psi_{\mQ^1\pm} , \\
 \sigma_1\gamma^0 = -i\sigma_2 , &  \psi_{\QS^1\pm} ,
 \end{cases} \nn
\ee 
where $\psi_{\mQ^1\pm}(x)$ and $\psi_{\QS^1\pm}(x)$ are referred to as {\it self-conjugated chiral uniqubit-spinor fields}. To be specific, we may take the following notations:
\be
\psi_{\QT-}(x) \equiv (\psi_{\QS^1-}(x), \psi_{\mQ^1-}(x) ) ,\quad \psi_{\QT+}(x) \equiv (\psi_{\QS^1+}(x), \psi_{\mQ^1+}(x) ), 
\ee 
to define self-conjugated chiral uniqubit-spinor fields.

With the introduction of new variables for coordinates,
\be
x^{\pm} = \frac{1}{2} ( x^0 \pm x^1), \quad \p_0 \pm \p_1 \equiv \p_{\pm}\equiv \frac{\p}{\p x^{\pm}}  , \nn
\ee 
the above action can simply be rewritten as follows:
\be \label{Waction}
 \cS_{\mQ^0_{\pm}} & = & \int d^2x \,  \frac{1}{2} \psi_{\mQ^0_{\pm}}(x)  i\p_{\pm} \psi_{\mQ^0_{\pm}}(x) .
\ee
Such an action is actually the same as the one given in Eq.(\ref{BSFA}) which is resulted directly from the simplest motion postulate and quadratic free motion postulate. 

It can be checked that the action of zeroqubit-spinor field possesses both homogeneous and non-homogeneous scaling symmetries as well as translation group symmetry,
\be
& & \psi_{\mQ^0_{\pm}}(x) \to \psi'_{\mQ^0_{\pm}}(x') =  e^{\varpi/2} \psi_{\mQ^0_{\pm}}(x), \quad x^{\mu} \to x'^{\mu} = e^{-\varpi} x^{\mu} ,  \nn \\
& &  \psi_{\mQ^0_{\pm}}(x) \to \psi'_{\mQ^0_{\pm}}(x' ) =  e^{\varpi_{\pm}/2} \psi_{\mQ^0_{\pm}}(x), \quad x^{\pm} \to x'^{\pm} = e^{-\varpi_{\pm}} x^{\pm} , \nn \\
& & x^{\mu} \to x^{\mu} + a^{\mu}, \quad \psi_{\mQ^0_{\pm}}(x) \to \psi_{\mQ^0_{\pm}}(x + a) ,  \nn \\
& & e^{-\varpi},\, e^{\varpi/2}  \in SC(1)\cong SG(1), \quad e^{\varpi_{\pm}/2}, e^{-\varpi_{\pm}}  \in SP(1,1)\cong SO(1,1) .
\ee

As there exists no scalar coupling to zeroqubit-spinor field, each zeroqubit-spinor field in two-dimensional Minkowski spacetime keeps a free motion. The equation of motion can easily be read off as follows:
\be
i\p_{\pm} \psi_{\mQ^0_{\pm}}(x) = 0 , 
\ee 
its general solution is found to be:
\be
& & \psi_{\mQ^0_{\pm}}(x) \equiv \psi_{\mQ^0_{\pm}}(x^{\mp}) = \int \frac{dp}{\sqrt{2\pi}} [ b_{\pm}(p) e^{-i p x^{\mp}} + b^{\dagger}_{\pm}(p) e^{i p x^{\mp}} ] ,  
\ee 
with the momentum $p_{\pm} = p_0 \pm p_1$ and definitions:
\be
& & \psi_{\mQ^0_{+}}(x^{-}): \quad p_{-} = 0, \quad p_0=p_1 = p, \nn \\
& &  \psi_{\mQ^0_{-}}(x^{+}): \quad p_{+} = 0, \quad p_0= -p_1 = p . \nn
\ee
Where $b_{\pm}(p)$ and $b^{\dagger}_{\pm}(p)$ are annihilation and creation operators which satisfy the following anti-commuting relation:
\be
\{ b_{\pm}(p), b^{\dagger}_{\pm}(p')\} = \delta(p-p')  .
\ee

In general, the zeroqubit-spinor fields $\psi_{\mQ^0_-}(x^{+} ) $ and $\psi_{\mQ^0_+}(x^{-} )$ can be regarded as left-handed and right-handed freely moving {\it self-conjugated chiral uniqubit-spinor fields} $\psi_{\QL^1}(x)$ and $\psi_{\QR^1}$, respectively.  

For a finite spatial dimension with proper boundary condition, the zeroqubit-spinor field or bit-spinor field may be viewed as {\it zeroqubit-spinor string} or {\it bit-spinor string} in two-dimensional Minkowski spacetime.


\subsection{ Periodic behavior of $\cQ_c$-spin charge for qubit-spinor fields with $\cQ_c$-spin charges $\CQc$=0,1,2,3,0 in corresponding to Hilbert space with dimensions $D_H$=1,2,4,8,16 and Minkowski spacetime with dimensions $D_h$=2,3,4,6,10 } 

By applying for the orthonormal qubit basis states, we have analyzed four types of local coherent states of qubits. They contain the basic qubit-basis states, 2-product qubit-basis states, 3-product qubit-basis states and 4-product qubit-basis states, which bring about the uniqubit-spinor, biqubit-spinor, triqubit-spinor and tetraqubit-spinor fields, respectively. The independent degrees of freedom of qubit basis states are determined by qubit number $Q_N$, which span $2^{Q_N}$-dimensional Hilbert space and bring on $\cM_c$-spin and $\cQ_c$-spin charges of qubit-spinor field based on the local coherent-qubits motion postulate with following along the maximum coherence motion principle. The $\cM_c$-spin charge determines the dimension of Minkowski spacetime and describes $\cM_c$-spin symmetry in association with Poincar\'e-type group symmetry (also called inhomogeneous Lorentz-type group symmetry). The $\cQ_c$-spin charge characterizes $\cQ_c$-spin symmetry in correlation to scalar couplings and meanwhile reflects a periodic behavior. 

The main properties of qubit-spinor fields and spacetime dimensions for $Q_N=0,1,2,3,4$ are summarized in Table 1. It is seen that there exists a simple relation between $\cQ_c$-spin charge $\CQc$ and qubit number $Q_N$ for $Q_N\le3$. For the case $Q_N = 4$, the $\cQ_c$-spin charge exhibits a periodic behavior with respect to the qubit number. To be specific, we can write down the simple relations as follows:
\be
& &  \CQc=q_c=Q_N = 0,1,2,3, \qquad Q_N \le 3 , \nn \\
& & \CQc=q_c=Q_N-4 = 0,  \quad \qquad Q_N = 4 ,
\ee
which indicates that $\CQc=q_c= 0,1,2,3$ should characterize {\it four categories of $\cQ_c$-spin charges}. We should consider the first five qubit numbers $Q_N=0,1,2,3,4$ as {\it basic qubit numbers}, where $Q_N=0$ is actually a trivial one. 
\\

\begin{table}[htp]
\caption{The properties of qubit-spinor fields and spacetime dimensions}
\begin{center}
\begin{tabular}{|c|c|c|c|c|c|c|}
\hline
Qubits $Q_N $ & $ 0$ & $ 1$ & $ 2$ & $ 3$ & $4$ \\
Qubit-spinor field &  $\psi_{\mQ^0}$ &  $\psi_{\mQ^1}$ &  $\psi_{\mQ^2}$ &  $\psi_{\mQ^3}$ & $\psi_{\mQ^4}$ \\
Qubit-spinor & bit-spinor & uniqubit-spinor & biqubit-spinor  & triqubit-spinor  & tetraqubit-spinor \\
Hilbert space $\cD_H$ & $ 2^0 =1 $ & $ 2^1=2 $ & $ 2^2=4 $ & $ 2^3=8 $ & $2^4=16 $ \\ 
$\cQ_c$-spin charge $\CQc$ & $ 0$ & $ 1$ & $  2 $ & $ 3$ & $0=4-4$ \\
$\cQ_c$-spin sym. & SP(0) & SP(1)$\cong$O(1) & SP(2)$\cong$U(1) & SP(3)$\cong$SU(2) & SP(0) \\
Spacetime $D_h$ & 2 & 3=2+$2^0$  & 4=2+$2^1$  &  6=2+$2^2$  & 10=2+$2^3$ \\
Trans. Dim. $D_h$-2 & 0 & 1  & 2  &  4  & 8 \\
Spinor structure   &  $\psi_{\QT\pm}$ &  $\psi_{\QS^1}$ &  $\psi_{\QS^2},\psi_{\QC^1}$ &  $\psi_{\QS^3}, \psi_{\QC^2}$ & $\psi_{\QS^4}$ \\
$\cM_c$-spin sym. & SP(1,1) & SP(1,2) & SP(1,3) & SP(1,5) & SP(1,9) \\
Isomorphic group &  & $\cong$ SL(2,$\mathbb{R}$)  & $\cong$ SL(2,$\mathbb{C}$)  & $\cong$ SL(2,$\mathbb{Q}$) &  $\cong$ SL(2,$\mathbb{O}$)  \\
Lorentz-type sym.  & SO(1,1) & SO(1,2) & SO(1,3) & SO(1,5) & SO(1,9) \\
Poincar\'e-type sym. & PO(1,1) & PO(1,2) & PO(1,3) & PO(1,5) & PO(1,9) \\
\hline
\end{tabular}
\end{center}
\label{Table 1}
\end{table}%

The dimension ($\cD_H$) of Hilbert space for qubit-spinor field is determined by its local coherent state of qubits with a given qubit number $Q_N$, namely, $\cD_H=2^{Q_N}$. The dimension ($D_h$) of Minkowski spacetime is determined from local coherent state of qubits via its maximally correlated motion with transverse dimension given by $D_h-2$. The local coherent state of qubits with qubit number $Q_N=4$ has zero $\cQ_c$-spin charge,  which is the same as the one of zeroqubit-spinor field or bit-spinor field and displays a periodic behavior of $\cQ_c$-spin charge. Similarly, the spacetime dimensions $D_h=2,3,4,6,10$ with respective to the basic qubit numbers $Q_N=0,1,2,3,4$ are regarded as {\it basic spacetime dimensions}. In general, there exist four categories with respective to $\cQ_c$-spin charges $\CQc=q_c=0,1,2,3$, which enables us to make a categorization for all hyperqubit-spinor fields and hyper-spacetime dimensions.


\section{ Categorization of hyperqubit-spinor fields and hyper-spacetimes with periodic $\cQ_c$-spin charges and the genesis of hyper-spacetime dimension from categoric qubit number }

The local coherent states of qubits built from 2-product and 3-product qubit-basis states are shown to  characterize Majorana and Dirac fermions as biqubit-spinor and triqubit-spinor fields with $\cQ_c$-spin symmetries SP(2)$\cong$U(1) and SP(3)$\cong$SU(2), respectively. The local coherent state of qubits formed from 4-product qubit-basis states brings on the tetraqubit-spinor field as hyperqubit-spinor field in 16-dimensional real Hilbert space and 10-dimensional Minkowski hyper-spacetime, which is found to have a zero $\cQ_c$-spin charge $\CQc=q_c=Q_N-4=0$ that is the same as the one of zeroqubit-spinor field in two-dimensional spacetime. By further analyzing local coherent states of qubits constructed from higher product qubit-basis states in this section, we are able to deduce a definite periodic property of $\cQ_c$-spin charge $\CQc$ for arbitrary qubit number $Q_N$. In general, we are going to show that any hyperqubit-spinor field in Hilbert space and any Minkowski hyper-spacetime dimension determined from the maximally correlated free motion can be categorized into four classes in correspondence to four categories of $\cQ_c$-spin charges $\CQc= q_c=0,1,2,3$, which can be carried out from classifying arbitrary qubit number $Q_N$ into categoric qubit number denoted as $Q_N^{(q_c,k)}$ in category-$q_c$ and $k$-th period.


\subsection{Local coherent state of five qubits with $\cQ_c$-spin charge $\CQc$=1 and the action of pentaqubit-spinor field as hyperqubit-spinor field in 11D hyper-spacetime}

Let us begin with considering local coherent states of five qubits, $\Psi_{\mQ^5}(x)$ and $\Psi_{\QS^5}(x)$, based on the two types of 5-product qubit-basis states,
\be
\Psi_{\mQ^5}(x) & = &  \sum_{s_1, \cdots, s_{5}}  \psi_{s_1\cdots s_{5}}(x)  \varsigma_{s_1\cdots s_{5}} \equiv \sum_{s_1, \cdots, s_{5}}  \psi_{s_1\cdots s_{5}}(x)\,  \varsigma_{s_1}\otimes \cdots \otimes \varsigma_{s_{5}}, \nn \\
 \Psi_{\QS^5}(x) & = &  \sum_{s_1, \cdots, s_{5}}  \psi_{s_1\cdots s_{5}}(x)  \zeta_{s_1\cdots s_{5}} \equiv \sum_{s_1, \cdots, s_{5}}  \psi_{s_1\cdots s_{5}}(x)\,  \zeta_{s_1}\otimes \cdots \otimes \zeta_{s_{5}}, \nn \\
 & \equiv & \sum_{s_1, \cdots, s_{5}}  \tilde{\psi}_{s_1\cdots s_{5}}(x)  \varsigma_{s_1\cdots s_{5}} \equiv \sum_{s_1, \cdots, s_{5}}  \tilde{\psi}_{s_1\cdots s_{5}}(x)\,  \varsigma_{s_1}\otimes \cdots \otimes \varsigma_{s_{5}},
\ee
which are regarded as hyperqubit-spinor fields with $\cD_H=2^{Q_N}=32$ local distribution amplitudes $\psi_{s_1\cdots s_{5}}(x)$ that span 32-dimensional Hilbert space. For convenience, we may refer to $\Psi_{\mQ^5}(x)$ and $\Psi_{\QS^5}(x)$ as {\it pentaqubit-spinor field} and {\it self-conjugated pentaqubit-spinor field}, respectively.

Based on the local coherent-qubits motion postulate with following along the maximum coherence motion principle, the action of pentaqubit-spinor field $\Psi_{\mQ^5}(x)$ as hyperqubit-spinor field is constructed to be:
 \be \label{actionQ5}
 \cS_{\mQ^5}  & = &  \int d^{11}x \{  \frac{1}{2}  \Psi_{\mQ^5}^{\dagger}(x)  \delta_{\mA}^{\; \mM}  \vUps^{\mA} i\p_{\mM} \Psi_{\mQ^5}(x)   - \frac{1}{2}\lambda_1 \phi(x) \Psi_{\mQ^5}^{\dagger}(x) \vtUps \Psi_{\mQ^5}(x) \}, 
\ee
with $\mA, \mM= 0, 1, \cdots, 10$. Where eleven $32\times 32$ real symmetric normalized $\cM_c$-matrices $\vUps^{\mA} = ( \vUps^0, \vUps^{\bmA}, \vUps^9, \vUps^{10} )$ ($\bmA=1,\cdots, 8$) and single $\cQ_c$-matrix are given by the following forms:
\be \label{UpsM5}
& & \vUps^{0} = \sigma_0 \otimes \Ups^0, \quad \vUps^{\bmA} = \sigma_3 \otimes \Ups^{\bA} ,  \nn \\
& & \vUps^{9} = \sigma_3 \otimes \Ups^{9}, \quad \vUps^{10} = \sigma_1 \otimes \Ups^0 , \nn \\
& & \vtUps = \sigma_2 \otimes \Ups^0, 
\ee
where $\Ups^0, \Ups^{\bA}, \Ups^9$ are $16\times 16$ real symmetric normalized $\cM_c$-matrices defined in Eq.(\ref{UpsM4}). $\vtUps$ is an imaginary antisymmetric normalized matrix which anticommutes with real symmetric normalized $\cM_c$-matrices except $\vUps^{0}$. 

It can be verified that $\vUps^{\mA} $ and $\vtUps$ satisfy the following Clifford algebra relations:
\be \label{CAQ5}
& & \{\vUps^{\mA}, \vUps^{\mB} \} =2 \delta^{\mA\mB}, \quad \{ \vUps^{0}, \vUps^{\mA} \} = 2 \vUps^{\mA}, \nn \\
& & \{ \vtUps, \vUps^{\mA}\} = 0, \quad \mA, \mB = 1, \cdots , 10 ,
\ee
which indicates that the $\cQ_c$-spin charge of pentaqubit-spinor field $\Psi_{\mQ^5}(x)$ is given by the following relation: 
\be
\CQc = q_c= Q_N-4 =1 .
\ee
Such a $\cQ_c$-spin charge reproduces the same one as that of uniqubit-spinor field in three-dimensional Minkowski spacetime. In structure, the three real symmetric normalized $\cM_c$-matrices $\vUps^0, \vUps^9, \vUps^{10}$ and the single antisymmetric $\cQ_c$-matrix $\vtUps$ are analogous to those for uniqubit-spinor field presented in Eqs.(\ref{UpsQ1}) and (\ref{tUpsQ1}). The remaining eight real symmetric normalized $\cM_c$-matrices $\vUps^{\bA}$ ($\bA=1,\cdots, 8$) are correlated to extra eight spatial dimensions. 

The action for self-conjugated pentaqubit-spinor field $\Psi_{\QS^5}(x)$ as hyperqubit-spinor field is constructed as follows:
 \be \label{actionM5}
 \cS_{\QS^5}  & = &  \int d^{11}x  \{ \frac{1}{2}  \bar{\Psi}_{\QS^5}(x)  \delta_{\mA}^{\; \mM}  \vGa^{\mA} i\p_{\mM} \Psi_{\QS^5}(x) - \frac{1}{2}\lambda_1 \phi(x) \bar{\Psi}_{\QS^5} (x) \Psi_{\QS^5}(x) \} , 
\ee
with $\mA, \mM= 0,1,2,3,5, \cdots, 11$. The $32\times 32$ $\vGa$-matrices  $\vGa^{\mA}= (\vGa^{0}, \vGa^{\tmA}, \vGa^{10}, \vGa^{11}) $ ($\tmA=  1,2,3,5, \cdots, 9$) get the following structures:
\be \label{GM5}
& & \vGa^{0} =\; \sigma_3 \otimes\Gamma^{0} ,  \quad \vGa^{\tmA} = \; \sigma_3 \otimes\Gamma^{\tA} , \nn \\
& & \vGa^{10} = i\sigma_2 \otimes \Gamma^{10}, \quad \vGa^{11} = i\sigma_1 \otimes \Gamma^{10}, 
\ee
with $16\times 16$ matrices ($\Gamma^{0}$, $\Gamma^{\tA}$, $\Gamma^{10}$) given in Eq.(\ref{TGM}). Obviously, three-dimensional spacetime in association with $\Gamma$-matrices $\Gamma^0$, $\Gamma^{10}$ and $\Gamma^{11}$ is formally in correspondence to that for self-conjugated uniqubit-spinor field shown in Eq.(\ref{GM1}). The other eight $\vGa$-matrices $\vGa^{\tmA}$ ($\tmA=1,2,3,5,\cdots, 9$) characterize the maximally correlated free motion in extra eight spatial dimensions.

The self-conjugated pentaqubit-spinor field $\Psi_{\QS^5}(x)$ as hyperqubit-spinor field gets the following explicit form:
\be
& & \Psi_{\QS^5}(x) =\binom{\psi_{\QC^4}}{\psi_{\hQC^4} } =  \begin{pmatrix} 
\psi_{\mQ^3_{\mC+}}(x) \\
\psi_{\mQ^3_{\mC-}}(x) \\
\psi_{\hat{\mQ}^3_{\mC  -}} (x) \\
\psi_{\hat{\mQ}^3_{\mC  +}} (x)
 \end{pmatrix}  =   \begin{pmatrix} 
\psi_{\mQ^2_{\mC++}}(x) \\
\psi_{\mQ^2_{\mC+-}}(x) \\
\psi_{\mQ^2_{\mC-+}}(x) \\
\psi_{\mQ^2_{\mC--}}(x) \\
\psi_{\hat{\mQ}^2_{\mC  --}} (x) \\
\psi_{\hat{\mQ}^2_{\mC  -+}} (x) \\
\psi_{\hat{\mQ}^2_{\mC  +-}} (x) \\
\psi_{\hat{\mQ}^2_{\mC  ++}} (x)
\end{pmatrix}  ,  \nn \\
& & \psi_{\hQC^4} = \sigma_1\otimes \sigma_1\otimes \sigma_1\otimes \sigma_1\psi_{\QC^4}^{\ast},
\ee
which satisfies the following self-conjugate condition:
\be
& & \Psi^c_{\QS^5}(x) = C_{\QS^5} \bar{\Psi}^T_{\QS^5}(x) = \Psi_{\QS^5}(x), \nn \\
& & C_{\QS^5} = -i \sigma_2 \otimes \sigma_2 \otimes\sigma_2 \otimes \sigma_2 \otimes \sigma_2 .
\ee

It can be verified that the actions given in Eqs.(\ref{actionQ5}) and (\ref{actionM5}) possess the following associated symmetry:
\be
G_S & = & SC(1)\ltimes P^{1,10}\ltimes SO(1,10)\adjoin SP(1, 10) \rtimes SG(1) \times SP(q_c=1)  \nn \\
& = & SC(1)\ltimes  PO(1, 10) \adjoin SP(1,10) \rtimes SG(1) \times SP(1) , 
\ee
where SP(1,10) represents maximal $\cM_c$-spin symmetry as hyperspin symmetry and SP(1)$\cong$O(1) denotes $\cQ_c$-spin symmetry with $\cQ_c$-spin charge $\CQc=q_c= 1$. PO(1,10) = P$^{1,10}\ltimes$ SO(1,10) is Poincar\'e-type/inhomogeneous Lorentz-type group symmetry with translation group symmetry P$^{1,10}$ in 11-dimensional Minkowski hyper-spacetime of coordinates. SC(1)$\adjoin$SG(1) labels an associated symmetry for global scaling transformations.

\subsection{Local coherent state of six qubits with $\cQ_c$-spin charge $\CQc$=2 and the action of hexaqubit-spinor field as hyperqubit-spinor field in 12D hyper-spacetime }

We now examine {\it local coherent state of six qubits} as hyperqubit-spinor field which is formed from 6-product qubit-basis states. The explicit forms are constructed as follows: 
\be
\Psi_{\mQ^6}(x) & = &  \sum_{s_1, \cdots, s_{6}}  \psi_{s_1\cdots s_{6}}(x)  \varsigma_{s_1\cdots s_{6}} \equiv \sum_{s_1, \cdots, s_{6}}  \psi_{s_1\cdots s_{6}}(x)\,  \varsigma_{s_1}\otimes \cdots \otimes \varsigma_{s_{6}}, \nn \\
 \Psi_{\QS^6}(x) & = &  \sum_{s_1, \cdots, s_{6}}  \psi_{s_1\cdots s_{6}}(x)  \zeta_{s_1\cdots s_{6}} \equiv \sum_{s_1, \cdots, s_{6}}  \psi_{s_1\cdots s_{6}}(x)\,  \zeta_{s_1}\otimes \cdots \otimes \zeta_{s_{6}}, \nn \\
 & \equiv & \sum_{s_1, \cdots, s_{6}}  \tilde{\psi}_{s_1\cdots s_{6}}(x)  \varsigma_{s_1\cdots s_{6}} \equiv \sum_{s_1, \cdots, s_{6}}  \tilde{\psi}_{s_1\cdots s_{6}}(x)\,  \varsigma_{s_1}\otimes \cdots \otimes \varsigma_{s_{6}},
\ee
where $\Psi_{\mQ^6}(x)$ is referred to as {\it hexaqubit-spinor field} and $\Psi_{\QS^6}(x)$ as {\it self-conjugated hexaqubit-spinor field} with qubit number $Q_N=6$. The local distribution amplitudes $\psi_{s_1\cdots s_{6}}(x)$ contain $\cD_H=2^{Q_N}=64$ independent degrees of freedom, which spans a 64-dimensional Hilbert space.

Again based on the local coherent-qubits motion postulate with following along the maximum coherence motion principle, the action of hexaqubit-spinor field $\Psi_{\mQ^6}(x)$ as hyperqubit-spinor field is found to be: 
 \be \label{actionQ6}
 \cS_{\mQ^6}  & = &  \int d^{12}x \{  \frac{1}{2}  \Psi_{\mQ^6}^{\dagger}(x)  \delta_{\mA}^{\; \mM}  \vUps^{\mA} i\p_{\mM} \Psi_{\mQ^6}(x)  - \frac{1}{2}\lambda_2 \phi_p(x) \Psi_{\mQ^6}^{\dagger}(x) \vtUps^p \Psi_{\mQ^6}(x) \} , 
\ee
with $\mA, \mM= 0, 1, \cdots, 11$. The $64\times 64$ real symmetric normalized $\cM_c$-matrices $\vUps^{\mA}$ with the notation $\vUps^{\mA} = ( \vUps^0, \vUps^{\bmA}, \vUps^9, \vUps^{10}, \vUps^{11} )$ ($\bmA=1,\cdots, 8$) can be expressed into the following general form: 
\be \label{UpsM6}
& & \vUps^{0} =  \sigma_0 \otimes \sigma_0 \otimes \Ups^0, \quad  \vUps^{\bmA} =  \sigma_0 \otimes \sigma_3 \otimes \Ups^{\bA},  \nn \\
& & \vUps^{9} = \sigma_0 \otimes \sigma_3 \otimes \Ups^{9}, \quad \vUps^{10} =  \sigma_0 \otimes \sigma_1 \otimes \Ups^0 , \nn \\
& & \vUps^{11}   =   \sigma_2 \otimes \sigma_2 \otimes \Ups^0  , \nn \\
& & \vtUps^1 = \sigma_3 \otimes  \sigma_2 \otimes \Ups^0, \quad \vtUps^2  =  \sigma_1 \otimes  \sigma_2 \otimes \Ups^0,
\ee
with $\Ups^0, \Ups^{\bA}, \Ups^9$ the $16\times 16$ real symmetric normalized $\cM_c$-matrices given in Eq.(\ref{UpsM4}). $\vtUps^p$ ($p=1,2$) are two imaginary antisymmetric normalized $\cQ_c$-matrices which anticommute with real symmetric normalized $\cM_c$-matrices except the unit matrix $\vUps^{0}$. 
 
The matrices $\vUps^{\mA} $ and $\vtUps^p$ satisfy the following Clifford algebra relations:
\be \label{CAQ6}
& & \{\vUps^{\mA}, \vUps^{\mB} \} = 2\delta^{\mA\mB}, \quad \{ \vUps^{0}, \vUps^{\mA} \} = 2 \vUps^{\mA}, \quad \{ \vtUps^p, \vtUps^{q}\} = 2\delta^{pq},  \nn \\ 
& & \{ \vtUps^p, \vUps^{\mA}\} = 0,  \quad \mA, \mB = 1, \cdots , 11, \quad p, q =1, 2 ,
\ee
which implies that the $\cQ_c$-spin charge of hexaqubit-spinor field $\Psi_{\mQ^6}(x)$ has the following relation with qubit number $Q_N$:
\be
\CQc=q_c=Q_N-4=2,
\ee 
which is equal to the $\cQ_c$-spin charge of biqubit-spinor field in four-dimensional Minkowski spacetime.

Actually, four real symmetric normalized $\cM_c$-matrices ($\vUps^0, \vUps^9, \vUps^{10}, \vUps^{11}$) and two imaginary antisymmetric normalized $\cM_c$-matrices ($\vtUps^1, \vtUps^2$) are analogous to those for biqubit-spinor field as shown in Eqs.(\ref{UpsQ2}) and (\ref{tUpsQ2}). The remaining eight real symmetric normalized $\cM_c$-matrices $\vUps^{\bA}$ ($\bA=1,\cdots, 8$) characterize extra eight spatial dimensions. 

In an analogous way, the action of self-conjugated hexaqubit-spinor field $\Psi_{\QS^6}(x)$ as hyperqubit-spinor field can be written as follows:
 \be \label{actionM6}
 \cS_{\QS^6}  & = &  \int d^{12}x \, \{ \frac{1}{2}  \bar{\Psi}_{\QS^6}(x)  \delta_{\mA}^{\; \mM}  \vGa^{\mA} i\p_{\mM} \Psi_{\QS^6}(x) - \frac{1}{2}\lambda_2 \phi_p(x) \bar{\Psi}_{\QS^6}(x)\tvGa^p \Psi_{\QS^6}(x) \} , 
\ee
with $\mA, \mM= 0,1,2,3,5, \cdots, 12$. The $64\times 64$ $\vGa$-matrices  $\vGa^{\mA}= (\vGa^{0},  \vGa^{\tmA}, \vGa^{10}, \vGa^{11},  \vGa^{12}) $ ($\tmA=  1,2,3,5, \cdots, 9$) and $\tvGa^p$ ($p=1,2$) can be constructed as follows: 
\be \label{GM6}
& &  \vGa^{0} = \sigma_2 \otimes \sigma_3 \otimes \Gamma^{0} , \quad  \vGa^{\tmA} = \sigma_2 \otimes \sigma_3 \otimes\Gamma^{\tA}  , \nn \\
& &  \vGa^{10} = -i\sigma_2 \otimes \sigma_2 \otimes \Gamma^{10}, \quad \vGa^{11} = i \sigma_2 \otimes \sigma_1 \otimes I_{16},  \nn \\
& & \vGa^{12} = -i \sigma_1 \otimes \sigma_0 \otimes I_{16},   \nn \\
& & \tvGa^1 = i\sigma_3 \otimes \sigma_0 \otimes I_{16}, \quad \tvGa^2= \sigma_0 \otimes \sigma_0 \otimes I_{16} , 
\ee
with ($ \Gamma^{0}, \Gamma^{\tA}, \Gamma^{10}$) presented in Eq.(\ref{TGM}). Where four $\vGa$-matrices ($\vGa^0, \vGa^9, \vGa^{10}, \vGa^{11}$) and two $\tvGa$-matrices ($\tvGa^1, \tvGa^2$) are similar to those for self-conjugated biqubit-spinor field in four-dimensional Minkowski spacetime as shown in Eq.(\ref{GM2}).

The self-conjugated hexaqubit-spinor field $\Psi_{\QS^6}(x)$ as hyperqubit-spinor field has the following spinor structuce:
\be
& & \Psi_{\QS^6}(x) =\binom{\psi_{\QC^5}}{\psi_{\hQC^5} } =  \begin{pmatrix} 
\psi_{\mQ^4_{\mC+}}(x) \\
\psi_{\mQ^4_{\mC-}}(x) \\
\psi_{\hat{\mQ}^4_{\mC  -}} (x) \\
\psi_{\hat{\mQ}^4_{\mC  +}} (x)
 \end{pmatrix}  =   \begin{pmatrix} 
\psi_{\mQ^3_{\mC++}}(x) \\
\psi_{\mQ^3_{\mC+-}}(x) \\
\psi_{\mQ^3_{\mC-+}}(x) \\
\psi_{\mQ^3_{\mC--}}(x) \\
\psi_{\hat{\mQ}^3_{\mC  --}} (x) \\
\psi_{\hat{\mQ}^3_{\mC  -+}} (x) \\
\psi_{\hat{\mQ}^3_{\mC  +-}} (x) \\
\psi_{\hat{\mQ}^3_{\mC  ++}} (x)
\end{pmatrix}  ,  \nn \\
& & \psi_{\hQC^5} = \sigma_1\otimes \sigma_1\otimes \sigma_1\otimes \sigma_1\otimes \sigma_1\psi_{\QC^5}^{\ast} ,
\ee
which satisfies the self-conjugate condition,
\be
& & \Psi^c_{\QS^6}(x) = C_{\QS^6} \bar{\Psi}^T_{\QS^6}(x) = \Psi_{\QS^6}(x),  \nn \\
& & C_{\QS^6} =   \sigma_3 \otimes \sigma_2 \otimes \sigma_2 \otimes\sigma_2 \otimes \sigma_2 \otimes \sigma_2 .
\ee

The actions presented in Eq.(\ref{actionQ6}) and (\ref{actionM6}) get the following associated symmetry:
\be
G_S & = & SC(1)\ltimes P^{1,11}\ltimes SO(1,11)\adjoin SP(1,11)  \times SG(1) \times SP(2) \nn \\
& = & SC(1)\ltimes  PO(1,11) \adjoin SP(1,11) \times SG(1)\times SP(2), 
\ee
where SP(1,11) denotes maximal $\cM_c$-spin symmetry as hyperspin symmetry. SP(2)$\cong$U(1) is the $\cQ_c$-spin symmetry with respect to $\cQ_c$-spin charge $\CQc=q_c=2$. PO(1,11) = P$^{1,11}\times$ SO(1,11) represents Poincar\'e-type/inhomogeneous Lorentz-type group symmetry with SO(1,11) the Lorentz-type group symmetry and P$^{1,11}$ the translation group symmetry in 12-dimensional Minkowski hyper-spacetime.


\subsection{Local coherent state of seven qubits with $\cQ_c$-spin charge $\CQc$=3 and the action of heptaqubit-spinor field as hyperqubit-spinor field in 14D hyper-spacetime }

We now come to consider a local coherent state which is form by 7-product qubit-basis states with qubit number $Q_N=7$:  
\be
\Psi_{\mQ^7}(x) & = &  \sum_{s_1, \cdots, s_{7}}  \psi_{s_1\cdots s_{7}}(x)  \varsigma_{s_1\cdots s_{7}} \equiv \sum_{s_1, \cdots, s_{7}}  \psi_{s_1\cdots s_{7}}(x)\,  \varsigma_{s_1}\otimes \cdots \otimes \varsigma_{s_{7}}, \nn \\
\Psi_{\QS^7}(x) & = &  \sum_{s_1, \cdots, s_{7}}  \psi_{s_1\cdots s_{7}}(x)  \zeta_{s_1\cdots s_{7}} \equiv \sum_{s_1, \cdots, s_{7}}  \psi_{s_1\cdots s_{7}}(x)\,  \zeta_{s_1}\otimes \cdots \otimes \zeta_{s_{7}}, \nn \\
 & \equiv & \sum_{s_1, \cdots, s_{7}}  \tilde{\psi}_{s_1\cdots s_{7}}(x)  \varsigma_{s_1\cdots s_{7}} \equiv \sum_{s_1, \cdots, s_{7}}  \tilde{\psi}_{s_1\cdots s_{7}}(x)\,  \varsigma_{s_1}\otimes \cdots \otimes \varsigma_{s_{7}} ,
\ee
we may refer to hyperqubit-spinor fields $\Psi_{\mQ^7}(x)$ and $\Psi_{\QS^7}(x)$ as {\it heptaqubit-spinor field} and {\it self-conjugated heptaqubit-spinor field}, respectively. They involve $\cD_H=2^{Q_N}= 128 $ local distribution amplitudes $\psi_{s_1\cdots s_{7}}(x)$ which span 128-dimensional Hilbert space.

The maximally correlated motion of heptaqubit-spinor field $\Psi_{\mQ^7}(x)$ brings on the following hermitian action in 14-dimensional Minkowski spacetime:
 \be \label{actionQ7}
 \cS_{\mQ^7}  & = &  \int d^{14}x \frac{1}{2}  \Psi_{\mQ^7}^{\dagger}(x)  \delta_{\mA}^{\; \mM}  \vUps^{\mA} i\p_{\mM} \Psi_{\mQ^7}(x) , 
\ee
with $\mA, \mM= 0, 1, \cdots, 13$. Where $\vUps^{\mA}$ are $128\times 128$ real symmetric normalized $\cM_c$-matrices, which can be expressed as follows with the notation $\vUps^{\mA} = ( \vUps^0, \vUps^{\bmA}, \vUps^9, \vUps^{10}, \vUps^{11} , \vUps^{12}, \vUps^{13})$ ($\bmA=1,\cdots, 8$):
\be \label{UpsM7}
& & \vUps^0 = \sigma_0 \otimes \sigma_0 \otimes \sigma_0 \otimes  \Ups^0 , \quad  \vUps^{\bmA} = \sigma_3\otimes  \sigma_0\otimes \sigma_3 \otimes  \Ups^A, \nn \\
& & \vUps^9 = \sigma_3\otimes  \sigma_0\otimes \sigma_3 \otimes  \Ups^9,\quad \vUps^{10} = \sigma_3\otimes  \sigma_0\otimes \sigma_1 \otimes \Ups^0 , \nn \\
& &  \vUps^{11} = \sigma_3\otimes  \sigma_2\otimes \sigma_2 \otimes \Ups^0, \quad \vUps^{12} = \sigma_1\otimes  \sigma_0\otimes \sigma_0 \otimes \Ups^0 ,  \nn \\
& & \vUps^{13} = \sigma_2\otimes  \sigma_2\otimes \sigma_0 \otimes \Ups^0,
\ee
with $\Ups^0, \Ups^{\bA}, \Ups^9$ the $16\times 16$ real symmetric normalized $\cM_c$-matrices given in Eq.(\ref{UpsM4}). The three imaginary antisymmetric normalized $\cQ_c$-matrices $\vtUps^p$ ($p=1,2,3$) are found to be:
\be \label{tUpsM3}
& & \vtUps^1 = \sigma_3\otimes \sigma_1\otimes \sigma_2 \otimes \Ups^0, \nn \\
& & \vtUps^2 = \sigma_3\otimes \sigma_3\otimes \sigma_2 \otimes \Ups^0, \nn \\
& & \vtUps^3 = \sigma_2\otimes \sigma_0\otimes \sigma_0 \otimes \Ups^0 ,
\ee
which are anti-commuting with real symmetric normalized $\cM_c$-matrices except two matrices $\vUps^0$ and $\vUps^{13}$. The matrices $\vUps^{\mA} $ and $\vtUps^p$ satisfy the following Clifford algebra relations:
\be \label{CAQ7}
& & \{\vUps^{\mA}, \vUps^{\mB} \} = 2\delta^{\mA\mB}, \quad \{ \vUps^{0}, \vUps^{\mA} \} =  2\vUps^{\mA}, \quad \{ \vtUps^p, \vtUps^{q}\} = 2\delta^{pq},  \nn \\ 
& &\mA, \mB = 1, \cdots , 13, \quad p, q =1, 2 , 3 ; \quad  \{ \vtUps^p, \vUps^{\mA}\} = 0, \; \mA\neq 0,13, 
\ee
where six real symmetric normalized $\cM_c$-matrices ($\vUps^0, \vUps^9, \vUps^{10}, \vUps^{11}, \vUps^{12}, \vUps^{13}$) and three antisymmetric normalized $\cQ_c$-matrices ($\vtUps^1, \vtUps^2, \vtUps^3$) are analogous to matrices for triqubit-spinor field in six-dimensional Minkowski spacetime as shown in Eqs.(\ref{CAQ3}) and (\ref{ACAQ3}). The other eight real symmetric normalized $\cM_c$-matrices $\vUps^{\bmA}$ ($\bmA=1,\cdots, 8$) correlate to extra eight spatial dimensions. The $\cQ_c$-spin charge of heptaqubit-spinor field is the same as that of triqubit-spinor field, i.e., $\CQc=q_c=3$.

The action of self-conjugated heptaqubit-spinor field as hyperqubit-spinor field $\Psi_{\QS^7}(x)$ is easily read:
 \be \label{actionM7}
 \cS_{\QS^7}  & = &  \int d^{14}x  \frac{1}{2}  \bar{\Psi}_{\QS^7}(x)  \delta_{\mA}^{\; \mM}  \vGa^{\mA} i\p_{\mM} \Psi_{\QS^7}(x) , 
\ee
with $\mA, \mM= 0,1,2,3,5, \cdots, 14$. The fourteen $128\times 128$ $\Gamma$-matrices $\vGa^{\mA} =  (\vGa^{0}, \vGa^{\tmA}, \vGa^{10}, \vGa^{11},  \vGa^{12} , \vGa^{13},  \vGa^{14} ) $ ($\tmA=  1,2,3,5, \cdots, 9$) are given by the following forms:  
\be \label{GM7}
& &  \vGa^{0} = \sigma_1 \otimes \sigma_2 \otimes \sigma_3 \otimes \Gamma^{0} , \quad \vGa^{10} = -i \sigma_1 \otimes \sigma_2\otimes \sigma_2 \otimes \Gamma^{10}, \nn \\
& &  \vGa^{11} = i \sigma_1 \otimes \sigma_2\otimes \sigma_1 \otimes I_{16}, \quad \vGa^{12} = -i \sigma_1 \otimes \sigma_1\otimes \sigma_0 \otimes I_{16}, \nn \\
& & \vGa^{13} = - i \sigma_2 \otimes \sigma_0\otimes \sigma_0 \otimes I_{16}, \quad \vGa^{14} = \sigma_2\otimes  \sigma_3\otimes \sigma_0  \otimes I_{16}, \nn \\
& & \vGa^{\tmA} = \sigma_1 \otimes \sigma_2 \otimes \sigma_3 \otimes\Gamma^{\tA} ,  \quad \tmA = 1,2,3,5, \cdots, 9,  
\ee
with ($ \Gamma^{0},  \Gamma^{\tA}, \Gamma^{10}$) presented in Eq.(\ref{TGM}). Where six $\vGa$-matrices ($\vGa^0, \vGa^{10}, \vGa^{11}, \vGa^{12}, \vGa^{13},  \vGa^{14}$) are analogous to those for self-conjugated triqubit-spinor field in six-dimensional Minkowski spacetime as shown in Eq.(\ref{GM3}).

The self-conjugated heptaqubit-spinor field  $\Psi_{\QS^7}(x)$ as hyperqubit-spinor field is defined as follows:
\be
& & \Psi_{\QS^7}(x) =\binom{\psi_{\QC^6}}{\psi_{\hQC^6} } =  \begin{pmatrix} 
\psi_{\mQ^5_{\mC+}}(x) \\
\psi_{\mQ^5_{\mC-}}(x) \\
\psi_{\hat{\mQ}^5_{\mC  -}} (x) \\
\psi_{\hat{\mQ}^5_{\mC  +}} (x)
 \end{pmatrix}  =   \begin{pmatrix} 
\psi_{\mQ^4_{\mC++}}(x) \\
\psi_{\mQ^4_{\mC+-}}(x) \\
\psi_{\mQ^4_{\mC-+}}(x) \\
\psi_{\mQ^4_{\mC--}}(x) \\
\psi_{\hat{\mQ}^4_{\mC  --}} (x) \\
\psi_{\hat{\mQ}^4_{\mC  -+}} (x) \\
\psi_{\hat{\mQ}^4_{\mC  +-}} (x) \\
\psi_{\hat{\mQ}^4_{\mC  ++}} (x)
\end{pmatrix}  ,  \nn \\
& & \psi_{\hQC^6} = \sigma_1\otimes \sigma_1\otimes \sigma_1\otimes \sigma_1\otimes \sigma_1\otimes \sigma_1\psi_{\QC^6}^{\ast} ,
\ee
which satisfies the following self-conjugate condition:
\be
& & \Psi^c_{\QS^7}(x) = C_{\QS^7} \bar{\Psi}^T_{\QS^7}(x) = \Psi_{\QS^7}(x), \nn \\
& & C_{\QS^7} =  \sigma_0 \otimes \sigma_3 \otimes \sigma_2 \otimes \sigma_2 \otimes\sigma_2 \otimes \sigma_2 \otimes \sigma_2 .
\ee

The actions given in Eq.(\ref{actionQ7}) and (\ref{actionM7}) turn out to have the following associated symmetry:
\be
G_S & = & SC(1)\ltimes P^{1,13}\ltimes SO(1,13)\adjoin SP(1,13)  \rtimes SG(1) \times SP(3) \nn \\
& = & SC(1)\ltimes  PO(1,13) \adjoin SP(1,13) \rtimes SG(1)\times SU(2), 
\ee
where SP(1,13) is hyperspin symmetry and PO(1,13) = P$^{1,13}\ltimes$ SO(1,13) is Poincar\'e-type/inhomogeneous Lorentz-type group symmetry in 14-dimensional Minkowski hyper-spacetime. SP(3)$\cong$SU(2) represents $\cQ_c$-spin symmetry which has the following group generators:
\be
& & \lambda_1= \frac{1}{2} \sigma_1 \otimes \sigma_3\otimes \sigma_0 \otimes I_{16} , \nn \\
& & \lambda_2= \frac{1}{2} \sigma_2 \otimes \sigma_3\otimes \sigma_0 \otimes I_{16} , \nn \\
& & \lambda_3 =  \frac{1}{2} \sigma_3 \otimes \sigma_0\otimes \sigma_0 \otimes I_{16} . \nn
\ee
It is obvious that the $\cQ_c$-spin charge is given by, 
\be
\CQc=q_c=Q_N-4=3 .
\ee


\subsection{Local coherent state of eight qubits and the action of octoqubit-spinor field as hyperqubit-spinor field in 18D hyper-spacetime in analogous to tetraqubit-spinor field in 10D hyper-spacetime and zeroqubit-spinor field in 2D spacetime} 

Let us analyze local coherent states of eight qubits formed from two types of 8-product qubit-basis states. They are hyperqubit-spinor fields $\Psi_{\mQ^8}(x)$ and $\Psi_{\QS^8}(x)$ in real qubit-basis and complex conjugated qubit-basis, respectively, with the following definitions:
\be \label{QS8}
\Psi_{\mQ^8}(x) & = &  \sum_{s_1, \cdots, s_{8}}  \psi_{s_1\cdots s_{8}}(x)  \varsigma_{s_1\cdots s_{8}} \equiv \sum_{s_1, \cdots, s_{8}}  \psi_{s_1\cdots s_{8}}(x)\,  \varsigma_{s_1}\otimes \cdots \otimes \varsigma_{s_{8}}, \nn \\
 \Psi_{\QS^8}(x) & = &  \sum_{s_1, \cdots, s_{8}}  \psi_{s_1\cdots s_{8}}(x)  \zeta_{s_1\cdots s_{8}} \equiv \sum_{s_1, \cdots, s_{8}}  \psi_{s_1\cdots s_{8}}(x)\,  \zeta_{s_1}\otimes \cdots \otimes \zeta_{s_{8}}, \nn \\
 & \equiv & \sum_{s_1, \cdots, s_{8}}  \tilde{\psi}_{s_1\cdots s_{8}}(x)  \varsigma_{s_1\cdots s_{8}} \equiv \sum_{s_1, \cdots, s_{8}}  \tilde{\psi}_{s_1\cdots s_{8}}(x)\,  \varsigma_{s_1}\otimes \cdots \otimes \varsigma_{s_{8}},
\ee
where $\Psi_{\mQ^8}(x)$ is referred to as {\it octoqubit-spinor field} and $\Psi_{\QS^8}(x)$ as {\it self-conjugated octoqubit-spinor field}. They contain $\cD_H=2^8=256$ local distribution amplitudes $\psi_{s_1\cdots s_{8}}(x)$ which span 256-dimensional Hilbert space.

Based on the local coherent-qubits motion postulate with following along the maximum coherence motion principle, we obtain the following action for octoqubit-spinor field $\Psi_{\mQ^8}(x)$:
 \be \label{actionQ8}
 \cS_{\mQ^8}  & = &  \int d^{18}x \frac{1}{2} \Psi_{\mQ^8}^{\dagger}(x)  \delta_{\mA}^{\; \mM}  \vUps^{\mA} i\p_{\mM} \Psi_{\mQ^8}(x) , 
\ee
with $\mA, \mM= 0, 1, \cdots, 17$. Where the eighteen $2^8\times 2^8$ real symmetric normalized $\cM_c$-matrices $\vUps^{\mA} = ( \vUps^0,  \vUps^{\bmA}, \vUps^{\bmA +8}, \vUps^{9+8} )$ ($\bmA=1,\cdots, 8$) have the following structures:
\be \label{UpsM8}
& & \vUps^{0} = \Ups^0 \otimes \Ups^0, \;\;  \vUps^{9+8} =  \Ups^9 \otimes \Ups^{9},  \nn \\
& & \vUps^{\bmA} = \Ups^{0} \otimes \Ups^{\bA} , \quad \vUps^{\bmA + 8} =  \Ups^{\bA}   \otimes \Ups^{9} ,  \; \; \bmA=1,\cdots, 8 ,
\ee
with $\Ups^0, \Ups^{\bA}, \Ups^9$ the $16\times 16$ real symmetric normalized $\cM_c$-matrices defined in Eq.(\ref{UpsM4}). The matrices $\vUps^{\mA} $ satisfy the following Clifford algebra:
\be \label{CAQ8}
\{\vUps^{\mA}, \vUps^{\mB} \} = 2\delta^{\mA\mB}, \quad \{ \vUps^{0}, \vUps^{\mA} \} = 2 \vUps^{\mA}, \quad \mA, \mB = 1, \cdots , 17.
\ee

As there exists no antisymmetric normalized $\cQ_c$-matrix which anticommutes with real symmetric normalized $\cM_c$-matrices, the octoqubit-spinor field $\Psi_{\mQ^8}(x)$ as hyperqubit-spinor field has zero $\cQ_c$-spin charge. Such a $\cQ_c$-spin charge is given by a relation with qubit number as follows:
\be
\CQc=q_c=Q_N-8= 0, 
\ee
which repeats the cases for tetraqubit-spinor field in ten-dimensional Minkowski hyper-spacetime and zeroqubit-spinor field in two-dimensional Minkowski spacetime. Where the ten real symmetric normalized $\cM_c$-matrices $\vUps^0, \vUps^{\bmA}, \vUps^{9+8}$ are similar to those of tetraqubit-spinor field in 10-dimensional Minkowski hyper-spacetime as shown in Eq.(\ref{UpsM4}), and the remaining eight real symmetric normalized $\cM_c$-matrices $\vUps^{\bmA + 8}$ ($\bmA=1,\cdots, 8$) involve additional eight spatial dimensions. Alternatively, the two real symmetric normalized $\cM_c$-matrices $\vUps^0, \vUps^{9+8}$ are similar to the ones for self-conjugated chiral uniqubit-spinor field or zeroqubit-spinor field in two-dimensional Minkowski spacetime as shown in Eq.(\ref{UpsQ1}), while the other sixteen real symmetric normalized $\cM_c$-matrices $ \vUps^{\bmA}$ and $\vUps^{\bmA + 8}$ ($\bmA=1,\cdots, 8$) concern extra sixteen spatial dimensions.

The action for self-conjugated octoqubit-spinor field $\Psi_{\QS^8}(x)$ as hyperqubit-spinor field can be expressed as follows:
 \be \label{actionM8}
 \cS_{\QS^8}  & = &  \int d^{18}x  \frac{1}{2}  \bar{\Psi}_{\QS^8}(x)  \delta_{\mA}^{\; \mM}  \vGa^{\mA} i\p_{\mM} \Psi_{\QS^8}(x) , 
\ee
with $\mA, \mM= 0,1,2,3,5, \cdots, 18$. Where $2^8\times 2^8$ $\Gamma$-matrices  $\vGa^{\mA}= (\vGa^{0}, \vGa^{\tmA}, \vGa^{\bmA+9}, \vGa^{10+8}) $ ($\tmA=  1,2,3,5, \cdots, 9$, $\bmA=  1,\cdots, 8$) have the following forms:
\be \label{GM8}
& & \vGa^{0} = \Gamma^0 \otimes\Gamma^{0} ,  \quad \vGa^{\tmA} = \Gamma^{0} \otimes\Gamma^{\tA} , \nn \\
& & \vGa^{\bmA+9} = \Gamma^{\tA} \otimes\Gamma^{10}, \quad \vGa^{10+8} = \Gamma^{10} \otimes \Gamma^{10}, 
\ee
with $\Gamma^{0}$, $\Gamma^{\tA}$ and $\Gamma^{10}$ the $16\times 16$ $\cM_c$-matrices given in Eq.(\ref{TGM}).  

It is noticed that the hyper-spacetime in association with $\vGa$-matrices $\vGa^0$, $\vGa^{\tmA}$ and $\vGa^{10+8}$ describes a free motion in ten-dimensional spacetime out of eighteen-dimensional Minkowski hyper-spacetime, which is in analogous to the case for self-conjugated tetraqubit-spinor field shown in Eq.(\ref{actionQ4}). The remaining eight $\cM_c$-matrices $\vGa^{\bmA +9 }$ characterize maximally correlated motion in extra eight spatial dimensions. On the other hand, we may regard two-dimensional spacetime in association with $\cM_c$-matrices $\vGa^0$ and $\vGa^{10+8}$ as basic spacetime for describing a longitudinal free motion, which is in analogous to a free motion of self-conjugated chiral uniqubit-spinor field or zeroqubit-spinor field shown in Eq.(\ref{actionQ0}), whereas the other sixteen $\cM_c$-matrices $\vGa^{\tmA}$ and $\vGa^{\bmA +9 }$ reflect the maximally correlated transverse motion in sixteen spatial dimensions. 

The self-conjugated octoqubit-spinor field $\Psi_{\QS^8}(x)$ as hyperqubit-spinor field is explicitly defined as follows:
\be
& & \Psi_{\QS^8}(x) =\binom{\psi_{\QC^7}}{\psi_{\hQC^7} } =  \begin{pmatrix} 
\psi_{\mQ^6_{\mC+}}(x) \\
\psi_{\mQ^6_{\mC-}}(x) \\
\psi_{\hat{Q}^6_{\mC  -}} (x) \\
\psi_{\hat{Q}^6_{\mC  +}} (x)
 \end{pmatrix}  =   \begin{pmatrix} 
\psi_{\mQ^5_{\mC++}}(x) \\
\psi_{\mQ^5_{\mC+-}}(x) \\
\psi_{\mQ^5_{\mC-+}}(x) \\
\psi_{\mQ^5_{\mC--}}(x) \\
\psi_{\hat{\mQ}^5_{\mC  --}} (x) \\
\psi_{\hat{\mQ}^5_{\mC  -+}} (x) \\
\psi_{\hat{\mQ}^5_{\mC  +-}} (x) \\
\psi_{\hat{\mQ}^5_{\mC  ++}} (x)
\end{pmatrix}  ,  \nn \\
& & \psi_{\hQC^7} = \sigma_1\otimes \sigma_1\otimes \sigma_1\sigma_1\otimes \sigma_1\otimes \sigma_1\otimes \sigma_1\psi_{\QC^7}^{\ast} ,
\ee
which satisfies the following self-conjugate condition:
\be
& & \Psi^c_{\QS^8}(x) = C_{\QS^8} \bar{\Psi}^T_{\QS^8}(x) = \Psi_{\QS^8}(x), \nn \\
& & C_{\QS^8} = \sigma_2 \otimes\sigma_2 \otimes \sigma_2 \otimes \sigma_2  \otimes \sigma_2 \otimes\sigma_2 \otimes \sigma_2 \otimes \sigma_2 .
\ee

The maximal associated symmetry of the actions in Eq.(\ref{actionQ8}) and (\ref{actionM8}) is found to be,
\be
G_S & = & SC(1)\ltimes P^{1,17}\ltimes SO(1,17)\adjoin SP(1,17)  \rtimes SG(1) \times SP(q_c=0) \nn \\
& = & SC(1)\ltimes  PO(1,17) \adjoin SP(1,17) \rtimes SG(1), 
\ee
where SP(1,17) is maximal $\cM_c$-spin symmetry as hyperspin symmetry. The semidirect group PO(1,17) = P$^{1,17}\ltimes$ SO(1,17) characterizes Poincer\'e-type/inhomogeneous Lorentz-type group symmetry with $P^{1,17}$ denoting the translation group symmetry in 18-dimensional Minkowski hyper-spacetime of coordinates. SP($q_c=0$) denotes in formal $\cQ_c$-spin symmetry to classify hyperqubit-spinor field, which indicates that there exist a class of hyperqubit-spinor field with zero $\cQ_c$-spin charge that gets no scalar coupling.


\subsection{Classifications of arbitrary qubit number and spacetime dimension with four categories of $\cQ_c$-spin charge and the genesis of spacetime dimension from categoric qubit number}

When looking back to $\cQ_c$-spin charge with a periodic behavior $\CQc = q_c = 0,1,2,3,0$ in correspondence to basic qubit-spinor fields, i.e., zeroqubit-spinor, uniqubit-spinor, biqubit-spinor, triqubit-spinor and tetraqubit-spinor fields, with respective to basic qubit numbers, i.e., $q_c=Q_N= 0,1,2,3,4$, we observe that local coherent states of qubits with high qubit number $Q_N\geq 4$ bring on a periodic $\cQ_c$-spin charge given by a simple relation with qubit number, i.e., $\CQc = Q_N-4=q_c = 0,1,2,3$, which characterize other four kinds of qubit-spinor fields: tetraqubit-spinor, pentaqubit-spinor, hexaqubit-spinor and heptaqubit-spinor fields with respective to four qubit numbers $Q_N=4,5,6,7$. 

To be manifest, let us summarize the main properties of hyperqubit-spinor fields and hyper-spacetime dimensions in table 2.

\begin{table}[htp]
\caption{The properties of local coherent state of qubits as hyperqubit-spinor field}
\begin{center}
\begin{tabular}{|c|c|c|c|c|c|}
\hline
Qubits $Q_N$  & 4 & 5 & 6 & 7 & 8  \\
Hyperqubit-spinor &  $\Psi_{\mQ^4}$ &  $\Psi_{\mQ^5}$ &  $\Psi_{\mQ^6}$ &  $\Psi_{\mQ^7}$ & $\Psi_{\mQ^8}$ \\
Qubit-spinor & tetraqubit- & pentaqubit-& hexaqubit-  & heptaqubit- & octoqubit-  \\
Hilbert space $\cD_H$ & $ 16=2^4 $ & $ 32=2^5 $ & $  64=2^6 $ & $  128=2^7 $ & $256=2^8 $ \\ 
$\cQ_c$-spin charge $\CQc$  & $Q_N$-4=0 & $Q_N$-4=1 & $Q_N$-4=2 & $Q_N$-4=3 & $Q_N$-4$\times$2=0 \\
$\cQ_c$-spin sym. & SP(0) & SP(1)$\cong$O(1) & SP(2)$\cong$U(1) & SP(3)$\cong$SU(2) & SP(0) \\
$\cM_c$-spin charge $\CMc$ & $ 2Q_N$+2=10   & $ 2Q_N$+1=11 & $ 2Q_N$=12  &  $ 2Q_N$=14  & $2Q_N$+2=18 \\
Hyper-spacetime $D_h$ & 2+$2^3$=2+8 & 3+8 & 4+8 &  6+8 & 10+8=2+16\\
Spinor structure &  $\Psi_{\QS^4} $ &  $\Psi_{\QS^5} $ &  $\Psi_{\QS^6},\Psi_{\QC^5} $ &  $\Psi_{\QS^7},\Psi_{\QC^6}$ &  $\Psi_{\QS^8} $ \\
Hyperspin sym. & SP(1,9) & SP(1,10) & SP(1,11) & SP(1,13) & SP(1,17) \\
Lorentz-type sym. & SO(1,9) & SO(1,10) & SO(1,11) & SO(1,13) & SO(1,17) \\
Poincar\'e-type sym. & PO(1,9) & PO(1,10) & PO(1,11) & PO(1,13) & PO(1,17) \\
\hline
\end{tabular}
\end{center}
\label{Table 2}
\end{table}

From the properties shown in table 1 and table 2 for basic qubit-spinor fields and hyperqubit-spinor fields, the $\cQ_c$-spin charge is resulted as the mod 4 qubit number, i.e.:
\be
 \CQc = q_c = 0,1,2,3 = Q_N (\mbox{mod 4}) ,
 \ee
which enables us to deduce a general theorem for the classification of qubit number.

{\it Qubit number classification theorem}:  any qubit number $Q_N$ used to form product qubit-basis states and build local coherent state of qubits can be classified into four categories as categoric qubit number $Q_N^{(q_c,k)}$, which is characterized by four $\cQ_c$-spin charges $\CQc=q_c=0,1,2,3$ and integer number $k=0,1,\cdots$.

As four categories are in correspondence to four basic qubit numbers, i.e., $\CQc=q_c=Q_N =0,1,2,3$, we are able to express any qubit number $Q_N$ in terms of $\cQ_c$-spin charge with the following general relation:
\be
 & & Q_N^{(q_c,k)} \equiv q_c + 4 k , \nn \\
 & & q_c  = 0,1,2,3 ,  \quad  k = 0,1,2, \cdots , 
\ee
which indicates that $\cQ_c$-spin charge $\CQc = q_c$ plays an essential role as qubit category characteristic number. The periodic feature of $\cQ_c$-spin charge $\CQc$ for any qubit number $Q_N^{(q_c,k)} = q_c + 4k$ can be expressed as follows: 
\be
& & \CQc\equiv\CQc^{(q_c,k)} = \CQc^{(q_c)}= q_c , \nn \\
& & q_c = 0,1,2, 3 , \quad k= 0, 1, 2, \cdots ,
\ee
where the integer number $k$ denotes as {\it cycle number of period} for $\cQ_c$-spin charge $\CQc$. Therefore, $\CQc^{(q_c,k)}$ is regarded as {\it categoric $\cQ_c$-spin charge} in category-$q_c$ and $k$-th period and $Q_N^{(q_c,k)}$ is considered as {\it categoric qubit number} in category-$q_c$ and $k$-th period. 

Following along the qubit number classification theorem and local coherent-qubits motion postulate, we are able to further deduce a general theorem for the classification of spacetime dimension. 

{\it Spacetime dimension classification theorem}: any spacetime dimension determined by the maximally correlated motion of qubit-spinor field with arbitrary categoric qubit number can be classified into four categories as categoric spacetime dimension $D_h^{(q_c,k)}$, which is characterized by four $\cQ_c$-spin charges $\CQc=q_c=0,1,2,3$ and periodic number $k=0,1,\cdots$.

As the basic spacetime dimensions generating from maximally correlated free motions of basic qubit-spinor fields with respective to four basic qubit numbers $Q_N =0,1,2,3$ are determined to be $D_{q_c}= 2,3,4,6$ which are characterized by four $\cQ_c$-spin charges $\CQc=q_c=Q_N =0,1,2,3$, we can express any spacetime dimension $D_h$ in terms of categoric basic dimension $D_{q_c}$ with the following general relation:
\be
& & D_h^{(q_c,k)}  \equiv D_{q_c} + 8k, \quad k= 0, 1, 2, \cdots ,  \nn \\
& &  D_{q_c}= 2,3,4,6, \quad q_c = 0,1,2, 3,  
\ee
which is referred to as {\it category-$q_c$ spacetime dimension in $k$-th period}. $D_{q_c}$ represents {\it basic spacetime dimension in category-$q_c$}, which is completely determined by $\cQ_c$-spin charge $\CQc=q_c$ with the following identity:
\be
 & & D_{q_c} = 2 + 2^{q_c-1} \theta(q_c-1) ,
\ee
which brings on four basic spacetime dimensions $D_{q_c} = 2, 3, 4, 6$ with respective to four $\cQ_c$-spin charges $q_c = 0, 1, 2, 3$. Where $\theta(z)$ is the $\theta$-function with $\theta(z) = 1$ for $z\ge 0 $ and $\theta(z) = 0$ for $z < 0$. 

Therefore, we are able to obtain the following general relation between categoric spacetime dimension $D_h^{(q_c,k)}$ and categoric qubit number $Q_N^{(q_c,k)}$,
\be
D_h^{(q_c,k)} = 2Q_N^{(q_c,k)} - 2(q_c - 1) +  2^{q_c-1} \theta(q_c-1) ,
\ee
which implies how the genesis of spacetime dimension is correlated to categoric qubit number and $\cQ_c$-spin charge.


\subsection{ Categorization of hyperqubit-spinor field and hyper-spacetime with four categories of $\cQ_c$-spin charge and the action of hyperqubit-spinor field in category-$q_c$ with hyperspin symmetry }

In analogous to qubit number classification theorem and spacetime dimension classification theorem, it is straightforward to deduce a general categorization theorem for qubit-spinor field.

{\it Qubit-spinor categorization theorem}: any qubit-spinor field as local coherent state of qubits with arbitrary qubit number can be categorized into four categories as categoric qubit-spinor field $\Psi_{Q_N^{(q_c,k)}}(x)$, which is characterized by four $\cQ_c$-spin charges $\CQc=q_c=0,1,2,3$ and periodic number $k=0,1,\cdots$. 

To make the categorizations for both hyperqubit-spinor field and hyper-spacetime through four $\cQ_c$-spin charges with respective to four basic qubit numbers and four basic spacetime dimensions, it is appropriate to build the action of hyperqubit-spinor field as local coherent state of qubits with categoric qubit number $Q_N^{(q_c,k)}=q_c + 4k$ in $k$-th period. Such a hyperqubit-spinor field is denoted by $\Psi_{Q_N^{(q_c,k)}}(x)$ and referred to as {\it category-$q_c$ hyperqubit-spinor field in $k$-th period}, which spans $\cD_{H}= 2^{Q_N^{(q_c,k)}}$ dimensional Hilbert space defined on $D_h^{(q_c,k)}$ dimensional Minkowski hyper-spacetime in category-$q_c$ and $k$-th period.

The hyperqubit-spinor field $\Psi_{\mQ^n}(x)$ with arbitrary qubit number $Q_N=n$ is defined by real $n$-product qubit-basis $\{ \varsigma_{s_1\cdots s_{n}} \}$ as follows: 
\be \label{QNn}
& & \Psi_{\mQ^n}(x) =  \sum_{s_1, \cdots, s_{n}}  \psi_{s_1\cdots s_{n}}(x)  \varsigma_{s_1\cdots s_{n}} \equiv \sum_{s_1, \cdots, s_{n}}  \psi_{s_1\cdots s_{n}}(x)\,  \varsigma_{s_1}\otimes \cdots \otimes \varsigma_{s_{n}} ,  \nn
\ee
with $\psi_{s_1\cdots s_{n}}(x) $ regarded as evolving local distribution amplitudes. Any hyperqubit-spinor field $\Psi_{\mQ^n}(x)$ can always be classified into four categories $\Psi_{\mQ^n}(x) \equiv \Psi_{Q_N^{(q_c,k)}}(x)$ with respective to $\cQ_c$-spin charges $q_c=0,1,2,3$ in $k$-th period.

For any qubit number in {\it Category-0}, i.e., $ Q_N^{(0,k)}=n=4k$, the action for {\it Category-0 hyperqubit-spinor field} $\Psi_{Q_N^{(0,k)}}(x)$ with zero $\cQ_c$-spin charge $\CQc=q_c=0$ in $k$-th period can generally be expressed as follows: 
 \be \label{actionQ_N^{(0)}}
 \cS_{Q_N^{(0)}}  & = &  \int [dx] \frac{1}{2} \Psi_{Q_N^{(0,k)}}^{\dagger}(x)  \delta_{\mA}^{\; \mM}  \vUps^{\mA} i\p_{\mM} \Psi_{Q_N^{(0,k)}}(x)   , 
\ee
with $\mA, \mM= 0, 1, \cdots, D_h^{(0,k)}-1$. The maximally correlated motion of Category-0 hyperqubit-spinor field leads to genesis of {\it Category-0 hyper-spacetime} with dimension $D_h^{(0,k)}= 2 + 8k$. For the case $k=0$, it provides the basic action for zeroqubit-spinor field or bit-spinor field $\psi_{\mQ^0_{\pm}}(x)$ in two-dimensional spacetime as shown in Eq.(\ref{actionQ0}). When $k\ge 1$, the $2^{4k}\times 2^{4k}$ real symmetric normalized $\cM_c$-matrices $\vUps^{\mA}$ can generally be expressed as $\vUps^{\mA} = ( \vUps^0, \vUps^{\bmA +8j}, \vUps^{9+8(k-1)} )$ ($\bmA=1,\cdots, 8$; $j=0,1,\cdots, k-1$). For the cases $k=1,2$, the $\cM_c$-matrices are presented in Eqs.(\ref{UpsM4}) and (\ref{UpsM8}), respectively. In the general case, we can express the $\cM_c$-matrices by the following explicit forms:
\be \label{UpsQ_N^{(0)}}
\vUps^{0} & = & \Ups^0_{(1)} \otimes \cdots \Ups^{0}_{(k-j)}  \cdots \otimes \Ups^{0}_{(k)}, \nn \\
\vUps^{\bmA+ 8\times0} & = & \Ups^9_{(1)} \otimes \cdots\Ups^{9}_{(k-j)}  \cdots \otimes \Ups^{\bA}_{(k)} , \nn \\
\vUps^{\bmA + 8j} & = &  \Ups^{9}_{(1)}  \otimes \cdots \Ups^{\bA}_{(k-j)}  \cdots \otimes \Ups^{9}_{(k)} , \nn \\
& &  j=1,\cdots, k-2, \nn \\ 
\vUps^{\bmA + 8(k-1)} & = &  \Ups^{\bA}_{(1)}  \otimes \cdots\Ups^{9}_{(k-j)}  \cdots \otimes \Ups^{9}_{(k)} , \nn \\
\vUps^{9+8(k-1)} & = &  \Ups^9_{(1)} \otimes \cdots\Ups^{9}_{(k-j)}  \cdots  \otimes \Ups^{9}_{(k)}, 
\ee
where the matrices $\Ups^0, \Ups^{\bA}, \Ups^9$ are $16\times 16$ real symmetric normalized $\cM_c$-matrices given in Eq.(\ref{UpsM4}). The subscript of $\cM_c$-matrices $\Ups^{A}_{(i)}$ ($i=1,2\cdots, k$) labels the column order in the product tensor.

For any qubit number in {\it Category-1}, i.e., $Q_N^{(1,k)}= 1+ 4k$, the action for {\it Category-1 hyperqubit-spinor field} $\Psi_{Q_N^{(1,k)}}(x)$ with $\cQ_c$-spin charge $\CQc=q_c=1$ in $k$-th period is generally obtained as follows:
 \be \label{actionQ_N^{(1)}}
 \cS_{Q_N^{(1)}} = \int [dx] \{ \frac{1}{2} \Psi_{Q_N^{(1,k)}}^{\dagger}(x)  \delta_{\mA}^{\; \mM}  \vUps^{\mA} i\p_{\mM} \Psi_{Q_N^{(1,k)}}(x) - \frac{1}{2}\lambda_1 \phi_{1}(x)\Psi_{Q_N^{(1,k)}}^{\dagger}(x)  \vtUps \Psi_{Q_N^{(1,k)}}(x) \}, 
\ee
with $\mA, \mM= 0, 1, \cdots, D_h^{(1,k)}-1$. The {\it Category-1 hyper-spacetime} gets dimensions $D_h^{(1,k)}= 3 + 8k$ ($k=0,1\cdots$) with three-dimensional spacetime as a basic one. In the simplest case with $k=0$, the action for uniqubit-spinor field $\psi_{\mQ^1}(x)$ in three-dimensional spacetime is presented in Eq.(\ref{actionB}). The $\cM_c$-matrices and $\cQ_c$-matrix for the case $k=1$ are given in Eq.(\ref{UpsM5}). 

In the general case $k\ge 1$, the $2^{4k+1}\times 2^{4k+1}$ real symmetric normalized $\cM_c$-matrices $\vUps^{\mA}$ are expressed as $\vUps^{\mA} = ( \vUps^0, \vUps^{\bmA +8j}, \vUps^{9+8(k-1)}, \vUps^{10+8(k-1)} )$ ($\bmA=1,\cdots, 8$; $j=0,1,\cdots, k-1$), which can be obtained by extending straightforwardly the case $k=1$ to the general case $k\ge 1$ with the following explicit structures:
\be \label{UpsQ_N^{(1)}}
\vUps^{0} & = & \sigma_0 \otimes \Ups^0_{(1)} \otimes \cdots \Ups^{0}_{(k-j)}  \cdots \otimes \Ups^{0}_{(k)}, \nn \\
\vUps^{\bmA + 8j} & = &  \sigma_3 \otimes  \Ups^{9}_{(1)}  \otimes \cdots \Ups^{\bA}_{(k-j)}  \cdots \otimes \Ups^{9}_{(k)} , \nn \\
& & j=0,1,\cdots, k-1, \nn \\
\vUps^{9+8(k-1)} & = &  \sigma_3 \otimes \Ups^9_{(1)} \otimes \cdots\Ups^{9}_{(k-j)}  \cdots  \otimes \Ups^{9}_{(k)}, \nn \\
\vUps^{10+8(k-1)} & = & \sigma_1 \otimes \Ups^0_{(1)} \otimes \cdots \Ups^{0}_{(k-j)}  \cdots \otimes \Ups^{0}_{(k)}, \nn \\
\vtUps & = & \sigma_2 \otimes \Ups^0_{(1)} \otimes \cdots \Ups^{0}_{(k-j)}  \cdots \otimes \Ups^{0}_{(k)} .
\ee

Similarly, the action for {\it Category-2 hyperqubit-spinor field} $\Psi_{Q_N^{(2,k)}}(x)$ with categoric qubit number $Q_N^{(2)}= 2+ 4k$ in the $k$-th period has the following general form:
 \be \label{actionQ_N^{(2)}}
 \cS_{Q_N^{(2)}} = \int [dx] \{  \frac{1}{2}\Psi_{Q_N^{(2,k)}}^{\dagger}(x)  \delta_{\mA}^{\; \mM}  \vUps^{\mA} i\p_{\mM} \Psi_{Q_N^{(2,k)}}(x) - \frac{1}{2}\lambda_2 \phi_p(x) \Psi_{Q_N^{(2,k)}}^{\dagger}(x) \vtUps^p \Psi_{Q_N^{(2,k)}}(x) \} , 
\ee
with $\mA, \mM= 0, 1, \cdots, D_h^{(2,k)}-1$. The dimension of {\it Category-2 hyper-spacetime} is given by $D_h^{(2,k)}= 4 + 8k$. The simplest case with $k=0$ corresponds to the action of biqubit-spinor field $\psi_{\mQ^2}(x)$ in four-dimensional spacetime as shown in Eq.(\ref{actionQ2}). For the case $k=1$, the $\cM_c$-matrices and $\cQ_c$-matrices are presented in Eq.(\ref{UpsM6}). 

For the general case $k\ge 1$, the $2^{4k+2}\times 2^{4k+2}$ real symmetric normalized $\cM_c$-matrices $\vUps^{\mA}$ can be written as $\vUps^{\mA} = ( \vUps^0, \vUps^{\bmA +8j}, \vUps^{9+8(k-1)}, \vUps^{10+8(k-1)}, \vUps^{11+8(k-1)} )$ ($\bmA=1,\cdots, 8$; $j=0,1,\cdots, k-1$). The explicit structures of $\cM_c$-matrices and $\cQ_c$-matrices in Category-2 can be expressed as follows:
\be \label{UpsQ_N^{(2)}}
\vUps^{0} & = & \sigma_0 \otimes \sigma_0 \otimes \Ups^0_{(1)} \otimes \cdots \Ups^{0}_{(k-j)}  \cdots \otimes \Ups^{0}_{(k)}, \nn \\
\vUps^{\bmA + 8j} & = &  \sigma_0 \otimes \sigma_3 \otimes \Ups^{9}_{(1)}  \otimes \cdots \Ups^{\bA}_{(k-j)}  \cdots \otimes \Ups^{9}_{(k)} , \nn \\
& &  j=0,1,\cdots, k-1, \nn \\
\vUps^{9+8(k-1)} & = &  \sigma_0 \otimes \sigma_3 \otimes \Ups^9_{(1)} \otimes \cdots\Ups^{9}_{(k-j)}  \cdots  \otimes \Ups^{9}_{(k)}, \nn \\
\vUps^{10+8(k-1)} & = & \sigma_0 \otimes \sigma_1 \otimes \Ups^0_{(1)} \otimes \cdots \Ups^{0}_{(k-j)}  \cdots \otimes \Ups^{0}_{(k)}, \nn \\
\vUps^{11+8(k-1)} & = & \sigma_2 \otimes \sigma_2 \otimes \Ups^0_{(1)} \otimes \cdots \Ups^{0}_{(k-j)}  \cdots \otimes \Ups^{0}_{(k)}, \nn \\
\vtUps^1 & = & \sigma_3 \otimes  \sigma_2 \otimes \Ups^0_{(1)} \otimes \cdots \Ups^{0}_{(k-j)}  \cdots \otimes \Ups^{0}_{(k)} , \nn \\
\vtUps^2 & = & \sigma_1 \otimes  \sigma_2 \otimes \Ups^0_{(1)} \otimes \cdots \Ups^{0}_{(k-j)}  \cdots \otimes \Ups^{0}_{(k)} .
\ee

In analogous, the action for {\it Category-3 hyperqubit-spinor field} $\Psi_{Q_N^{(3,k)}}(x)$ with categoric qubit number $Q_N^{(3,k)}= 3+ 4k$ in the $k$-th period can be written in the following general form:
 \be \label{actionQ_N^{(3)}}
 \cS_{Q_N^{(3)}} = \int [dx]\, \frac{1}{2}  \Psi_{Q_N^{(3,k)}}^{\dagger}(x)  \delta_{\mA}^{\; \mM}  \vUps^{\mA} i\p_{\mM} \Psi_{Q_N^{(3,k)}}(x) , 
 \ee
with $\mA, \mM= 0, 1, \cdots, D_h^{(3,k)}-1$. The spacetime dimension of {\it Category-3 hyper-spacetime} is given by $D_h^{(2,k)}= 6 + 8k$. The action with $k=0$ is presented in Eq.(\ref{actionQ3}) for triqubit-spinor field $\psi_{\mQ^3}(x)$ in six-dimensional spacetime. For the case $k=1$, the $\cM_c$-matrices and $\cQ_c$-matrices are shown in Eqs.(\ref{UpsM7}) and (\ref{tUpsM3}), respectively. 

For the general case $k\ge 1$, the $2^{4k+3}\times 2^{4k+3}$ real symmetric normalized $\cM_c$-matrices $\vUps^{\mA}$ can be written as $\vUps^{\mA} = ( \vUps^0,  \vUps^{\bmA +8j}, \vUps^{9+8(k-1)}, \vUps^{10+8(k-1)}, \vUps^{11+8(k-1)}, \vUps^{12+8(k-1)}, \vUps^{13+8(k-1)} )$ ($\bmA=1,\cdots, 8$; $j=0,1,\cdots, k-1$). The $\cM_c$-matrices in Category-3 can be expressed as follows:
\be \label{UpsQ_N^{(2)}}
\vUps^{0} & = & \sigma_0 \otimes \sigma_0 \otimes \sigma_0 \otimes \Ups^0_{(1)} \otimes \cdots \Ups^{0}_{(k-j)}  \cdots \otimes \Ups^{0}_{(k)}, \nn \\
\vUps^{\bmA + 8j} & = &  \sigma_3\otimes  \sigma_0\otimes \sigma_3 \otimes  \Ups^{9}_{(1)}  \otimes \cdots \Ups^{\bA}_{(k-j)}  \cdots \otimes \Ups^{9}_{(k)} , \nn \\
& &  j=0,1,\cdots, k-1, \nn \\
\vUps^{9+8(k-1)} & = &  \sigma_3\otimes  \sigma_0\otimes \sigma_3 \otimes \Ups^9_{(1)} \otimes \cdots\Ups^{9}_{(k-j)}  \cdots  \otimes \Ups^{9}_{(k)}, \nn \\
\vUps^{10+8(k-1)} & = & \sigma_3\otimes  \sigma_0\otimes \sigma_1 \otimes  \Ups^0_{(1)} \otimes \cdots \Ups^{0}_{(k-j)}  \cdots \otimes \Ups^{0}_{(k)}, \nn \\
\vUps^{11+8(k-1)} & = & \sigma_3\otimes  \sigma_2\otimes \sigma_2 \otimes  \Ups^0_{(1)} \otimes \cdots \Ups^{0}_{(k-j)}  \cdots \otimes \Ups^{0}_{(k)}, \nn \\
\vUps^{12+8(k-1)} & = & \sigma_1\otimes  \sigma_0\otimes \sigma_0 \otimes \Ups^0_{(1)} \otimes \cdots \Ups^{0}_{(k-j)}  \cdots \otimes \Ups^{0}_{(k)}, \nn \\
\vUps^{13+8(k-1)} & = &  \sigma_2\otimes  \sigma_2\otimes \sigma_0 \otimes  \Ups^0_{(1)} \otimes \cdots \Ups^{0}_{(k-j)}  \cdots \otimes \Ups^{0}_{(k)} .
\ee
The three imaginary antisymmetric normalized $\cQ_c$-matrices $\vtUps^p$ ($p=1,2,3$) in Category-3 are given by,
\be
\vtUps^1 & = & \sigma_3\otimes \sigma_1\otimes \sigma_2 \otimes \Ups^0_{(1)} \otimes \cdots 
\Ups^{0}_{(k-j)}  \cdots \otimes \Ups^{0}_{(k)} , \nn \\
\vtUps^2 & = &  \sigma_3\otimes \sigma_3\otimes \sigma_2 \otimes\Ups^0_{(1)} \otimes \cdots 
\Ups^{0}_{(k-j)}  \cdots \otimes \Ups^{0}_{(k)} , \nn \\
\vtUps^2 & = &  \sigma_2\otimes \sigma_0\otimes \sigma_0 \otimes\Ups^0_{(1)} \otimes \cdots 
\Ups^{0}_{(k-j)}  \cdots \otimes \Ups^{0}_{(k)} .
\ee
The matrices $\vUps^{\mA} $ and $\vtUps^p$ satisfy the following Clifford algebra relations:
\be \label{CAQ_N}
& & \{\vUps^{\mA}, \vUps^{\mB} \} = 2\delta^{\mA\mB}, \quad \{ \vUps^{0}, \vUps^{\mA} \} =  2\vUps^{\mA}, \quad \{ \vtUps^p, \vtUps^{q}\} = 2\delta^{pq},  \nn \\ 
& & \mA, \mB = 1, \cdots , D_h^{(3,k)}-1, \quad p, q =1, 2, 3, \nn \\
& & \{ \vtUps^p, \vUps^{\mA}\} = 0, \quad \mA \neq \vUps^{0}, \vUps^{13+8(k-1)} .
\ee

In general, the action for category-$q_c$ hyperqubit-spinor field $\Psi_{Q_N^{(q_c,k)}}(x)$ in $k$-th period possesses the following associated symmetry:
\be
G_S & = & SC(1)\ltimes P^{1, D_h^{(q_c,k)}-1}\ltimes SO(1, D_h^{(q_c,k)}-1) \adjoin SP(1, D_h^{(q_c,k)}-1) \rtimes SG(1) \times SP(q_c) \nn \\
& = & SC(1)\ltimes  PO(1, D_h^{(q_c,k)}-1) \adjoin SO(1, D_h^{(q_c,k)}-1) \rtimes SG(1)\times SP(q_c) .
\ee


\section{ String-like qubit-spinor field with free motion in two-dimensional spacetime and the self-conjugated chiral qubit-spinor field with full anti-commuting $\Gamma$-matrices }

The $\cQ_c$-spin charge as qubit category characteristic number enables us to categorize all hyperqubit-spinor fields by utilizing categoric qubit number $Q_N^{(q_c,k)}=q_c + 4k$ in the $k$-th period ($k=0,1,\cdots$) and meanwhile to construct category-$q_c$ hyperqubit-spinor field $\Psi_{Q_N^{(q_c,k)}}(x)$ in hyper-spacetime based on the periodic feature of $\cQ_c$-spin charge $\CQc^{(q_c,k)}=\CQc^{(q_c)}=q_c$. The dimension of hyper-spacetime is also classified into categoric spacetime dimension $D_h^{(q_c,k)}=D_{q_c} + 8k$ with $D_{q_c}$ representing four basic spacetime dimensions $D_{q_c}=2,3,4,6$ in correspondence to four $\cQ_c$-spin charges $Q_c =0,1,2,3$. In this section, we are going to investigate self-conjugated chiral qubit-spinor structure. On one hand, a self-conjugated chiral-like qubit-spinor structure allows us to construct string-like qubit-spinor field with free motion in two-dimensional spacetime from self-conjugated qubit-spinor field defined in $D_h$-dimensional spacetime. On the other hand, a self-conjugate qubit-spinor field with arbitrary qubit number $Q_N$ in category-$q_c$ can always be described by a self-conjugated chiral qubit-spinor field with one additional qubit number $Q_N+1$ but with the same category-$q_c$ in large Hilbert space. As a specific example, a massless Dirac fermion as category-3 self-conjugated triqubit-spinor field in six-dimensional spacetime is shown to be described by self-conjugated chiral tetraqubit-spinor field in category-3. In particular, we will demonstrate that in light of self-conjugated chiral qubit-spinor field, all $\Gamma$-matrices represented in large Hilbert space become anti-commuting.


\subsection{ Self-conjugated chiral-like spinor structure and string-like qubit-spinor field}

Based on the periodic feature of $\cQ_c$-spin charge $\CQc^{(q_c,k)}=\CQc^{(q_c)}=q_c$, all qubit-spinor fields $\Psi_{Q_N^{(q_c,k)}}(x)$ in the same category-$q_c$ but with different period ($k=1,2,\cdots$) should possess the same $\cQ_c$-spin symmetry, and the $\cM_c$-matrices should have a similar matrix structure in correlation to four basic spacetime dimensions $D_{q_ c}=2,3,4,6$.

Let us begin with examining the zeroqubit-spinor field $\psi_{\mQ^0}(x)$, tetraqubit-spinor field $\Psi_{\mQ^4}(x)$ and octoqubit-spinor field $\Psi_{\mQ^8}(x)$ in category-0. As the zeroqubit-spinor field can be regarded as a self-conjugated chiral uniqubit-spinor field in two-dimensional Minkowski spacetime shown in Eq.(\ref{actionQ0}), the relevant two $\cM_c$-matrices are in analogous to two $\cM_c$-matrices $\Ups^0$ and $\Ups^9$ for tetraqubit-spinor field $\Psi_{\mQ^4}(x)$ or $\Gamma$-matrices $\Gamma^0$ and $\Gamma^{10}$ for self-conjugated tetraqubit-spinor field $\Psi_{\QS^4}(x) $, such two $\cM_c$-matrices characterize longitudinal free motion in a relevant two-dimensional Minkowski spacetime. 
 
To make an explicit illustration, it is useful to express the tetraqubit-spinor field $\Psi_{\mQ^4}(x)$ as the sum of two {\it self-conjugated chiral-like tetraqubit-spinor fields} $\Psi_{\mQ^4 \pm}(x)$, i.e.:  
\be \label{MW4}
& & \Psi_{\mQ^4}(x) = \Psi_{\mQ^4-}(x) + \Psi_{\mQ^4+}(x), \nn \\
& & \Psi_{\mQ^4 \pm}(x) = \Ups_{\pm} \Psi_{\mQ^4}(x), \quad \Ups_{\pm} = \frac{1}{2} ( 1 \pm \Ups^{9} ). 
\ee
In a similar way, we make the following decomposition for $\Psi_{\QS^4}(x)$:
\be  \label{QSpm4}
& & \Psi_{\QS^4}(x) = \Psi_{\QS^4-}(x) + \Psi_{\QS^4+}(x), \nn \\
& &  \Psi_{\QS^4\pm}(x) = \Gamma_{\pm} \Psi_{\QS^4}(x) , \quad  \Gamma_{\pm} = \frac{1}{2} ( 1 \pm \Gamma^{0} ) ,
\ee
where the project operators $\Ups_{\pm}$ and $ \Gamma_{\pm}$ define a self-conjugated chiral-like structure for  tetraqubit-spinor field, which allows us to rewrite the actions in Eqs.(\ref{actionQ4}) and (\ref{actionM4}) into the following forms in ten-dimensional hyper-spacetime:
\be \label{actionQW4}
 \cS_{\mQ^4}  & = &  \int d^{10}x  \frac{1}{2} \{  \Psi_{\mQ^4+}^{\dagger}(x)  \delta_{a}^{\; \mu}  \Ups^{a} i\p_{\mu} \Psi_{\mQ^4+}(x) 
 +  \Psi_{\mQ^4-}^{\dagger}(x)  \delta_{a}^{\; \mu}  \Ups^{a} i\p_{\mu} \Psi_{\mQ^4-}(x) \nn \\
 & + & \Psi_{\mQ^4+}^{\dagger}(x)  \delta_{\bA}^{\; \bM}  \Ups^{\bA} i\p_{\mM} \Psi_{\mQ^4-}(x) + \Psi_{\mQ^4-}^{\dagger}(x)  \delta_{\bA}^{\; \mM}  \Ups^{\bA} i\p_{\bM} \Psi_{\mQ^4+}(x)\} , 
\ee
with $a, \mu = 0, 9$ and $\bA, \bM = 1, 2, \cdots, 8$, and  
\be \label{actionMW4}
 \cS_{\QS^4}& = & \int d^{10}x  \frac{1}{2} \{  \bar{\Psi}_{\QS^4+}(x)  \delta_{a}^{\; \mu}  \Gamma^{a} i\p_{\mu} \Psi_{\QS^4+}(x) 
 +  \bar{\Psi}_{\QS^4-}(x)  \delta_{a}^{\; \mu}  \Gamma^{a} i\p_{\mu} \Psi_{\QS^4-}(x) \nn \\
 & + & \bar{\Psi}_{\QS^4+}(x)  \delta_{\tA}^{\; \tM}  \Gamma^{\tA} i\p_{\tM} \Psi_{\QS^4-}(x) + \bar{\Psi}_{\QS^4-}(x)  \delta_{\tA}^{\; \tM}  \Gamma^{\tA} i\p_{\tM} \Psi_{\QS^4+}(x)\} , 
\ee
with $a, \mu = 0, 10$ and $\tA, \tM = 1, 2, 3, 5,\cdots, 9$. 

It is noticed from the above actions that each self-conjugated chiral-like tetraqubit-spinor field $\Psi_{\mQ^4 +}(x)$ ($\Psi_{\mQ^4-}(x)$) or $\Psi_{\QS^4+}(x)$ ($\Psi_{\QS^4-}(x)$) gets a free motion only in two-dimensional Minkowski spacetime. The free motion in other eight spatial dimensions only occurs when two self-conjugated chiral-like tetraqubit-spinor fields with opposite chirality $\Psi_{\mQ^4 +}(x)$ and $\Psi_{\mQ^4-}(x)$ or $\Psi_{\QS^4+}(x)$ and $\Psi_{\QS^4-}(x)$ present simultaneously and obey the maximum coherence motion principle. Namely, their interferential terms for the derivative operators in association with eight $\cM_c$-matrices $\Ups^{\bA}$ ($\bA=1,2,\cdots, 8$)  or eight $\Gamma$-matrices $\Gamma^{\tA}$ ($\tA=1,2,3,5,\cdots, 9$), describe the maximally correlated motion in eight spatial dimensions. Therefore, the action for each self-conjugated chiral-like tetraqubit-spinor field can be expressed as follows in two-dimensional Minkowski spacetime:
\be \label{actionQM4}
 \cS_{\mQ^4\pm}  & = &  \int d^{2}x\,  \frac{1}{2} \Psi_{\mQ^4\pm}^{\dagger}(x)  \delta_{a}^{\; \mu}  \Ups^{a} i\p_{\mu} \Psi_{\mQ^4\pm}(x) , \nn \\
 \cS_{\QS^4\pm}  & = &  \int d^{2}x \, \frac{1}{2} \bar{\Psi}_{\QS^4\pm}(x)  \delta_{a}^{\; \mu}  \Gamma^{a} i\p_{\mu} \Psi_{\QS^4\pm}(x) , 
\ee
with $a,\mu =0,9$ for $\Psi_{\mQ^4\pm}(x)$ and $a,\mu =0,10$ for $\Psi_{\QS^4\pm}(x)$.

Similarly, it can be verified that the {\it self-conjugated chiral-like octoqubit-spinor fields} $\Psi_{\mQ^8\pm}(x)$ and $\Psi_{\QS^8\pm}(x)$ get a free motion in two-dimensional Minkowski spacetime. Their actions can be presented as follows:
\be \label{actionQM8}
 \cS_{\mQ^8\pm}  & = &  \int d^{2}x\,  \frac{1}{2} \Psi_{\mQ^8\pm}^{\dagger}(x)  \delta_{a}^{\; \mu}  \vUps^{a} i\p_{\mu} \Psi_{\mQ^8\pm}(x) , \nn \\
 \cS_{\QS^8\pm}  & = &  \int d^{2}x \, \frac{1}{2} \bar{\Psi}_{\QS^8\pm}(x)  \delta_{a}^{\; \mu}  \vGa^{a} i\p_{\mu} \Psi_{\QS^8\pm}(x) , 
\ee
with $a,\mu =0,17$ for $\Psi_{\mQ^8\pm}(x)$ and $a,\mu =0,18$ for $\Psi_{\QS^8\pm}(x)$. Where we have introduced the following definitions:
\be
& & \Psi_{\mQ^8 \pm}(x) = \vUps_{\pm} \Psi_{\mQ^8}(x), \quad  \vUps_{\pm} = \frac{1}{2} ( 1 \pm \vUps^{9+8} ), \nn \\
& & \Psi_{\QS^8 \pm}(x) = \vGa_{\pm} \Psi_{\QS^8}(x), \quad  \vGa_{\pm} = \frac{1}{2} ( 1 \pm \vGa^{0} ).
\ee
with $\vUps^{9+8}\equiv \vUps^{17}$ given in Eq.(\ref{UpsM8}) and $\vGa^{0}$ presented in Eq.(\ref{GM8}). Where $\Psi_{\mQ^8}(x)$ and $\Psi_{\QS^8}(x)$ are octoqubit-spinor field and self-conjugated octoqubit-spinor field defined in Eq.(\ref{QS8}), respectively, which are shown in Eqs.(\ref{actionQ8}) and (\ref{actionM8}) to have a free motion in 18-dimensional hyper-spacetime. 

It is easy to check that the above actions possess $\cM_c$-spin symmetry SP(1,1) and global scaling symmetry SG(1) in association with Lorentz-type group symmetry SO(1,1) and translation group symmetry $P^{1,1}$ as well as conformal scaling symmetry SC(1). Meanwhile, the $\cM_c$-spin symmetry concerning $D_h-2$ spatial dimensions becomes an intrinsic hyperspin symmetry SP($D_h$-2).

For the self-conjugated chiral-like tetraqubit-spinor field $\Psi_{\mQ^4\pm}(x)$ ($\Psi_{\QS^4\pm}(x)$), the associated symmetry is given by, 
\be
G_S & = & SC(1)\ltimes P^{1,1}\ltimes SO(1,1)\adjoin SP(1,1)\rtimes SG(1)\times SP(8) \nn \\
& = &  SC(1)\ltimes PO(1,1)\adjoin SP(1,1) \times SG(1)\times SP(8) ,  
\ee 
where SP(8)$\cong$SO(8) provides a maximal internal hyperspin symmetry. 

For the self-conjugated chiral-like octoqubit-spinor field $\Psi_{\mQ^8\pm}(x)$ ($\Psi_{\QS^8\pm}(x)$), the associated symmetry reads as follows:
\be 
G_S & = & SC(1)\ltimes P^{1,1}\ltimes SO(1,1)\adjoin SP(1,1)\rtimes SG(1) \times SP(16) \nn \\
& = &  SC(1)\ltimes PO(1,1)\adjoin SP(1,1) \rtimes SG(1) \times SP(16), 
\ee 
which brings on the maximal internal hyperspin symmetry SP(16)$\cong$SO(16).


It is straightforward to generalize the above analyses to arbitrary hyperqubit-spinor field in category-0 with {\it self-conjugated chiral-like hyperqubit-spinor structure}. A self-conjugated chiral-like hyperqubit-spinor field in category-0 can generally be defined as follows: 
\be \label{SCCLHQSF0}
& & \Psi_{Q_N^{(0,k)}\pm}(x) = \vUps_{\pm} \Psi_{Q_N^{(0,k)}}(x), \quad  \vUps_{\pm} = \frac{1}{2} ( 1 \pm \Ups^{9+8(k-1)} ), 
\ee
with $\vUps_{\pm}$ the project operators and $k$ the periodic number ($k=0,1,\cdots$). For the cases with $k=0,1,2$, they are already analyzed above. For a general case $k\ge 3$, the action for self-conjugated chiral-like hyperqubit-spinor field $\Psi_{Q_N^{(0,k)}\pm}(x)$ in category-0 gets the following general form: 
\be \label{actionQMk0}
 \cS_{Q_N^{(0)}\pm}  & = &  \int d^{2}x\,  \frac{1}{2} \Psi_{Q_N^{(0,k)}\pm}^{\dagger}(x)  \delta_{a}^{\; \mu}  \vUps^{a} i\p_{\mu} \Psi_{Q_N^{(0,k)}\pm}(x) , 
\ee
which has a free motion only in two-dimensional Minkowski spacetime with $a,\mu =0, (1+ 8k)$. Such an action possesses the following associated symmetry:
\be 
G_S & = & SC(1)\ltimes P^{1,1}\ltimes SO(1,1)\adjoin SP(1,1)\rtimes SG(1) \times SP(8k) \nn \\
& = &  SC(1)\ltimes PO(1,1)\adjoin SP(1,1) \rtimes SG(1) \times SP(8k) ,
\ee 
where SP(8k) represents internal hyperspin symmetry.

As a self-conjugated chiral-like qubit-spinor field has a free motion only in two-dimensional spacetime, we are naturally inspired to associate it with string theory\cite{ST1,ST2}. So, the self-conjugated chiral-like tetraqubit-spinor field $\Psi_{\mQ^4\pm}(x)$ or $\Psi_{\QS^4\pm}(x)$ and self-conjugated chiral-like octoqubit-spinor field $\Psi_{\mQ^8\pm}(x)$ or $\Psi_{\QS^8\pm}(x)$ may be regarded as a {\it string-like tetraqubit-spinor field} and {\it string-like octoqubit-spinor field}, respectively. In general, a self-conjugated chiral-like hyperqubit-spinor field in category-0 $\Psi_{Q_N^{(0,k)}\pm}(x)$ with a free motion in two-dimensional spacetime is viewed as a {\it string-like hyperqubit-spinor field in category-0}, which is in analogous to the zeroqubit-spinor field shown in Eq.(\ref{actionQ0}). 

The above discussions can be extended to any hyperqubit-spinor field in category-$q_c$ with self-conjugated chiral-like hyperqubit-spinor structure, i.e., $\Psi_{Q_N^{(q_c,k)}\pm}(x)$, its action can be written into the following form:  
\be \label{actionQMkq}
 \cS_{Q_N^{(q_c,k)}\pm}  & = &  \int d^{2}x\,  \frac{1}{2} \Psi_{Q_N^{(q_c,k)}\pm}^{\dagger}(x) \delta_{a}^{\; \mu}  \vUps^{a} i\p_{\mu} \Psi_{Q_N^{(q_c,k)}\pm}(x) , 
\ee
with $a,\mu =0, (1+ 8k)$ for the $k$-th period of $\cQ_c$-spin charge and the definition,
\be
& & \Psi_{Q_N^{(q_c,k)}\pm}(x) = \vUps_{\pm} \Psi_{Q_N^{(q_c,k)}}(x), \quad  \vUps_{\pm} = \frac{1}{2} ( 1 \pm \vUps^{9+8(k-1)} ) .
\ee

To be more general, we can define a {\it string-like hyperqubit-spinor field} in category-$q_c$ with any self-conjugated chiral-like hyperqubit-spinor structure, i.e.:
\be
& & \Psi_{Q_N^{(q_c,k)}\pm}^{(j)}(x) = \vUps_{\pm}^j \Psi_{Q_N^{(q_c,k)}}^{(j)}(x), \quad  \vUps_{\pm}^j = \frac{1}{2} ( 1 \pm \vUps^{j} ) , 
\ee
where $\vUps^{j}$ can be any $\vUps$-matrix with $j=1,2,\cdots, 1+8k$ for the $k$-th period of $\cQ_c$-spin charge $\CQc^{(q_c,k)}$. The action for such a general string-like hyperqubit-spinor field is presented as follows:
\be \label{actionString}
 \cS_{Q_N^{(q_c,k)}\pm}  & = &  \int d^{2}x\,  \frac{1}{2} \Psi_{Q_N^{(q_c,k)}\pm}^{(j) \dagger}(x) \delta_{a}^{\; \mu}  \vUps^{a} i\p_{\mu} \Psi_{Q_N^{(q_c,k)}\pm}^{(j)}(x) , 
\ee
with $a,\mu =0, j$. 

The action for a general string-like hyperqubit-spinor field possesses the following associated symmetry:
\be 
G_S & = & SC(1)\ltimes P^{1,1}\ltimes SO(1,1)\adjoin SP(1,1) \rtimes SG(1) \times SP(D_{q_c}-2+8k) \times SP(q_c)  \nn \\
& = &  SC(1)\ltimes PO(1,1)\adjoin SP(1,1) \rtimes SG(1) \times SP(D_{q_c}-2+8k)\times SP(q_c) ,
\ee 
with $D_{q_c} = 3,4,6$ for $q_c=1,2,3$. SP$(D_{q_c}-2+8k)\times$SP($q_c$) represents the maximal internal symmetry for self-conjugated chiral-like hyperqubit-spinor field in category-$q_c$ and $k$-th period.

Therefore, we can always construct a {\it self-conjugated chiral-like qubit-spinor field} $\Psi_{Q_N^{(q_c,k)}\pm}(x)$ from a general qubit-spinor field in category-$q_c$ with categoric qubit number $Q_N^{(q_c,k)}$, so that it moves freely only in two-dimensional spacetime. Such a self-conjugated chiral-like qubit-spinor field $\Psi_{Q_N^{(q_c,k)}\pm}(x)$ is referred to as {\it string-like qubit-spinor field}.


\subsection{ Self-conjugated chiral hyperqubit-spinor field and full anti-commuting $\Gamma$-matrices in an extended Hilbert space }

When the dimension $\cD_H$ of Hilbert space spanned by product qubit basis states is larger than the dimension $D_h$ of Minkowski spacetime determined from the maximally correlated motion of qubit-spinor field, the motion correlation $\vGa$-matrices are not all satisfying anti-commuting relations, which indicates that such a Hilbert space formed by product qubit basis states is not large enough. To make all $\vGa$-matrices becoming anti-commuting, it is appropriate to consider self-conjugated chiral qubit-spinor structure for qubit-spinor field represented in an enlarged Hilbert space. A self-conjugated chiral qubit-spinor field can generally be constructed from self-conjugated qubit-spinor field with one additional qubit number in chiral qubit-spinor representation so as to keep the same independent degrees of freedom with equal $\cM_c$-spin charge and $\cQ_c$-spin charge.

Let us begin with examining the tetraqubit-spinor field. As shown in Eq.({\ref{SP10G}), not all $\Gamma$-matrices $\Gamma^{\mA}$ defined in Eq.({\ref{TGM}) are anti-commuting. The tetraqubit-spinor field in 16-dimensional Hilbert space belongs to a qubit-spinor representation in vector space of eight dimensions, i.e., $\cD_H=2^{D/2}=16$ with $D=8$. Whereas its maximally correlated motion brings on ten-dimensional Minkowski hyper-spacetime $D_h=10$. It naturally inspires us to represent such a tetraqubit-spinor field with 16 independent degrees of freedom in a qubit-spinor representation of vector space with the same dimension as that of Minkowski hyper-spacetime, i.e., $D=D_h$.

It is straightforward to construct a self-conjugated chiral hyperqubit-spinor field $\Psi_{\mQ^5-}(x)$ in terms of self-conjugated chiral-like hyperqubit-spinor fields $\Psi_{\mQ^4\pm}(x)$ defined in Eq.(\ref{MW4}), its explicit form is given as follows:
\be
& & \Psi_{\mQ^5-}(x) = \binom{\Psi_{\mQ^4-}(x) }{\Psi_{\mQ^4+}(x)} \equiv \vGa_{-}  \Psi_{\mQ^5}(x), \nn \\
& & \vGa_{-} = \frac{1}{2} ( 1 - \gamma_{\mQ^5} ), \quad  \gamma_{\mQ^5} = \sigma_3\otimes \Ups^{9} ,
\ee
where the 16 independent degrees of freedom of tetraqubit-spinor field are represented as a self-conjugated chiral hyperqubit-spinor field with chiral qubit-spinor representation defined in 10 dimensions $D=10$. The action in Eq.(\ref{actionQW4}) can be rewritten into the following form:
\be \label{actionQW5}
 \cS_{\mQ^5-}  & \equiv &  \cS_{\mQ^4} = \int d^{10}x\,  \frac{1}{2} \bar{\Psi}_{\mQ^5-}(x)  \delta_{\mA}^{\; \mM}  \vGa^{\mA} i\p_{\mM} \Psi_{\mQ^5-}(x) , 
 \ee
 with $\bar{\Psi}_{\mQ^5-}(x) = \Psi^{\dagger}_{\mQ^5-}\vGa^0$. The $\vGa$-matrices $\vGa^{\mA}= (\vGa^0, \vGa^{\tA}, \vGa^{10})$  ($\tA = 1,2,3,5, \cdots, 9$) are explicitly given as follows:
 \be \label{TGMMtW5}
& & \vGa^0 = \; \sigma_2 \otimes \Ups^0, \qquad  \vGa^{10} = - i\sigma_1 \otimes \Ups^{0}, \nn \\
& &  \vGa^{\tA} = - i\sigma_3 \otimes \Ups^{\bA}, \quad \tA=1,2,3=\bA, \nn \\
& & \vGa^{\tA} = - i\sigma_3 \otimes \Ups^{\bA+1},  \;\; \tA = 5, \cdots, 9= \bA + 1, 
\ee
where $\Ups^{0}$ and $\Ups^{\bA}$ ($\bA=1,\cdots, 8$) are presented in Eq.(\ref{UpsM4}). 

Similarly,  the action in Eq.(\ref{actionMW4}) can be expressed into the following form:
 \be
 \cS_{\QS^5-}  & \equiv &  \cS_{\QS^4} = \int d^{10}x \, \frac{1}{2} \bar{\Psi}_{\QS^5-}(x)  \delta_{\mA}^{\; \mM}  \vGa^{\mA} i\p_{\mM} \Psi_{\QS^5-}(x) , 
\ee
with $\bar{\Psi}_{\QS^5-}(x) = \Psi^{\dagger}_{\QS^5-}\vGa^0$. The $\vGa$-matrices $\vGa^{\mA}= (\vGa^0, \vGa^{10}, \vGa^{\tA})$  ($\tA = 1,2,3,5, \cdots, 9$) get the following structure:
\be \label{TGMMhW5}
& & \vGa^0 = \sigma_1 \otimes \Gamma^0, \quad  \vGa^{10} = -i\sigma_2 \otimes \Gamma^{0}, \nn \\
& & \vGa^{\tA} = \sigma_0 \otimes \Gamma^{\tA}, \quad \tA =1,2,3, 5, \cdots, 9, 
\ee
where $ \Gamma^0$ and $\Gamma^{\tA}$ ($\tA= 1,2,3,5, \cdots, 9$) are defined in Eq.(\ref{TGM}).  $\Psi_{\QS^5-}(x)$ is prsented as a self-conjugated chiral hyperqubit-spinor field with the following explicit form: 
\be
& & \Psi_{\QS^5-}(x) = \binom{\Psi_{\QS^4-}(x) }{\Psi_{\QS^4+}(x)} \equiv \vGa_{-}  \Psi_{\QS^5}(x), \nn \\
& & \vGa_{-} = \frac{1}{2} ( 1 -\gamma_{\QS^5} ), \quad \gamma_{\QS^5} = \sigma_3\otimes \Gamma^{0},
\ee
where $\Psi_{\QS^4\pm}(x)$ are the self-conjugated chiral-like hyperqubit-spinor fields defined in Eq.(\ref{QSpm4}).

It can be verified that both $\Psi_{\mQ^5-}(x)$ and $\Psi_{\QS^5-}(x)$ satisfy the following self-conjugated chiral conditions:
\be
& & \Psi^c_{\mQ^5-}(x) = C_{\mQ^5-} \bar{\Psi}^T_{\mQ^5-}(x) = \Psi_{\mQ^5-}(x), \quad \gamma_{\mQ^5}\Psi_{\mQ^5-}(x) =  -\Psi_{\mQ^5-}(x), \nn \\
& & \Psi^c_{\QS^5-}(x) = C_{\QS^5-} \bar{\Psi}^T_{\QS^5-}(x) = \Psi_{\QS^5-}(x), \quad \gamma_{\QS^5} \Psi_{\mQ^5-}(x) = -\Psi_{\QS^5-}(x), \nn \\
& & C_{\mQ^5-} = \vGa^{0}, \quad C_{\QS^5-} = \sigma_1 \otimes C_{\QS^4} = \sigma_1 \otimes  \sigma_2 \otimes \sigma_2 \otimes \sigma_2 \otimes \sigma_2, 
\ee
with $C_{\QS^4}$ defined in Eq.(\ref{CCQS4}). 

The self-conjugated chiral hyperqubit-spinor fields $\Psi_{\mQ^5-}(x)$ and $\Psi_{\QS^5-}(x)$ are referred to as {\it self-conjugated chiral pentaqubit-spinor fields}. In such a chiral qubit-spinor representation, all $\vGa$-matrices $\vGa^{\mA}$ defined in Eqs.(\ref{TGMMtW5}) and (\ref{TGMMhW5}) become anti-commuting, 
\be
\{\vGa^{\mA},  \vGa^{\mB} \} = 2\eta^{\mA\mB} , 
\ee
with $\eta^{\mA\mB} = \diag.(1, -1, \cdots, -1)$. Thus the group generators of hyperspin symmetry SP(1,9) can simply be given by the commutators of $\vGa$-matrices, 
\be
\Sigma^{\mA\mB} = \frac{i}{4}[\vGa^{\mA},  \vGa^{\mB} ], 
\ee
with $\mA, \mB= 0,1,2,3,5, \cdots, 10$.

The above analysis can be generalized to build the action for any hyperqubit-spinor field in category-$q_c$. 
For the general hyperqubit-spinor field in category-$0$, the action can be rewritten into the following form: 
\be \label{actionQWC0}
 \cS_{Q_N^{(1)}-}  & \equiv &  \cS_{Q_N^{(0)}} = \int d^{D_h^{(0,k)}}x\,  \frac{1}{2} \bar{\Psi}_{Q_N^{(1,k)}-}(x)  \delta_{\mA}^{\; \mM}  \vGa^{\mA} i\p_{\mM} \Psi_{Q_N^{(1,k)}-}(x) , 
 \ee
with $\bar{\Psi}_{Q_N^{(1,k)}-}(x) = \Psi^{\dagger}_{Q_N^{(1,k)}-}\vGa^0$. Where $\Psi_{Q_N^{(1,k)}-}(x)$ defines a self-conjugated chiral hyperqubit-spinor field with the following form:
\be \label{SCCC0}
& & \Psi_{Q_N^{(1,k)}-}(x) = \binom{\Psi_{Q_N^{(0,k)}-}(x)}{\Psi_{Q_N^{(0,k)}+}(x)} \equiv \vGa_{-}  \Psi_{Q_N^{(1,k)}}(x) , \nn \\
& & \vGa_{-} = \frac{1}{2} ( 1 - \gamma_{D_h^{(0,k)}+1} ), \quad  \gamma_{D_h^{(0,k)}+1} = \sigma_3\otimes \vUps^{9+8(k-1)} ,
\ee
where $\Psi_{Q_N^{(0,k)}\pm}(x)$ are the so-called self-conjugated chiral-like hyperqubit-spinor field in category-0 defined in Eq.(\ref{SCCLHQSF0}), and $\Psi_{Q_N^{(1,k)}}(x)$ is a hyperqubit-spinor field in category-1. The $\vGa$-matrices $\vGa^{\mA}= (\vGa^0, \vGa^{\tA + 8j}, \vGa^{10+8(k-1)})$  ($\tA = 1,2,3,5, \cdots, 9$;  $j=1,\cdots, k-1$; $k=1, 2, \cdots, $) have the following explicit structure:
 \be \label{GMC0}
& & \vGa^0 = \; \sigma_2 \otimes \vUps^0, \nn \\
& &  \vGa^{\tA + 8j } = - i\sigma_3 \otimes \vUps^{\bA + 8j}, \nn \\
& & \vGa^{10+8(k-1)} = - i\sigma_1 \otimes \vUps^{0}, 
\ee
where $\vUps^{0}$, $\vUps^{9+8(k-1)}$ and $\vUps^{\bA+8j}$ ($\bA=1,\cdots, 8$) are presented in Eq.(\ref{UpsQ_N^{(0)}}). 

Analogously, for the general hyperqubit-spinor field in category-$1$,  its action can be expressed as follows:
\be \label{actionQWC1}
 \cS_{Q_N^{(2)}-}  & \equiv &  \cS_{Q_N^{(1)}} = \int d^{D_h^{(1,k)}}x\, \{ \frac{1}{2} \bar{\Psi}_{Q_N^{(2,k)}-}(x)  \delta_{\mA}^{\; \mM}  \vGa^{\mA} i\p_{\mM} \Psi_{Q_N^{(2,k)}-}(x)  \nn \\
 & - & \frac{1}{2}\lambda_1 \phi_{1}(x)\bar{\Psi}_{Q_N^{(2,k)}-}(x)  \tilde{\vGa} \Psi_{Q_N^{(2,k)}-}(x) \}, 
 \ee
with $\bar{\Psi}_{Q_N^{(2,k)}-}(x) = \Psi^{\dagger}_{Q_N^{(2,k)}-}\vGa^0$. $\Psi_{Q_N^{(2,k)}-}(x)$ is defined as a self-conjugated chiral hyperqubit-spinor field,
\be \label{SCCC1}
& & \Psi_{Q_N^{(2,k)}-}(x) = \binom{0}{\Psi_{Q_N^{(1,k)}}(x)} \equiv \vGa_{-}  \Psi_{Q_N^{(2,k)}}(x) , \nn \\
& & \vGa_{-} = \frac{1}{2} ( 1 - \gamma_{D_h^{(1,k)}+1} ), \quad  \gamma_{D_h^{(1,k)}+1} = \sigma_3\otimes \sigma_0\otimes \vUps^{0} ,
\ee
where $\Psi_{Q_N^{(2,k)}}(x)$ is a hyperqubit-spinor field in category-2. The $\vGa$-matrices $\vGa^{\mA}= (\vGa^0, \vGa^{\tA + 8j}, \vGa^{10+8(k-1)}, \vGa^{11+8(k-1)}, \tilde{\vGa})$  ($\tA = 1,2,3,5, \cdots, 9$;  $j=1,\cdots, k-1$; $k=1,\cdots, $) get the following structure:
 \be \label{GMC1}
& & \vGa^0 = \; \sigma_1 \otimes \sigma_2 \otimes \vUps^0, \nn \\
&&  \vGa^{\tA + 8j } = i\sigma_1 \otimes \sigma_1 \otimes \vUps^{\bA + 8j}, \nn \\
& & \vGa^{10+8(k-1)} =  i\sigma_1 \otimes \sigma_1 \otimes \vUps^{9+8(k-1)},  \nn \\ 
& & \vGa^{11+8(k-1)} = - i\sigma_1 \otimes \sigma_3 \otimes \vUps^{0}, \nn \\
& & \tvGa = -i \sigma_2 \otimes \sigma_0 \otimes \vUps^0.
\ee

Similarly, we are able to rewrite the action for the general hyperqubit-spinor field in category-$2$ as follows:
\be \label{actionQWC2}
 \cS_{Q_N^{(3)}-}  & \equiv &  \cS_{Q_N^{(2)}} = \int d^{D_h^{(2,k)}}x\, \{ \frac{1}{2} \bar{\Psi}_{Q_N^{(3,k)}-}(x)  \delta_{\mA}^{\; \mM}  \vGa^{\mA} i\p_{\mM} \Psi_{Q_N^{(3,k)}-}(x)  \nn \\
 & - & \frac{1}{2}\lambda_2 \tilde{\phi}_p(x)\bar{\Psi}_{Q_N^{(3,k)}-}(x)  \tilde{\vGa}^p \Psi_{Q_N^{(3,k)}-}(x) \}, 
 \ee
with $\bar{\Psi}_{Q_N^{(3,k)}-}(x) = \Psi^{\dagger}_{Q_N^{(3,k)}-}\vGa^0$. $\Psi_{Q_N^{(3,k)}-}(x)$ is defined as a self-conjugated chiral hyperqubit-spinor field,
\be \label{SCCC2}
& & \Psi_{Q_N^{(3,k)}-}(x) = \binom{0}{\Psi_{Q_N^{(2,k)}}(x)} \equiv \vGa_{-}  \Psi_{Q_N^{(3,k)}}(x) , \nn \\
& & \vGa_{-} = \frac{1}{2} ( 1 - \gamma_{D_h^{(2,k)}+1} ), \quad  \gamma_{D_h^{(2,k)}+1} = \sigma_3\otimes \sigma_0\otimes \sigma_0\otimes \vUps^{0} ,
\ee
where $\Psi_{Q_N^{(3,k)}}(x)$ is a hyperqubit-spinor field in category-3. The $\vGa$-matrices $\vGa^{\mA}= (\vGa^0, \vGa^{\tA + 8j}, \vGa^{10+8(k-1)}, \vGa^{11+8(k-1)}, \vGa^{12+8(k-1)}$)  ($\tA = 1,2,3,5, \cdots, 9$;  $j=1,\cdots, k-1$; $k=1,2,\cdots, $) and $\tilde{\vGa}_p$ ($p=1,2$) have the following form:
 \be \label{GMC2}
& & \vGa^0 = \; \sigma_2 \otimes \sigma_0 \otimes \sigma_0 \otimes \vUps^0, \nn \\
& & \vGa^{\tA + 8j } = i\sigma_1 \otimes \sigma_0 \otimes \sigma_3 \otimes \vUps^{\bA + 8j}, \nn \\
& & \vGa^{10+8(k-1)} =  i\sigma_1 \otimes \sigma_0 \otimes \sigma_3 \otimes \vUps^{9+8(k-1)},  \nn \\
& & \vGa^{11+8(k-1)} = i\sigma_1 \otimes \sigma_0 \otimes \sigma_1 \otimes \vUps^{0}, \nn \\
& & \vGa^{12+8(k-1)} = i\sigma_1 \otimes \sigma_2 \otimes \sigma_2 \otimes \vUps^{0}, \nn \\
& & \tvGa^1 = i \sigma_1 \otimes \sigma_1 \otimes \sigma_2 \otimes \vUps^0, \nn \\
& & \tvGa^2 = i \sigma_1 \otimes \sigma_3 \otimes \sigma_2 \otimes \vUps^0 .
\ee

It can be verified that $\Psi_{Q_N^{(q_c,k)}-}(x)$ ($q_c=1,2,3$) satisfies the following self-conjugated chiral condition:
\be
& & \Psi^c_{Q_N^{(q_c,k)}-}(x)= C_{Q_N^{(q_c,k)}} \bar{\Psi}^T_{Q_N^{(q_c,k)}-}(x) = \Psi_{Q_N^{(q_c,k)}-}(x), \quad C_{Q_N^{(q_c,k)}} = \vGa^{0},  \nn \\
& & \gamma_{D_h^{(q_c-1,k)}+1}\Psi_{Q_N^{(q_c,k)}-}(x) = -\Psi_{Q_N^{(q_c,k)}-}(x) =  \gamma_{D_h^{(q_c-1, k)}+1}\Psi^c_{Q_N^{(q_c,k)}-}(x), 
\ee
where $\Psi_{Q_N^{(q_c,k)}-}(x)$ ($q_c=1,2,3$) is referred to as self-conjugated chiral hyperqubit-spinor field in category-($q_c$-1) and $k$-th period. In such a self-conjugated chiral qubit-spinor representation, all $\vGa$-matrices become anti-commuting, 
\be
& & \{\vGa^{\mA},  \vGa^{\mB} \} = 2\eta^{\mA\mB}, \quad \{\vGa^{\mA},  \tilde{\vGa}^{p} \} = 0, \quad \{\tilde{\vGa}^{p},  \tilde{\vGa}^{q} \} = 2\eta^{pq}, 
\ee
with $\mA, \mB= 0,1,2,3,5, \cdots, D_h^{(q_c-1,k)}$ and $p, q = 1, \cdots, q_c-1$. The group generators of hyperspin symmetry SP(1, $D_h^{(q_c-1,k)}$-1) and $\cQ_c$-spin symmetry SP($q_c-1$) are directly defined by the commutators of $\vGa$-matrices as follows: 
\be
\Sigma^{\mA\mB} = \frac{i}{4}[\vGa^{\mA},  \vGa^{\mB} ] , \quad \tilde{\Sigma}^{pq} = \frac{i}{4}[\tilde{\vGa}^{p},  \tilde{\vGa}^{p} ] . 
\ee

Therefore, when the categoric hyperqubit-spinor field $\Psi_{Q_N^{(q_c,k)}}(x)$ with categoric qubit number $Q_N^{(q_c,k)}$ in category-$q_c$ and $k$-th period is represented by a self-conjugated chiral hyperqubit-spinor field $\Psi_{Q_N^{(q_c+1,k)}-}(x)$ constructed from the hyperqubit-spinor field $\Psi_{Q_N^{(q_c+1,k)}}(x)$ in category-($q_c$+1) with one additional qubit number $Q_N^{(q_c+1,k)} = Q_N^{(q_c,k)}+1$, all $\vGa$-matrices that reflect both $\cM_c$-spin charge of hyperspin symmetry and $\cQ_c$-spin charge of $\cQ_c$-spin symmetry become full anti-commuting. The categoric hyperqubit-spinor field $\Psi_{Q_N^{(q_c,k)}}(x)$ and self-conjugated chiral hyperqubit-spinor field $\Psi_{Q_N^{(q_c+1,k)}-}(x)$ are defined in Hilbert space with dimensions $\cD_H= 2^{Q_N^{(q_c,k)}}$ and $\cD_H=2^{Q_N^{(q_c,k)} + 1}$, respectively, but they contain the same independent degrees of freedom with equal $\cM_c$-spin charge and $\cQ_c$-spin charge.


\subsection{ Massless Dirac fermion as self-conjugated chiral tetraqubit-spinor field in category-$3$ with full anti-commuting $\cM_c$-matrices and $\cQ_c$-matrices in 6D spacetime }

The actions in Eqs.(\ref{actionD2P}) and (\ref{actionDC}) indicate that a massless Dirac fermion can be described by a self-conjugated triqubit-spinor field $\psi_{\QS^3}(x)$ given in Eq.(\ref{MD23}) or equivalently by a self-conjugated triqubit-spinor field $\psi_{\fQS^3}(x)$ presented in Eq.(\ref{fQS3}) via a unitary transformation. Nevertheless, in such qubit-spinor representations, the $\Gamma$-matrices $\Gamma^{\ha}$ presented in Eq.(\ref{GM3}) or Eq.(\ref{DGM}) are not all anti-commuting, which motivates us to consider an extended qubit-spinor representation with self-conjugated chiral spinor structure. In fact, leptons and quarks in the electroweak interaction of SM behave as Weyl fermions with a chiral spinor structure in four-dimensional spacetime.

The self-conjugated triqubit-spinor field $\psi_{\QS^3}(x)$ can generally be rewritten into the sum of two 
{\it self-conjugated chiral triqubit-spinor fields} as follows:
\be
& & \psi_{\QS^3}(x) = \psi_{\hQS^3-}(x) + \psi_{\hQS^3+}(x) , \nn \\
& & \psi_{\hQS^3\pm}(x) = \hat{\Gamma}_{\pm} \psi_{\QS^3}(x), \nn \\
& &  \hat{\Gamma}_{\pm} = \frac{1}{2} ( 1 \pm \hat{\gamma}_7), \quad \hat{\gamma}_7= \sigma_3\otimes \gamma_5, 
\ee
with the definitions,
\be \label{QSW3}
& &  \psi_{\hQS^3-}(x) \equiv \binom{\psi_{\QC^2-}(x)}{\psi_{\QC^2-}^{\; \bc}(x)} \equiv \binom{\psi_{\QWn^2}(x)}{\psi_{\QWn^2}^{\; \bc}(x)}, \nn \\\
& & \psi_{\hQS^3+}(x) \equiv \binom{\psi_{\QC^2+(x)}}{\psi_{\QC^2+}^{\; \bc}(x)} \equiv \binom{\psi_{\QWp^2}(x)}{\psi_{\QWp^2}^{\; \bc}(x)}, \nn \\
& &  \psi_{\QC^2\pm}^{\; \bc} \equiv C_D \bar{\psi}_{\QC^2\pm}^{T}, \quad C_D = -i \sigma_3\otimes \sigma_2 ,
\ee
where $\psi_{\QC^2-}(x)$ and $\psi_{\QC^2+}(x)$ are chiral complex biqubit-spinor fields, 
\be \label{QC2PM}
& & \psi_{\QC^2\pm} \equiv \gamma_{\pm} \psi_{\QC^2}  \equiv \psi_{\QWpn^2} \equiv \gamma_{\pm} \psi_{D} , \nn \\
& & \gamma_{\pm} = \frac{1}{2} ( 1 \pm \gamma_5) , \quad \gamma_5 = \sigma_3\otimes \sigma_0, 
\ee
which are regarded as the left-handed and right-handed Weyl fermions defined from Dirac fermion in the spinor representation of four dimensions, 

In light of {\it self-conjugated chiral triqubit-spinor fields} $\psi_{\hQS^3\pm}(x)$ which satisfy the following self-conjugated chiral conditions:
\be
& &  \psi_{\hQS^3\pm}^c(x) = C_{\QS^3} \bar{\psi}_{\hQS^3\pm}^T(x) =   \psi_{\hQS^3\pm}(x) , \nn \\
& & \hat{\gamma}_7  \psi_{\hQS^3\pm}(x) = \hat{\gamma}_7 \psi_{\hQS^3\pm}^c(x) =  \pm  \psi_{\hQS^3\pm}(x) , \nn \\
& & C_{\QS^3}  = - i \sigma_1\otimes   \sigma_3\otimes  \sigma_2 , 
\ee
the action in Eq.(\ref{actionM3P}) can be rewritten into the following form in six-dimensional spacetime:
\be  \label{actionQS3}
\cS_{\QS^3}   & = & \int d^6x \,  \frac{1}{2}  [ \bar{\psi}_{\hQS^3+}(x)  \delta_{a}^{\;\mu}  \Gamma^a i\p_{\mu} \psi_{\hQS^3+}(x) + \bar{\psi}_{\hQS^3-}(x)  \delta_{a}^{\;\mu}  \Gamma^a i\p_{\mu} \psi_{\hQS^3-}(x)  \nn \\
& + & \bar{\psi}_{\hQS^3+}(x)  \delta_{k}^{\; m}  \Gamma^k i\p_{m} \psi_{\hQS^3-}(x) + \bar{\psi}_{\hQS^3-}(x)  \delta_{k}^{\; m}  \Gamma^k i\p_{m} \psi_{\hQS^3+}(x) ] , 
\ee
with $a,\mu=0,1,2,3$ and $k,m=5,6$. Where the correlation motions in the fifth and sixth spatial dimensions can only appear between two self-conjugated chiral triqubit-spinor fields $\psi_{\hQS^3-}(x)$ and $\psi_{\hQS^3+}(x)$. 

By introducing an alternative qubit-spinor field $\psi_{\hQS^4-}$ with satisfying the following self-conjugated chiral conditions:
\be
& & \psi_{\hQS^4-} = \binom{ \psi_{\hQS^3-}}{\psi_{\hQS^3+}}, \nn \\
& &  \psi_{\hQS^4-}^{c}(x)  =  C_{\hQS^4} \bar{\psi}^T_{\hQS^4-}(x) =  \psi_{\hQS^4-}, \nn \\
& & \hat{\gamma}_9 \psi_{\hQS^4-} = \hat{\gamma}_9 \psi^c_{\hQS^4-} = -\psi_{\hQS^4-}, \quad \hat{\gamma}_9 =  \sigma_3\otimes \hat{\gamma}_7, \nn \\
& & C_{\hQS^4} = \sigma_0\otimes C_{\QS^3}  = - i \sigma_0\otimes \sigma_1\otimes   \sigma_3\otimes  \sigma_2 , 
\ee
we can express the above action into the following simple form:
\be  \label{actionhQS4-}
\cS_{\hQS^4-} \equiv \cS_{\QS^3}   & = &  \int d^6x \,  \frac{1}{2}  \bar{\psi}_{\hQS^4-}(x)  \delta_{\ha}^{\;\hm}  \vGa^{\ha} i\p_{\hm} \psi_{\hQS^4-}(x) , 
\ee
with $\bar{\psi}_{\hQS^4-}(x) =\psi_{\hQS^4-}^{\dagger}(x)\vGa^0$. We may refer to $\psi_{\hQS^4-}(x)$ as {\it self-conjugated chiral tetraqubit-spinor field}. In such a chiral spinor representation, the $\vGa$-matrices $\vGa^{\ha}\equiv (\vGa^a, \vGa^k)$ ($a=0,1,2,3$; $k=5,6$) get following explicit form:
\be \label{GMQS4}
& & \vGa^a = \sigma_0\otimes \sigma_0\otimes \gamma^a, \nn \\
& &  \vGa^5 = i \sigma_2\otimes \sigma_0\otimes \gamma_5, \nn \\
& & \vGa^6 = i \sigma_1\otimes \sigma_3\otimes \gamma_5, 
\ee
with $\gamma^a$ defined in Eq.(\ref{GM2}). The $\cQ_c$-matrices $\tGa^p = \vGa^{6+p}$($p=1,2,3$) are found to be, 
\be
& & \vGa^7 = i \sigma_1\otimes \sigma_1\otimes \gamma_5 \equiv \tilde{\Gamma}^1, \nn \\
& &  \vGa^8 = i \sigma_1\otimes \sigma_2\otimes \gamma_5 \equiv \tilde{\Gamma}^2, \nn \\
& &  \vGa^9 = i \sigma_3\otimes \sigma_0\otimes \gamma_5 \equiv \tilde{\Gamma}^3.
\ee

It can be verified that all $\Gamma$-matrices become anti-commuting,  
\be
& & \{\Gamma^{\ha}, \Gamma^{\hb} \} = 2 \eta^{\ha\hb} , \quad  \{\tilde{\Gamma}^{p}, \tilde{\Gamma}^{q} \} = 2 \eta^{pq} , \quad  \{\Gamma^{\ha}, \tilde{\Gamma}^{p} \} = 0 .
\ee
The group generators of $\cM_c$-spin symmetry SP(1,5) and $\cQ_c$-spin symmetry SP(3) can simply be given by the commutators of $\vGa$-matrices, 
\be
& & \Sigma^{\ha\hb} = \frac{i}{4}  [\vGa^{\ha}, \vGa^{\hb} ] , \quad \tilde{\Sigma}^{pq} = \frac{i}{4}[ \tvGa^p , \tvGa^q ] ,
\ee 
where $\cQ_c$-spin symmetry SP(3)$\cong$SU(2) reflects an U(1) charge-spin between the chiral complex biqubit-spinor field $\psi_{\QC^2\pm}(x)$ and its charge conjugated one $\psi_{\QC^2\pm}^{\; \bc}$ as well as a rotational-spin property between two self-conjugated chiral triqubit-spinor fields $\psi_{\hQS^3-}(x)$ and $\psi_{\hQS^3+}(x)$.


\section{ Entangled hyperqubit-spinor field from locally entangled-qubits motion postulate and inhomogeneous hyperspin symmetry as basic symmetry of lepton-quark states }

It is shown that in light of self-conjugated chiral qubit-spinor structure, all $\Gamma$-matrices become anti-commuting and the group generators of $\cM_c$-spin and $\cQ_c$-spin symmetries are directly presented by the commutators of $\Gamma$-matrices, which provides an appropriate qubit-spinor representation in Hilbert space for building the action of qubit-spinor field. In this section, we are going to investigate further the intrinsic property of self-conjugated chiral qubit-spinor field, which allows us to extend the motion-correlation $\cM_c$-spin symmetry to be an {\it inhomogeneous $\cM_c$-spin symmetry} in association with inhomogeneous Lorentz-type/Poincar\'e-type group symmetry. In particular, we are led to make a {\it locally entangled-qubits motion postulate} by which a locally entangled state of qubits is proposed to be as an entangled qubit-spinor field, which enables us to show how inhomogeneous hyperspin symmetry of entangled hyperqubit-spinor field reflect basic symmetries of lepton-quark states.


\subsection{Entangled state in quantum theory and locally entangled-qubits motion postulate }

It is known in quantum physics that multiple qubits can exhibit quantum entanglement, which is regarded as the distinguishing feature between a qubit and a bit.  For instance, the entanglement of quantum state of two qubits can be denoted either by Dirac ``bra-ket" notation or by spin-like ``column-row" notation as follows:
\be
& & |\Psi_E\rangle = \alpha_{00} |0\rangle \otimes |0\rangle + \alpha_{11} |1\rangle \otimes |1\rangle \equiv  \alpha_{00} |00\rangle + \alpha_{11} |11\rangle , \nn \\ 
& & \Psi_E =  \alpha_{++}\binom{1}{0}\otimes\binom{1}{0} +  \alpha_{--} \binom{0}{1}\otimes\binom{0}{1} =  \begin{pmatrix} 
\alpha_{++}\\
0\\
0 \\
\alpha_{--}
 \end{pmatrix}, \nn \\
 & &  |\alpha_{00}|^2 + |\alpha_{11}|^2 = 1, \quad |\alpha_{++}|^2 + |\alpha_{--}|^2 = 1, 
\ee
where $|\alpha_{00}|^2$ or $|\alpha_{++}|^2$ and $|\alpha_{11}|^2$ or $|\alpha_{--}|^2$ are probabilities of measuring $ |0\rangle \otimes |0\rangle$ or $\binom{1}{0}\otimes\binom{1}{0}$ and $|1\rangle \otimes |1\rangle$ or $\binom{0}{1}\otimes\binom{0}{1}$, respectively.  For $\alpha_{00} = \alpha_{11} = 1/{\sqrt {2}} $ or $\alpha_{++} = \alpha_{--} = 1/{\sqrt {2}}$,  it is called equal superposition with equal probability $|1/{\sqrt {2}}|^{2}=1/2$. 

In general, an entangled system is considered to be the system in which its quantum state cannot be factored as a product of pure states. 

For a more formal analysis, let us consider two noninteracting systems $S$ and $S'$ with Hilbert spaces H$_{S}$ and H$_{S'}$, respectively. The Hilbert space of their composite system is defined by the tensor product H$_{S}\otimes$H$_{S'}$. If two systems are in states $\Psi_S$ and $\Psi_{S'}$, the state of the composite system is given by the tensor product $\Psi_S\otimes\Psi_{S'}$. If states of the composite system can be represented in such a form in the tensor product, they are called product states or separable states. Whereas not all states can be expressed as product states or separable states. 

Given a basis $\{\xi_i\}$ for H$_{S}$  and a basis  $\{\xi'_i\}$ for H$_{S'}$, a general state in H$_{S}\otimes$H$_{S'}$ can be expressed as follows:
\be
\Psi = \sum_{i,j} \psi_{ij}\, \xi_i \otimes \xi'_j  , \nn
\ee 
where $\Psi$ is called separable if
\be
& & \psi_{ij} = \psi_i\, \psi'_j,  \quad \Psi = \psi \otimes \psi', \nn \\
& &  \psi = \sum_{i} \psi_{i}\, \xi_i, \quad \psi' = \sum_{j} \psi'_{j}\, \xi'_j, \nn
\ee
with $\psi$ and $\psi'$ representing two vectors. 

When $\Psi$ becomes inseparable if 
\be 
\psi_{ij} \neq \psi_i\, \psi'_j , \quad \Psi \neq \psi \otimes \psi', \nn
\ee
hold for all vectors $\psi_i$ and $\psi'_j$. 

If a state is inseparable, it is referred to as {\it entangled state}. As an example, let us consider the simplest case with two basis states $\varsigma_s= \{ |0\rangle, |1\rangle\} $ or $\varsigma_s = \{\binom{1}{0}, \binom{0}{1} \}$
of Hilbert space H$_{S}$ and two basis states $\varsigma_{s'}= \{ |0\rangle, |1\rangle \} $ or $\varsigma_{s'} = \{ \binom{1}{0}, \binom{0}{1}\}$ of Hilbert space H$_{S'}$, an entangled state can be constructed as
\be
& & |\psi_{E_-}\rangle = \frac{1}{\sqrt{2}} \left( |0\rangle \otimes |1\rangle + |1\rangle \otimes |0\rangle \right) \equiv  \frac{1}{\sqrt{2}} \left( |01\rangle + |10\rangle \right) \nn
\ee
with the Dirac ``bra-ket" notation, which can also be expressed equivalently by using the spin-like ``column-row" notation as follows:
\be
& & \psi_{E_-} = \frac{1}{\sqrt{2}} \left( \binom{1}{0}\otimes\binom{0}{1} +  \binom{0}{1}\otimes\binom{1}{0} \right) = \frac{1}{\sqrt{2}}   \begin{pmatrix} 
0\\
1\\
1 \\
0
 \end{pmatrix} ,
\ee
which is actually one of four Bell states. In general, Bell states are maximally entangled pure states in Hilbert space H$_{S}\otimes$H$_{S'}$, but they cannot be separated into pure states of each H$_{S}$ and H$_{S'}$.

Let us make a detailed discussion on the maximally entangled pure states in H$_{S}\otimes$H$_{S'}$ based on the real basic qubit-basis $\{\varsigma_s\}$ defined in Eq.(\ref{QB}). Applying for {\it 2-product qubit-basis states} $\varsigma_{s_1s_2}$ defined in Eq.(\ref{RPQB}), we can express a biqubit-spinor state into the following general form:
\be \label{4BS}
& & \Psi = \sum_{s_1s_2}\, \psi_{s_1s_2} \varsigma_{s_1s_2} \equiv  \sum_{s_1s_2} \psi_{s_1s_2} \, \varsigma_{s_1}\otimes \varsigma_{s_2} = \begin{pmatrix} 
\psi_{++}\\
\psi_{+-}\\
\psi_{-+} \\
\psi_{--}
 \end{pmatrix},  
\ee
with $s_1,s_2= \pm $. where $\psi_{s_1s_2}$ denote four constant amplitudes. The four Bell states as maximally entangled pure states in such a spinor representation can be read from the above biqubit-spinor state with the following conventions:
\be
& & \psi_{E_{\pm}} = \hat{\gamma}_{\pm} \Psi, \quad \psi_{E'_{\pm}} = \gamma_5\hat{\gamma}_{\pm} \Psi, \quad \psi_{s_1s_2} = \frac{1}{\sqrt{2}}, \nn \\
& & \hat{\gamma}_{\pm} = \frac{1}{2} ( 1 \pm \hat{\gamma}_5) , \quad  \hat{\gamma}_5 = \sigma_3\otimes \sigma_3, \quad  \gamma_5 = \sigma_3\otimes \sigma_0 ,
\ee 
where $\psi_{E_{\pm}}$ and $\psi_{E'_{\pm}}$ are regarded as chiral biqubit-spinor states.

To be more explicit, let us write down four Bell states in the spinor representation as follows:
\be
& & \psi_{E_+} = \psi_{++} \varsigma_{+}\otimes \varsigma_{+} +   \psi_{--} \varsigma_{-}\otimes \varsigma_{-} = \frac{1}{\sqrt{2}} \begin{pmatrix} 
1\\
0\\
0 \\
1
 \end{pmatrix} , \nn \\
& &  \psi_{E_-} = \psi_{-+} \varsigma_{-}\otimes \varsigma_{+} +   \psi_{+-} \varsigma_{+}\otimes \varsigma_{-} = \frac{1}{\sqrt{2}}   \begin{pmatrix} 
0\\
1\\
1 \\
0
 \end{pmatrix}, \nn \\
& &  \psi_{E'_+} = \psi_{++} \varsigma_{+}\otimes \varsigma_{+} -   \psi_{--} \varsigma_{-}\otimes \varsigma_{-} = \frac{1}{\sqrt{2}} \begin{pmatrix} 
1\\
0\\
0 \\
-1
 \end{pmatrix} , \nn \\
 & & \psi_{E'_-} = \psi_{-+} \varsigma_{-}\otimes \varsigma_{+} -   \psi_{+-} \varsigma_{+}\otimes \varsigma_{-} = \frac{1}{\sqrt{2}} \begin{pmatrix} 
0\\
1\\
-1 \\
0
 \end{pmatrix} . 
 \ee
So that the general entangled states can be represented as follows:
\be
& & \Psi_{E_{+}} = \hat{\gamma}_{+} \Psi = \frac{1}{\sqrt{2}}(\psi_{++} + \psi_{--} )\, \psi_{E_+} +   \frac{1}{\sqrt{2}}(\psi_{++} - \psi_{--} )\, \psi_{E'_+} = \begin{pmatrix} 
\psi_{++}\\
0\\
0 \\
\psi_{--}
 \end{pmatrix},  \nn \\
& & \Psi_{E_-} = \hat{\gamma}_{-} \Psi  =  \frac{1}{\sqrt{2}}(\psi_{+-} + \psi_{-+} )\, \psi_{E_-} +  \frac{1}{\sqrt{2}}(\psi_{+-} + \psi_{-+} )\, \psi_{E'_-} = \begin{pmatrix} 
0\\
\psi_{+-}\\
\psi_{-+} \\
0
 \end{pmatrix},   \nn \\
& &  |\psi_{++}|^2 +  |\psi_{--}|^2 = 1, \quad |\psi_{+-}|^2 +  |\psi_{-+}|^2 = 1 ,
\ee 
which indicates that a general entangled state for two qubits can be expressed as a superposition of Bell states and meanwhile constructed by a chiral qubit-spinor state with global amplitudes constrained by the total probability.

From the above analyses, we come to the statement that by adopting qubit basis states in the spinor representation, an entangled state in quantum physics can formally be expressed into a chiral qubit-spinor state. Therefore, a general entangled state in quantum physics can be regarded as an entangled state of qubits with the probability amplitudes satisfying the normalization condition constrained from the total probability.

It is noticed that all above analyses and discussions about entangled state as an inseparable state in the tensor product Hilbert space are actually applicable to local states of qubits rather than only global states. Namely, they should hold for both global and local coherent states of qubits. 
 
Let us now turn to investigate local coherent state of qubits as qubit-spinor field. It is demonstrated in the previous section that a self-conjugated chiral qubit-spinor field becomes more appropriate to build the action, so that all $\Gamma$-matrices satisfy the anticommutative relations. Meanwhile, the group algebra of $\cM_c$-spin and $\cQ_c$-spin symmetries can simply be formed from the commutators of $\Gamma$-matrices. In analogous to the above analyses and discussions on entangled states in quantum physics as global coherent states of qubits, we come to a similar statement that a self-conjugated chiral qubit-spinor field is conceptually regarded as an {\it entangled qubit-spinor field}. For that, we are led to make the following postulate.

{\it Locally entangled-qubits motion postulate}: a locally entangled state of qubits as a column vector field of local distribution amplitudes obeying the maximum coherence motion principle brings on an {\it entangled qubit-spinor field} which is postulated to be as {\it fundamental building block of nature}. Such an entangled qubit-spinor field can generally be expressed as a {\it self-conjugated chiral qubit-spinor field}.


\subsection{ Majorana-Weyl type fermions as entangled qubit-spinor fields in category-2 $\And$ category-3 and the inhomogeneous $\cM_c$-spin symmetries associating with inhomogeneous Lorentz group symmetries in four- and six-dimensional Minkowski spacetimes }

The actions in Eqs.(\ref{actionDC}) and (\ref{actionfQS3}) indicate that the self-conjugated triqubit-spinor field $\psi_{\fQS^3}(x)$ in category-3 defined in Eq.(\ref{fQS3}) characterizes a massless Dirac fermion $\psi_D(x)$ in six-dimensional spacetime. Whereas in such a triqubit-spinor representation, the eight-dimensional Hilbert space spanned by 3-product qubit basis states is not large enough to make all $\Gamma$-matrices anti-commuting. As shown in Eq.(\ref{actionhQS4-}), it is more appropriate to extend eight-dimensional Hilbert space to 16-dimensional Hilbert space spanned by 4-product qubit basis states with a chiral structure so as to keep the same independent degrees of freedom. 

Let us begin with introducing {\it entangled triqubit-spinor fields} $\psi_{\QEpn^3}(x)$, which are defined by the following self-conjugated chiral triqubit-spinor fields $\psi_{\fQS^3\pm}(x)$: 
\be
& &  \psi_{\QEn^3}(x) \equiv \psi_{\fQS^3-}(x) \equiv \Gamma_{-}\psi_{\fQS^3}(x) = \binom{\psi_{L}(x)}{\psi_{L}^{\; \bc}(x)}, \nn \\
& & \psi_{\QEp^3}(x) \equiv \psi_{\fQS^3+}(x) \equiv \Gamma_{+}\psi_{\fQS^3}(x) = \binom{\psi_{R}(x)}{\psi_{R}^{\; \bc}(x)} , \nn \\
& &  \Gamma_{\pm} = \frac{1}{2} ( 1 \pm \hat{\gamma}_7), \quad \hat{\gamma}_7= \sigma_3\otimes \gamma_5, 
\ee
where $\psi_{L}$ and $\psi_{R}$ are the usual left-handed and right-handed Weyl fermions defined by the Dirac fermion $\psi_D(x)$ as follows:
\be
& & \psi_{L,R}  = \gamma_{\mp} \psi_D , \quad \gamma_{\pm} = \frac{1}{2} ( 1 \pm \gamma_5) , \nn \\
& &  \psi_{L, R}^{\; \bc} =C_D \bar{\psi}_{L, R}^{T}, \quad C_D = -i\sigma_3\otimes \sigma_2 ,
\ee
with $\psi_{L, R}^{\; \bc}$ being complex charge-conjugated ones. In general, $\psi_D$ can be regarded as complex biqubit-spinor field $\psi_D\equiv \psi_{\QC^2}$, and $\psi_{L,R}$ are viewed as chiral complex biqubit-spinor fields $\psi_{L,R}\equiv \psi_{\fQWpn^2}$.

The self-conjugated triqubit-spinor field $\psi_{\fQS^3}(x)$ can be expressed as the superposition of two entangled triqubit-spinor fields, i.e.:
\be
& & \psi_{\fQS^3}(x)   \equiv  \psi_{\QEn^3}(x) +  \psi_{\QEp^3}(x)
= \binom{\psi_{D}(x) } { \psi_{D}^{\bc}(x) }, 
\ee
where $\psi_{\QEpn^3}(x)$ satisfy the following self-conjugated chiral conditions:
\be \label{QESCC}
& & \psi_{\QEpn^3}^{c}(x) = C_{\QE^3} \bar{\psi}_{\QEpn^3}^T(x) = \psi_{\QEpn^3}(x), \quad C_{\QE^3} = \sigma_1 \otimes  C_{D} , \nn \\
& & \hat{\gamma}_7 \psi_{\QEpn^3}(x) = \pm  \psi_{\QEpn^3}(x), \quad \hat{\gamma}_7 \psi_{\QEpn^3}^{c}(x)=\pm \psi_{\QEpn^3}^{c}(x) .
\ee

Such defined entangled triqubit-spinor fields $\psi_{\QEpn^3}(x)$ are also considered as {\it Majorana-Weyl type fermions}. To be distinguishable with ordinary {\it left-handed and right-handed Weyl fermions} defined in four-dimensional spacetime via the project operators $ \gamma_{\mp} = \frac{1}{2}( 1\mp \gamma_5) \equiv \gamma_{L,R}$, the entangled qubit-spinor fields defined with the project operators $\Gamma_{\mp} = \frac{1}{2}( 1\mp \hat{\gamma}_{D+1})$ in high dimensional spacetime are referred to as {\it westward qubit-spinor field and eastward qubit-spinor field}, respectively. So that, $\psi_{\QEn^3}(x)$ is called as {\it westward entangled triqubit-spinor field} and $\psi_{\QEp^3}(x)$ as {\it eastward entangled triqubit-spinor field}.

In terms of the westward and eastward entangled triqubit-spinor fields $\psi_{\QEpn^3}(x)$, the actions in Eqs.(\ref{actionfQS3}) and (\ref{actionDM}) in category-3 can be rewritten into the following form:
\be  \label{actionQE4}
\cS_{\QE^4} \equiv \cS_{\fQS^3}  & = & \int d^6x \,  \frac{1}{2}  [ \bar{\psi}_{\QEp^3}(x)  \delta_{a}^{\;\mu}  \Gamma^a i\p_{\mu} \psi_{\QEp^3}(x) + \bar{\psi}_{\QEn^3}(x)  \delta_{a}^{\;\mu}  \Gamma^a i\p_{\mu} \psi_{\QEn^3}(x) ] \nn \\
& + & \bar{\psi}_{\QEp^3}(x)  \delta_{k}^{\; m}  \Gamma^k i\p_{m} \psi_{\QEn^3}(x) + \bar{\psi}_{\QEn^3}(x)  \delta_{k}^{\; m}  \Gamma^k i\p_{m} \psi_{\QEp^3}(x) ] , \nn \\
& \equiv &  \int d^6x \,   \bar{\psi}_{\QE^4}(x)  \delta_{\ha}^{\;\hm}  \Sigma_{-}^{\ha} i\p_{\hm} \psi_{\QE^4}(x) , 
\ee
with $a,\mu=0,1,2,3$, $\, k,m=5,6$ and $\ha, \hm = 0,1,2,3,5,6$ labeling six-dimensional spacetime. We have introduced in the second equality the matrices $\Sigma_{-}^{\ha}$ defined by $\Gamma$-matrices as follows:
\be
& & \Sigma_{-}^{\ha} = \frac{1}{2}\Gamma^{\ha}\Gamma_{-}, \quad \Gamma_{-} = \frac{1}{2} ( 1- \hat{\gamma}_9 ), \nn \\
& & \hat{\gamma}_9 =  \sigma_3\otimes \hat{\gamma}_7 = \sigma_3\otimes \sigma_3\otimes \gamma_5 . 
\ee
Where $\psi_{\QE^4}$ represents a westward entangled qubit-spinor field satisfying the following self-conjugated chiral conditions:
\be
& & \psi_{\QE^4} = \binom{ \psi_{\QEn^3}}{\psi_{\QEp^3}} = \begin{pmatrix} 
 \psi_{L}(x)   \\  \psi_{L}^{\; \bc}(x)  \\  \psi_{R}(x)  \\  \psi_{R}^{\; \bc}(x)
\end{pmatrix} \equiv \Gamma_{-}\psi_{\fQS^4}, \nn \\ 
& & \psi_{\QE^4}^{c}(x)  =  C_{\QE^4} \bar{\psi}^T_{\QE^4}(x) = \psi_{\QE^4}, \nn \\
& & \hat{\gamma}_9 \psi_{\QE^4} = -\psi_{\QE^4}, \quad \hat{\gamma}_9 \psi^c_{\QE^4} = -\psi^c_{\QE^4}, \nn \\
& & C_{\QE^4} = \sigma_0\otimes C_{\QE^3}  = - i \sigma_0\otimes \sigma_1\otimes   \sigma_3\otimes  \sigma_2 . 
\ee
For convenience, $\psi_{\QE^4}$ is referred to as {\it entangled tetraqubit-spinor field}, which is also regarded as Majorana-Weyl type fermion. 

In such a westward spinor representation, both $\Gamma$-matrices $\Gamma^{\ha}\equiv (\Gamma^a, \Gamma^k)$ ($a=0,1,2,3$; $k=5,6$) and $\cQ_c$-matrices $\tGa^p$ ($p=1,2,3$) can be expressed as the following forms:
\be \label{GMQE4}
& & \Gamma^a = \sigma_0\otimes \sigma_0\otimes \gamma^a, \nn \\
& &  \Gamma^5 = i \sigma_1\otimes \sigma_3\otimes \gamma_5, \nn \\
& & \Gamma^6 = i \sigma_2\otimes \sigma_0\otimes \gamma_5, \nn \\
& & \Gamma^7 = i \sigma_1\otimes \sigma_1\otimes \gamma_5  \equiv \tGa^1 , \nn \\
& & \Gamma^8 = i \sigma_1\otimes \sigma_2\otimes \gamma_5 \equiv \tGa^2 , \nn \\
& &  \Gamma^9 = i \sigma_3\otimes \sigma_0\otimes \gamma_5 \equiv \tGa^3 ,
\ee
with $\gamma^a$ ($a=0,1,2,3$) defined in Eq.(\ref{DGM}). In terms of the entangeled tetraqubit-spinor field $\psi_{\QE^4}$, it is easy to check that there exists no scalar coupling term in associating to $\tGa^p$ matrices. This is because the matrices $\Gamma^{\ha}$ and $\tGa^p$ have different commutation relations with $\hat{\gamma}_9$, i.e.:
\be
[ \tGa^p, \hat{\gamma}_9] = 0, \quad  \{ \Gamma^{\ha}, \hat{\gamma}_9 \} = 0  .   
\ee

It is noticed from the action in Eq.(\ref{actionQE4}) that the correlation motion in the fifth and sixth spatial dimensions can occur only between two entangled triqubit-spinor fields $\psi_{\QEp^3}(x)$ and $\psi_{\QEn^3}(x)$, which indicates that each entangled triqubit-spinor field $\psi_{\QEp^3}(x)$ (or $\psi_{\QEn^3}(x)$) has a free motion only in four-dimensional spacetime. So that the action of entangled triqubit-spinor field $\psi_{\QE^3}(x)$ can present an equivalent action for category-2 biqubit-spinor field in four-dimensional spacetime, i.e.: 
\be \label{actionQE3}
\cS_{\QE^3}   & \equiv & \int d^4x \,  \{ \bar{\psi}_{\QE^3}(x)  \delta_{a}^{\;\mu}  \Sigma_{-}^{a} i\p_{\mu} \psi_{\QE^3}(x)  -   \lambda_{2}  \phi_p(x) \bar{\psi}_{\QE^3}(x) \tSi_{-}^p \psi_{\QE^3}(x)    \nn \\
 & + & \frac{1}{2}\eta^{\mu\nu} \p_{\mu}\phi_p(x)\p_{\nu}\phi^p(x) -  \frac{1}{4}\lambda_Q (\phi_p(x)\phi^p(x))^2 \} ,
\ee
with $a,\mu = 0,1,2,3$ and $p=1,2$. Where $\Sigma_{-}^{a}$ and $\tSi_{-}^p$ matrices are given by the $\Gamma$-matrices $\Gamma^{\ha}$ through the following explicit forms:
\be  \label{GMQE3}
& & \Sigma_{-}^{a} = \frac{1}{2}\Gamma^{a}\Gamma_{-}, \quad \tSi_{-}^{p} = \frac{1}{2}\tGa^{p}\Gamma_{-}, \nn \\
& & \Gamma^a =  \sigma_0\otimes \gamma^{a} ,  \nn \\
& &   \Gamma^5 = i\sigma_1\otimes \sigma_3\otimes \sigma_0 \equiv \tGa^1 ,  \nn \\
& & \Gamma^6 = i\sigma_2\otimes \sigma_3\otimes \sigma_0  \equiv \tGa^2 , \nn \\
& & \Gamma_{-} = \frac{1}{2} ( 1- \hat{\gamma}_7 ), 
\ee
with $\gamma^a$ ($a=0,1,2,3$) given in Eq.(\ref{DGM}).

In the entangled qubit-spinor representation, all $\Gamma$-matrices given in Eqs.(\ref{GMQE4}) and (\ref{GMQE3}) become anti-commuting,  
\be
& & \{\Gamma^{\ha}, \Gamma^{\hb} \} = 2 \eta^{\ha\hb} , \quad  \{\Gamma^{\ha}, \tilde{\Gamma}^{p} \} = 0, \quad \{\tilde{\Gamma}^{p}, \tilde{\Gamma}^{q} \} = 2 \eta^{pq} ,
\ee
with $\ha, \hb =0,1,2,3,5,6$, $p,q =1,2,3$, and 
\be
\{\Gamma^{a}, \Gamma^{b} \} = 2 \eta^{a b}, \quad \{\Gamma^{a}, \tilde{\Gamma}^{p} \} = 0, \quad \{\tilde{\Gamma}^{p}, \tilde{\Gamma}^{q} \} = 2 \eta^{pq} ,
\ee
with $a, b =0,1,2,3$, $p,q=1,2$. 

The commutators of $\Gamma$-matrices provide directly the group generators of extended $\cM_c$-spin symmetry, i.e.:
\be
& & \Sigma^{\ha\hb} = \frac{i}{4}  [\Gamma^{\ha}, \Gamma^{\hb} ] ,  \quad \Sigma_{-}^{\ha} = \frac{1}{2}\Gamma^{\ha}\Gamma_{-}, \quad \Gamma_{-} = \frac{1}{2} ( 1- \hat{\gamma}_9 ), 
\ee 
which satisfy the following group algebra:
\be
& & [\Sigma^{\ha\hb}, \Sigma^{\hc\hd}] =  i (\Sigma^{\ha\hd}\eta^{\hb\hc} -\Sigma^{\hb\hd}  \eta^{\ha\hc} - \Sigma^{\ha\hc} \eta^{\hb\hd} + \Sigma^{\hb\hc} \eta^{\ha\hd}) , \nn \\
& &  [\Sigma^{\ha\hb}, \Sigma_{-}^{\hc}] = i( \Sigma_{-}^{\ha}\eta^{\hb\hc} -\Sigma_{-}^{\hb}  \eta^{\ha\hc}  ) , \quad [\Sigma_{-}^{\ha}, \Sigma_{-}^{\hb}] = 0 .  
\ee
A similar group algebra holds for $\Sigma^{a b}$ and $\Sigma_{-}^{a}$,
\be
& & \Sigma^{a b} = \frac{i}{4}  [\Gamma^{a}, \Gamma^{b} ] ,  \quad \Sigma_{-}^{a} = \frac{1}{2}\Gamma^{a}\Gamma_{-}, \quad \Gamma_{-} = \frac{1}{2} ( 1- \hat{\gamma}_7 ) .
\ee

The group algebra for the group generators $\Sigma^{a b}$ or $\Sigma^{\ha\hb}$ characterize the motion-correlation $\cM_c$-spin symmetry SP(1,$D_h$-1) for $D_h=4$ or $D_h=6$. The commutating group algebra for the group generators $\Sigma_{-}^{a}$ or $\Sigma_{-}^{\ha}$ brings on an Abelian-type group symmetry for the action of entangled qubit-spinor field. 

Let us study explicitly such an Abelian-type group symmetry for both westward entangled triqubit-spinor field and westward entangled tetraqubit-spinor field. They transform under the group operation as follows:
\be
& &  \psi_{\QE^3}(x) \to  \psi'_{\QE^3}(x) =  S(\varpi)  \psi_{\QE^3}(x) , \quad S(\varpi) = e^{i \varpi_{a}\Sigma_{-}^{a}/2 } , \nn \\
& &  \psi_{\QE^4}(x) \to  \psi'_{\QE^4}(x) =  S(\varpi)  \psi_{\QE^4}(x) , \quad S(\varpi) = e^{i \varpi_{\ha}\Sigma_{-}^{\ha}/2 } , 
\ee
with the explicit forms,
\be \label{WES}
& & \psi'_{\QE^3}(x) = \psi_{\QE^3}^{(w)}(x)  + \tilde{\psi}_{\QE^3}^{(e)}(x)  , \quad S(\varpi) = 1 + i \varpi_{a} \Sigma_{-}^{a}/2, \nn \\
& &  \psi_{\QE^3}^{(w)}(x) \equiv  \psi_{\QE^3}(x), \quad \tilde{\psi}_{\QE^3}^{(e)}(x) \equiv i \frac{1}{2}\varpi_{a} \Sigma_{-}^{a} \psi_{\QE^3}(x), 
\ee
and
\be
& & \psi'_{\QE^4}(x) = \psi_{\QE^4}^{(w)}(x) + \tilde{\psi}_{\QE^4}^{(e)}(x), \quad S(\varpi) = 1 + i \varpi_{\ha} \Sigma_{-}^{\ha}/2 , \nn \\
& &  \psi_{\QE^4}^{(w)}(x) \equiv  \psi_{\QE^4}(x), \quad \tilde{\psi}_{\QE^4}^{(e)}(x) \equiv i \frac{1}{2}\varpi_{\ha} \Sigma_{-}^{\ha} \psi_{\QE^4}(x), 
\ee  
where $\tilde{\psi}_{\QE^3}^{(e)}(x)$ and $\tilde{\psi}_{\QE^4}^{(e)}(x)$ represent the shifted new parts of entangled qubit-spinor fields after such Abelian-type group transformations. The superscripts `$w$' and `$e$' on the entangled qubit-spinor fields label their westward and eastward chirality properties, respectively, 
\be \label{WECP}
& & \hat{\gamma}_7 \psi_{\QE^3}^{(w)}(x) = -  \psi_{\QE^3}^{(w)}(x), \quad \hat{\gamma}_7 \tilde{\psi}_{\QE^3}^{(e)}(x) = + \tilde{\psi}_{\QE^3}^{(e)}(x), \nn \\
& & \hat{\gamma}_9 \psi_{\QE^4}^{(w)}(x) = -  \psi_{\QE^4}^{(w)}(x), \quad \hat{\gamma}_9 \tilde{\psi}_{\QE^4}^{(e)}(x) = + \tilde{\psi}_{\QE^4}^{(e)}(x) .
\ee

Such an Abelian-type group symmetry reflects a translation-like spin group symmetry of  entangled qubit-spinor field in Hilbert space. It is noted that the shifted new part of entangled qubit-spinor field gets a sign flip in chirality as shown in Eq.(\ref{WECP}). Namely, when the entangled qubit-spinor field is westward, its shifted new part after the transformation becomes eastward, which holds the other way round. 

This enlarged translation-like Abelian spin group symmetry is analogous to the translation group symmetry P$^{1,D_h-1}$ of coordinates in Minkowski spacetime. For convenience, we may use the notation W$^{1,D_h-1}$ to denote such a translation-like Abelian spin group symmetry since it brings on a sign flip in chirality for the shifted new part after such a group transformation. As such a translation-like Abelian spin group symmetry W$^{1,D_h-1}$ arises initially from the locally entangled-qubits motion postulate, it is regarded as an {\it entanglement-correlated Abelian-type spin group symmetry} in the entangled qubit-spinor representation of Hilbert space. 

As such an entanglement-correlated translation-like spin group symmetry W$^{1,D_h-1}$ is associated with the sign flip in chirality of westward and eastward entangled qubit-spinor fields for the initial and shifted new parts, we may refer to W$^{1,D_h-1}$ as {\it $\cW_e$-spin symmetry} for short. The extended $\cM_c$-spin symmetry is referred to as {\it inhomogeneous $\cM_c$-spin symmetry} which is denoted as follows:
\be
WS(1,D_h-1)  \equiv  SP(1,D_h-1)\rtimes W^{1,D_h-1} ,
\ee
where the symbol `$\rtimes$' is used to indicate that the inhomogeneous $\cM_c$-spin symmetry WS(1,$D_h$-1) is a semidirect product group with $\cM_c$-spin symmetry group SP(1, $D_h$-1) and $\cW_e$-spin symmetry group W$^{1,D_h-1}$.

The group generators of $\cQ_c$-spin symmetry is simply given by
\be
\Sigma^{p q} = \frac{i}{4}  [\Gamma^{D_h +p}, \Gamma^{D_h +q} ]   = \frac{i}{4}[ \tGa^p , \tGa^q ] \equiv  \tilde{\Sigma}^{pq}, 
\ee
with $\cQ_c$-spin charges $\CQc=q_c= 2$ and $\CQc=q_c= 3$ in correspondence to the entangled qubit-spinor fields $\psi_{\QE^3}(x)$ and $\psi_{\QE^4}(x) $. 

The actions in Eqs.(\ref{actionQE3}) and (\ref{actionQE4}) possess the following inhomogeneous $\cM_c$-spin symmetry} and $\cQ_c$-spin symmetry:
\be
& & G_S = WS(1, D_h-1)\times  SP(q_c) \equiv  SP(1,D_h-1)\rtimes W^{1,D_h-1} \times  SP(q_c) , 
\ee
with $D_h=4$ and $D_h=6$ for $q_c=2$ and $q_c=3$, respectively. Such an inhomogeneous $\cM_c$-spin symmetry is in association with inhomogeneous Lorentz-type/Poincar\'e-type group symmetry in Minkowski spacetime,
\be
 PO(1, D_h-1) = P^{1,D_h-1}\ltimes SO(1,D_h-1) , \nn
\ee
which is a semidirect product group symmetry with P$^{1,D_h-1}$ representing the translation group symmetry of coordinates in Minkowski spacetime. 

In general, the maximal symmetry of the actions in Eqs.(\ref{actionQE3}) and (\ref{actionQE4}) can be expressed as the following associated symmetry with coincidental transformations:
\be
G_S & = &  SC(1)\ltimes P^{1,D_h-1}\ltimes SO(1,D_h-1)\adjoin SP(1,D_h-1)\rtimes W^{1,D_h-1} \times SG(1) \times SP(q_c)  
\nn \\
& = & SC(1)\ltimes PO(1, D_h-1) \adjoin WS(1,D_h-1)  \times SG(1) \times  SP(q_c) ,  
\ee
with $D_h=4, 6$ for $q_c=2, 3$, respectively. Where the symbol ``$\adjoin$" is used to indicate the associated symmetry in which the transformation of $\cM_c$-spin symmetry SP(1,$D_h$-1) in Hilbert space must be coincidental to that of isomorphic Lorentz-type group symmetry SO(1,$D_h$-1) in globally flat Minkowski spacetime. SP($q_c$) represents the $\cQ_c$-spin symmetry and SC(1)$\adjoin$SG(1) denotes the associated symmetry for global scaling transformations.


\subsection{ Entangled pentaqubit-spinor field as entangled hyperqubit-spinor field in category-0 and inhomogeneous hyperspin symmetry WS(1,9) as unified spin-color symmetry of chiral lepton-quark state in 10D hyper-spacetime}

A massless Dirac fermion as triqubit-spinor field in category-3 is shown to be characterized by an entangled tetraqubit-spinor field with inhomogeneous $\cM_c$-spin symmetry WS(1,5) in six-dimensional spacetime. Based on the locally entangled-qubits motion postulate, let us further examine an {\it entangled pentaqubit-spinor field} constructed from a chiral spinor structure of tetraqubit-spinor field in Category-0 with zero $\cQ_c$-spin charge $\CQc=q_c=0$ in ten-dimensional hyper-spacetime. Such an entangled pentaqubit-spinor field is verified to unify chiral spin-color lepton-quark states into an {\it entangled hyperqubit-spinor field}.

In light of chiral qubit-spinor structure of tetraqubit-spinor field, we are able to construct a {\it self-conjugated chiral pentaqubit-spinor field} as follows:
\be \label{QE5}
& & \Psi_{\QE^5}(x) = \binom{\Psi_{\fQW^4}(x) }{\Psi^{\, \bc}_{\fQW^4}(x) } , 
\ee
where $\Psi_{\fQW^4}(x)$ is a {\it chiral complex tetraqubit-spinor filed} with $\Psi^{\;\; \bc}_{\fQW^4}(x)$ as the complex charge-conjugated one. They have the following explicit forms: 
\be  
& & \Psi_{\fQW^4}(x) \equiv \begin{pmatrix} 
 \psi_{\fQ_{\mW++}^2} \\   \psi_{\fQ_{\mW+-}^2} \\  \psi_{\fQ_{\mW-+}^2} \\  \psi_{\fQ_{\mW--}^2}
\end{pmatrix}  \equiv \begin{pmatrix} 
Q^r_{L}(x) \\  Q^b_{L}(x) \\ Q^g_{L}(x) \\ Q^w_{L}(x) 
\end{pmatrix}, \nn \\
& &  \Psi^{\, \bc}_{\fQW^4}(x) \equiv \begin{pmatrix} 
 \psi^{\bc}_{\fQ_{\mW++}^2} \\   \psi^{\bc}_{\fQ_{\mW+-}^2} \\  \psi^{\bc}_{\fQ_{\mW-+}^2} \\  \psi^{\bc}_{\fQ_{\mW--}^2}
\end{pmatrix} \equiv 
\begin{pmatrix} 
Q^{r\, \bc}_{L}(x) \\  Q^{b\, \bc}_{L}(x) \\ Q^{g\, \bc}_{L}(x) \\ Q^{w\, \bc}_{L}(x) 
\end{pmatrix} ,  
\ee
with the definitions,
\be
& & \Psi^{\, \bc}_{\fQW^4}(x) = C_{\fQW^4} \bar{\Psi}^{\, T}_{\fQW^4}(x), \quad  C_{\fQW^4} = \sigma_0\otimes \sigma_0\otimes C_{D}, \nn \\
& & \psi_{\fQ_{\mW s_1s_2}^2} \equiv Q^{\alpha}_{L}(x) \equiv \gamma_{-} Q^{\alpha}(x), \quad \gamma_{-} = \frac{1}{2} ( 1-\gamma_5 ), \nn \\
& & \psi_{\fQ_{\mW s_1s_2 }^2}^{\bc} \equiv Q^{\alpha\, \bc}_{L}(x) = C_D \bar{Q}^{\alpha\; T}_{L }(x) , \quad C_D = - i \sigma_3\otimes \sigma_2 , \nn \\
& & s_1,s_2 = \pm , \quad \alpha = \{ s_1s_2 \} \equiv (r, b, g, w), 
\ee
where $Q^{\alpha}_{L}(x)$ with $\alpha = r, b, g, w$ represent the left-handed Dirac fermions with unified four color-spin charges in correspondence to the `red', `blue', `green' and `white'. The first three color-spin charges `red', `blue' and `green' characterize three chromo-quarks and the fourth color-spin charge `white' reflects a relevant lepton. $Q^{\alpha}_{L}(x)$ and $Q^{\alpha\, \bc}_{L}(x)$ are defined as left-handed lepton-quark state and its complex charge-conjugated one. Each $Q^{\alpha}_{L}(x)$ is regarded as a {\it chiral complex biqubit-spinor field} $\psi_{\fQ_{\mW s_1s_2}^2}$.

Based on the locally entangled-qubits motion postulate, the self-conjugated chiral pentaqubit-spinor field $\Psi_{\QE^5}(x)$ is referred to as {\it entangled pentaqubit-spinor field}, its action in ten-dimensional hyper-spacetime is built to be as follows:
\be \label{actionQE5}
 \cS_{\QE^5}  & = &  \int d^{10}x  \bar{\Psi}_{\QE^5}(x)  \delta_{\mA}^{\; \mM}  \Sigma_{-}^{\mA} i\p_{\mM} \Psi_{\QE^5}(x) , 
\ee
where $\Sigma_{-}^{\mA}$ ($\mA=0,1,2,3,5,\cdots,10$) matrices are defined with $\Gamma$-matrices $\Gamma^{\mA}= (\Gamma^a, \Gamma^A) $ ($a=0,1,2,3, A= 5,\cdots,10$), which have the following explicit forms:
\be \label{GMQE5}
& & \Sigma_{-}^{\mA}= \frac{1}{2}\Gamma^{\mA}\Gamma_{-}, \quad 
\Gamma_{-} = \frac{1}{2} ( 1-\hat{\gamma}_{11} ), \nn \\
& & \hat{\gamma}_{11} = \sigma_3\otimes \sigma_0 \otimes \sigma_0 \otimes \gamma_5, \nn \\
& & \Gamma^0 = \, \sigma_0 \otimes \sigma_0 \otimes\sigma_0 \otimes \sigma_1 \otimes \sigma_0, \nn \\
& &  \Gamma^1 = i \sigma_0 \otimes \sigma_0 \otimes \sigma_0\otimes \sigma_2\otimes \sigma_1, \nn \\
& & \Gamma^2 = i \sigma_0 \otimes  \sigma_0 \otimes\sigma_0\otimes  \sigma_2\otimes \sigma_2, \nn \\
& & \Gamma^3 = i \sigma_0 \otimes \sigma_0 \otimes\sigma_0\otimes  \sigma_2\otimes \sigma_3, \nn \\
& &  \Gamma^5 = i \sigma_1 \otimes \sigma_0 \otimes \sigma_2\otimes  \sigma_3\otimes \sigma_0, \nn \\
& &  \Gamma^6 =i \sigma_2 \otimes  \sigma_3 \otimes\sigma_2\otimes  \sigma_3\otimes \sigma_0, \nn \\
& & \Gamma^7 = i \sigma_1 \otimes \sigma_2 \otimes\sigma_3\otimes  \sigma_3\otimes \sigma_0 , \nn \\
& & \Gamma^8 = i \sigma_2 \otimes \sigma_2 \otimes  \sigma_0\otimes \sigma_3\otimes \sigma_0, \nn \\
& &  \Gamma^9 = i \sigma_1 \otimes  \sigma_2 \otimes \sigma_1\otimes \sigma_3\otimes \sigma_0 , \nn \\
& &  \Gamma^{10} =  i \sigma_2 \otimes \sigma_1 \otimes \sigma_2\otimes \sigma_3\otimes \sigma_0 ,
\ee
where all $\Gamma$-matrices become anti-commuting,
\be
\{ \Gamma^{\mA} , \Gamma^{\mB} \} = 2\eta^{\mA\mB} .  
\ee

In general, $\Psi_{\QE^5}(x)$ is also referred to as {\it entangled hyperqubit-spinor field} for chiral lepton-quark state, which satisfies the following self-conjugated chiral conditions:
\be
& & \Psi_{\QE^5}^{c}(x) = C_{\QE^5} \bar{\Psi}_{\QE^5}^T(x) = \Psi_{\QE^5}(x), \nn \\
& &  \hat{\gamma}_{11} \Psi_{\QE^5}(x) = \hat{\gamma}_{11} \Psi_{\QE^5}^{c}(x) = - \Psi_{\QE^5}(x),  \nn \\
& &  C_{\QE^5} = \sigma_1 \otimes \sigma_0 \otimes \sigma_0 \otimes C_D , 
\ee

It can be verified in the entangled pentaqubit-spinor representation that besides the $\Gamma$-matrices $\Gamma^{\mA}$ ($\mA = 0,1,2,3,5,\cdots, 10$) and $\hat{\gamma}_{11}$, there exists no any other anti-commuting $\Gamma$-matrix that can bring on scalar coupling term with the entangled pentaqubit-spinor field $\Psi_{\QE^5}(x)$. Therefore, the entangled pentaqubit-spinor field $\Psi_{\QE^5}(x)$ has zero $\cQ_c$-spin charge $\cC_{Q_c} = q_c =0$ and the action in Eq.(\ref{actionQE5}) gets no scalar coupling.

The action in Eq.(\ref{actionQE5}) possesses an associated symmetry in which the inhomogeneous hyperspin symmetry WS(1,9) is in association with inhomogeneous Lorentz-type/Poincar\'e-type group symmetry PO(1,9) in ten-dimensional Minkowski hyper-spacetime. Such an associated symmetry is presented as follows:
\be
G_S & = & SC(1)\ltimes PO(1, 9) \adjoin WS(1,9) \rtimes SG(1) \times SP(q_c=0) \nn \\
& = & SC(1)\ltimes P^{1,9}\ltimes SO(1,9) \adjoin SP(1,9)\rtimes  W^{1,9} \rtimes SG(1),  
\ee
where we have used the symbol ``$\adjoin$" to indicate the associated symmetry which requires the transformation of hyperspin symmetry SP(1,9) be coincidental to that of Lorentz-type group symmetry SO(1,9) $\cong$ SP(1,9). The group generators of WS(1,9) can directly be presented by the commutators of $\Gamma$-matrices:
\be
& & \Sigma^{\mA\mB}= \frac{i}{4}[\Gamma^{\mA},  \Gamma^{\mB} ], \quad  \Sigma_{-}^{\mA}= \frac{1}{2}\Gamma^{\mA}\Gamma_{-},  
 \ee
with $\mA, \mB= 0,1,2,3,5, \cdots, 10$. Where $\Sigma^{\mA\mB}$ represent the group generators of hyperspin symmetry SP(1,9) and $\Sigma_{-}^{\mA}$ the group generators of $\cW_e$-spin symmetry W$^{1,9}$.

The hyperspin symmetry SP(1,9) reflects a unified spin $\&$ color charge symmetry for chiral lepton-quark state. To be more explicit, it is natural to decompose the hyperspin symmetry SP(1,9) into the following subgroups:
\be
SP(1,9) & \supset &  SP(1,3)\times SP(6)  \cong SO(1,3) \times SO(6)\nn \\
 & \cong & SO(1,3) \times SU_C(4),
\ee
with the corresponding group generators,
\be
& & \Sigma^{\mA\mB}  ( \mA, \mB= 0,1,2,3 ) \in sp(1,3)\cong so(1,3)  \nn \\
& &  \Sigma^{\mA\mB}  ( \mA, \mB= 5,6,7,8,9,10 ) \in sp(6)\cong so(6)\cong su_C(4) .
\ee
Where the subgroup symmetry SP(1,3) $\cong$ SO(1,3) in connection with ordinary four-dimensional Minkowski spacetime represents the usual spin symmetry, and the subgroup symmetry SP(6)$\cong$SO(6) $\cong$ SU$_C$(4) in correlation with extra six spacial dimensions characterizes a unified four color-spin charge symmetry of chiral lepton-quark state.


\subsection{Entangled hexaqubit-spinor field in Category-1 as entangled hyperqubit-spinor field and inhomogeneous hyperspin symmetry as unified spin-color symmetry of lepton-quark state in 11D hyper-spacetime}

It is known that quarks in the strong interaction of SM behave as vector-like Dirac fermions, which indicates that the entangled pentaqubit-spinor field as self-conjugated chiral pentaqubit-spinor field is not an appropriate entangled hyperqubit-spinor field for characterizing quarks in SM. We should consider an entangled hyperqubit-spinor field that contains both left-handed and right-handed lepton-quark states.

Let us introduce an entangled hyperqubit-spinor field with the following hyperqubit-spinor structure:
\be
& & \Psi_{\QE^6}(x) = \binom{\Psi_{\QEn^5}(x)}{\Psi_{\QEp^5}(x)}, \nn \\
& & \Psi_{\QEn^5}(x) = \binom{\Psi_{\fQWn^4}(x) }{\Psi^{\; \; \bc}_{\fQWn^4}(x) } , \quad \Psi_{\QEp^5}(x) = \binom{\Psi_{\fQWp^4}(x) }{\Psi^{\; \; \bc}_{\fQWp^4}(x) } , 
\ee
where $\Psi_{\QEpn^5}(x)$ are entangled pentaqubit-spinor fields defined as follows:
\be \label{QS5}
& & \Psi_{\QEpn^5}(x) \equiv \Gamma_{\pm} \Psi_{\fQS^5}(x), \nn \\
& & \Psi_{\fQS^5}(x) = \binom{\Psi_{\fQC^4}(x) }{\Psi^{\; \bc}_{\fQC^4}(x) } , \nn \\
& & \Gamma_{\pm} = \frac{1}{2} ( 1 \pm \hat{\gamma}_{11} ) , \;\; \hat{\gamma}_{11} = \sigma_3\otimes \sigma_0 \otimes \sigma_0 \otimes \gamma_5
\ee
with $\Psi_{\QEn^5}(x)\equiv \Psi_{\QE^5}(x)$ the entangled pentaqubit-spinor field as shown in Eq.(\ref{QE5}). $\Psi_{\fQS^5}(x)$ is a self-conjugated pentaqubit-spinor field in Category-1 with $\cQ_c$-spin charge $\CQc=q_c=1$, which is formed from the following {\it complex tetraqubit-spinor field}: 
\be \label{QC4}
& & \Psi_{\fQC^4}(x) = \begin{pmatrix} 
 \psi_{\fQ_{\mC++}^2} \\   \psi_{\fQ_{\mC+-}^2} \\  \psi_{\fQ_{\mC-+}^2} \\  \psi_{\fQ_{\mC--}^2}
\end{pmatrix} \equiv \begin{pmatrix} 
Q^r(x) \\  Q^b(x) \\ Q^g(x) \\ Q^w(x) 
\end{pmatrix}, \nn \\
& &  \Psi^{\; \bc}_{\fQC^4}(x) = \begin{pmatrix} 
 \psi^{\bc}_{\fQ_{\mC++}^2} \\   \psi^{\bc}_{\fQ_{\mC+-}^2} \\  \psi^{\bc}_{\fQ_{\mC-+}^2} \\  
 \psi^{\bc}_{\fQ_{\mC--}^2}
\end{pmatrix}  \equiv 
\begin{pmatrix} 
Q^{r\, \bc}(x) \\  Q^{b\, \bc}(x) \\ Q^{g\, \bc}(x) \\ Q^{w\, \bc}(x) 
\end{pmatrix} ,  \nn \\
& & \psi_{\fQ_{\mC s1s2}^2}\equiv Q^{\alpha}(x) , \;\; s_1,s_2 = \pm , \;\; \alpha \equiv \{ s_1s_2 \} \equiv (r, b, g, w), \nn \\
& & \psi_{\fQ_{\mC s1s2}^2}^{\bc} \equiv Q^{\alpha\, \bc}(x) = C_D \bar{Q}^{\alpha\, T}(x) \equiv  \sigma_2\otimes \sigma_2 \psi_{\fQ_{\mC s1s2}^2}^{\ast}, 
\ee
where the complex tetraqubit-spinor field $\Psi_{\fQC^4}(x)$ is composed of {\it complex biqubit-spinor fields} $\psi_{\fQ_{\mC s1s2 }^2}$ that are regarded as Dirac fermion states $Q^{\alpha}(x)$.  Such Dirac fermion states $Q^{\alpha}(x)$ labeled with four color-spin charges `red', `blue', `green' and `white'  form a lepton-quark state. $Q^{\alpha\, \bc}(x)$ is defined as the complex charge conjugation state of lepton-quark state $Q^{\alpha}(x)$.

The complex tetraqubit-spinor field $\Psi_{\fQC^4}(x)$ can generally be written as the sum of chiral complex tetraqubit-spinor fields $\Psi_{\fQWpn^4}(x)$,
\be
\Psi_{\fQC^4}(x) \equiv \Psi_{\fQWn^4}(x) + \Psi_{\fQWp^4}(x) ,
\ee
where $\Psi_{\fQWpn^4}(x)$ and their complex charge-conjugated ones $ \Psi^{\; \bc}_{\fQWpn^4}(x)$ are given by,
\be \label{fQWpn4}
& & \Psi_{\fQWpn^4}(x) \equiv \vGa^5_{\pm} \Psi_{\fQC^4}(x) = \begin{pmatrix} 
Q^r(x) \\  Q^b(x) \\ Q^g(x) \\ Q^w(x) 
\end{pmatrix}_{R, L}, \nn \\
& &  \Psi^{\; \bc}_{\fQWpn^4}(x) \equiv  \vGa^5_{\mp} \Psi^{\; \bc}_{\fQC^4}(x) = \begin{pmatrix} 
Q^{r\, \bc}(x) \\  Q^{b\, \bc}(x) \\ Q^{g\, \bc}(x) \\ Q^{w\, \bc}(x) 
\end{pmatrix}_{L, R} , \nn \\
& & \vGa^5_{\pm} = \frac{1}{2}(1\pm \tilde{\gamma}_5) , \quad  \tilde{\gamma}_5 = \sigma_0\otimes \sigma_0\otimes \gamma_5, 
\ee
which are related to the left-handed and right-handed lepton-quark states $Q^{\alpha}_{L, R}(x)$.

The action of entangled hyperqubit-spinor field $\Psi_{\QE^6}(x)$ can be written as follows:
\be \label{actionQE6}
 \cS_{\QE^6}  & = &  \int d^{11}x \, \{ \bar{\Psi}_{\QE^6}(x)  \delta_{\mA}^{\; \mM}  \vSi_{-}^{\mA} i\p_{\mM} \Psi_{\QE^6}(x)  - \lambda_{1} \phi_1(x) \bar{\Psi}_{\QE^6} (x) \tvSi_{-} \Psi_{\QE^6}(x) \}, 
\ee
where $\vSi_{-}^{\mA}$ and $\tvSi_{-}$ matrices are defined with $\vGa$-matrices $\vGa^{\hat{\mA}}= (\vGa^a, \vGa^A, \vGa^{12}) $ ($\hat{\mA} =0,1,2,3,5,\cdots, 12$, $a=0,1,2,3, A= 5,\cdots, 11$). Their explicit forms are given by,
\be \label{QE6}
& &  \vSi^{\mA}= \frac{1}{2}\vGa^{\mA}\vGa_{-}, \quad \tvSi_{-}= \frac{1}{2}\tvGa\vGa_{-},  \nn \\
& & \vGa^{\mA}= \; \sigma_0\otimes \Gamma^{\mA},  \quad  \mA=0,1,2,3, 5,\cdots,10, \nn \\
& & \vGa^{11} = \, i\sigma_1\otimes \hat{\gamma}_{11} , \nn \\
& &   \vGa^{12} = - i\sigma_2\otimes \hat{\gamma}_{11} \equiv \tvGa , \nn \\
& &  \vGa_{-} = \frac{1}{2} ( 1-\hat{\gamma}_{13} ), \quad \hat{\gamma}_{13} = \sigma_3 \otimes \hat{\gamma}_{11}, 
\ee
with $\Gamma^{\mA}$ ($\mA=0,1,2,3, 5,\cdots, 10$) defined in Eq.(\ref{GMQE5}). The entangled hyperqubit-spinor field $\Psi_{\QE^6}(x)$ satisfies the following self-conjugated chiral conditions:
\be
& & \Psi_{\QE^6}^{c}(x) = C_{\QE^6} \bar{\Psi}_{\QE^6}^T(x) = \Psi_{\QE^6}(x), \nn \\
& &  \hat{\gamma}_{13}\Psi_{\QE^6}(x) = \hat{\gamma}_{13} \Psi_{\QE^6}^{c}(x) = - \Psi_{\QE^6}(x), \nn \\
& & C_{\QE^6} = \sigma_0 \otimes C_{\QE^5} = \sigma_0 \otimes \sigma_1 \otimes \sigma_0 \otimes \sigma_0 \otimes C_D . 
\ee
We may refer to $\Psi_{\QE^6}(x)$ as {\it entangled hexaqubit-spinor field}.

The action in Eq.(\ref{actionQE6}) possesses an associated symmetry in which the inhomogeneous hyperspin symmetry WS(1,10) is in association with inhomogeneous Lorentz-type/Poincar\'e-type group symmetry PO(1,10). The whole associated symmetry of the action can be expressed as follows:
\be
G_S & = & SC(1)\ltimes PO(1,10)\adjoin WS(1,10) \rtimes SG(1) \times SP(1) \nn \\
& = & SC(1)\ltimes P^{1,10} \ltimes SO(1,10)\adjoin SP(1,10) \rtimes W^{1,10} \rtimes SG(1) \times O(1),  
\ee
where the group transformations of hyperspin symmetry SP(1,10) and Lorentz-type group symmetry SO(1,10) must be coincidental as indicated by the symbol ``$\adjoin$".  The group generators of WS(1,10) are simply presented by the commutators of anti-commuting $\vGa$-matrices,
\be
& & \varSigma^{\mA\mB}= \frac{i}{4}[\vGa^{\mA},  \vGa^{\mB} ], \quad  \vSi_{-}^{\mA}= \frac{1}{2}\vGa^{\mA}\vGa_{-} , \nn
 \ee
with $\mA, \mB= 0,1,2,3,5, \cdots, 11$. The $\cQ_c$-spin symmetry SP(1)$\cong$O(1) becomes manifest with transformation $\Psi_{\QE^6}(x) \to - \Psi_{\QE^6}(x)$, which is coincidental to the chirality property of entangled hyperqubit-spinor field $\hat{\gamma}_{13}\Psi_{\QE^6}(x)=  - \Psi_{\QE^6}(x)$.

The hyperspin symmetry SP(1,10) can be decomposed into the following subgroups:
\be
& & SP(1,10) \supset  SP(1,3)\times SP(7)  \cong SO(1,3) \times SO(7) \nn \\
& &   \supset SO(1,3) \times SU_C(4) \supset SO(1,3) \times SU_C(3),
\ee
with the subgroup generators,
\be
& & \Sigma^{\mA\mB}  ( \mA, \mB= 0,1,2,3 ) \in sp(1,3)\cong so(1,3)  \nn \\
& &  \Sigma^{\mA\mB}  ( \mA, \mB= 5,\cdots, 10 ) \in sp(6)\cong so(6) \cong su_C(4). 
\ee
The unitary symmetry SU$_C$(4) is regarded as color-spin charge symmetry of lepton-quark state, which is correlated to extra six dimensions in Minkowski hyper-spacetime of coordinates. The subgroup symmetry SU$_C$(3) of SU$_C$(4) produces the strong interaction symmetry of three vector-like chromo-quarks in SM. 

To show explicitly the vector-like property of color-spin symmetry SU$_C$(3), it is useful to express the action in Eq.(\ref{actionQE6}) into the following equivalent one by using the self-conjugated pentaqubit-spinor field $\Psi_{\QS^5}(x)$:
\be \label{actionQS5}
 \cS_{\fQS^5} \equiv \cS_{\QE^6}  & = &  \int d^{11}x \, \{ \frac{1}{2}  \bar{\Psi}_{\fQS^5}(x)  \delta_{\mA}^{\; \mM}  \Gamma^{\mA} i\p_{\mM} \Psi_{\fQS^5}(x)  - \frac{1}{2}\lambda_1 \phi(x) \bar{\Psi}_{\fQS^5} (x)\tGa \Psi_{\fQS^5}(x) \}, 
\ee
where the $\Gamma$-matrices are given by,
\be \label{GMQS5}
& & \Gamma^{\mA}= \Gamma^{\mA},  \; \; \mA=0,1,2,3,5,\cdots, 10 , \nn \\
& &  \Gamma^{11} =  i \hat{\gamma}_{11}, \quad \tGa =  I_{32}, 
\ee
with $\Gamma^{\mA}$ ($\mA=0,1,2,3,5,\cdots, 10$) defined in Eq.(\ref{GMQE5}) and $I_{32}$ the $32\times32$ unit matrix.  In obtaining the above action, we have used the following identity resulting from the definitions given in Eq.(\ref{QS5}):
\be
& & \Psi_{\fQS^5}(x) \equiv \Psi_{\QEn^5 }(x) + \Psi_{\QEp^5 }(x), \nn \\
& & \Psi_{\fQS^5}^{c}(x) = C_{\QE^5} \bar{\Psi}_{\fQS^5}^T(x) = \Psi_{\fQS^5}(x) ,
\ee
where $\Psi_{\fQS^5}(x)$ is defined in Eqs.(\ref{QS5}) and (\ref{QC4}) with lepton-quark state formed by Dirac fermions.


\subsection{ Entangled Heptaqubit-spinor field in Category-2 as entangled hyperqubit-spinor field and inhomogeneous hyperspin symmetry as unified spin-color-flavor symmetry of lepton-quark states in 12D hyper-spacetime}

The lepton-quark state appearing in category-1 entangled hexaqubit-spinor field does characterize the vector-like strong interaction in SM, but it cannot reflect the flavor property of leptons and quarks in the weak interaction of SM, which brings us to consider the next category-2 entangled hyperqubit-spinor field with $\cQ_c$-spin charge $\CQc=q_c=2$. Such an entangled hyperqubit-spinor field is referred to as {\it entangled heptaqubit-spinor field} with the following qubit-spinor structure: 
\be \label{QE7}
& & \Psi_{\QE^7}(x)  \equiv \binom{\Psi_{\fQW^6}(x)}{\Psi_{\fQW^6}^{\, c}(x)}\equiv \binom{\Psi_{\fQS^6}^{L}(x)}{\Psi_{\fQS^6}^{L\, c}(x)} , 
\ee
where $\Psi_{\fQW^6}(x)$ is regarded as a {\it chiral hexaqubit-spinor field} with $\Psi_{\fQW^6}^{\, c}(x)$ as its complex charge-conjugated one. $\Psi_{\fQW^6}(x)$ and $\Psi_{\fQW^6}^{\, c}(x)$ can be provided from a {\it self-conjugated hexaqubit-spinor field} $\Psi_{\fQS^6}(x)$ via the following definitions:
\be
& &  \Psi_{\fQW^6}(x) \equiv \Psi_{\fQS^6}^{L}(x) \equiv \vGa_{-}^5 \Psi_{\fQS^6}(x) , \nn \\
& & \Psi_{\fQW^6}^{\, c}(x) = C_{\fQW^6} \bar{\Psi}_{\fQW^6}^{T}(x) \equiv \Psi_{\fQS^6}^{L\, c}(x) =  \vGa_{+}^5 \Psi_{\fQS^6}(x), \nn \\
& & \Psi_{\fQS^6}^{\, c}(x) = C_{\fQS^6} \bar{\Psi}_{\fQS^6}^{T}(x) = \Psi_{\fQS^6}(x), \nn \\
& & \vGa_{\pm}^5 = \frac{1}{2} ( 1 \pm \tilde{\gamma}_5 ), \quad \tilde{\gamma}_5 = \sigma_0 \otimes \sigma_0 \otimes  \sigma_0 \otimes \sigma_0\otimes \gamma_5 , \nn \\
& & C_{\fQW^6} = C_{\fQS^6} = \sigma_1 \otimes \sigma_1 \otimes  \sigma_0 \otimes \sigma_0\otimes C_D .
\ee
which have the following explicit forms:
\be \label{fQS6}
& &  \Psi_{\fQS^6}^{L}(x) \equiv \begin{pmatrix}
\Psi_{\fQC^4}(x) \\ \Psi_{\fQC^4}^{'\; \bc}(x) \\ \Psi'_{\fQC^4}(x) \\ \Psi_{\fQC^4}^{\; \bc}(x) 
\end{pmatrix}_{L}, \quad  \Psi_{\fQS^6}^{L\; c}(x) \equiv \begin{pmatrix}
\Psi_{\fQC^4}(x) \\ \Psi_{\fQC^4}^{'\; \bc}(x) \\ \Psi'_{\fQC^4}(x) \\ \Psi_{\fQC^4}^{\; \bc}(x) 
\end{pmatrix}_{R} ,
\ee
where $\Psi_{\fQC^4}(x)$ and $\Psi'_{\fQC^4}(x)$ are two complex tetraqubit-spinor fields with definitions presented in Eq.(\ref{QC4}). They are formed from two flavors of lepton-quark states $Q^{\alpha}(x)$ and $Q^{'\alpha}(x)$ with superscript $\alpha = (r,b,g,w)$ labeling four color-spin charges. The projection operators $\vGa_{\mp}^5$ give rise to definitions for the left-handed and right-handed lepton-quark states $Q^{\alpha}_{L, R}(x)$.

The entangled heptaqubit-spinor field $\Psi_{\QE^7}(x)$ satisfies the following self-conjugated chiral conditions:
\be
& & \Psi_{\QE^7}^{c}(x) = C_{\QE^7} \bar{\Psi}_{\QE^7}^{T}(x)= \Psi_{\QE^7}(x) , \nn \\
& & \hat{\gamma}_{15}\Psi_{\QE^7}(x) = \hat{\gamma}_{15}\Psi_{\QE^7}^{c}(x) = - \Psi_{\QE^7}(x), \nn \\
& & C_{\QE^7}= \sigma_1 \otimes  C_{\fQW^6} = \sigma_1 \otimes \sigma_1 \otimes \sigma_1 \otimes  \sigma_0 \otimes \sigma_0\otimes C_D  , \nn \\
& & \hat{\gamma}_{15} = \sigma_3 \otimes \sigma_0 \otimes \sigma_0 \otimes  \sigma_0 \otimes \sigma_0\otimes \gamma_5 .
\ee
The action for such an entangled heptaqubit-spinor field $\Psi_{\QE^7}(x)$ is constructed to be,
\be \label{actionQE7}
 \cS_{\QE^7}  =   \int d^{12}x \,  \{ \bar{\Psi}_{\QE^7}(x)  \delta_{\mA}^{\; \mM}  \vSi_{-}^{\mA} i\p_{\mM} \Psi_{\QE^7}(x) -  \lambda_{2} \phi_p(x) \bar{\Psi}_{\QE^7} (x) \tvSi_{-}^p \Psi_{\QE^7}(x) \}, 
\ee
with $\mA, \mM= 0,1,2,3,5, \cdots, 12$. The $\vSi_{-}^{\mA}$ and $\tvSi_{-}^p$ matrices are defined with $\vGa$-matrices $\vGa^{\hat{\mA}} = (\vGa^a, \vGa^A, \vGa^{13}, \vGa^{14}) $ ($\hat{\mA} = 0,1,2,3,5,\cdots, 14$, $a=0,1,2,3, A= 5,\cdots, 12$) as follows: 
\be \label{GMQE7}
& &  \vSi_{-}^{\mA} = \frac{1}{2} \vGa^{\mA}\vGa_{-}, \quad  \tvSi_{-}^{p} = \frac{1}{2} \tvGa^{p}\vGa_{-} , \nn \\
& &\vGa^{a} = \; \sigma_0 \otimes \sigma_0 \otimes \Gamma^{a},  \quad a=0,1,2,3, \nn \\
& & \vGa^{A} = \; \sigma_1 \otimes \sigma_0 \otimes \Gamma^{A}, \quad A=5, \cdots, 10,
 \nn \\
& & \vGa^{11} =  i\sigma_1 \otimes \sigma_1 \otimes \hat{\gamma}_{11} , \nn \\
& & \vGa^{12} =  i\sigma_1 \otimes \sigma_2 \otimes \hat{\gamma}_{11} , \nn \\
& & \vGa^{13} =  i\sigma_1 \otimes \sigma_3 \otimes  \hat{\gamma}_{11} 
 \equiv  \tvGa^{1} , \nn \\
& & \vGa^{14} =  i\sigma_2 \otimes \sigma_0 \otimes \tilde{\gamma}_5 \; \equiv  \tvGa^{2} , \nn \\
& & \hat{\gamma}_{15} = i\sigma_3 \otimes \sigma_0 \otimes \tilde{\gamma}_5 , \quad 
 \vGa_{-} = \frac{1}{2} ( 1-\hat{\gamma}_{15} ) , \nn \\
& & \hat{\gamma}_{11} = \sigma_3\otimes \sigma_0 \otimes \sigma_0 \otimes \gamma_5 , \nn \\ & & \tilde{\gamma}_5 = \sigma_0 \otimes  \sigma_0 \otimes \sigma_0\otimes \gamma_5, 
\ee
with $\Gamma^{\mA}$ ($A=0,1,2,3, 5,\cdots, 10$) defined in Eq.(\ref{GMQE5}). It is easy to check that all $\vGa$-matrices $\vGa^{\hat{\mA}}$ become anti-commuting.

The action in Eq.(\ref{actionQE7}) possesses an associated symmetry in which the inhomogeneous hyperspin symmetry WS(1,11) is in association with inhomogeneous Lorentz-type/Poincar\'e-type group symmetry PO(1,11) together with $\cQ_c$-spin symmetry SP(2) as well as global scaling symmetry, i.e.:
\be
G_S & = & SC(1)\ltimes PO(1,11) \adjoin WS(1,11) \rtimes SG(1)\times SP(2) \nn \\
& = & SC(1)\ltimes P^{1,11}\ltimes SO(1,11) \adjoin  SP(1,11)\rtimes W^{1,11} \rtimes SG(1)\times SP(2) , 
\ee
where the symbol ``$\adjoin$" is used to indicate the associated symmetry with coincidental transformations between symmetry groups SP(1,11) and SO(1,11), which is ensured by the isomorphic property SP(1,11)$\cong$ SO(1,11). 

The group generators of inhomogeneous hyperspin symmetry WS(1,11) and $\cQ_c$-spin symmetry SP(2) are directly given by the commutators of $\vGa$-matrices, 
\be
& & \varSigma^{\mA\mB} = \frac{i}{4} [ \vGa^{\mA}, \vGa^{\mB} ], \quad \vSi_{-}^{\mA} = \frac{1}{2} \vGa^{\mA}\vGa_{-}, \nn \\
& &  \varSigma^{D_h+p\, D_h+q} = \frac{i}{4} [ \vGa^{12+p}, \vGa^{12+q} ]=  \frac{i}{4} [ \tvGa^{p}, \tvGa^{q} ] \equiv \tvSi^{p q}, 
\ee
with $\mA, \mB= 0,1,2,3,5,\cdots, 12$ and $p, q=1,2$. The $\cQ_c$-spin symmetry characterizes the chiral flavor-spin symmetry SP(2)$\cong$U(1) between two flavors of lepton-quark states $\Psi_{\fQC^4}(x)$ and $\Psi'_{\fQC^4}(x)$. 

The hyperspin symmetry SP(1,11) can generally be decomposed into the following subgroup structure: 
\be
& & SP(1,11) \supset  SP(1,3)\times SP(8)  \supset SP(1,3) \times SP(6) \times SP(2) \nn \\
& & \cong SO(1,3) \times SO(6) \times SO(2) \cong SO(1,3)\times SU_C(4) \times U(1), 
\ee
where the symmetry SU$_C$(4)$\times$U(1) characterizes the vector-like four color-spin symmetry SU$_C$(4) and flavor-spin symmetry U(1) for two type of flavors of lepton-quark states. The corresponding subgroup algebras are given by,
\be
& & \varSigma^{\mA\mB}  ( \mA, \mB= 0,1,2,3 ) \in sp(1,3)\cong so(1,3)  \nn \\
& & \varSigma^{\mA\mB}  (\mA, \mB= 5,6,7,8,9,10 )  \in sp(6) \cong so(6)\cong su_C(4) , \nn \\
& &  \varSigma^{\mA\mB}  ( \mA, \mB= 11,12) \in sp(2)\cong so(2)\cong u(1) .  \nn 
\ee

To show the vector-like color-spin charge symmetry in SM, the action in Eq.(\ref{actionQE7}) for the entangled heptaqubit-spinor field $ \Psi_{\QE^7}(x)$ can equivalently be expressed into the following form by using the self-conjugated hexaqubit-spinor field $\Psi_{\QS^6}(x)$: 
\be \label{actionQS6}
 \cS_{\QS^6}  & = &  \int d^{12}x \, \{  \frac{1}{2}  \bar{\Psi}_{\QS^6}(x)  \delta_{\mA}^{\; \mM}  \vGa^{\mA} i\p_{\mM} \Psi_{\QS^6}(x)  - \frac{1}{2}\lambda_2 \phi_p(x) \bar{\Psi}_{\QS^6} (x) \tvGa^p \Psi_{\QS^6}(x) \}, 
\ee
with $\mA, \mM= 0,1,2,3,5, \cdots, 12$. The $\Gamma$-matrices $\vGa^{\mA} = (\vGa^a, \vGa^A) $ ($a=0,1,2,3, A= 5,\cdots, 12$) and $\tvGa$-matrices $\tvGa^p$ (p=1,2) are explicitly  given as follows: 
\be \label{GMQS6}
& &  \vGa^{\mA} = \sigma_0 \otimes \Gamma^{\mA}, \quad \mA=0,1,2,3, 5, \cdots, 10, \nn \\
& & \vGa^{11} =  i\sigma_1 \otimes \hat{\gamma}_{11} , \nn \\
& & \vGa^{12} =  i\sigma_2 \otimes \hat{\gamma}_{11} , \nn \\
& & \vGa^{13} =  i\sigma_3 \otimes \hat{\gamma}_{11} = i\hat{\gamma}_{13} \equiv \tvGa^{1} , \nn \\
& & \vGa^{14} = \; \sigma_0 \otimes I_{64} = \, \, I_{128} \equiv  \tvGa^{2}.
\ee
with $\Gamma^{\mA}$ ($\mA=0,1,2,3, 5, \cdots, 10$) defined in Eq.(\ref{GMQE5}). The $\cQ_c$-spin symmetry reflects the chiral flavor-spin symmetry SP(2) $\cong$ U(1) with the group generator,
\be
\tilde{\Sigma}^{12}_0 = -\frac{1}{2} \sigma_3 \otimes \sigma_3 \otimes  \sigma_0 \otimes \sigma_0\otimes \gamma_5. \nn
\ee

It is shown that the entangled heptaqubit-spinor field in 128-dimensional Hilbert space spanned by 7-product qubit basis states unifies two type of flavors of lepton-quark states into a single entangled hyperqubit-spinor field. Nevertheless, the maximal subgroup symmetry SU$_C$(4)$\times$U$_V$(1)$\times$U$_A$(1) deduced from hyperspin symmetry SP(1,11) and $\cQ_c$-spin symmetry SP(2) cannot produce the basic symmetries SU$_C$(3)$\times$SU$_L$(2)$\times$U$_Y$(1) of leptons and quarks in SM. So that the entangled heptaqubit-spinor field as entangled hyperqubit-spinor field of leptons and quarks remains incomplete to realize the SM.


\section{ Entangled octoqubit-spinor field as ultra-grand unified qubit-spinor field in 14D hyper-spacetime with inhomogeneous hyperspin symmetry as ultra-grand unified symmetry and the comprehension on lepton-quark state with one more family and the observed universe with 4D spacetime }

Three entangled hyperqubit-spinor fields with respective to $\cQ_c$-spin charges $\CQc=q_c=0,1,2$ and hyper-spacetime dimensions $D_h=10, 11, 12$ have been analyzed to show how they characterize the basic properties of lepton-quark states, such as color-spin and flavor-spin charges. Nevertheless, their maximal inhomogeneous hyperspin symmetry and $\cQ_c$-spin symmetry cannot yet produce the basic symmetries of chiral type leptons and quarks in SM though the entangled heptaqubit-spinor field contains $\cD_H=64$ independent degrees of freedom which is equal to the degrees of freedom involved in each family of leptons and quarks in SM. To realize the basic symmetries of leptons and quarks in SM as subgroup symmetries of hyperspin symmetry, it is natural to take into account, from the qubit-spinor categorization theorem, the next entangled hyperqubit-spinor field in category-3 based on the locally entangled-qubits motion postulate. Such an entangled hyperqubit-spinor field is referred to as entangled octoqubit-spinor field which spans 256-dimensional Hilbert space and has a free motion in 14-dimensional Minkowski hyper-spacetime. Its action will be shown to possess the inhomogeneous hyperspin symmetry WS(1,13) which contains the subgroup symmetry SO(10) as grand unified group symmetry \cite{GUT1,GUT2}, so that the basic symmetries of leptons and quarks in SM can be reproduced as subgroup symmetries of hyperspin symmetry SP(1,13). The $\cQ_c$-spin symmetry SP(3)$\cong$SU(2) is found to characterize the family-spin symmetry for two families of lepton-quark states in SM, which enables us to comprehend the basic issue on {\it why there exist more than one family of leptons and quarks in nature}. Meanwhile, we are able to provide a natural explanation on the longstanding open question {\it why our observed universe is only four-dimensional spacetime}. 


\subsection{ Entangled octoqubit-spinor field in 14D hyper-spacetime with inhomogeneous hyperspin and $\cQ_c$-spin symmetries in producing basic symmetries of leptons and quarks in SM and comprehension on lepton-quark state beyond one family in nature  }

In the weak interaction of SM, either leptons or quarks as left-handed Weyl fermions always form weak isospin doublet. In the strong interaction of SM, quarks appear as vector-like Dirac fermions. In the electromagnetic interaction of SM, both charged leptons and quarks behave as vector-like Dirac fermions. In the entangled qubit-spinor representation, the hyperspin symmetry of entangled pentaqubit-spinor field in category-0 reflects a unified color-spin symmetry of chiral lepton-quark state, which is not suitable to characterize either the electromagnetic interaction of vector-like charged leptons and quarks in SM or the strong interaction of vector-like quarks in SM. Though the hyperspin symmetries for both entangled hexaqubit-spinor field in category-1 and entangled heptaqubit-spinor field in category-2 can describe the vector-like properties of charged leptons and quarks in the electromagnetic and/or strong interactions, but they cannot realize the weak isospin symmetry of chiral leptons and quarks in SM. Following along the qubit-spinor categorization theorem, we are going to present a detailed analysis and discussion on the entangled octoqubit-spinor field as hyperqubit-spinor field in category-3. 

To characterize the chiral weak isospin property, let us begin with a category-3 {\it chiral complex heptaqubit-spinor field} $\Psi_{\fQW^7}(x)$ defined as follows:
\be \label{QW7}
& & \Psi_{\fQW^7}(x) \equiv \binom{\Psi_{\fQWn^6}(x)}{\Psi_{\fQWp^6}(x)} \equiv \binom{\Psi_{\fQC^6}^{L}(x)}{\Psi_{\fQC^6}^{R}(x)} , \nn \\ 
& & \Psi_{\fQWpn^6}(x) \equiv \vGa_{\pm}^5  \Psi_{\fQC^6}(x) \equiv \Psi_{\fQC^6}^{R, L}(x), \nn \\
& &  \vGa_{\pm}^5 = \frac{1}{2} ( 1 \pm \tilde{\gamma}_5 ), \quad \tilde{\gamma}_5 = \sigma_0 \otimes  \sigma_0 \otimes  \sigma_0 \otimes \sigma_0\otimes \gamma_5  , 
\ee
where $\Psi_{\fQWpn^6}(x)$ are {\it chiral complex hexaqubit-spinor fields} defined from the {\it complex hexaqubit-spinor field} $\Psi_{\fQC^6}(x)$. They have the following explicit forms:
\be \label{QC6}
& & \Psi_{\fQC^6}(x)  \equiv \begin{pmatrix}
\Psi^{u}_{\fQC^4}(x) \\  \Psi^{d' \,\bc}_{\fQC^4}(x) \\ \Psi^{d}_{\fQC^4}(x)  \\ -\Psi^{u' \,\bc}_{\fQC^4}(x)
\end{pmatrix},  \quad \Psi_{\fQC^6}^{R, L}(x)  \equiv \begin{pmatrix}
\Psi^{u}_{\fQC^4}(x) \\  \Psi^{d' \,\bc}_{\fQC^4}(x) \\ \Psi^{d}_{\fQC^4}(x)  \\ -\Psi^{u' \,\bc}_{\fQC^4}(x)
\end{pmatrix}_{R, L} \equiv \Psi_{\fQWpn^6}(x),  
\ee
with $\Psi^{q}_{\fQC^4}(x)$ ($q=u,u',d,d'$) regarded as {\it complex tetraqubit-spinor fields} and $\Psi^{q\; \bc }_{\fQC^4}(x)$ as complex charge-conjugated ones. Their explicit forms are given as follows:
\be \label{qQC4}
& & \Psi^{q}_{\fQC^4}(x)  \equiv \begin{pmatrix}
Q^{r}(x) \\ Q^{b}(x) \\ Q^{g}(x) \\ Q^{w}(x) 
\end{pmatrix}^{q}, \quad  \Psi^{q\; \bc }_{\fQC^4}(x)  \equiv \begin{pmatrix}
Q^{r\, \bc}(x) \\ Q^{b\, \bc}(x) \\ Q^{g, \bc}(x) \\ Q^{w\, \bc}(x) 
\end{pmatrix}^{q}, 
\ee
where the superscripts $q=u, u', d, d'$ represent four kinds of lepton-quark states with respective to the notations $\left(Q^{\alpha}\right)^{q} = (U^{\alpha}, U^{'\alpha}, D^{\alpha}, D^{'\alpha})$. The projection operators $\vGa^5_{\mp}$ define the left-handed and right-handed lepton-quark states, respectively. There are two kinds of up-type and down-type lepton-quark states with unified four color-spin charges $\alpha=(r, b, g, w)$ corresponding to `red', `blue', `green' and `white'. 

The action for such a chiral heptaqubit-spinor field $\Psi_{\fQW^7}(x)$ in category-3 is constructed as follows:
\be \label{actionQW7}
\cS_{\fQW^7}  & = &  \int d^{14}x \,  \frac{1}{2} [ \bar{\Psi}_{\fQW^7}(x) \delta_{\mA}^{\; \mM}  \vGa^{\mA} i\p_{\mM} \Psi_{\fQW^7}(x) + H.c. ] , 
\ee
which gets a free motion in 14-dimensional hyper-spacetime with $\mA, \mM= 0,1,2,3,5, \cdots, 14$ by following along the maximum coherence motion principle. The $\vGa$-matrices $\vGa^{\mA} $ ($\mA=0,1,2,3, 5,\cdots, 14$) have the following explicit forms:
\be \label{GMQW7}
& & \vGa^0 = \;\, \sigma_0 \otimes \sigma_0 \otimes \sigma_0 \otimes \sigma_0 \otimes\sigma_0 \otimes \sigma_1 \otimes \sigma_0, \nn \\
& &  \vGa^1 = \; i \sigma_0 \otimes \sigma_0 \otimes \sigma_0 \otimes \sigma_0 \otimes \sigma_0\otimes \sigma_2\otimes \sigma_1, \nn \\
& & \vGa^2 =  \;i \sigma_0 \otimes \sigma_0 \otimes \sigma_0 \otimes  \sigma_0 \otimes\sigma_0\otimes  \sigma_2\otimes \sigma_2, \nn \\
& & \vGa^3 =  \;i \sigma_0 \otimes \sigma_0 \otimes \sigma_0 \otimes \sigma_0 \otimes\sigma_0\otimes  \sigma_2\otimes \sigma_3, \nn \\
& &  \vGa^5 =  \;i \sigma_1 \otimes \sigma_0 \otimes \sigma_1 \otimes \sigma_0 \otimes \sigma_2\otimes  \sigma_3\otimes \sigma_0, \nn \\
& & \vGa^6 = \; i \sigma_1 \otimes \sigma_0 \otimes \sigma_2 \otimes  \sigma_3 \otimes\sigma_2\otimes  \sigma_3\otimes \sigma_0, \nn \\
& & \vGa^7 =  \; i \sigma_1 \otimes \sigma_0 \otimes \sigma_1 \otimes \sigma_2 \otimes\sigma_3\otimes  \sigma_3\otimes \sigma_0 , \nn \\
& &  \vGa^8 = \; i \sigma_1 \otimes \sigma_0 \otimes \sigma_2 \otimes \sigma_2 \otimes  \sigma_0\otimes \sigma_3\otimes \sigma_0, \nn \\
& &  \vGa^9 = \; i \sigma_1 \otimes \sigma_0 \otimes \sigma_1 \otimes  \sigma_2 \otimes \sigma_1\otimes \sigma_3\otimes \sigma_0 , \nn \\
& &  \vGa^{10} =  i \sigma_1 \otimes \sigma_0 \otimes \sigma_2 \otimes \sigma_1 \otimes \sigma_2\otimes \sigma_3\otimes \sigma_0 , \nn \\
& & \vGa^{11} =  i\sigma_2 \otimes \sigma_0 \otimes \sigma_0 \otimes  \sigma_0 \otimes \sigma_0\otimes \sigma_3\otimes \sigma_0 , \nn \\
& & \vGa^{12} =  i\sigma_1 \otimes \sigma_1 \otimes \sigma_3 \otimes  \sigma_0 \otimes \sigma_0\otimes \sigma_3\otimes \sigma_0 , \nn \\
& & \vGa^{13} =  i\sigma_1 \otimes \sigma_2 \otimes \sigma_3 \otimes  \sigma_0 \otimes \sigma_0\otimes \sigma_3\otimes \sigma_0 , \nn \\
& & \vGa^{14} =  i\sigma_1 \otimes \sigma_3 \otimes  \sigma_3 \otimes  \sigma_0 \otimes \sigma_0\otimes \sigma_3\otimes \sigma_0 . 
\ee
It can be verified that the hermitian action in Eq.(\ref{actionQW7}) can be expressed as follows:
\be \label{actionQW7}
\cS_{\fQW^7}  & = &  \int d^{14}x \,  \frac{1}{2} \{ \bar{\Psi}_{\fQW^7}(x) \delta_{\mA}^{\; \mM}  \vGa^{\mA} i\p_{\mM} \Psi_{\fQW^7}(x) +  \bar{\Psi}_{\fQW^7}^{c}(x) \delta_{a}^{\; \mu}  \vGa^{a} i\p_{\mu} \Psi_{\fQW^7}^{c}(x) \nn \\
& - & \bar{\Psi}_{\fQW^7}^{c}(x) \delta_{A}^{\; M}  \vGa^{A} i\p_{M} \Psi_{\fQW^7}^{c}(x)  \} , \nn
\ee
where we have used the notations $\mA=(a, A), \mM= (\mu, M)$ and $\vGa^{\mA} = (\vGa^a, \vGa^A)$ with $a, \mu =0,1,2,3; \; A, M = 5,\cdots, 14$. $\Psi_{\fQW^7}^{c}(x)$ is defined as the complex charge-conjugation of $\Psi_{\fQW^7}(x)$ and has the following properties:
\be
& &  \Psi_{\fQW^7}^{c}(x) = C_{\fQW^7} \bar{\Psi}^{T}_{\fQW^7}(x) , \nn \\
& & \hat{\gamma}_{15} \Psi_{\fQW^7}(x) = - \Psi_{\fQW^7}(x), \;\; \hat{\gamma}_{15} \Psi_{\fQW^7}^{c}(x) = - \Psi_{\fQW^7}^{c}(x), \nn \\
& & C_{\fQW^7}  = \vGa_2\vGa_0\vGa_6\vGa_8\vGa_{10}\vGa_{12}\vGa_{14} \gamma_5 \nn \\ & & \qquad \; =   \sigma_1 \otimes \sigma_2 \otimes  \sigma_2 \otimes  \sigma_0 \otimes \sigma_0\otimes C_D, \nn \\
& & \hat{\gamma}_{15}  = \sigma_3 \otimes \sigma_0 \otimes  \sigma_0 \otimes  \sigma_0 \otimes \sigma_0\otimes \gamma_5 .  
\ee

The above formalism of the action indicates that it is appropriate to introduce an entangled hyperqubit-spinor field with the following qubit-spinor structure: 
\be \label{QE8}
& & \Psi_{\QE^8}(x) = \binom{\Psi_{\fQW^7}(x) } { \Psi_{\fQW^7}^{c}(x) }, \nn \\
& & \Psi_{\fQW^7}(x) \equiv \binom{\Psi_{\fQWn^6}(x)}{\Psi_{\fQWp^6}(x)} \equiv \binom{\Psi_{\fQC^6}^L(x)}{\Psi_{\fQC^6}^R(x)} , \nn \\
& & \Psi_{\fQW^7}^{c}(x)  \equiv \binom{\Psi_{\fQWp^6}^{c}(x)}{\Psi_{\fQWn^6}^{c}(x)}  \equiv \binom{\Psi_{\fQC^6}^{R\, c}(x)}{\Psi_{\fQC^6}^{L\, c}(x)} , 
\ee
where $\Psi_{\fQWpn^6}^{c}(x)$ are the complex charge-conjugation of chiral complex hexaqubit-spinor fields $\Psi_{\fQWpn^6}(x)$ with the following definition:
\be \label{QC6c}
& & \Psi_{\fQWpn^6}^{c}(x) = C_{\fQWpn^6} \bar{\Psi}^{T}_{\fQWpn^6}(x) \equiv \vGa_{\mp}^5  \Psi_{\fQC^6}^{c}(x)  , \nn \\
& &  \Psi_{\fQC^6}^{c}(x) = C_{\fQC^6} \bar{\Psi}^{T}_{\fQC^6}(x), \nn \\
& & C_{\fQC^6}  = C_{\fQWpn^6}= \sigma_2 \otimes  \sigma_2 \otimes  \sigma_0 \otimes \sigma_0\otimes C_D .
\ee
The projection operators $\vGa^5_{\mp}$ given in Eq.(\ref{QW7}) define the left-handed and right-handed lepton-quark states, respectively. $\Psi_{\fQC^6}^{c}(x)$ is the complex charge-conjugation of $\Psi_{\fQC^6}(x)$ with the following explicit form:
\be
 \Psi_{\fQC^6}^{c}(x)  \equiv \begin{pmatrix}
\Psi^{u'}_{\fQC^4}(x) \\  \Psi^{d\; \bc}_{\fQC^4}(x) \\ \Psi^{d'}_{\fQC^4}(x)  \\ -\Psi^{u\; \bc}_{\fQC^4}(x)
\end{pmatrix} , \quad  \Psi_{\fQC^6}^{L, R\; c}(x)  \equiv \begin{pmatrix}
\Psi^{u'}_{\fQC^4}(x) \\  \Psi^{d\; \bc}_{\fQC^4}(x) \\ \Psi^{d'}_{\fQC^4}(x)  \\ -\Psi^{u\; \bc}_{\fQC^4}(x)
\end{pmatrix}_{R, L} .
\ee

In terms of the entangled hyperqubit-spinor field $\Psi_{\QE^8}(x)$, the action in Eq.(\ref{actionQW7}) can be rewritten into the following simple form: 
\be  \label{actionQE8}
\cS_{\QE^8}  & \equiv & \int d^{14}x \,   \bar{\Psi}_{\QE^8}(x) \delta_{\mA}^{\;\mM}  \vSi_{-}^{\mA} i\p_{\mM} \Psi_{\QE^8}(x).
\ee
with $\mA, \mM= 0,1,2,3,5, \cdots, 14$. $\vSi_{-}^{\mA}$ matrices are defined by $\vGa$-matrices $\vGa^{\mA}= (\vGa^a, \vGa^A) $ ($a=0,1,2,3, A= 5,\cdots, 14$). All $\vGa$-matrices $\vGa^{\hat{\mA}} =(\vGa^{\mA}, \vGa^{D_h+p}) $ ($\mA=0,1,2,3, 5, \cdots, 14$, $p=1,2,3$) and $\cQ_c$-matrices $\tvGa^p\equiv \vGa^{D_h+p} $ ($p=1,2,3$) are found to have the following structures:
\be \label{GMQE8}
& & \vSi_{-}^{\mA} = \frac{1}{2} \vGa^{\mA}\vGa_{-}, \quad  \vGa_{-} = \frac{1}{2} ( 1 - \hat{\gamma}_{17} ),  \nn \\
& & \vGa^a = \,  \sigma_0 \otimes \vGa^a, \quad a=0,1,2,3, \nn \\
& & \vGa^{\mA} = \,  \sigma_3 \otimes \vGa^A, \quad A= 5, \cdots, 14, \nn \\
& & \vGa^{15} = i \sigma_1 \otimes \tilde{\gamma}_5 \equiv \tvGa^1 , \nn \\
& & \vGa^{16} = i \sigma_2 \otimes \tilde{\gamma}_5 \equiv \tvGa^2 , \nn \\
& & \vGa^{17} = i \sigma_3 \otimes \hat{\gamma}_{15}  \equiv \tvGa^3 , \nn \\
& & \tilde{\gamma}_5 = \sigma_0 \otimes \sigma_0 \otimes  \sigma_0 \otimes  \sigma_0 \otimes \sigma_0\otimes \gamma_5, \nn \\
& & \hat{\gamma}_{17} = \sigma_0\otimes \sigma_3 \otimes \sigma_0 \otimes  \sigma_0 \otimes  \sigma_0 \otimes \sigma_0\otimes \gamma_5 \
\equiv \sigma_0\otimes  \hat{\gamma}_{15}, 
\ee
where $\vGa^a$ and $\vGa^A$ ($a=0,1,2,3,  A= 5, \cdots, 14$) on the right-hand side of equality are defined in Eq.(\ref{GMQW7}).

The entangled hyperqubit-spinor field $\Psi_{\QE^8}(x)$ is composed of four kinds of lepton-quark states and satisfies the following self-conjugated chiral conditions: 
\be \label{CCQE8}
& & \Psi^{c}_{\QE^8}(x) = C_{\QE^8} \bar{\Psi}^T_{\QE^8}(x) =  \Psi_{\QE^8}(x), \nn \\
& & C_{\QE^8} =  i\vGa_2\vGa_0\vGa_6\vGa_8\vGa_{10}\vGa_{12}\vGa_{14} \vGa_{16} = \sigma_1\otimes C_{\fQW^7} \nn \\
& & \quad \quad = \sigma_1 \otimes \sigma_1 \otimes \sigma_2 \otimes  \sigma_2 \otimes  \sigma_0 \otimes \sigma_0\otimes C_D , \nn \\
& & \hat{\gamma}_{17} \Psi_{\QE^8}(x) = - \Psi_{\QE^8}(x), \quad  \hat{\gamma}_{17} = \sigma_0\otimes  \hat{\gamma}_{15} ,
\ee
where $\Psi_{\QE^8}(x)$ is referred to as {\it entangled octoqubit-spinor field} in category-3.

All $\vGa$-matrices are anti-commuting and satisfy special commutation relations with the matrix $\hat{\gamma}_{17}$, 
\be
& & \{ \vGa^{\mA}, \vGa^{\mB} \} = 2 \eta^{\mA\mB}, \qquad \qquad \mA, \mB = 0,1,2,3,5,\cdots, 17, \nn \\
& & \{ \vGa^{\mA}, \hat{\gamma}_{17} \} = 0, \qquad \qquad \quad \; \mA = 0,1,2,3,5,\cdots, 14, \nn \\
& &  [ \vGa^{\mA}, \hat{\gamma}_{17} ]  \equiv  [ \vGa^{D_h+p}, \hat{\gamma}_{17} ]  \equiv  [ \tvGa^{p}, \hat{\gamma}_{17} ] = 0, \quad \mA=15,16,17, \; p = 1,2,3, 
\ee
where the commutator property in the last equality results in the absence of scalar coupling term in association with the $\cQ_c$-matrices $\tvGa^p$ ($p=1,2,3$), which becomes a manifest feature for entangled octoqubit-spinor field $\Psi_{\QE^8}(x)$ in category-3.

The action in Eq.(\ref{actionQE8}) possesses an associated symmetry in which the inhomogeneous hyperspin symmetry WS(1,13)  and $\cQ_c$-spin symmetry SP(3) in association with inhomogeneous Lorentz-type/Poincar\'e group symmetry PO(1,13) together with the scaling symmetries SG(1) and SC(1), i.e.:
\be
G_S & = & SC(1)\ltimes PO(1,13) \adjoin WS(1,13) \rtimes SG(1)\times SP(3) , \nn \\
 & = & SC(1)\ltimes P^{1,13}\ltimes SO(1,13) \adjoin SP(1,13)\rtimes W^{1,13} \rtimes SG(1)\times SP(3) ,
\ee
which indicates from the symbol `$\adjoin$' that the group transformation of hyperspin symmetry SP(1,13) in Hilbert space should be coincidental to that of Lorentz-type group symmetry SO(1,13) in Minkowski hyper-spacetime via the isomorphic property of two group symmetries, i.e., SP(1,13)$\cong$ SO(1,13). 

The group generators of inhomogeneous hyperspin symmetry WS(1,13) and $\cQ_c$-spin symmetry SP(3) can directly be expressed by the commutators of $\vGa$-matrices, 
\be
& & \varSigma^{\mA\mB} = \frac{i}{4} [ \vGa^{\mA}, \vGa^{\mB} ], \quad \vSi_{-}^{\mA} = \frac{1}{2} \vGa^{\mA}\vGa_{-} , \nn \\
& & \varSigma^{D_h+p\, D_h+q} = \frac{i}{4} [ \vGa^{14+p}, \vGa^{14+q} ]  = \frac{i}{4} [ \tvGa^{p}, \tvGa^{q} ]\equiv  \tvSi^{p,q},  
\ee
with $\mA, \mB= 0,1,2,3,5,\cdots, 14$ and $p, q=1,2,3$. 

By decomposing the hyperspin symmetry SP(1,13) into possible subgroup symmetries, we arrive at the following appropriate subgroup symmetry:
\be
 SP(1,13) & \supset &  SP(1,3)\times SP(10)  \supset SP(1,3) \times SP(6) \times SP(4) \nn \\
&  \cong & SO(1,3)\times SO(10) \supset SO(1,3) \times SO(6)\times SO(4) \nn \\
& \cong & SO(1,3) \times SU_C(4)\times SU_L(2)\times SU_R(2) \nn \\
& \supset & SO(1,3) \times SU_C(3)\times SU_L(2)\times U_Y(1),
\ee
where the direct product subgroup symmetry $SO(1,3)\times SU_C(3)\times SU_L(2)\times U_Y(1)$ in the last equality produces the basic symmetries in SM, which reflects the Lorentz-spin charge, color-spin charge, isospin charge and hyper-electric charge in SM. The direct product subgroup algebra is in correspondence to the following group generators:
\be
& & \varSigma^{\mA\mB}  ( \mA, \mB= 0,1,2,3 ) \in sp(1,3)\cong so(1,3)  \nn \\
& & \varSigma^{\mA\mB}  (\mA, \mB= 5,6,7,8,9,10 )  \in sp(6) \cong so(6)\cong su_C(4) , \nn \\
& &  \varSigma^{\mA\mB}  ( \mA, \mB= 11,12,13,14) \in sp(4)\cong so(4)\cong su_L(2)\times su_R(2).  \nn 
\ee

It is clear that the hyperspin symmetry SP(1,13) of entangled octoqubit-spinor field $\Psi_{\QE^8}(x)$ as entangled hyperqubit-spinor field in category-3 does bring on the basic symmetries of leptons and quarks in SM as subgroup symmetries. From the entangled qubit-spinor representation of $\Psi_{\QE^8}(x)$ and the definitions of $\Psi_{\fQW^7}(x)$ and $\Psi_{\fQW^7}^{c}(x)$, it can be verified that the $\cQ_c$-spin symmetry SP(3) describes the {\it family-spin symmetry} SU$_F$(2)$\cong$SP(3) for two kinds of lepton-quark states, i.e., $Q^{\alpha}=(U^{\alpha}, D^{\alpha})$ and  $Q^{' \alpha}=(U^{' \alpha}, D^{' \alpha})$.

So far we come to the observation that when a locally entangled state of qubits as entangled hyperqubit-spinor field is considered as fundamental building block of nature with following along the locally entangled-qubits motion postulate, the resulting inhomogeneous hyperspin symmetry of entangled hyperqubit-spinor field naturally reflects the fundamental symmetry of nature. To realize the basic symmetries of leptons and quarks in SM as subgroup symmetries of hyperspin symmetry, the entangled octoqubit-spinor field $\Psi_{\QE^8}(x)$ in category-3 is found to be the {\it minimal entangled hyperqubit-spinor field}, which leads the hyperspin symmetry SP(1,13) $\cong$ SO(1,13) as {\it minimal unified symmetry} to produce the basic symmetries of SM as subgroup symmetries presented via a direct product group symmetry, i.e., SO(1,3)$\times$SU$_C$(3)$\times$SU$_L$(2)$\times$ U$_Y$(1). Meanwhile, the intrinsic $\cQ_c$-spin symmetry SP(3) as family-spin symmetry SU$_F$(2) indicates the simultaneous existence of two families of leptons and quarks in nature. 

It becomes manifest that the category-3 entangled octoqubit-spinor field $\Psi_{\QE^8}(x)$ in 14-dimensional hyper-spacetime brings about the prediction on the simultaneous presence of two families of chiral type leptons and quarks in nature, which provides a natural answer to the best-known question ``who ordered that?" exclaimed when I.I. Rabi was informed in 1936 the discovery of the muon as a new particle with an unexpected surprising property\footnote{In the middle of 1930s, all the subatomic particles of nature were thought to be composed of the electron, proton and neutron, it was surprised to find that the muon was not the hadron predicted by theoretical particle physicist, but a new and entirely unexpected type of charged lepton, a heavier relative to the electron, which made I.I. Rabi posed his famous question in pondering the existence of the muon: ``who ordered that?" . }

Therefore, it is the entangled octoqubit-spinor field that provides a minimal entangled hyperqubit-spinor field to unify simultaneously two families of chiral type leptons and quarks in SM into a single entangled hyperqubit-spinor field and meanwhile produce exactly the basic symmetries of SM SO(1,3)$\times$SU$_C$(3)$\times$SU$_L$(2)$\times$U$_Y$(1) as a direct product subgroup symmetry of hyperspin symmetry SP(1,13)$\cong$SO(1,13). It becomes a necessity to merge automatically two families of chiral type leptons and quarks in SM into such a minimal entangled hyperqubit-spinor field in 14-dimensional hyper-spacetime. The two families of chiral type lepton-quark states are characterized by the intrinsic $\cQ_c$-spin symmetry SP(3) as family-spin symmetry SU$_F$(2), which brings on a natural explanation on why there exist leptons and quarks beyond one family in SM. It is clear that the hyperspin symmetry SP(1,13)$\cong$SO(1,13) for the unification model \cite{UM} contains a maximal subgroup symmetry SP(10)$\cong$SO(10) as grand unified symmetry\cite{GUT1,GUT2} which unifies all internal symmetries SU$_C$(3)$\times$SU$_L$(2)$\times$U$_Y$(1) of leptons and quarks in SM.


\subsection{Ultra-grand unified qubit-spinor field in 14D hyper-spacetime with inhomogeneous hyperspin symmetry as ultra-grand unified symmetry and comprehension on the observed universe with only 4-dimensional spacetime} 

The entangled octoqubit-spinor field is shown to be the minimal entangled hyperqubit-spinor field, which realizes all basic symmetries in SM as a direct product subgroup symmetry of hyperspin symmetry and obtains the action with a free motion in 14-dimensional hyper-spacetime. Nevertheless, our currently observed universe is only four-dimensional spacetime. How to understand such a universe with four-dimensional spacetime from 14-dimensional hyper-spacetime becomes an inevitable question to be answered in exploring the foundation of the hyperunified field theory. In fact, the question why our observed universe is only four-dimensional spacetime has been a longstanding and essential issue. Following along the locally entangled-qubits motion postulate and noticing the fact that our observed universe is composed of the atomic matter as basic constituent of observable matter world, we are able to provide a natural explanation on such a basic question. 

Let us first rewrite the chiral complex heptaqubit-spinor field $\Psi_{\fQW^7}(x)$ in Eq.(\ref{QW7}) into the sum of two {\it typical self-conjugated chiral hyperqubit-spinor fields},
\be \label{fQWQE7}
& &\Psi_{\fQW^7}(x) \equiv  \frac{1}{\sqrt{2}} [ \Psi_{\fQEnf}(x) + i \Psi_{\fQEns}(x) ] , \nn \\
& & \Psi_{\fQW^7}^{c}(x) \equiv  \frac{1}{\sqrt{2}} [ \Psi_{\fQEnf}(x) - i \Psi_{\fQEns}(x) ],  
\ee 
where $\Psi_{\fQEni}(x) $ ($i=1,2$) have the following qubit-spinor structure:
\be \label{fQE7}
& & \Psi_{\fQEni}(x) \equiv \binom{\Psi_{\fQWni}(x)}{\Psi_{\fQWni}^{c}(x)} \equiv 
\binom{\Psi_{\fQ_{\mS_i}^6}^{L}(x)}{\Psi_{\fQ_{\mS_i}^6}^{L\, c}(x)} .
\ee
$\Psi_{\fQWni}(x)$ ($i=1,2$) are two {\it chiral complex hexaqubit-spinor fields} defined from two {\it typical self-conjugated hexaqubit-spinor fields} $\Psi_{\fQ_{\mS_i}^6}(x)$ as follows:
\be
& & \Psi_{\fQWni}(x) \equiv \Psi_{\fQ_{\mS_i}^6}^{L}(x) = \vGa^5_{-} \Psi_{\fQ_{\mS_i}^6}(x), \nn \\
& & \Psi_{\fQWni}^{c}(x) = C_{\fQ_{\mW_i}^6} \bar{\Psi}_{\fQWni}^{T}(x) \equiv  \Psi_{\fQ_{\mS_i}^6}^{L\, c}(x) = \vGa^5_{+} \Psi_{\fQ_{\mS_i}^6}(x), \nn \\
& & \Psi_{\fQ_{\mS_i}^6}^{c} (x) = C_{\fQ_{\mS_i}^6} \bar{\Psi}_{\fQ_{\mS_i}^6}^{T}(x) = \Psi_{\fQ_{\mS_i}^6}(x),  \nn \\
& & C_{\fQ_{\mW_i}^6} = C_{\fQ_{\mS_i}^6} = \sigma_2 \otimes \sigma_2 \otimes \sigma_0 \otimes \sigma_0 \otimes C_D , \nn \\
& &  \vGa^5_{\pm}  = \frac{1}{2} (1 \pm \tilde{\gamma}_5 ) , \quad  \tilde{\gamma}_5  = \sigma_0 \otimes \sigma_0 \otimes \sigma_0 \otimes \sigma_0 \otimes \gamma_5 ,
\ee
where $\Psi_{\fQ_{\mS_i}^6}(x)$ ($i=1,2$) have the following explicit structures: 
\be
& &   \Psi_{\fQ_{\mS_i}^6}^{L}(x)  \equiv \begin{pmatrix}
\Psi_{\fQ_{\mC_i}^4}^{u}(x) \\  \Psi_{\fQ_{\mC_i}^4}^{d\, \bc}(x)  \\ \Psi_{\fQ_{\mC_i}^4}^{d}(x) \\ - \Psi_{\fQ_{\mC_i}^4}^{u\, \bc}(x)  
\end{pmatrix}_{L} , \quad  \Psi_{\fQ_{\mS_i}^6}^{L, c}(x)  \equiv \begin{pmatrix}
\Psi_{\fQ_{\mC_i}^4}^{u}(x) \\  \Psi_{\fQ_{\mC_i}^4}^{d\, \bc}(x)  \\ \Psi_{\fQ_{\mC_i}^4}^{d}(x) \\ - \Psi_{\fQ_{\mC_i}^4}^{u\, \bc}(x)  
\end{pmatrix}_{R} . 
\ee

It is noted that $\Psi_{\fQ_{\mS_i}^6}(x)$ ($i=1,2$) should be distinguished from the self-conjugated hexaqubit-spinor field $\Psi_{\fQ_{\mS}^6}(x)$ defined in Eq.(\ref{fQS6}) as they exhibit a new kind of twisted hyperqubit-spinor structure. Where $\Psi_{\fQ_{\mC_i}^4}^{q}(x)$ ($q=u, d$) are defined as {\it complex tetraqubit-spinor fields} with $\Psi_{\fQ_{\mC_i}^4}^{q\, \bc}(x)$ as complex charge-conjugated ones, they are composed of lepton-quark states with the following explicit forms:
\be
& &  \Psi_{\fQ_{\mC_i}^4}^{q}(x) \equiv \begin{pmatrix}
Q_i^{r}(x) \\ Q_i^{b}(x) \\ Q_i^{g}(x) \\ Q_i^{w}(x) 
\end{pmatrix}^q, \quad  \Psi_{\fQ_{\mC_i}^4}^{q\, \bc}(x) \equiv \begin{pmatrix}
Q^{r\, \bc}_{i }(x) \\ Q^{b\, \bc}_{i}(x) \\ Q^{g\, \bc}_{i}(x) \\ Q^{w\, \bc}_{i}(x) 
\end{pmatrix}^q . 
\ee
where the superscripts $q=u$ and $q=d$ label up-type and down-type lepton-quark states, i.e., $Q^{\alpha\, u}_i(x) \equiv U_i^{\alpha}(x)$ and $Q^{\alpha\, d}_i(x) \equiv D_i^{\alpha}(x)$, with four color-spin charges $\alpha = r, b, g, w$ in correspondence to `red', `blue', `green' and `white'. The subscripts $i=1,2$ reflect two classes of lepton-quark states.  The projection operators $\vGa^5_{\mp}$ define the left-handed and right-handed lepton-quark states, i.e., $Q_{i\, L,R}^{\alpha\, u}(x) \equiv U_{i\, L, R}^{\alpha}(x)$ and $Q_{i\, L,R}^{\alpha\, d}(x) \equiv D_{i\, L, R}^{\alpha}(x)$.

From the definitions presented in Eqs.(\ref{fQWQE7}) and (\ref{fQE7}), it can be checked that $\Psi_{\fQEni}(x)$ satisfy the following self-conjugated chiral conditions:
\be
& & \Psi_{\fQEni}^{c}(x) \equiv  C_{\fQ_{\mE_i}^7}\bar{\Psi}_{\fQEni}^{T}(x) = \Psi_{\fQEni}(x),
 \nn \\
& & \hat{\gamma}_{15} \Psi_{\fQEni}(x) =  \hat{\gamma}_{15} \Psi_{\fQEni}^{c}(x) = - \Psi_{\fQEni}(x) , 
 \nn \\
& & C_{\fQ_{\mE_i}^7} = C_{\fQW^7} =   \sigma_1 \otimes \sigma_2 \otimes  \sigma_2 \otimes  \sigma_0 \otimes \sigma_0\otimes C_D . 
\ee
So that the action in Eq.(\ref{actionQW7}) can be rewritten into the following form:
\be  \label{actionfQE7}
\cS_{\fQW^7}   & \equiv & \int d^{14}x \,  \frac{1}{2}  \{  \bar{\Psi}_{\fQEnf}(x) \delta_{a}^{\;\mu}  \vGa^{a} i\p_{\mu}  \Psi_{\fQEnf}(x) + \bar{\Psi}_{\fQEns}(x) \delta_{a}^{\;\mu}  \vGa^{a} i\p_{\mu}  \Psi_{\fQEns}(x) \nn \\
& + & i\bar{\Psi}_{\fQEnf}(x) \delta_{A}^{\; M}  \vGa^{A} i\p_{M}  \Psi_{\fQEns}(x) -i \bar{\Psi}_{\fQEns}(x) \delta_{A}^{\; M}  \vGa^{A} i\p_{M}  \Psi_{\fQEnf}(x) \} \nn \\
& + & \frac{1}{2} i\p_{M}[  \bar{\Psi}_{\fQEnf}(x) \delta_{A}^{\; M}  \vGa^{A}  \Psi_{\fQEnf}(x)  + \bar{\Psi}_{\fQEns}(x) \delta_{A}^{\; M}  \vGa^{A}  \Psi_{\fQEns}(x) ] , 
\ee
with $a, \mu = 0,1,2,3$ and $A, M = 5,\cdots, 14$. Where we have used the following identity:
\be
& & \bar{\Psi}_{\fQEni}(x) \delta_{A}^{\; M}  \vGa^{A} i\p_{M} \Psi_{\fQEni}(x) \equiv i\p_{M}[  \bar{\Psi}_{\fQEni}(x) ] \delta_{A}^{\; M}  \vGa^{A}  \Psi_{\fQEni}(x) \nn \\ 
& & \; = \frac{1}{2} i\p_{M}[  \bar{\Psi}_{\fQEni}(x) \delta_{A}^{\; M}  \vGa^{A}  \Psi_{\fQEni}(x) ] ,  \quad A, M = 5, \cdots ,14,  \nn
\ee
due to the self-conjugated feature of $\Psi_{\fQEni}(x)$, and also the following property of $\vGa$-matrices:
\be
C_{\fQ_{\mE_i}^7} \vGa^a C_{\fQ_{\mE_i}^7}^{-1} = - (\vGa^a)^T, \quad C_{\fQ_{\mE_i}^7} \vGa^A C_{\fQ_{\mE_i}^7}^{-1} = (\vGa^A)^T,  \nn
\ee
with $a=0,1,2,3$ and $A = 5, \cdots ,14$. 

The total derivative appearing in the last term of the action in Eq.(\ref{actionfQE7}) indicates that the free motion of $\Psi_{\fQEni}(x)$($i=1,2$) in extra ten spatial dimensions can only occur through the cross term between $\Psi_{\fQEnf}(x)$ and $\Psi_{\fQEns}(x)$. Such a feature is attributed to the specific {\it twisted qubit-spinor structure} of $\Psi_{\fQEni}(x)$($i=1,2$), which distinguishes from the structure of entangled heptaqubit-spinor field $\Psi_{\QE^7}(x)$ given in Eq.(\ref{QE7}). It is noted that the letter in bold type `$\fQ_{\mE_i}$' has been used to distinguish them. For convenience of mention in terminology, we may refer to such typical self-conjugated chiral hyperqubit-spinor fields $\Psi_{\fQEni}(x)$ as {\it twisting entangled hyperqubit-spinor fields}.

In general, the action in Eq.(\ref{actionfQE7}) can be expressed into the following simple form:
\be  \label{actionfQE8}
\cS_{\fQE^8} & \equiv & \int d^{14}x \,  \bar{\Psi}_{\fQE^8}(x) \delta_{\mA}^{\;\mM}  \vSi_{-}^{\mA} i\p_{\mM} \Psi_{\fQE^8}(x) ,
\ee
with $\vSi_{-}^{\mA}$ ($\mA=0,1,2,3, 5, \cdots, 14$) matrices given by $\vGa$-matrices $\vGa^{\mA}= (\vGa^a, \vGa^A) $ ($a=0,1,2,3, A= 5,\cdots, 14$). The $\vGa$-matrices $\vGa^{\mA} $ ($\mA=0,1,2,3, 5, \cdots, 14$) and $\cQ_c$-matrices $\tvGa^p$ ($p=1,2,3$) can be combined to form the full $\vGa$-matrices  $\vGa^{\hat{\mA}} \equiv (\vGa^{\mA}, \vGa^{D_h+p})\equiv(\vGa^{\mA}, \tvGa^{p}) $ ($\hat{\mA}=0,1,2,3, 5, \cdots, 17$) with the following forms: 
\be \label{GMfQE8}
& & \vSi_{-}^{\mA} = \frac{1}{2} \vGa^{\mA}\vGa_{-}, \quad  
\vGa_{-} = \frac{1}{2} ( 1 - \hat{\gamma}_{17} ), \nn \\
& & \vGa^a = \;\;\; \, \sigma_0\otimes \vGa^a,  \qquad a=0,1,2,3 , \nn \\
& &  \vGa^A =-\sigma_2\otimes \vGa^A,     \qquad A=5,\cdots,14, \nn \\
& &  \vGa^{15} =\; i\sigma_1\otimes \tilde{\gamma}_5\equiv \tvGa^1, \nn \\
& &  \vGa^{16} =\; i\sigma_2 \otimes \hat{\gamma}_{15} \equiv \tvGa^2, \nn \\
& &  \vGa^{17} =\; i\sigma_3\otimes \tilde{\gamma}_5 \equiv \tvGa^3, \nn \\
& & \hat{\gamma}_{17} = \sigma_0\otimes  \hat{\gamma}_{15} , 
\ee
where the matrices $\vGa^a$ and $\vGa^A$ on the right-hand side of equality are presented in Eq.(\ref{GMQW7}). 

The entangled hyperqubit-spinor field $\Psi_{\fQE^8}(x)$ in Eq.(\ref{actionfQE8}) is defined as follows:
\be \label{fQE8}
& & \Psi_{\fQE^8}(x) \equiv \binom{\Psi_{\fQEnf}(x)} { \Psi_{\fQEns}(x)} \equiv 
 \begin{pmatrix}
\Psi_{\fQ_{\mS1}^6}^{L}(x) \\  \Psi_{\fQ_{\mS_1}^6}^{L\, c}(x)  \\ \Psi_{\fQ_{\mS_2}^6}^{L}(x) \\ \Psi_{\fQ_{\mS_2}^6}^{L\, c}(x)
\end{pmatrix} , 
\ee
which satisfies the following self-conjugated chiral conditions:
\be
& &  \Psi_{\fQE^8}^{c}(x) =  C_{\fQE^8} \bar{\Psi}_{\fQE^8}^{T}(x) =  \Psi_{\fQE^8}(x) , \quad \hat{\gamma}_{17} \Psi_{\fQE^8}(x)  = - \Psi_{\fQE^8}(x), \nn \\
& &  C_{\fQE^8} = \sigma_0\otimes C_{\fQ_{\mE_i}^7}  = \sigma_0\otimes \sigma_1 \otimes \sigma_2 \otimes \sigma_2 \otimes \sigma_0 \otimes \sigma_0 \otimes C_D .
\ee
Note that $\Psi_{\fQE^8}(x)$ defines an alternative {\it entangled octoqubit-spinor field}, which has a different qubit-spinor structure from that of entangled octoqubit-spinor field $\Psi_{\QE^8}(x)$ given in Eq.(\ref{QE8}). $\Psi_{\fQE^8}(x)$ is distinguished symbolically by the letter in bold type `$\fQE$'.

As all $\vGa$-matrices in Eq.(\ref{GMfQE8}) become anti-commuting, the group generators of inhomogeneous hyperspin symmetry WS(1,13) and $\cQ_c$-spin symmetry SP(3) for the action in Eq.(\ref{actionfQE8}) can directly be given by the commutators of $\vGa$-matrices, 
\be
& & \varSigma^{\mA\mB} = \frac{i}{4} [ \vGa^{\mA}, \vGa^{\mB} ], \quad \vSi_{-}^{\mA} = \frac{1}{2} \vGa^{\mA}\vGa_{-} , \nn \\
& &   \varSigma^{D_h+p\, D_h+q} \equiv \frac{i}{4} [ \vGa^{14 + p}, \vGa^{14+q} ] = \frac{i}{4} [ \tvGa^{p}, \tvGa^{q} ] \equiv \tvSi^{p,q} ,  
\ee
with $\mA, \mB= 0,1,2,3,5,\cdots, 14$ and $p, q=1,2,3$. 

It is seen from the action in Eq.(\ref{actionfQE7}) that each twisting entangled hyperqubit-spinor field $\Psi_{\fQEni}(x)$ has a free motion only in four-dimensional spacetime, which motivates us to build the action for each $\Psi_{\fQEni}(x)$ in four dimensional spacetime. The action for each twisting entangled hyperqubit-spinor field is found to have the following general form: 
\be \label{actionfQE74D}
\cS_{\fQ_{\mE_i}^7}   & \equiv & \int d^4x \,  \{  \bar{\Psi}_{\fQEni}(x)  \delta_{a}^{\;\mu}  \varSigma_{-}^{a} i\p_{\mu} \Psi_{\fQEni}(x) \nn \\
& + & \bar{\Psi}_{\fQEni}(x)  [ H^{A}(x) + i\gamma_5 H_5^{A}(x) ] \varSigma_{-}^{A} \Psi_{\fQEni}(x) \}, 
\ee
with $a,\mu = 0,1,2,3$, and the $\vGa$-matrices, 
\be
& & \varSigma_{-}^{a} = \frac{1}{2} \vGa^{a}\vGa_{-}, \quad  \varSigma_{-}^{A} =  \frac{1}{2} \vGa^{A}\vGa_{-}, \nn \\
& &  \vGa_{-} = \frac{1}{2} ( 1 - \hat{\gamma}_{15} ) , \quad a = 0,1,2,3; \; A=5, \cdots, 14 .
\ee
The vector fields $H^{A}(x)$ and $H_5^{A}(x)$ in Hilbert space are regarded as Higgs-like scalar fields. 

It can be verified that the action in Eq.(\ref{actionfQE74D}) possesses the following associated symmetry:
\be
G_S & = & SC(1)\ltimes PO(1,3)\adjoin WS(1,3) \rtimes SG(1) \times SP(10) , \nn \\
& = &  SC(1)\ltimes P^{1,3}\ltimes SO(1,3)\adjoin SP(1,3)\rtimes W^{1,3}\rtimes SG(1) \times SO(10), 
\ee
which is considered as a subgroup symmetry of the whole associated symmetry in which the inhomogeneous hyperspin symmetry and Poincar\'e-type group symmetry together with global scaling symmetry, i.e., SC(1) $\ltimes$ PO(1,13) $\adjoin$ WS(1,13) $\rtimes$ SG(1).

It is interesting to notice that each twisting entangled hyperqubit-spinor field $\Psi_{\fQEni}(x)$ contains 64 independent degrees of freedom, which characterizes exactly all intrinsic features for each family of chiral type leptons and quarks in SM. So that the hyperspin symmetry SP(10) brings on the grand unified symmetry with the following appropriate subgroup symmetries: 
\be
G & = & SP(10) \cong SO(10) \supset SU_C(4)\times SU_L(2)\times SU_R(2) \supset SU_C(3)\times SU_L(2)\times U_Y(1), \nn \\
G & = & SP(10) \cong SO(10)\supset SU(5)\times U(1) \supset SU_C(3)\times SU_L(2)\times U_Y(1), 
\ee
which brings about the grand unified theory\cite{GUT1,GUT2}  for leptons and quarks in SM with including right-handed neutrinos. Where SU(5) is regarded as a minimal grand unified theory\cite{SU5}, and SU$_C$(4)$\times$SU$_L$(2)$\times$SU$_R$(2) as a lepton-quark unified theory\cite{PS} which gives rise to the unified four color-spin symmetry and left-right isospin symmetry. 

As each twisting entangled hyperqubit-spinor field $\Psi_{\fQEni}(x)$ unifies each family of chiral type leptons and quarks in SM into a single qubit-spinor field with grand unified symmetry SP(10)$\cong$SO(10), it is appropriate to call $\Psi_{\fQEni}(x)$ as {\it grand unified qubit-spinor field} denoted as follows:
\be
\Psi_{\fQ_{\mG_i}^{-}}(x) \equiv \Psi_{\fQEni}(x).
\ee
To be more explicit, the sixty four independent degrees of freedom in each family of chiral type leptons and quarks in SM can be expressed into a single westward entangled hyperqubit-spinor field\cite{HUFT}:
\be
& & \Psi_{W\1 i}^{T}(x) \equiv  \Psi_{\fQ_{\mG_i}^{-}}^{T}(x) \equiv \Psi_{\fQEni}^{T}(x)   \nn \\
& & =[ (U_i^{r}, U_i^{b}, U_i^{g}, U_i^{w}, D^{r}_{ic}, D^{b}_{ic}, D^{g}_{ic}, D^{w}_{ic}, D_i^{r}, D_i^{b}, D_i^{g}, D_i^{w}, -U^{r}_{ic}, -U^{b}_{ic}, -U^{g}_{ic}, -U^{w}_{ic})_L\, ,   \nn \\
& & \;\; (U_i^{r}, U_i^{b}, U_i^{g}, U_i^{w}, D^{r}_{ic}, D^{b}_{ic}, D^{g}_{ic}, D^{w}_{ic}, D_i^{r},  D_i^{b}, D_i^{g}, D_i^{w}, -U^{r}_{ic}, -U^{b}_{ic}, -U^{g}_{ic}, -U^{w}_{ic})_R ]^T ,
\ee
with $i=1,2$ for two families. The superscript $T$ denotes the transposition of a column matrix.  $U_i^{\alpha}$ and $D_i^{\alpha}$ are up-type and down-type lepton-quark states in SM with $\alpha= (r,\, b\, , g\, , w)$ representing the trichromatic (red, blue, green) and white colors, respectively. $Q_{i c}^{\alpha} =(U_{i c}^{\alpha},  D_{i c}^{\alpha})$ are defined as complex charge-conjugated Dirac spinors, 
i.e., $Q_{i\1 c}^{\alpha} = C_D \bar{Q}_i^T= C_D \gamma_0 Q_i^{\ast}$ with $C_D$ the charge-conjugation matrix $C_D = -i\sigma_3\otimes\sigma_2$. The subscripts ``L" and ``R" in $Q_{i L,R}^{\alpha} = (U_i^{\alpha}, D_i^{\alpha})_{L,R}$ denote the left-handed and right-handed Dirac spinors, respectively, i.e., $Q_{i L,R}^{\alpha} = \frac{1}{2}(1\mp \gamma_5) Q_i^{\alpha}$ with $\gamma_5   Q_{i L,R}^{\alpha}  = \mp Q_{i L,R}^{\alpha}$.


As the entangled octoqubit-spinor field $\Psi_{\fQE^8}(x)$ is composed of two grand unified qubit-spinor fields, it can be verified that the intrinsic $\cQ_c$-spin symmetry SP(3) characterizes the family-spin symmetry SU$_F$(2)$\cong$SP(3) for two families of chiral type leptons and quarks in SM, i.e., $Q_1^{\alpha}=(U_1^{\alpha}, D_1^{\alpha})$ and  $Q_2^{\alpha}=(U_2^{\alpha}, D_2^{\alpha})$, with respective to two grand unified qubit-spinor fields $\Psi_{\fQ_{\mG_1}^{-}}(x)$ and $\Psi_{\fQ_{\mG_2}^{-}}(x)$.

From the above analyses, we come to the observation that if there exists only one family of chiral type leptons and quarks in nature, their maximally correlated motion brings on a free motion just in four-dimensional spacetime. Whereas the locally entangled-qubits motion postulate and maximum coherence motion principle inevitably lead to the prediction on the simultaneous existence of two families of chiral type leptons and quarks in nature with free motion in 14-dimensional hyper-spacetime. Let us suppose that all chiral type leptons and quarks become massive with different masses, and meanwhile the heavy leptons and quarks decay into light leptons and quarks. Such a situation does occur in the real case of SM, all heavy massive charged leptons and quarks decay eventually into the lightest massive electron and neutrinos as well as up quark and down quark. All the lightest leptons and quarks belong to a single family as the first family in SM. In fact, the observed matter in our living universe consists of the atomic matter which is just composed of the lightest up and down quarks together with electron. When combining such a fact with the observation that each twisting entangled hyperqubit-spinor field as grand unified qubit-spinor field ($\Psi_{\fQ_{\mG_i}^{-}}(x) \equiv \Psi_{\fQEni}(x)$) contains exactly a single family of chiral type lepton-quark state and has a free motion only in four-dimensional spacetime, we arrive at the conclusion that the lightest up and down quarks together with electron can only keep a free motion in four-dimensional spacetime since they belong to a single family of lepton-quark state as the first family in SM and constitute the observed atomic matter in our universe. In other word, all existing matter in our living universe can only move in four-dimensional Minkowski spacetime, which brings about a natural explanation on the longstanding issue {\it why our observed universe is only four-dimensional spacetime}. 

Therefore, the entangled octoqubit-spinor field spanned in 256-dimensional Hilbert space is regarded as an {\it ultra-grand unified qubit-spinor field} which is composed of two grand unified qubit-spinor fields. Such an ultra-grand unified qubit-spinor field keeps a free motion in 14-dimensional hyper-spacetime determined by $\cM_c$-spin charge, which may generally be denoted as follows:
\be
& & \Psi_{\fQUn}(x) \equiv \Psi_{\fQE^8}(x) \equiv \binom{\Psi_{\fQGnf}(x)}{\Psi_{\fQGns}(x)} \equiv  \binom{\Psi_{W\1 1}(x)}{\Psi_{W\1 2}(x)}  , \nn \\ 
& & \Psi_{\QUn}(x)\equiv  \Psi_{\QE^8}(x) = \binom{\Psi_{\fQW^7}(x) } { \Psi_{\fQW^7}^{c}(x) },
\ee
where the superscript `-' of the symbols $\fQU$ and $\QU$ characterizes the chirality property, which is referred to as {\it U-parity} of grand unified qubit-spinor field for each family of lepton-quark state in SM. The inhomogeneous hyperspin symmetry and $\cQ_c$-spin symmetry of ultra-grand unified qubit-spinor field $\Psi_{\fQUn}(x)$ ($\Psi_{\QUn}(x)$) with {\it negative U-parity} is regarded as {\it ultra-grand unified symmetry} denoted as follows:
\be
G_U = WS(1,13)\times SP(3) \cong SO(1,13)\rtimes W^{1,13} \times SU_F(2), \nn
\ee 
which is in association with inhomogeneous Lorentz-type/Poincar\'e-type group symmetry PO(1,13) in 14-dimensional Minkowski hyper-spacetime to provide an associated symmetry. The subgroup symmetry SP(10) of ultra-grand unified symmetry SP(1,13)$\cong$SO(1,13) brings about grand unified symmetry SO(10)$\cong$SP(10)  for each family of chiral type leptons and quarks in SM. Each grand unified qubit-spinor field $\Psi_{\fQGni}(x)$ defined by the twisting entangled hyperqubit-spinor field $\Psi_{\fQEni}(x)$ characterizes exactly a single family of chiral type leptons and quarks in SM, which keeps a free motion only in four-dimensional spacetime.


\section{ Hyperunified qubit-spinor field as fundamental building block of nature and inhomogeneous hyperspin symmetry as hyperunified symmetry from maximum locally entangled-qubits motion principle }

Based on the locally entangled-qubits motion postulate, four categories of entangled hyperqubit-spinor fields with respective to categoric $\cQ_c$-spin charges $\CQc^{(q_c,k)} = q_c=0,1,2, 3$ in the first period  ($k=1$) have been investigated in detail, which are shown to have free motions in hyper-spacetime with respective to dimensions $D_h=10, 11, 12, 14$. The $\cQ_c$-spin charges not only provide the categorization for entangled hyperqubit-spinor fields as unified qubit-spinor fields of lepton-quark states, but also reflect the flavor-spin and family-spin charges of lepton-quark states via the intrinsic $\cQ_c$-spin symmetries. Unlike the hyperspin symmetry that reflects a motion-correlation spin symmetry, the $\cQ_c$-spin symmetry characterizes a motion-irrelevance intrinsic symmetry. For the entangled octoqubit-spinor field in category-3, the hyperspin symmetry SP(1,13)$\cong$SO(1,13) as ultra-grand unified symmetry in 14-dimensional hyper-spacetime is verified to describe consistently a minimal unified symmetry of helicity-spin, boost-spin, color-spin and isospin charges of lepton-quark states in SM, while the intrinsic $\cQ_c$-spin symmetry SP(3) presents as a direct product group symmetry to hyperspin symmetry SP(1,13) and describes the family-spin symmetry SU$_F$(2) for two families of chiral type leptons and quarks in SM. From the unification point of view, it appears a necessity to attribute such an intrinsic family-spin symmetry into a motion-correlation hyperspin symmetry so as to obtain a single hyperunified symmetry in hyper-spacetime, which comes to the main purpose in this section. For that, we are led to make the {\it least $\cQ_c$-spin postulate} and propose the {\it maximum locally entangled-qubits motion principle} as guiding principle, which enables us to construct a hyperunified qubit-spinor field of lepton-quark states as fundamental building block of nature. 



\subsection{Least $\cQ_c$-spin postulate and maximum locally entangled-qubits motion principle} 

The entangled hyperqubit-spinor fields in category-2 and category-3 possess the $\cQ_c$-spin symmetries SP(2) and SP(3), which indicates that the maximal symmetry of entangled hyperqubit-spinor field with $\cQ_c$-spin charges $\CQc=q_c=2,3$ is not yet characterized by a single inhomogeneous hyperspin symmetry as the intrinsic $\cQ_c$-spin symmetry SP($q_c$) appears as a direct product group symmetry to inhomogeneous hyperspin symmetry WS(1,$D_h$-1). To get inhomogeneous hyperspin symmetry as a single hyperunified symmetry, we are motivated to make a {\it least $\cQ_c$-spin postulate} and propose a {\it maximum locally entangled-qubits motion principle} as guiding principle, so that it brings on an entangled hyperqubit-spinor field as a hyperunified qubit-spinor field.  

{\it Least $\cQ_c$-spin postulate}: To make inhomogeneous hyperspin symmetry as a single hyperunified symmetry of entangled hyperqubit-spinor field, a hyperunified qubit-spinor field is postulated to have the least $\cQ_c$-spin charge in building hyperunified field theory. This  postulate indicates that such a hyperunified qubit-spinor field should belong to either category-0 or category-1 entangled hyperqubit-spinor field with $\cQ_c$-spin charge $\CQc^{(q_c,k)} =q_c = 0$ or $1$.  With such a least $\cQ_c$-spin postulate, we are led to propose the following guiding principle.

{\it Maximum locally entangled-qubits motion principle}: when combing the least $\cQ_c$-spin postulate and locally entangled-qubits motion postulate together with local coherent-qubits motion postulate, we arrive at a maximum locally entangled-qubits motion principle as guiding principle. We should demonstrate that it is such a maximum locally entangled-qubits motion principle together with the maximum coherence motion principle that brings about an entangled hyperqubit-spinor field with least $\cQ_c$-spin charge as hyperunified qubit-spinor field. Such a hyperunified qubit-spinor field is regarded as fundamental building block of nature.

Following along the maximum locally entangled-qubits motion principle and noticing the categorization theorems on both qubit-spinor field with categoric qubit number and Minkowski spacetime with categoric dimension, we should come to investigate locally entangled states of nine qubits ($Q_N=9$) and ten qubits ($Q_N=10$) with respective to $\cQ_c$-spin charges $\CQc^{(q_c,k)}=q_c=0$ and $\CQc^{(q_c,k)}=q_c=1$ in the second period $k=2$, they may be referred to as {\it entangled enneaqubit-spinor field} in category-0 and {\it entangled decaqubit-spinor field} in cetegory-1.


\subsection{ Entangled enneaqubit-spinor field with zero $\cQ_c$-spin charge as minimal hyperunified qubit-spinor field and inhomogeneous hyperspin symmetry as minimal hyperunified symmetry in 18D hyper-spacetime }

The entangled octoqubit-spinor field as ultra-grand unified qubit-spinor field has a maximal $\cQ_c$-spin charge $\CQc^{(q_c,k)}=q_c =3$ in the first period $k=1$, which leads the maximum $\cQ_c$-spin symmetry SP(3) to characterize the family-spin symmetry SU$_F$(2)$\cong$SP(3) of two families of chiral type leptons and quarks in SM. To embed such a $\cQ_c$-spin symmetry into a motion-correlation hyperspin symmetry in hyper-spacetime, it is a necessity to explore the next entangled hyperqubit-spinor field with least $\cQ_c$-spin charge. Following along the periodic feature of $\cQ_c$-spin charge, we come to study the locally entangled state of nine qubits with zero $\cQ_c$-spin charge in the second period $k=2$, which is referred to as {\it entangled enneaqubit-spinor field} in category-0.

Let us begin with the construction of entangled enneaqubit-spinor field $\Psi_{\QE^9}(x)$ with zero $\cQ_c$-spin charge $\CQc^{(q_c,k)}=q_c=0$ in the second period $k=2$. It is straightforward to build $\Psi_{\QE^9}(x)$ via the following qubit-spinor structure:
\be \label{QE9}
& & \Psi_{\QE^9}(x) = \binom{\Psi_{\QE^8}(x) } { \tPsi_{\QE^8}(x) } \equiv \binom{\Psi_{\QUn}(x) } { \Psi_{\QUp}(x) } \equiv \Psi_{\QHz}(x), 
\ee
where $\Psi_{\QE^8}(x)\equiv \Psi_{\QUn}(x)$ is an entangled octoqubit-spinor field which is regarded as ultra-grand unified qubit-spinor field for two families of chiral type lepton-quark states with negative U-parity as shown in Eq.(\ref{QE8}). $\tPsi_{\QE^8}(x)\equiv \Psi_{\QUp}(x) $ is an additional entangled octoqubit-spinor field with positive U-parity, which is defined as follows:
\be
& & \tPsi_{\QE^8}(x) \equiv \binom{\tPsi_{\fQW^7}(x) } { \tPsi_{\fQW^7}^{c}(x) } \equiv  \Psi_{\QUp}(x) , \nn \\
& & \tPsi_{\fQW^7}(x) \equiv \binom{\tPsi_{\fQWp^6}(x)}{\tPsi_{\fQWn^6}(x)} \equiv \binom{\tPsi_{\fQC^6}^R(x)}{\tPsi_{\fQC^6}^L(x)} , \nn \\
& & \tPsi_{\fQW^7}^{c}(x)  \equiv \binom{\tPsi_{\fQWn^6}^{c}(x)}{\tPsi_{\fQWp^6}^{c}(x)}  \equiv \binom{\tPsi_{\fQC^6}^{L\, c}(x)}{\tPsi_{\fQC^6}^{R\, c}(x)} ,  
\ee
where $\tPsi_{\fQW^7}(x)$ is a chiral complex heptaqubit-spinor field with $\tPsi_{\fQW^7}^{c}(x)$ as its complex charge-conjugated one. They all have positive U-parity,  
\be
& &  \tPsi_{\fQW^7}^{c}(x) = C_{\fQW^7} \bar{\tPsi}^{T}_{\fQW^7}(x) , \nn \\
& & C_{\fQW^7} = \sigma_1 \otimes \sigma_2 \otimes  \sigma_2 \otimes  \sigma_0 \otimes \sigma_0\otimes C_D, \nn \\
& & \hat{\gamma}_{15} \tPsi_{\fQW^7}(x) = + \tPsi_{\fQW^7}(x), \quad \hat{\gamma}_{15} \tPsi_{\fQW^7}^{c}(x) = + \tPsi_{\fQW^7}^{c}(x) .
\ee
$\tPsi_{\QE^8}(x)\equiv \Psi_{\QUp}(x)$ with positive U-parity satisfies the following self-conjugated chiral conditions: 
\be
& &  \tPsi^{c}_{\QE^8}(x) = C_{\QE^8} \bar{\tPsi}^T_{\QE^8}(x) =  \tPsi_{\QE^8}(x)\equiv \Psi_{\QUp}^{c}(x) = C_{\QU} \bar{\Psi}^T_{\QUp}(x)  , \nn \\
& & \hat{\gamma}_{17}\Psi_{\QUp}(x) \equiv \hat{\gamma}_{17} \tPsi_{\QE^8}(x)  = + \tPsi_{\QE^8}(x) \equiv + \Psi_{\QUp}(x), \nn \\
& &  C_{\QU} \equiv C_{\QE^8}\equiv \sigma_1\otimes C_{\fQW^7} , \quad  \hat{\gamma}_{17} = \sigma_0\otimes  \hat{\gamma}_{15} .
\ee

It is noticed that $ \Psi_{\QUp}(x)\equiv \tPsi_{\QE^8}(x)$ and $\Psi_{\QUn}(x) \equiv\Psi_{\QE^8}(x)$ have a similar qubit-spinor structure but with opposite U-parity. The chiral complex heptaqubit-spinor field $\tPsi_{\fQW^7}(x)$ are composed of {\it chiral complex hexaqubit-spinor fields} $\tPsi_{\fQWpn^6}(x)$, and similarly for their complex charge conjugated ones $\tPsi_{\fQW^7}^{c}(x)$ and $\tPsi_{\fQWpn^6}^{c}(x)$, i.e.:
\be \label{tQC6c}
& & \tPsi_{\fQWpn^6}(x) \equiv \tPsi_{\fQC^6}^{R, L}(x) \equiv \vGa_{\pm}^5  \tPsi_{\fQC^6}(x) , \nn \\
& & \tPsi_{\fQWpn^6}^{c}(x) = C_{\fQWpn^6} \bar{\tPsi}^{T}_{\fQWpn^6}(x)  \equiv \vGa_{\mp}^5  \tPsi_{\fQC^6}^{c}(x), \nn \\
& &  \tPsi_{\fQC^6}^{c}(x) = C_{\fQC^6} \bar{\tPsi}^{T}_{\fQC^6}(x), \nn \\
& & C_{\fQC^6}  = C_{\fQWpn^6}= \sigma_2 \otimes  \sigma_2 \otimes  \sigma_0 \otimes \sigma_0\otimes C_D, \nn \\
& &  \vGa_{\pm}^5 = \frac{1}{2} ( 1 \pm \tilde{\gamma}_5 ), \quad \tilde{\gamma}_5 = \sigma_0 \otimes  \sigma_0 \otimes  \sigma_0 \otimes \sigma_0\otimes \gamma_5 , 
\ee
where $\tPsi_{\fQC^6}^{R, L} (x)$ are {\it chiral complex hexaqubit-spinor fields} with $\tPsi_{\fQC^6}^{R, L\, c}(x)$ their complex charge-conjugated ones. They have the following explicit structures:
\be
\tPsi_{\fQC^6}^{R, L}(x)  \equiv \begin{pmatrix}
\tPsi^{u}_{\fQC^4}(x) \\  \tPsi^{d' \,\bc}_{\fQC^4}(x) \\ \tPsi^{d}_{\fQC^4}(x)  \\ -\tPsi^{u' \,\bc}_{\fQC^4}(x)
\end{pmatrix}_{R, L}, \quad  \tPsi_{\fQC^6}^{R, L\, c}(x)  \equiv \begin{pmatrix}
\tPsi^{u'}_{\fQC^4}(x) \\  \tPsi^{d\; \bc}_{\fQC^4}(x) \\ \tPsi^{d'}_{\fQC^4}(x)  \\ -\tPsi^{u\; \bc}_{\fQC^4}(x)
\end{pmatrix}_{L, R} ,
\ee
where $\tPsi^{q\, R, L}_{\fQC^4}(x)$ ($q=u, u', d, d'$) are {\it chiral complex tetraqubit-spinor fields} with $\tPsi^{q\,\bc\; L, R }_{\fQC^4}(x)$ the complex charge-conjugated ones. Their explicit forms can be expressed as follows:
\be \label{tqQC4}
& & \tPsi^{q\, R, L}_{\fQC^4}(x)  \equiv \begin{pmatrix}
\tQ^{r}(x) \\ \tQ^{b}(x) \\ \tQ^{g}(x) \\ \tQ^{w}(x) 
\end{pmatrix}^{q}_{R, L}, \quad  \tPsi^{q\,\bc\, L, R }_{\fQC^4}(x)  \equiv \begin{pmatrix}
\tQ^{r\, \bc}(x) \\ \tQ^{b\, \bc}(x) \\ \tQ^{g, \bc}(x) \\ \tQ^{w\, \bc}(x) 
\end{pmatrix}^{q}_{L, R} , 
\ee
with $q=u, u', d, d'$ labeling four kinds of lepton-quark states presented by the following notations:
\be
\left(\tQ^{\alpha}\right)^{u,u',d,d'}_{R, L} \equiv  \tU^{\alpha}_{R, L}, \tU^{'\alpha}_{R, L}, \tD^{\alpha}_{R, L}, \tD^{'\alpha}_{R, L} .
\ee 
Note that the lepton-quark states $\left(\tQ^{\alpha}\right)^{q}_{R, L}$ always appear to have opposite chirality to $\left(Q^{\alpha}\right)^{q}_{L, R}$. We may refer to $\left(\tQ^{\alpha}\right)^{q}_{R, L}$ as {\it mirror lepton-quark states}. They involve two kinds of up-type ($q=u,u'$) and down-type ($q=d,d'$) mirror lepton-quark states with four color-spin charges $\alpha=r, b, g, w$ in correspondence to `red', `blue', `green' and `white'.  The projection operators $\vGa^5_{\pm}$ define the right-handed and left-handed mirror lepton-quark states notated by the subscripts `R' and `L', respectively. 

Therefore, $\Psi_{\QUp}(x) \equiv \tPsi_{\QE^8}(x)$ with positive U-parity is regarded as {\it mirror entangled octoqubit-spinor field} for two families of {\it mirror lepton-quark states}. It will be demonstrated below that the $\cQ_c$-spin symmetry SP(3) as family-spin symmetry SU$_F$(2) of entangled octoqubit-spinor field does turn into inhomogeneous hyperspin symmetry of entangled enneaqubit-spinor field $\Psi_{\QE^9}(x)$, so that we will refer to $\Psi_{\QE^9}(x)$ as {\it minimal hyperunified qubit-spinor field} denoted as $\Psi_{\QHz}(x)\equiv \Psi_{\QE^9}(x)$. 

From the above analyses, we are able to obtain the following action for the minimal hyperunified qubit-spinor field $\Psi_{\QHz}(x)\equiv \Psi_{\QE^9}(x)$ in 18-dimensional hyper-spacetime:
\be  \label{actionQE9}
\cS_{\QHz} \equiv \cS_{\QE^9}  & \equiv & \int d^{18}x \,   \bar{\Psi}_{\QE^9}(x) \delta_{\mA}^{\;\mM}  \vSi_{-}^{\mA} i\p_{\mM} \Psi_{\QE^9}(x) \nn \\
& \equiv & \int d^{18}x \,   \bar{\Psi}_{\QHz}(x) \delta_{\mA}^{\;\mM}  \vSi_{-}^{\mA} i\p_{\mM} \Psi_{\QHz}(x).
\ee
with $\mA, \mM= 0,1,2,3,5, \cdots, 18$. The matrices $\vSi_{-}^{\mA}$ are defined from the $\vGa$-matrices $\vGa^{\mA}= (\vGa^a, \vGa^A) $ ($a=0,1,2,3, A= 5,\cdots, 18$) with the following explicit structures: 
\be \label{GMQE9}
& & \vSi_{-}^{\mA} = \frac{1}{2} \vGa^{\mA}\vGa_{-}, \quad  \vGa_{-} = \frac{1}{2} ( 1 - \hat{\gamma}_{19} ),  \nn \\
& & \vGa^a = \,  \sigma_0 \otimes \sigma_0 \otimes \vGa^a, \quad a=0,1,2,3, \nn \\
& & \vGa^{\mA} = \,  \sigma_0 \otimes \sigma_3 \otimes \vGa^A, \quad A= 5, \cdots, 14, \nn \\
& & \vGa^{15} = i \sigma_2 \otimes \sigma_1 \otimes \tilde{\gamma}_5 , \nn \\
& & \vGa^{16} = i \sigma_2 \otimes \sigma_2 \otimes \tilde{\gamma}_5 , \nn \\
& & \vGa^{17} = i \sigma_2 \otimes \sigma_3 \otimes \hat{\gamma}_{15} , \nn \\
& & \vGa^{18} = i \sigma_1 \otimes \sigma_0 \otimes \hat{\gamma}_{15} , \nn \\
& &  \hat{\gamma}_{19} = \sigma_3\otimes \sigma_0\otimes  \hat{\gamma}_{15} =  \sigma_3\otimes \hat{\gamma}_{17} , \nn \\
& & \tilde{\gamma}_5 = \sigma_0 \otimes \sigma_0 \otimes  \sigma_0 \otimes  \sigma_0 \otimes \sigma_0\otimes \gamma_5, 
\ee
with $\vGa^a$ and $\vGa^A$ ($a=0,1,2,3,  A= 5, \cdots, 14$) on the right-hand side of equality defined in Eq.(\ref{GMQW7}). 

It can be checked that $\Psi_{\QE^9}(x)\equiv \Psi_{\QHz}(x)$ satisfies the following self-conjugated chiral conditions:
\be
& & \Psi_{\QE^9}^{c}(x) = C_{\QE^9} \bar{\Psi}_{\QE^9}^{T}(x) = \Psi_{\QE^9}(x)\equiv \Psi_{\QHz}^{c}(x) = C_{\QH^0} \bar{\Psi}_{\QHz}^{T}(x),   \nn \\
& &  \hat{\gamma}_{19} \Psi_{\QHz}(x) \equiv \hat{\gamma}_{19} \Psi_{\QE^9}(x)  = - \Psi_{\QE^9}(x)\equiv - \Psi_{\QHz}(x), \nn \\
& & C_{\QH^0} \equiv C_{\QE^9} = \sigma_0\otimes  C_{\QE^8} =  \sigma_0\otimes \sigma_1\otimes C_{\fQW^7},
\ee
where $\Psi_{\QE^9}(x)\equiv \Psi_{\QHz}(x)$ has a negative chirality in 18-dimensional hyper-spacetime, which may be referred to as {\it negative H-parity}.


To characterize explicitly the family-spin charge of lepton-quark states in SM, let us consider an alternative qubit-spinor structure of entangled enneaqubit-spinor field based on the twisting entangled hyperqubit-spinor field as grand unified qubit-spinor field $\Psi_{\fQEni}(x)\equiv  \Psi_{\fQGni}(x)$.  Such an entangled enneaqubit-spinor field $\Psi_{\fQE^9}(x)$ as minimal hyperunified qubit-spinor field $\Psi_{\fQHz}(x)$ is simply defined as follows:
\be \label{fQE9}
& & \Psi_{\fQE^9}(x) = \binom{\Psi_{\fQE^8}(x) } { \tPsi_{\fQE^8}(x) } \equiv \binom{\Psi_{\fQUn}(x) } { \Psi_{\fQUp}(x) } \equiv  \Psi_{\fQHz}(x) , 
\ee
where $\Psi_{\fQE^8}(x)\equiv \Psi_{\fQUn}(x)$ is an entangled octoqubit-spinor field, which is viewed as an ultra-grand unified qubit-spinor field with negative U-parity as shown in Eq.(\ref{fQE8}). $\tPsi_{\fQE^8}(x) \equiv \Psi_{\fQUp}(x)$ with positive U-parity is defined as follows with a similar qubit-spinor structure:  
\be \label{tfQE8}
& & \tPsi_{\fQE^8}(x) \equiv \binom{\Psi_{\fQ_{\mE_3^+}^7}(x)} { \Psi_{\fQ_{\mE_4^+}^7}(x)} \equiv 
 \begin{pmatrix}
\tPsi_{\fQ_{\mS3}^6}^{R}(x) \\  \tPsi_{\fQ_{\mS_3}^6}^{R\, c}(x)  \\ \tPsi_{\fQ_{\mS_4}^6}^{R}(x) \\ \tPsi_{\fQ_{\mS_4}^6}^{R\, c}(x)
\end{pmatrix} \equiv \binom{\Psi_{\fQGpt}(x)} { \Psi_{\fQGpft}(x)} \equiv  \Psi_{\fQUp}(x), 
\ee
which satisfies the following self-conjugated chiral conditions: 
\be
& &\tPsi_{\fQE^8}^{c}(x) =  C_{\fQE^8} \bar{\tPsi}_{\fQE^8}^{T}(x) =  \tPsi_{\fQE^8}(x) \equiv \Psi_{\fQUp}^{c}(x) =  C_{\fQU} \bar{\Psi}_{\fQUp}^{T}(x), \nn \\
&& \Psi_{\fQ_{\mE_j^+}^7}^{c}(x) =  C_{\fQ_{\mE_j}^7} \bar{\Psi}_{\fQ_{\mE_j^+}^7}^{T}(x) =  \Psi_{\fQ_{\mE_j^+}^7}(x) \equiv \Psi_{\fQGpj}^{c}(x) =  C_{\fQG} \bar{\Psi}_{\fQGpj}^{T}(x) ,
 \nn \\
& & \hat{\gamma}_{17} \Psi_{\fQUp}(x) \equiv \hat{\gamma}_{17} \tPsi_{\fQE^8}(x)  = + \tPsi_{\fQE^8}(x) \equiv + \Psi_{\fQUp}(x) , \nn \\
& &  C_{\fQU}\equiv C_{\fQE^8} = \sigma_0\otimes C_{\fQ_{\mE_i}^7} = \sigma_0\otimes C_{\fQW^7}\equiv  \sigma_0\otimes C_{\fQG} , \nn \\
& & C_{\fQG} =  C_{\fQ_{\mE_j}^7} \equiv C_{\fQW^7} =  C_{\fQ_{\mE_i}^7} =  \sigma_1 \otimes \sigma_2 \otimes \sigma_2 \otimes \sigma_0 \otimes \sigma_0 \otimes C_D ,
\ee
with $j=3,4$.

It is noticed that $\Psi_{\fQUp}(x)\equiv \tPsi_{\fQE^8}(x)$ has an opposite chirality to $\Psi_{\fQUn}(x)\equiv\Psi_{\fQE^8}(x)$ and characterizes mirror lepton-quark states with positive U-parity. So that the entangled octoqubit-spinor field with positive U-parity $\tPsi_{\fQE^8}(x)$ is regarded as {\it mirror ultra-grand unified qubit-spinor field} $\Psi_{\fQUp}(x)$. Correspondingly, the twisting entangled hyperqubit-spinor field $\Psi_{\fQEpj}(x)$ with positive U-parity is regarded as {\it mirror grand unified qubit-spinor field} $\Psi_{\fQGpj}(x)$ ($j=3,4$), which is defined as follows:
\be \label{fQG7}
& & \Psi_{\fQGpj}(x) \equiv \Psi_{\fQEpj}(x) \equiv \binom{\Psi_{\fQWpj}(x)}{\Psi_{\fQWpj}^{c}(x)} \equiv 
\binom{\tPsi_{\fQ_{\mS_j}^6}^{R}(x)}{\tPsi_{\fQ_{\mS_j}^6}^{R\, c}(x)} . 
\ee
where $\Psi_{\fQWpj}(x)$ ($j=3,4$) are {\it mirror chiral complex hexaqubit-spinor fields} and $\tPsi_{\fQ_{\mS_j}^6}^{R}(x)$ ($j=3,4$) are {\it mirror self-conjugated hexaqubit-spinor field} with the following explicit forms:
\be
& &   \tPsi_{\fQ_{\mS_j}^6}^{R}(x)  \equiv \begin{pmatrix}
\tPsi_{\fQ_{\mC_j}^4}^{u}(x) \\  \tPsi_{\fQ_{\mC_j}^4}^{d\, \bc}(x)  \\ \tPsi_{\fQ_{\mC_j}^4}^{d}(x) \\ - \tPsi_{\fQ_{\mC_j}^4}^{u\, \bc}(x)  
\end{pmatrix}_{R} , \quad  \tPsi_{\fQ_{\mS_j}^6}^{R\; c}(x)  \equiv \begin{pmatrix}
\tPsi_{\fQ_{\mC_j}^4}^{u}(x) \\  \tPsi_{\fQ_{\mC_j}^4}^{d\, \bc}(x)  \\ \tPsi_{\fQ_{\mC_j}^4}^{d}(x) \\ - \tPsi_{\fQ_{\mC_j}^4}^{u\, \bc}(x)  
\end{pmatrix}_{L} .  
\ee
$\tPsi_{\fQ_{\mC_j}^4}^{q}(x)$ ($j=3,4$) are regarded as {\it mirror complex tetraqubit-spinor fields} with the complex charge-conjugated ones $\tPsi_{\fQ_{\mC_j}^4}^{q\, \bc}(x)$ ($j=3,4$). Their explicit forms are given as follows:
\be
& &  \tPsi_{\fQ_{\mC_j}^4}^{q}(x) \equiv \begin{pmatrix}
\tQ_j^{r}(x) \\ \tQ_j^{b}(x) \\ \tQ_j^{g}(x) \\ \tQ_j^{w}(x) 
\end{pmatrix}^q, \quad  \tPsi_{\fQ_{\mC_j}^4}^{q\, \bc}(x) \equiv \begin{pmatrix}
\tQ^{r\, \bc}_{j }(x) \\ \tQ^{b\, \bc}_{j}(x) \\ \tQ^{g\, \bc}_{j}(x) \\ \tQ^{w\, \bc}_{j}(x) 
\end{pmatrix}^q  
\ee
where $q=u, d$ represent the up-type and down-type mirror lepton-quark states $\tQ^{\alpha\, u}_j(x) = \tU_j^{\alpha}(x)$ and $\tQ^{\alpha\, u}_j(x) = \tD_j^{\alpha}(x)$ with four color-spin charges $\alpha = r, b, g, w$ corresponding to `red', `blue', `green' and `white'. The subscripts $j=3,4$ denote two families of mirror lepton-quark states.  The subscripts `R' and `L' label the right-handed and left-handed mirror lepton-quark states $\tQ_{j\, R,L}^{\alpha\, u, d}(x) = (\tU_{j\, R, L}^{\alpha}(x), \tD_{j\, R, L}^{\alpha}(x))$. 

In light of entangled octoqubit-spinor fields as ultra-grand unified qubit-spinor fields with the mirror pair of negative and positive U-parities, i.e., $\Psi_{\fQUn}\equiv \Psi_{\fQE^8}(x)$ and $\Psi_{\fQUp}(x)\equiv \tPsi_{\fQE^8}(x)$, it is easy to verify that $\Psi_{\fQHz}(x)\equiv \Psi_{\fQE^9}(x)$ with {\it negative H-parity} satisfies the following self-conjugated chiral conditions:
\be
& &\Psi_{\fQE^9}^{c}(x) = C_{\fQE^9} \bar{\Psi}_{\fQE^9}^{T}(x) = \Psi_{\QE^9}(x) \equiv \Psi_{\fQHz}^{c}(x) = C_{\fQH^0} \bar{\Psi}_{\fQHz}^{T}(x), \nn \\
& & \hat{\gamma}_{19} \Psi_{\fQHz}(x) \equiv \hat{\gamma}_{19} \Psi_{\fQE^9}(x)  = - \Psi_{\fQE^9}(x) \equiv - \Psi_{\fQHz}(x) , \nn \\
& & C_{\fQH^0} \equiv C_{\fQE^9} = \sigma_0\otimes  C_{\fQE^8} = \sigma_0\otimes  \sigma_0\otimes  C_{\fQ_{\mE_i}^7} \equiv  \sigma_0\otimes  \sigma_0\otimes  C_{\fQG} .  
\ee
So that the action in Eq.(\ref{actionQE9}) can be rewritten into the following form:
\be  \label{actionfQE9}
\cS_{\fQHz} \equiv \cS_{\fQE^9}  & \equiv & \int d^{18}x \,   \bar{\Psi}_{\fQE^9}(x) \delta_{\mA}^{\;\mM}  \vSi_{-}^{\mA} i\p_{\mM} \Psi_{\fQE^9}(x), \nn \\
& \equiv & \int d^{18}x \,   \bar{\Psi}_{\fQHz}(x) \delta_{\mA}^{\;\mM}  \vSi_{-}^{\mA} i\p_{\mM} \Psi_{\fQHz}(x)
\ee
with $\mA, \mM= 0,1,2,3,5, \cdots, 18$. The matrices $\vSi_{-}^{\mA}$ are defined from $\vGa$-matrices $\vGa^{\mA}= (\vGa^a, \vGa^A) $ ($a=0,1,2,3, A= 5,\cdots, 18$) with the following structures:
\be \label{GMfQE9}
& & \vSi_{-}^{\mA} = \frac{1}{2} \vGa^{\mA}\vGa_{-}, \quad  \vGa_{-} = \frac{1}{2} ( 1 - \hat{\gamma}_{19} ),  \nn \\
& & \vGa^a = \;\;\; \,  \sigma_0 \otimes \sigma_0 \otimes \vGa^a, \quad a=0,1,2,3, \nn \\
& & \vGa^{\mA} = \,  -\sigma_0 \otimes \sigma_2 \otimes \vGa^A, \quad A= 5, \cdots, 14, \nn \\
& &  \vGa^{15} =\; i\sigma_2 \otimes \sigma_1\otimes \tilde{\gamma}_5 , \nn \\
& &  \vGa^{16} =\; i\sigma_2 \otimes \sigma_2 \otimes \hat{\gamma}_{15}, \nn \\
& &  \vGa^{17} =\; i\sigma_2 \otimes \sigma_3\otimes \tilde{\gamma}_5 , \nn \\
& & \vGa^{18} = \; i \sigma_1 \otimes \sigma_0 \otimes \hat{\gamma}_{15} , \nn \\
& &  \hat{\gamma}_{19} = \sigma_3\otimes \sigma_0\otimes  \hat{\gamma}_{15} =  \sigma_3\otimes \hat{\gamma}_{17} ,  \nn \\
& & \tilde{\gamma}_5 = \sigma_0 \otimes \sigma_0 \otimes  \sigma_0 \otimes  \sigma_0 \otimes \sigma_0\otimes \gamma_5, 
\ee
where $\vGa^a$ and $\vGa^A$ ($a=0,1,2,3,  A= 5, \cdots, 14$) on the right-hand side of equality are given in Eq.(\ref{GMQW7}).

As there exists no other $\vGa$-matrix that can be anti-commuting with all $\vGa$ matrices $\vGa^{\mA}= (\vGa^a, \vGa^A) $ ($a=0,1,2,3, A= 5,\cdots, 18$) and couple with scalar fields, the entangled enneaqubit-spinor field as {\it minimal hyperunified qubit-spinor field}, $\Psi_{\fQE^9}(x)\equiv \Psi_{\fQHz}(x)$ or $\Psi_{\QE^9}(x)\equiv \Psi_{\QHz}(x)$, is a category-0 entangled hyperqubit-spinor field with $\cQ_c$-spin charge $\CQc=q_c=0$. 

It can be verified that the actions in Eqs.(\ref{actionQE9}) and (\ref{actionfQE9}) possess an associated symmetry in which the inhomogeneous hyperspin symmetry WS(1,17) is in association with inhomogeneous Lorentz-type/Poincar\'e-type group symmetry PO(1,17) as well as global scaling symmetries SG(1) and SC(1), i.e.:
\be
G_S & = & SC(1)\ltimes PO(1,17) \adjoin WS(1,17) \rtimes SG(1) \nn \\
& = & SC(1)\ltimes P^{1,17}\ltimes SO(1,17) \adjoin  SP(1,17) \rtimes W^{1,17}\rtimes SG(1) ,
\ee
where the symbol ``$\adjoin$" is adopted to indicate the associated symmetry so that the transformation of hyperspin symmetry SP(1,17) should be coincidental to that of Lorentz-type group symmetry SO(1,17). 

The group generators of inhomogeneous hyperspin symmetry WS(1,17) are directly given by the commutators of $\vGa$-matrices as follows: 
\be
& & \varSigma^{\mA\mB} = \frac{i}{4} [ \vGa^{\mA}, \vGa^{\mB} ], \quad \vSi_{-}^{\mA} = \frac{1}{2} \vGa^{\mA}\vGa_{-}, 
\ee
with $\mA, \mB= 0,1,2,3,5,\cdots, 18$. The hyperspin symmetry SP(1,17) can be decomposed into the following subgroup symmetries:
\be
 SP(1,17) & \supset &  SP(1,3)\times SP(14)  \supset SP(1,3) \times SP(10) \times SP(4) \nn \\
&  \cong & SO(1,3)\times SO(14) \supset SO(1,3) \times SO(10)\times SO(4) \nn \\
& \supset & SO(1,3)\times SO(6) \times SO(4)\times SO(4) \nn \\
& \cong & SO(1,3) \times SU_C(4)\times SU_L(2)\times SU_R(2) \times SU_F(2)\times SU_M(2) \nn \\
& \supset & SO(1,3) \times SU_C(3)\times SU_L(2)\times U_Y(1) ,
\ee
with the group generators of subgroup symmetries given as follows:
\be
& & \varSigma^{\mA\mB}  ( \mA, \mB= 0,1,2,3 ) \in sp(1,3)\cong so(1,3)  \nn \\
& &  \varSigma^{\mA\mB}  (\mA, \mB= 5,6,7,8,9,10,11,12,13,14 )  \in sp(10) \cong so(10), \nn \\
& &  \varSigma^{\mA\mB}  ( \mA, \mB= 15,16,17,18) \in sp(4)\cong so(4)\cong su_F(2)\times su_M(2) , \nn \\
& & \varSigma^{\mA\mB}  (\mA, \mB= 5,6,7,8,9,10 )  \in sp(6) \cong so(6)\cong su_C(4) , \nn \\
& &  \varSigma^{\mA\mB}  ( \mA, \mB= 11,12,13,14) \in sp(4)\cong so(4)\cong su_L(2)\times su_R(2) .  \nn 
\ee
The subgroup symmetry SP(10)$\cong$SO(10) leads to the grand unified symmetry and the subgroup symmetry SP(4)$\cong$SO(4)$\cong$SU$_F$(2)$\times$SU$_M$(2) characterizes the family-spin symmetry SU$_F$(2) and {\it mirror family-spin symmetry} SU$_M$(2) for two families of chiral type lepton-quark states and mirror lepton-quark states, $Q_i^{\alpha}=(U_i^{\alpha}, D_i^{\alpha})$ ($i=1,2$) and  $\tQ_j^{\alpha}=(\tU_j^{\alpha}, \tD_j^{\alpha})$ ($j=3,4$).

It is clear that the motion-irrelevant intrinsic $\cQ_c$-spin symmetry SP(3) as family-spin symmetry SU$_F$(2) of entangled octoqubit-spinor field $\Psi_{\fQE^8}(x)$ ($\Psi_{\QE^8}(x)(x)$) in 14-dimensional hyper-spacetime does transmute into a motion-correlation hyperspin symmetry SP(1,17) of entangled enneaqubit-spinor field $\Psi_{\fQE^9}(x)$ ($\Psi_{\QE^9}(x)$) in 18-dimensional hyper-spacetime. 

Therefore, the least $\cQ_c$-spin postulate brings us to construct entangled enneaqubit-spinor field $\Psi_{\fQE^9}(x)$ (or $\Psi_{\QE^9}(x)$) which provides a minimal hyperunified qubit-spinor field $\Psi_{\fQHz}(x)\equiv \Psi_{\fQE^9}(x)$ (or $\Psi_{\QHz}(x)\equiv \Psi_{\QE^9}(x)$) with negative H-parity. Its maximal inhomogeneous hyperspin symmetry WS(1,17) leads to a minimal hyperunified symmetry. Such a minimal hyperunified qubit-spinor field brings about the prediction for the existence of mirror lepton-quark states. Specifically, two twisting entangled hyperqubit-spinor fields as grand unified qubit-spinor fields $\Psi_{\fQGni}(x)\equiv \Psi_{\fQEni}(x)$ ($i=1,2$) with negative U-parity describe two families of chiral type lepton-quark states in SM, whereas two mirror twisting entangled hyperqubit-spinor fields as mirror grand unified qubit-spinor fields $\Psi_{\fQGpj}(x)\equiv \Psi_{\fQEpj}(x)$ ($j=3,4$) with positive U-parity characterize two families of mirror lepton-quark states beyond the SM.


\subsection{ Entangled decaqubit-spinor field in 19D hyper-spacetime as hyperunified qubit-spinor field for fundamental building block of nature and inhomogeneous hyperspin symmetry WS(1,18) as hyperunified fundamental symmetry of nature}

The entangled enneaqubit-spinor field as minimal hyperunified qubit-spinor field $\Psi_{\fQHz}(x) \equiv \Psi_{\fQE^9}(x)$ (or $\Psi_{\QHz}(x) \equiv \Psi_{\QE^9}(x)$ ) provides a unified description on two families of chiral type leptons and quarks in SM together with two families of mirror leptons and quarks beyond SM. Nevertheless, the experimental observation in high energy physics indicates that there exist at least three families of chiral type leptons and quarks as basic constituents of nature. To unify all discovered leptons and quarks into a single hyperunified qubit-spinor field, we are led to consider further the next entangled hyperqubit-spinor field. Such an entangled hyperqubit-spinor field should be a locally entangled state of ten qubits with $\cQ_c$-spin charge $\CQc=1$ based on the periodic feature of $\cQ_c$-spin charge and categorization of qubit-spinor field, which is referred to as {\it entangled decaqubit-spinor field} in category-1. 

The entangled decaqubit-spinor field can be constructed in a straightforward way by taking entangled enneaqubit-spinor field as a minimal hyperunified qubit-spinor field $\Psi_{\QE^9}(x)\equiv \Psi_{\QHz}(x)$ (or $\Psi_{\fQE^9}(x)\equiv \Psi_{\fQHz}(x)$). Let us first construct entangled decaqubit-spinor field with the following qubit-spinor structure:
\be
\Psi_{\QE^{10}}(x) & \equiv & \binom{ \Psi_{\QE^9}(x)}{ \hPsi_{\QE^9}(x) } = \begin{pmatrix}
\Psi_{\QE^8}(x) \\ \tPsi_{\QE^8}(x) \\ \tPsi'_{\QE^8}(x) \\ \Psi'_{\QE^8}(x) 
\end{pmatrix} \equiv \Psi_{\QH}(x) , 
\ee
with the identical definition notated as follows:
\be
\Psi_{\QH}(x) & \equiv & \binom{ \Psi_{\QHn}(x)}{ \Psi_{\QHp}(x) } = \begin{pmatrix}
\Psi_{\QUn}(x) \\ \Psi_{\QUp}(x) \\ \hPsi_{\QUp}(x) \\ \hPsi_{\QUn}(x) ,
\end{pmatrix}, 
\ee
which is regarded as {\it hyperunified qubit-spinor field} denoted by $\Psi_{\QH}(x)\equiv \Psi_{\QE^{10}}(x)$. We have introduced two entangled enneaqubit-spinor fields as minimal hyperunified qubit-spinor fields, $\Psi_{\QHn}(x) \equiv \Psi_{\QE^9}(x) \equiv \Psi_{\QHz}(x) $ and $\Psi_{\QHp}(x)\equiv \hPsi_{\QE^9}(x)$, with {\it negative and positive H-parities}, respectively, i.e.:
\be
& & \hat{\gamma}_{19} \Psi_{\QHn}(x)  = -\Psi_{\QHn}(x) , \quad \hat{\gamma}_{19} \Psi_{\QHp}(x)  = +\hPsi_{\QHp}(x)  \nn  \\
& &  \hat{\gamma}_{19} = \sigma_3\otimes \sigma_0\otimes  \hat{\gamma}_{15} =  \sigma_3\otimes \hat{\gamma}_{17} . \nn
\ee 

The action for such an entangled decaqubit-spinor field as hyperunified qubit-spinor field is obtained, based on the maximum coherence motion Principe, as follows:
\be  \label{actionQH}
\cS_{\QH} & \equiv & \int d^{19}x \,  \{ \bar{\Psi}_{\QH}(x)  \delta_{\mA}^{\;\mM}  \vSi_{-}^{\fA} i\p_{\mM} \Psi_{\QH}(x) -  \lambda_{1} \phi_1(x)  \bar{\Psi}_{\QH}(x)  \tvSi_{-} \Psi_{\QH}(x) \nn \\
& \equiv & \int d^{19}x \,  \{ \bar{\Psi}_{\QE^{10}}(x)  \delta_{\mA}^{\;\mM}  \vSi_{-}^{\fA} i\p_{\mM} \Psi_{\QE^{10}}(x) - \lambda_{1} \phi_1(x)  \bar{\Psi}_{\QE^{10}}(x)  \tvSi_{-} \Psi_{\QE^{10}}(x) \}, 
\ee
which has free motion in 19-dimensional hyper-spacetime with $\mA, \mM= 0,1,2,3,5, \cdots, 19$. The  matrices $\vSi_{-}^{\fA}$ and $\tvSi_{-}$ are defined from $\vGa$-matrices $\vGa^{\hat{\mA}} \equiv (\vGa^{\mA}, \vGa^{20}) \equiv (\vGa^a, \vGa^A, \tvGa) $ ($\hat{\mA}=0,1,2,3, 5,\cdots, 20$) with the following explicit forms:
\be \label{GMQH}
& & \varSigma_{-}^{\fA} = \frac{1}{2} \vGa^{\fA}\vGa_{-}, \quad  \mA = 0,1,2,3,5, \cdots 19, \nn \\
& & \tvSi_{-} \equiv \varSigma_{-}^{20} = \frac{1}{2} \vGa^{20}\vGa_{-}  = \frac{1}{2} \tvGa \vGa_{-}, \quad  \vGa_{-} = \frac{1}{2} ( 1 - \hat{\gamma}_{21} ),  \nn \\
& & \vGa^a = \;\; \,  \sigma_0 \otimes \sigma_0 \otimes \sigma_0 \otimes \vGa^a, \quad a=0,1,2,3, \nn \\
& & \vGa^{\mA} = \;\;\,  \sigma_0 \otimes \sigma_0 \otimes \sigma_3 \otimes \vGa^A, \quad A= 5, \cdots, 14, \nn \\
& &  \vGa^{15} =\; i\sigma_0 \otimes \sigma_2 \otimes \sigma_1\otimes \tilde{\gamma}_5 , \nn \\
& &  \vGa^{16} =\; i\sigma_0 \otimes \sigma_2 \otimes \sigma_2 \otimes \tilde{\gamma}_5 , \nn \\
& &  \vGa^{17} =\; i\sigma_0 \otimes \sigma_2 \otimes \sigma_3\otimes \hat{\gamma}_{15}  , \nn \\
& & \vGa^{18} = \; i \sigma_0 \otimes \sigma_1 \otimes \sigma_0 \otimes \hat{\gamma}_{15} , \nn \\
& & \vGa^{19} = \; i \sigma_1 \otimes \sigma_3 \otimes \sigma_0 \otimes \hat{\gamma}_{15} , \nn \\
& & \vGa^{20} = \; i \sigma_2 \otimes \sigma_3 \otimes \sigma_0 \otimes \hat{\gamma}_{15} \equiv \tvGa, \nn \\
& &  \hat{\gamma}_{21} = \sigma_3\otimes \sigma_3\otimes \sigma_0\otimes  \hat{\gamma}_{15} =  \sigma_3\otimes \hat{\gamma}_{19} , \nn \\
& & \tilde{\gamma}_5 = \sigma_0 \otimes \sigma_0 \otimes  \sigma_0 \otimes  \sigma_0 \otimes \sigma_0\otimes \gamma_5,  
\ee
where $\vGa^a$ and $\vGa^A$ ($a=0,1,2,3,  A= 5, \cdots, 14$) on the right-hand side of equality are defined in Eq.(\ref{GMQW7}). 

The entangled decaqubit-spinor field $\Psi_{\QE^{10}}(x)$ as hyperunified qubit-spinor field, $\Psi_{\QH}(x)\equiv \Psi_{\QE^{10}}(x)$, satisfies the following self-conjugated chiral conditions:
\be
& &  \Psi_{\QH}^{c}(x) = C_{\QH}\bar{\Psi}_{\QH}^{T}(x) =  \Psi_{\QH}(x) \equiv \Psi_{\QE^{10}}^{c}(x) = \Psi_{\QE^{10}}(x), \nn \\
& & C_{\QH} = \sigma_0\otimes C_{\QHz} \equiv \sigma_0\otimes C_{\QE^9} = \sigma_0\otimes \sigma_0\otimes C_{\QE^8}, \nn \\
& & \hat{\gamma}_{21} \Psi_{\QH}(x) \equiv \hat{\gamma}_{21} \Psi_{\QE^{10}}(x) = -\Psi_{\QE^{10}}(x) \equiv - \Psi_{\QH}(x) ,
\ee
where $\Psi_{\QH}(x)$ is defined by convention to have negative chirality in the spinor representation of twenty dimensions.

The scalar coupling with a single $\tvGa$ matrix indicates that the entangled decaqubit-spinor field as hyperunified qubit-spinor field belongs to category-1 hyperqubit-spinor field with $\cQ_c$-spin charge $\CQc=q_c=1$. As the hyperunified qubit-spinor field $\Psi_{\QH}(x)$ is composed of two minimal hyperunified qubit-spinor fields $\Psi_{\QHn}(x)$ and $\Psi_{\QHp}(x)$, it concerns four families of chiral type lepton-quark states in two ultra-grand unified hyperqubit-spinor fields $\Psi_{\QUn}(x)$ and $\hPsi_{\QUn}(x)$ with negative U-parity and also four families of mirror lepton-quark states in two ultra-grand unified hyperqubit-spinor fields $\Psi_{\QUp}(x)$ and $\hPsi_{\QUp}(x)$ with positive U-parity.


Let us now turn to analyze alternative entangled decaqubit-spinor field as hyperunified qubit-spinor field, which provides a distinguished qubit-spinor structure so as to characterize explicitly four families of chiral type lepton-quark states and mirror lepton-quark states. Such an entangled decaqubit-spinor field as hyperunified qubit-spinor field $\Psi_{\fQH}(x)\equiv \Psi_{\fQE^{10}}(x)$ is constructed straightforwardly from grand unified qubit-spinor fields which are composed of twisting entangled hyperqubit-spinor fields with both negative and positive U-parities.  

In general, such an entangled decaqubit-spinor field $\Psi_{\fQE^{10}}(x)$ contains two entangled enneaqubit-spinor fields $\Psi_{\fQE^9}(x)$ and $\hPsi_{\fQE^9}(x)$ which are regarded as minimal hyperunified qubit-spinor fields with negative and positive H-parities, i.e., $\Psi_{\fQHn}(x)$ and $\Psi_{\fQHp}(x)$. Explicitly, $\Psi_{\fQE^{10}}(x)$ has the following structure:
\be
\Psi_{\fQE^{10}}(x) & \equiv & \binom{ \Psi_{\fQE^9}(x)}{ \hPsi_{\fQE^9}(x) } = \begin{pmatrix}
\Psi_{\fQE^8}(x) \\ \tPsi_{\fQE^8}(x) \\ \tPsi'_{\fQE^8}(x) \\ \Psi'_{\fQE^8}(x) 
\end{pmatrix} = \begin{pmatrix}
\Psi_{\fQEnf}(x) \\ \Psi_{\fQEns}(x) \\ \Psi_{\fQEpt}(x) \\ \Psi_{\fQEpft}(x)  \\ 
\Psi_{\fQEpf}(x) \\ \Psi_{\fQEps}(x) \\ \Psi_{\fQEnt}(x) \\ \Psi_{\fQEnft}(x) 
\end{pmatrix} \equiv \Psi_{\fQH}(x) , 
\ee
with the identical expression notated as follows:
\be
\Psi_{\fQH}(x) & \equiv & \binom{ \Psi_{\fQHn}(x)}{ \Psi_{\fQHp}(x) } = \begin{pmatrix}
\Psi_{\fQUn}(x) \\ \Psi_{\fQUp}(x) \\ \hPsi_{\fQUp}(x) \\ \hPsi_{\fQUn}(x) ,
\end{pmatrix} = \begin{pmatrix}
\Psi_{\fQGnf}(x) \\ \Psi_{\fQGns}(x) \\ \Psi_{\fQGpt}(x) \\ \Psi_{\fQGpft}(x)  \\ 
\Psi_{\fQGpf}(x) \\ \Psi_{\fQGps}(x) \\ \Psi_{\fQGnt}(x) \\ \Psi_{\fQGnft}(x) 
\end{pmatrix} ,
\ee
where the subscripts with letters in bold style `$\fQE$' and `$\fQH$' are adopted to distinguish from the hyperunified qubit-spinor field discussed in previous analysis. The two mirror pairs of twisting entangled hyperqubit-spinor fields as grand unified qubit-spinor fields with negative and positive U-parity, $\Psi_{\fQGpni}(x)\equiv \Psi_{\fQEpni}(x)$ ($i=1,2$) and $\Psi_{\fQGpnj}(x)\equiv \Psi_{\fQEpnj}(x) $ ($j=3,4$), are defined as follows:
\be \label{fQGpn}
& & \Psi_{\fQGpni}(x) \equiv \Psi_{\fQEpni}(x) \equiv \binom{\Psi_{\fQWpni}(x)}{\Psi_{\fQWpni}^{c}(x)} \equiv \binom{\Psi_{\fQ_{\mS_i}^6}^{\tR, L}(x)}{\Psi_{\fQ_{\mS_i}^6}^{\tR, L\; c}(x)} , \nn \\
& & \Psi_{\fQGpnj}(x) \equiv \Psi_{\fQEpnj}(x) \equiv \binom{\Psi_{\fQWpnj}(x)}{\Psi_{\fQWpnj}^{c}(x)} \equiv \binom{\Psi_{\fQ_{\mS_j}^6}^{\tR, L}(x)}{\Psi_{\fQ_{\mS_j}^6}^{\tR, L\; c}(x)} ,
\ee
where $\Psi_{\fQWpni}(x)$ ($i=1,2$) and $\Psi_{\fQWpnj}(x)$ ($j=3,4$) are two mirror pairs of chiral complex hexaqubit-spinor fields. They are given by two mirror pairs of self-conjugated hexaqubit-spinor fields $\Psi_{\fQ_{\mS_i}^6}^{\tR, L}(x)$ ($i=1,2$) and $\Psi_{\fQ_{\mS_j}^6}^{\tR, L}(x)$ ($j=3,4$) with the following explicit forms:
\be
& &   \Psi_{\fQ_{\mS_f}^6}^{L}(x)  \equiv \begin{pmatrix}
\Psi_{\fQ_{\mC_f}^4}^{u}(x) \\  \Psi_{\fQ_{\mC_f}^4}^{d\, \bc}(x)  \\ \Psi_{\fQ_{\mC_f}^4}^{d}(x) \\ - \Psi_{\fQ_{\mC_f}^4}^{u\, \bc}(x)  
\end{pmatrix}_{L} , \quad  
\Psi_{\fQ_{\mS_f}^6}^{L\; c}(x)  \equiv \begin{pmatrix}
\Psi_{\fQ_{\mC_f}^4}^{u}(x) \\  \Psi_{\fQ_{\mC_f}^4}^{d\, \bc}(x)  \\ \Psi_{\fQ_{\mC_f}^4}^{d}(x) \\ - \Psi_{\fQ_{\mC_f}^4}^{u\, \bc}(x)  
\end{pmatrix}_{R}, \nn \\
& & \Psi_{\fQ_{\mS_f}^6}^{\tR}(x)  \equiv \begin{pmatrix}
\tPsi_{\fQ_{\mC_f}^4}^{u}(x) \\  \tPsi_{\fQ_{\mC_f}^4}^{d\, \bc}(x)  \\ \tPsi_{\fQ_{\mC_f}^4}^{d}(x) \\ - \tPsi_{\fQ_{\mC_f}^4}^{u\, \bc}(x)  
\end{pmatrix}_{R} , \quad 
\Psi_{\fQ_{\mS_f}^6}^{\tR\; c}(x)  \equiv \begin{pmatrix}
\tPsi_{\fQ_{\mC_f}^4}^{u}(x) \\  \tPsi_{\fQ_{\mC_f}^4}^{d\, \bc}(x)  \\ \tPsi_{\fQ_{\mC_f}^4}^{d}(x) \\ - \tPsi_{\fQ_{\mC_f}^4}^{u\, \bc}(x)  
\end{pmatrix}_{L}  
\ee
with $f= i, j =1,2,3,4$ ($i= 1,2$, $j= 3,4$).  Where $\Psi_{\fQ_{\mC_f}^4}^{q}(x)$ and $\tPsi_{\fQ_{\mC_f}^4}^{q}(x)$ ($q=u, d$) are the mirror pairs of complex tetraqubit-spinor fields,
\be \label{fQC4}
& &  \Psi_{\fQ_{\mC_f}^4}^{q\, L, R}(x) \equiv \begin{pmatrix}
Q_f^{r}(x) \\ Q_f^{b}(x) \\ Q_f^{g}(x) \\ Q_f^{w}(x) 
\end{pmatrix}^q_{L, R}, \quad  
\tPsi_{\fQ_{\mC_f}^4}^{q\, L, R}(x) \equiv \begin{pmatrix}
\tQ_f^{r}(x) \\ \tQ_f^{b}(x) \\ \tQ_f^{g}(x) \\ \tQ_f^{w}(x) 
\end{pmatrix}^q_{L, R},
\ee
with their complex charge-conjugated ones $\Psi_{\fQ_{\mC_f}^4}^{q\, \bc}(x)$ and $\tPsi_{\fQ_{\mC_f}^4}^{q\, \bc}(x)$,
\be \label{fQC4C}
& & \Psi_{\fQ_{\mC_f}^4}^{q\, \bc\, L, R}(x) \equiv \begin{pmatrix}
Q^{r\, \bc}_{f}(x) \\ Q^{b\, \bc}_{f}(x) \\ Q^{g\, \bc}_{f}(x) \\ Q^{w\, \bc}_{f}(x) 
\end{pmatrix}^q_{L, R} ,  \quad  \tPsi_{\fQ_{\mC_f}^4}^{q\, \bc\, L, R}(x) \equiv \begin{pmatrix}
\tQ^{r\, \bc}_{f}(x) \\ \tQ^{b\, \bc}_{f}(x) \\ \tQ^{g\, \bc}_{f}(x) \\ \tQ^{w\, \bc}_{f}(x) 
\end{pmatrix}^q_{L, R} , \nn \\
& & Q^{\alpha\, \bc}_{f}(x) = C_D \bar{Q}^{\alpha\, T}_{f}(x), \quad \tQ^{\alpha\, \bc}_{f}(x)= C_D \bar{\tQ}^{\alpha\, T}_{f}(x), \quad C_D = - i\sigma_3\otimes \sigma_2 ,
\ee
where $q=u, d$ represent the up-type and down-type lepton-quark states in SM, i.e., $Q^{\alpha\, u}_{f\, L, R}(x) = U_{f\, L, R}^{\alpha}(x)$ and $Q^{\alpha\, d}_{f\, L, R}(x) = D_{f\, L, R}^{\alpha}(x)$, and also the up-type and down-type mirror lepton-quark states, i.e., $\tQ^{\alpha\, u}_{f\, L, R}(x) = \tU_{f\, L, R}^{\alpha}(x)$ and $\tQ^{\alpha\, d}_{f\, L, R}(x) = \tD_{f\, L, R}^{\alpha}(x)$. The four color-spin charges $\alpha = r, b, g, w$ correspond to `red', `blue', `green' and `white', and the subscripts $f=1,2, 3,4$ denote four families of chiral type lepton-quark states and mirror lepton-quark states.  The letters `L' and `R' label the left-handed and right-handed lepton-quark states and the corresponding mirror lepton-quark states with opposite chirality. 

In such a representation of qubit-spinor structure, the action of hyperunified qubit-spinor field $\Psi_{\fQH}(x)$ can be expressed as follows:
\be  \label{actionfQH}
\cS_{\fQH} & \equiv & \int d^{19}x \,  \{ \bar{\Psi}_{\fQH}(x)  \delta_{\mA}^{\;\mM}  \vSi_{-}^{\fA} i\p_{\mM} \Psi_{\fQH}(x) - \lambda_{1} \phi_1(x)  \bar{\Psi}_{\fQH}(x)   \tvSi_{-} \Psi_{\fQH}(x),
\ee
with $\mA, \mM= 0,1,2,3,5, \cdots, 19$. The matrices $\vSi_{-}^{\mA}$ and $\tvSi_{-}$ are given by $\vGa$-matrices $\vGa^{\hat{\mA}} \equiv (\vGa^{\mA}, \vGa^{20}) \equiv (\vGa^a, \vGa^A, \tvGa) $ ($\hat{\mA}=0,1,2,3, 5,\cdots, 20$). They have the following explicit forms: 
\be \label{GMfQH}
& & \varSigma_{-}^{\mA} = \frac{1}{2} \vGa^{\mA}\vGa_{-}, \quad \mA = 0,1,2,3,5, \cdots 19, \nn \\
& & \varSigma_{-}^{20} = \frac{1}{2} \vGa^{20}\vGa_{-}  = \frac{1}{2} \tvGa \vGa_{-}\equiv \tvSi_{-}, \quad  \vGa_{-} = \frac{1}{2} ( 1 - \hat{\gamma}_{21} ),  \nn \\
& & \vGa^a = \;\; \,  \sigma_0 \otimes \sigma_0 \otimes \sigma_0 \otimes \vGa^a, \quad a=0,1,2,3, \nn \\
& & \vGa^{\mA} = - \sigma_0 \otimes \sigma_0 \otimes \sigma_2 \otimes \vGa^A, \quad A= 5, \cdots, 14, \nn \\
& &  \vGa^{15} =\; i\sigma_0 \otimes \sigma_2 \otimes \sigma_1\otimes \tilde{\gamma}_{5} , \nn \\
& &  \vGa^{16} =\; i\sigma_0 \otimes \sigma_2 \otimes \sigma_2 \otimes \hat{\gamma}_{15} , \nn \\
& &  \vGa^{17} =\; i\sigma_0 \otimes \sigma_2 \otimes \sigma_3\otimes \tilde{\gamma}_{5}  , \nn \\
& & \vGa^{18} = \; i \sigma_0 \otimes \sigma_1 \otimes \sigma_0 \otimes \hat{\gamma}_{15} , \nn \\
& & \vGa^{19} = \; i \sigma_1 \otimes \sigma_3 \otimes \sigma_0 \otimes \hat{\gamma}_{15} , \nn \\
& & \vGa^{20} = \; i \sigma_2 \otimes \sigma_3 \otimes \sigma_0 \otimes \hat{\gamma}_{15} \equiv \tvGa, \nn \\
& &  \hat{\gamma}_{21} = \sigma_3\otimes \sigma_3\otimes \sigma_0\otimes  \hat{\gamma}_{15} =  \sigma_3\otimes \hat{\gamma}_{19} , \nn \\
& & \tilde{\gamma}_5 = \sigma_0 \otimes \sigma_0 \otimes  \sigma_0 \otimes  \sigma_0 \otimes \sigma_0\otimes \gamma_5,  
\ee
which have a different structure in comparison to those in Eq.(\ref{GMQH}). Where $\vGa^a$ and $\vGa^A$ ($a=0,1,2,3,  A= 5, \cdots, 14$) on the right-hand side of equality are defined in Eq.(\ref{GMQW7}).  

The entangled decaqubit-spinor field as hyperunified qubit-spinor field $\Psi_{\fQE^{10}}(x)\equiv \Psi_{\fQH}(x)$ satisfies the following self-conjugated chiral conditions:
\be
& &  \Psi_{\fQH}^{c}(x) = C_{\fQH}\bar{\Psi}_{\fQH}^{T}(x) =  \Psi_{\fQH}(x) \equiv \Psi_{\fQE^{10}}^{c}(x) = \Psi_{\fQE^{10}}(x), \nn \\
& & C_{\fQH} = \sigma_0\otimes C_{\fQHz} \equiv \sigma_0\otimes C_{\fQE^9} = \sigma_0\otimes \sigma_0\otimes C_{\fQE^8}, \nn \\
& & C_{\fQE^8} = \sigma_0\otimes C_{\fQ_{\mE_i}^7}  = \sigma_0\otimes \sigma_1 \otimes \sigma_2 \otimes \sigma_2 \otimes \sigma_0 \otimes \sigma_0 \otimes C_D , \nn \\
& & \hat{\gamma}_{21} \Psi_{\fQH}(x) \equiv \hat{\gamma}_{21} \Psi_{\fQE^{10}}(x) = -\Psi_{\fQE^{10}}(x) \equiv - \Psi_{\fQH}(x) ,
\ee
which indicates that the entangled decaqubit-spinor field as hyperunified qubit-spinor field has only a trivial $\cQ_c$-spin symmetry SP(1)$\cong$O(1) with transformation $\Psi_{\fQH}(x)\to - \Psi_{\fQH}(x)$ which coincides with the chiral condition $\hat{\gamma}_{21} \Psi_{\fQH}(x) = - \Psi_{\fQH}(x)$. 

It is easy to check that the actions in Eqs.(\ref{actionfQH}) and (\ref{actionQH}) possess an associated symmetry characterized by the inhomogeneous hyperspin symmetry WS(1,18) associated with inhomogeneous Lorentz-type/Poincar\'e-type group symmetry PO(1,18) together with global scaling symmetry SG(1) and SC(1), which is given as follows:
\be
G_S & = & SC(1)\ltimes PO(1,18) \adjoin WS(1,18) \times SG(1)  \nn \\
& = & SC(1)\ltimes P^{1,18}\ltimes SO(1,18) \adjoin  SP(1,18) \rtimes W^{1,18}\times SG(1) ,
\ee
where the transformation of hyperspin symmetry SP(1,18) should be coincidental to that of Lorentz-type group symmetry SO(1,18) as expressed by the symbol ``$\adjoin$" which is adopted to indicate the associated symmetry. The group generators of inhomogeneous hyperspin symmetry WS(1,18) are provided directly by the following commutators of $\vGa$-matrices: 
\be
& & \varSigma^{\mA\mB} = \frac{i}{4} [ \vGa^{\mA}, \vGa^{\mB} ], \quad \vSi_{-}^{\mA} = \frac{1}{2} \vGa^{\mA}\vGa_{-}, 
\ee
with $\mA, \mB= 0,1,2,3,5,\cdots, 19$. 

In general, the hyperspin symmetry SP(1,18) gets a natural decomposition into the following subgroup symmetries:
\be
 SP(1,18) & \supset & SP(1,3)\times SP(15)  \supset SP(1,3) \times SP(10) \times SP(5) \nn \\
&  \cong & SO(1,3)\times SO(15) \supset SO(1,3) \times SO(10)\times SO_F(5) \nn \\
& \supset & SO(1,3)\times SO(6) \times SO(4)\times SO_F(4) \nn \\
& \cong & SO(1,3) \times SU_C(4)\times SU_L(2)\times SU_R(2) \times SU_F(2)\times SU_M(2),
\ee
with the subgroup algebras given by the following group generators:
\be
& & \varSigma^{\fA\fB}  ( \fA, \fB= 0,1,2,3 ) \in sp(1,3)\cong so(1,3)  \nn \\
& &  \varSigma^{\fA\fB}  (\fA, \fB= 5,\cdots, 19)  \in sp(15) \cong so(15), \nn \\
& &  \varSigma^{\fA\fB}  (\fA, \fB= 5,\cdots, 14 )  \in sp(10) \cong so(10), \nn \\
& &  \varSigma^{\fA\fB}  (\fA, \fB= 15,\cdots, 19 )  \in sp(5) \cong so_F(5), \nn \\
& & \varSigma^{\fA\fB}  (\fA, \fB= 5,\cdots,10 )  \in sp(6) \cong so(6)\cong su_C(4) , \nn \\
& &  \varSigma^{\fA\fB}  ( \fA, \fB= 11,\cdots14) \in sp(4)\cong so(4)\cong su_L(2)\times su_R(2) , 
\ee
where the subgroup symmetry SP$_F$(5)$\cong$SO$_F$(5) characterizes the family-spin symmetry for four families of mirror pairs of lepton-quark states.

Therefore, the entangled decaqubit-spinor field $\Psi_{\fQE^{10}}(x)$ (or $\Psi_{\QE^{10}}(x)$) enables us to unify all discovered leptons and quarks into a single hyperunified qubit-spinor field, i.e., $\Psi_{\fQH}(x)\equiv \Psi_{\fQE^{10}}(x)$ (or $\Psi_{\QH}(x)\equiv \Psi_{\QE^{10}}(x)$), which is regarded as the fundamental building block of nature. The maximal inhomogeneous hyperspin symmetry WS(1,18) in association with inhomogeneous Lorentz-type/Poincar\'e-type group symmetry PO(1,18) together with global scaling symmetry SG(1) and SC(1) brings on a hyperunified symmetry as fundamental symmetry of nature.


\subsection{ Conservation of H-parity and U-parity in hyperunified field theory with vector-like lepton-quark state and comprehension on P-parity violation in SM } 

The category-1 entangled decaqubit-spinor field as hyperunified qubit-spinor field $\Psi_{\fQH}(x)$ brings about four families of chiral type lepton-quark states with negative U-parity and four families of mirror lepton-quark states with positive U-parity, i.e., $\Psi_{\fQ_{\mG_f}^-}(x)$ and $\Psi_{\fQ_{\mG_f}^+}(x)$ ($f=1,2,3,4$), which motivates us to express such a hyperunified qubit-spinor field $\Psi_{\fQH}(x)$ into four families of vector-like lepton-quark states in 19-dimensional hyper-spacetime.

To be explicit and specific, let us reconstruct the entangled decaqubit-spinor field $\Psi_{\fQE^{10}}(x)\equiv \Psi_{\fQH}(x)$ into a {\it self-conjugated enneaqubit-spinor field}, which can be expressed as a superposition of two entangled enneaqubit-spinor fields given by minimal hyperunified qubit-spinor fields with the following form:
\be
\Psi_{\fQS^9}(x) & \equiv & \Psi_{\fQHn}(x) + \Psi_{\fQHp}(x) \equiv  \binom{ \Psi_{\fQU}(x)}{ \hPsi_{\fQU}(x) } \equiv  \begin{pmatrix}
\Psi_{\fQGst}(x) \\ \Psi_{\fQGnd}(x) \\ \Psi_{\fQGth}(x) \\ \Psi_{\fQGft}(x) 
\end{pmatrix} , 
\ee
with the definitions,
\be
\Psi_{\fQU}(x) & \equiv &  \Psi_{\fQUn}(x)  + \hPsi_{\fQUp}(x) \equiv \Psi_{\fQS^8}(x)  \equiv  \binom{\Psi_{\fQ_{\mS_1}^7}(x)}{\Psi_{\fQ_{\mS_2}^7}(x) }, \nn \\
\hPsi_{\fQU}(x) & \equiv & \hPsi_{\fQUn}(x) + \Psi_{\fQUp}(x) \equiv \hPsi_{\fQS^8}(x)  \equiv  \binom{\Psi_{\fQ_{\mS_3}^7}(x)}{\Psi_{\fQ_{\mS_4}^7}(x) },
\ee
and
\be
& & \Psi_{\fQ_{\mS_f}^7}(x) \equiv \Psi_{\fQGf}(x) 
\equiv  \binom{\Psi_{\fQ_{\mC_f}^6}(x)}{\Psi^c_{\fQ_{\mC_f}^6}(x)} , \; \; f=1,2,3,4 .
\ee
All above defined hyperqubit-spinor fields $\Psi_{\fQS^9}(x)$, $\Psi_{\fQS^8}(x) \equiv \Psi_{\fQU}(x)$, $\hPsi_{\fQS^8}(x) \equiv \hPsi_{\fQU}(x)$ and $\Psi_{\fQ_{\mS_f}^7}(x) \equiv \Psi_{\fQGf}(x)$ ($f=1,2,3,4$) are self-conjugated hyperqubit-spinor fields and satisfy the following self-conjugation conditions:
\be
& & \Psi_{\fQS^9}^c(x) = C_{\fQS^9}\bar{\Psi}_{\fQS^9}^T(x) = \Psi_{\fQS^9}(x) , \nn \\
& & \Psi_{\fQS^8}^c(x) = C_{\fQS^8}\bar{\Psi}_{\fQS^8}^T(x) = \Psi_{\fQS^8}(x) , \nn \\
& & \Psi_{\fQ_{\mS_f}^7}^c(x) = C_{\fQS^7}\bar{\Psi}_{\fQ_{\mS_f}^7}^T(x) = \Psi_{\fQ_{\mS_f}^7}(x) , \nn \\
& & C_{\fQS^9}\equiv \sigma_0\otimes  C_{\fQS^8}, \quad  C_{\fQS^8} \equiv \sigma_0\otimes  C_{\fQS^7} , \nn \\
& & C_{\fQS^7} =  \sigma_1 \otimes \sigma_2 \otimes \sigma_2 \otimes \sigma_0 \otimes \sigma_0 \otimes C_D ,
\ee
where the self-conjugated enneaqubit-spinor field $\Psi_{\fQS^9}(x)$ provides a hyperunified qubit-spinor field with different spinor structure. The four self-conjugated hyperqubit-spinor fields $\Psi_{\fQ_{\mS_f}^7}(x)$ ($f=1,2,3,4$) characterize four families of grand unified qubit-spinor fields $\Psi_{\fQGf}(x)\equiv \Psi_{\fQ_{\mS_f}^7}(x)$ ($f=1,2,3,4$), which are described by four {\it complex hexaqubit-spinor fields} $\Psi_{\fQ_{\mC_f}^6}(x)$ together with their complex charge-conjugated ones $\Psi_{\fQ_{\mC_f}^6}^{c}(x)$.  Explicitly, $\Psi_{\fQ_{\mC_f}^6}(x)$ ($\Psi_{\fQ_{\mC_f}^6}^{c}(x)$) are composed of four families of vector-like lepton-quark states which are given by the following forms:
\be \label{QC6}
& & \Psi_{\fQ_{\mC_f}^6}(x)  \equiv \begin{pmatrix}
\Psi^{\hu}_{\fQ_{\mC_f}^4}(x) \\  \Psi^{\ckd \,\bc}_{\fQ_{\mC_f}^4}(x) \\ 
\Psi^{\hd}_{\fQ_{\mC_f}^4}(x)  \\ -\Psi^{\cku \,\bc}_{\fQ_{\mC_f}^4}(x)
\end{pmatrix},  \quad
\Psi_{\fQ_{\mC_f}^6}^{c}(x)  \equiv \begin{pmatrix}
\Psi^{\cku}_{\fQ_{\mC_f}^4}(x) \\  \Psi^{\hd\; \bc}_{\fQ_{\mC_f}^4}(x) \\ 
\Psi^{\ckd}_{\fQ_{\mC_f}^4}(x)  \\ -\Psi^{\hu\; \bc}_{\fQ_{\mC_f}^4}(x)
\end{pmatrix} , 
\ee
where $\Psi^{\hq}_{\fQ_{\mC_f}^4}(x)$ ($\hq=\hu, \hd, \cku, \ckd $ and $f=1,2,3,4$) are complex tetraqubit-spinor fields with $\Psi^{\hq\; \bc }_{\fQ_{\mC_f}^4}(x)$ their complex charge-conjugated ones. Their explicit forms are given as follows:
\be \label{qQC4}
& & \Psi^{\hq}_{\fQ_{\mC_f}^4}(x)  \equiv \begin{pmatrix}
Q_f^{r}(x) \\ Q_f^{b}(x) \\ Q_f^{g}(x) \\ Q_f^{w}(x) 
\end{pmatrix}^{\hq}, \quad  \Psi^{\hq\; \bc }_{\fQ_{\mC_f}^4}(x)  \equiv \begin{pmatrix}
Q_f^{r\, \bc}(x) \\ Q_f^{b\, \bc}(x) \\ Q_f^{g, \bc}(x) \\ Q_f^{w\, \bc}(x) 
\end{pmatrix}^{\hq}, 
\ee
where $\left(Q_f^{\alpha}\right)^{\hq}$ denote vector-like lepton-quark states in four families $\Psi_{\fQGf}(x) \equiv \Psi_{\fQ_{\mS_f}^7}(x)$ with $f=1,2,3,4$. The superscripts $\hq=(\hu, \hd, \cku, \ckd)$ label four kinds of vector-like lepton-quark states in each family and $\alpha=(r, b, g, w)$ denote four color-spin charges in correspondence to `red', `blue', `green' and `white' for each vector-like lepton-quark state. 

The four kinds of vector-like lepton-quark states are denoted as $\left(Q_f^{\alpha}\right)^{\hq} = (\hU_f^{\alpha}, \hD_f^{\alpha}, \ckU_f^{\alpha}, \ckD_f^{\alpha})$ with the following definitions:
\be \label{VLQS}
& & \hU_f^{\alpha} \equiv U_{f L}^{\alpha} + \tU_{f R}^{\alpha} , \quad  \ckU_f^{\alpha} \equiv  U_{f R}^{\alpha} + \tU_{f L}^{\alpha} , \nn \\
& & \hD_f^{\alpha} \equiv D_{f L}^{\alpha} + \tD_{f R}^{\alpha} , \quad  \ckD_f^{\alpha} \equiv  D_{f R}^{\alpha} + \tD_{f L}^{\alpha} , 
\ee 
where $U_{f L}^{\alpha}$, $U_{f R}^{\alpha}$, $D_{f L}^{\alpha}$ and $D_{f R}^{\alpha}$ represent the left-handed and right-handed up-type and down-type lepton-quark states, and $\tU_{f L}^{\alpha}$, $\tU_{f R}^{\alpha}$, $\tD_{f L}^{\alpha}$ and $\tD_{f R}^{\alpha}$ denote the corresponding left-handed and right-handed mirror lepton-quark states. 

In general, each family of chiral type lepton-quark states in SM and mirror lepton-quark states beyond SM as well as vector-like lepton-quark states can be expressed by the so-called westward and eastward entangled hyperqubit-spinor fields\cite{HUFT} as well as their superposition, i.e.:
\be \label{WEQH}
& & \Psi_{W\1 f}^{T}(x) \equiv  \Psi_{\fQ_{\mG_f}^{-}}^{T}(x)  \nn \\
& & \;\; =[ (U_f^{r}, U_f^{b}, U_f^{g}, U_f^{w}, D^{r}_{fc}, D^{b}_{fc}, D^{g}_{fc}, D^{w}_{fc}, D_f^{r}, D_f^{b}, D_f^{g}, D_f^{w}, -U^{r}_{fc}, -U^{b}_{fc}, -U^{g}_{fc}, -U^{w}_{fc})_L,   \nn \\
& & \;\; (U_f^{r}, U_f^{b}, U_f^{g}, U_f^{w}, D^{r}_{fc}, D^{b}_{fc}, D^{g}_{fc}, D^{w}_{fc}, D_f^{r},  D_f^{b}, D_f^{g}, D_f^{w}, -U^{r}_{fc}, -U^{b}_{fc}, -U^{g}_{fc}, -U^{w}_{fc})_R ]^T , \nn \\
& & \Psi_{E\1 f}^{T}(x) \equiv  \Psi_{\fQ_{\mG_f}^{+}}^{T}(x)  \nn \\
& & \;\; =[ (\tU_f^{r}, \tU_f^{b}, \tU_f^{g}, \tU_f^{w}, \tD^{r}_{fc}, \tD^{b}_{fc}, \tD^{g}_{fc}, \tD^{w}_{fc}, \tD_f^{r}, \tD_f^{b}, \tD_f^{g}, \tD_f^{w}, -\tU^{r}_{fc}, -\tU^{b}_{fc}, -\tU^{g}_{fc}, -\tU^{w}_{fc})_R ,   \nn \\
& & \;\; (\tU_f^{r}, \tU_f^{b}, \tU_f^{g}, \tU_f^{w}, \tD^{r}_{fc}, \tD^{b}_{fc}, \tD^{g}_{fc}, \tD^{w}_{fc}, \tD_f^{r},  \tD_f^{b}, \tD_f^{g}, \tD_f^{w}, -\tU^{r}_{fc}, -\tU^{b}_{fc}, -\tU^{g}_{fc}, -\tU^{w}_{fc})_L ]^T , \nn \\
& & \Psi_{\fQGf}(x) = \Psi_{W\1 f}^{T}(x) +  \Psi_{E\1 f}^{T}(x) \equiv \Psi_{\fQ_{\mS_f}^7}(x) \nn \\
& & \;\; =[ \hU_f^{r}, \hU_f^{b}, \hU_f^{g}, \hU_f^{w}, \hD^{r}_{fc}, \hD^{b}_{fc}, \hD^{g}_{fc}, \hD^{w}_{fc}, \hD_f^{r}, \hD_f^{b}, \hD_f^{g}, \hD_f^{w}, -\hU^{r}_{fc}, -\hU^{b}_{fc}, -\hU^{g}_{fc}, -\hU^{w}_{fc} ,   \nn \\
& & \;\; \; \ckU_f^{r}, \ckU_f^{b}, \ckU_f^{g}, \ckU_f^{w}, \ckD^{r}_{fc}, \ckD^{b}_{fc}, \ckD^{g}_{fc}, \ckD^{w}_{fc}, \ckD_f^{r},  \ckD_f^{b}, \ckD_f^{g}, \ckD_f^{w}, -\ckU^{r}_{fc}, -\ckU^{b}_{fc}, -\ckU^{g}_{fc}, -\ckU^{w}_{fc} ]^T ,
\ee
with $f=1,2,3,4$. The superscript $T$ denotes the transposition of column matrix.

In terms of the self-conjugated enneaqubit-spinor field $\Psi_{\fQS^9}(x)$ as hyperunified qubit-spinor field, we are able to express the action in Eq.(\ref{actionfQH}) into the following form: 
\be  \label{actionfQS9}
\cS_{\fQS^9}  \equiv \cS_{\fQH} = \int d^{19}x \, \frac{1}{2} \{ \bar{\fPsi}_{\fQS^{9}}(x)  \delta_{\mA}^{\;\mM}  \vGa^{\fA} i\p_{\mM} \fPsi_{\fQS^{9}}(x) - \lambda_{1} \phi_1(x)  \bar{\fPsi}_{\fQS^{9}}(x)\fPsi_{\fQS^{9}}(x) \}, 
\ee
with $\mA, \mM= 0,1,2,3,5, \cdots, 19$. The $\vGa$-matrices $\vGa^{\hat{\mA}} \equiv (\vGa^a, \vGa^A) $ ($a=0,1,2,3; A=5,\cdots, 19$) are given by, 
\be \label{GMfQS9}
& & \vGa^a = \;\; \,   \sigma_0 \otimes \sigma_0 \otimes \vGa^a, \quad a=0,1,2,3, \nn \\
& & \vGa^{\mA} = -  \sigma_0 \otimes \sigma_2 \otimes \vGa^A, \quad A= 5, \cdots, 14, \nn \\
& &  \vGa^{15} =\; i \sigma_2 \otimes \sigma_1\otimes \tilde{\gamma}_{5} , \nn \\
& &  \vGa^{16} =\; i \sigma_2 \otimes \sigma_2 \otimes \hat{\gamma}_{15} , \nn \\
& &  \vGa^{17} =\; i \sigma_2 \otimes \sigma_3\otimes \tilde{\gamma}_{5}  , \nn \\
& & \vGa^{18} = \; i \sigma_1 \otimes \sigma_0 \otimes \hat{\gamma}_{15} , \nn \\
& & \vGa^{19} = \; i \sigma_3 \otimes \sigma_0 \otimes \hat{\gamma}_{15} , \nn \\
& &  \hat{\gamma}_{15} =  \sigma_3 \otimes \sigma_0 \otimes  \sigma_0 \otimes  \sigma_0 \otimes \sigma_0\otimes \gamma_5, \nn \\
& & \tilde{\gamma}_5 = \sigma_0 \otimes \sigma_0 \otimes  \sigma_0 \otimes  \sigma_0 \otimes \sigma_0\otimes \gamma_5,  
\ee
with $\vGa^a$ and $\vGa^A$ ($a=0,1,2,3,  A= 5, \cdots, 14$) on the right-hand side of equality given in Eq.(\ref{GMQW7}). 

Such a formalism of the action in Eq.(\ref{actionfQS9})  indicates that both U-parity for grand unified qubit-spinor field and ordinary P-parity defined in SM should be conserved in the hyperunified field theory.

It is noted that the vector-like lepton-quark states $\left(Q_f^{\alpha}\right)^{\hq} = (\hU_f^{\alpha}, \hD_f^{\alpha}, \ckU_f^{\alpha}, \ckD_f^{\alpha})$ defined in Eq.(\ref{VLQS}) as Dirac fermions should be distinguished from Dirac-type leptons and quarks $Q_f^{\alpha} = (U_f^{\alpha}, D_f^{\alpha})$ appearing in the electromagnetic and strong interactions of SM, which are defined as follows:
\be
& & U_f^{\alpha} \equiv U_{f L}^{\alpha} + U_{f R}^{\alpha} , \quad D_f^{\alpha} \equiv D_{f L}^{\alpha} + D_{f R}^{\alpha} ,
\ee 
which behave as Dirac fermions only in the electromagnetic and strong interactions without considering the weak interaction in SM.

From the above analyses and discussions, we come to the observation that the hyperunified field theory for hyperunified qubit-spinor field in 19-dimensional hyper-spacetime preserves H-parity and U-parity as well as P-parity. Whereas the observed leptons and quarks in SM are actually chiral fermions in correspondence to negative U-parity in hyperunified qubit-spinor field. Therefore, to realize the SM with three families of chiral type leptons and quarks, both H-parity and U-parity as well as family-spin symmetry must be broken down. Therefore, the maximum P-parity violation\cite{PV1,PV2,PV3} and also CP-violation\cite{KM} in the weak interaction of SM should be caused from spontaneous symmetry breaking mechanism\cite{SCPV1,SCPV2} and/or geometrical symmetry breaking mechanism associated with the reduction of spacetime dimensions. A study on symmetry breaking mechanism is beyond the scope of present paper, we will present a general discussion in another paper.


\section{Summaries and conclusions with remarks}

We have provided a detailed and systematic investigation on the foundation of the hyperunified field theory laid from the maximum coherence motion principle and maximum locally entangled-qubits motion principle as guiding principles. Our main observations and conclusions are summarized as follows. 

Following along the action principle with path integral formulation by treating basic field as a function, and starting with thinking of an intuitive notion that the universe is composed of fundamental building block which always keeps a continuous motion in a simple and coherent way, we have paid on the motion nature of basic constituent rather than symmetry property in the usual consideration and made a proposal that the fundamental building block of nature is characterized by a real column vector field $\Psi(x)$ expressed as a set of real fields $\psi_i(x)$ ($i=1,\cdots, \cD_H$), i.e., $\Psi^{T}(x) = (\psi_1(x), \psi_2(x), \cdots, \psi_{\cD_H}(x) )$. Such a real column vector field $\Psi(x)$ is considered as a continuous and differentiable column function of real variables $\{x\} \equiv \{ x^{\fM}; \fM=0,1,\cdots, D_h-1\}$ in the path integral formulation. The general action of real column vector field $\Psi(x)$ has been built by making initially the simplest motion postulate which states that the action keeps merely a bilinear form of $\Psi(x)$ with involving only the first order derivative with respect to the variables $ x^{\fM}$ and meanwhile satisfies the hermiticity requirement. Then the maximum coherence motion principle has been introduced as guiding principle from further proposing the maximum correlation motion postulate and quadratic free motion postulate together with the simplest motion postulate, which brings the real column vector field $\Psi(x)$ to be basic spinor field with the natural appearance of canonical anticommutation relation and leads the real variables $x^{\fM}$ to span $D_h$-dimensional Minkowski spacetime with only one temporal dimension. The hermiticity requirement of the action allows only a pure imaginary coupling constant to meet with anti-commuting feature of real column vector field $\Psi(x)$ and brings on the first order derivative associated with the pure imaginary factor `$i$' to form a linear self-adjoint operator $i \partial_{\fM}$. Such an operator describes the free motion of basic spinor field as fundamental building block. The maximally correlated motion of column vector field $\Psi(x)$ is found to be characterized by the real symmetric $\cM_c$-spin matrices with the number of matrices designated as $\cM_c$-spin charge $\cC_{\cM_c}=D_h$. Meanwhile, the antisymmetric matrices are shown to reflect the motion-irrelevance intrinsic property of column vector field $\Psi(x)$ and referred to as $\cQ_c$-spin matrices with the number of matrices designated as $\cQ_c$-spin charge $\CQc=q_c$. As the maximum coherence motion principle as guiding principle leads $\cM_c$-spin matrices and $\cQ_c$-spin matrices to satisfy Clifford algebra, the action for such a spinor field $\Psi(x)$ has been demonstrated to bring about the natural emergence of associated symmetry in which the motion-correlation $\cM_c$-spin/hyperspin symmetry SP(1,$D_h$-1) and global scaling symmetry SG(1) together with intrinsic $\cQ_c$-spin symmetry SP($q_c$) in Hilbert space are in association with the Poincar\'e-type group symmetry PO(1,$D_h$-1) and global scaling symmetry SC(1) in Minkowski spacetime. For the simplest case of single component spinor field with zero $\cQ_c$-spin charge $\CQc =0$, it is verified to have a free motion only in two-dimensional spacetime.

By proposing the local coherent-qubits motion postulate which states that a local coherent state of qubits as a column vector field of local distribution amplitudes, by obeying the maximum coherence motion principle, brings on a qubit-spinor field as basic constituent of matter, we have presented a detailed analysis on the basic properties of qubit-spinor field with arbitrary qubit number $Q_N$. 

Begin with examining the simplest local coherent state of single qubit ($Q_N=1$) with two local distribution amplitudes which obeying the maximum coherence motion principle, we have demonstrated that such a local coherent state provides a Taijion spinor field referred to as real/self-conjugated uniqubit-spinor field in two-dimensional Hilbert space. Meanwhile, its maximally correlated motion brings about three real symmetric $\cM_c$-spin matrices with $\cM_c$-spin charge $\cC_{\cM_c}=3$, which results in the appearance of three-dimensional Minkowski spacetime. The single antisymmetric matrix brings on a unit $\cQ_c$-spin charge $\CQc=q_c=Q_N=1$ and leads to the scalar coupling with one scalar field. It has been shown that Majorana fermion with four independent degrees of freedom is described by a local coherent state of 2-qubit ($Q_N$=2) with $\cQ_c$-spin charge $\CQc=q_c=Q_N=2$, which is referred to as real/self-conjugated biqubit-spinor field in four-dimensional real/self-conjugated Hilbert space. While massless Dirac fermion with eight independent degrees of freedom is characterized by a local coherent state of 3-qubit ($Q_N$=3) with $\cQ_c$-spin charge $\CQc=q_c=Q_N=3$, which is referred to as real/self-conjugated triqubit-spinor field in eight-dimensional real/self-conjugated Hilbert space. In general, Majorana and Dirac fermions are verified to be characterized equivalently by the corresponding complex uniqubit-spinor field and complex biqubit-spinor field with respective to two-dimensional complex Hilbert space and four-dimensional complex Hilbert space. The maximally correlated motions of biqubit-spinor and triqubit-spinor fields bring on the appearance of four-dimensional and six-dimensional Minkowski spacetimes, respectively. It is seen that the appearance of spatial dimensions is in connection to the rotational $\cM_c$-spin symmetry of qubit-spinor field, while the appearance of temporal dimension is related to the non-homogeneous scaling symmetry of qubit-spinor field. Such a feature is attributed to the invariance principle of the action under the associated symmetry that the $\cM_c$-spin symmetry transformation in Hilbert space must be coincidental to the Lorentz-type symmetry transformation in Minkowski spacetime. When regarding a single component real spinor field as zeroqubit-spinor field or bit-spinor field with $\cQ_c$-spin charge $\CQc=q_c=Q_N=0$, we have arrived at a simple relation between $\cQ_c$-spin charge and qubit number, $\CQc=q_c=Q_N=0,1,2,3$, for four basic qubit-spinor fields with qubit number $Q_N\le 3$.  

For the local coherent states of qubits with qubit number $Q_N\ge 4$, we have presented a detailed analysis on tetraqubit-spinor, pentaqubit-spinor, hexaqubit-spinor, heptaqubit-spinor and octoqubit-spinor fields with respective to the qubit numbers $Q_N=4, 5, 6, 7, 8$,  which correspond to Hilbert spaces with dimensions $\cD_H= 2^4, 2^5, 2^6, 2^7, 2^8$. They are generally referred to as hyperqubit-spinor fields. The motion-correlation $\cM_c$-spin matrices for both $\Ups$-matrices and $\Gamma$-matrices have explicitly been constructed to characterize the maximally correlated motions of hyperqubit-spinor fields, which brings on the appearance of Minkowski hyper-spacetime with dimensions determined from $\cM_c$-spin charge to be $D_h\equiv \CMc=10, 11, 12, 14, 18$ with respective to qubit numbers $Q_N=4, 5, 6, 7, 8$. The intrinsic $\cQ_c$-spin matrices have been demonstrated to reflect the basic properties of hyperqubit-spinor fields, which displays a periodic feature indicated from $\cQ_c$-spin charges $\CQc = q_c = 0,1,2,3,0$ in correspondence to qubit numbers $Q_N=4, 5, 6, 7, 8$. In comparison to four basic qubit-spinor fields which exhibit a simple relation between $\cQ_c$-spin charges and qubit numbers $\CQc=q_c=Q_N=0,1,2,3$, we come to the observation that the $\cQ_c$-spin charges of qubit-spinor fields reflect a periodic feature characterized by four categoric $\cQ_c$-spin charges $\CQc = q_c=0,1,2,3$.

We have deduced a periodic law of $\cQ_c$-spin charge to make categorizations for all qubit-spinor fields with arbitrary qubit number and also for all Minkowski spacetimes determined from the maximally correlated motions of qubit-spinor fields. It has been verified that there exist four categories with respective to four $\cQ_c$-spin charges, which can be  presented as the mod 4 qubit number.  A general relation between the qubit number $Q_N$ and $\cQ_c$-spin charge $q_c$ has been obtained to be that, $Q_N^{(q_c,k)}= \CQc^{(q_c,k)} + 4k = q_c + 4k$, where four $\cQ_c$-spin charges $\CQc^{(q_c,k)}= q_c = 0,1,2,3$ characterize four categories and the integer number $k=0,1,\cdots$ reflects periodic behavior. So that any qubit-spinor field as local coherent state of qubits with categoric qubit number $Q_N^{(q_c,k)}$ can be classified into the categoric qubit-spinor field denoted as $\Psi_{Q_N^{(q_c,k)}}(x)$, which belongs to the spinor representation in $2^{Q_N^{(q_c,k)}}$-dimensional Hilbert space and has free motion in Minkowski spacetime with categoric dimension $D_h^{(q_c,k)}$. Such a categoric dimension $D_h^{(q_c,k)}$ is given by the general relation $D_h^{(q_c,k)}=D_{q_c} + 8k$, where $k=0,1,\cdots$ denotes the $k$-th period and $D_{q_c}$ stands for the basic spacetime dimension in category-$q_c$. In general, there are four basic spacetime dimensions $D_{q_c} = 2 + 2^{q_c-1} \theta(q_c-1) = 2, 3, 4, 6$ with respective to four $\cQ_c$-spin charges $\CQc=q_c =0,1,2,3$, which correspond to four basic qubit numbers $Q_N=0,1,2,3$ including the case of zero qubit.

When the independent degrees of freedom $\cD_H=2^{Q_N}$ of self-conjugated qubit-spinor field $\Psi_{Q_N^{(q_c,k)}}(x)$ in category-$q_c$ and $k$-th period are more than twice the qubit number, i.e., $2^{Q_N} > 2Q_N$ or $Q_N>2$, the $2^{Q_N}$-dimensional Hilbert space spanned by $Q_N$-product qubit-basis states appears to be not big enough to bring all motion-correlation $\Gamma$-matrices to be anti-commuting, which leads us to construct a self-conjugated chiral qubit-spinor field in the same category-$q_c$ but in $2^{Q_N+1}$-dimensional Hilbert space spanned by $(Q_N+1)$-product qubit-basis states. Such a self-conjugated chiral qubit-spinor structure makes all motion-correlation $\Gamma$-matrices and intrinsic $\cQ_c$-matrices to become anti-commuting, and meanwhile the group generators of $\cM_c$-spin/hyperspin symmetry and $\cQ_c$-spin symmetry can directly be presented by the commutators of $\Gamma$-matrices and $\cQ_c$-matrices. From the concept of quantum entanglement of qubits, which states that an inseparable state in tensor product Hilbert space is referred to as entangled state, we have made a locally entangled-qubits motion postulate as such a statement is applicable to both global and local states. The locally entangled-qubits motion postulate indicates that a locally entangled state of qubits as a column vector field of local distribution amplitudes with obeying the maximum coherence motion principle brings on an entangled qubit-spinor field which is supposed to provide the fundamental building block of nature. In formal, a self-conjugated chiral qubit-spinor field with $2^{Q_N}$ independent local distribution amplitudes is shown to be an inseparable state in $2^{Q_N+1}$-dimensional tensor product Hilbert space formed from $(Q_N+1)$-product qubit-basis states, which is referred to as entangled qubit-spinor field.

Based on the locally entangled-qubits motion postulate, we have presented a detailed analysis on the basic properties of entangled hyperqubit-spinor fields as unified qubit-spinor fields of lepton-quark states in hyper-spacetime, which include the entangled hyperqubit-spinor fields with respective to $\cQ_c$-spin charges $\CQc^{(q_c,k)} = q_c = 0,1,2,3$ ($k=1$) in correspondence to categoric hyper-spacetime dimensions $D_h^{(q_c,k)}=10, 11, 12, 14$. The corresponding hyperspin symmetry has been shown to reflect the basic properties of lepton-quark states, such as color-spin, isospin, flavor-spin as well as family-spin charges. The relevant $\cQ_c$-spin charge allows us not only to classify the categories of entangled hyperqubit-spinor fields as unified qubit-spinor fields of lepton-quark states, but also to characterize the flavor-spin and family-spin charges of lepton-quark states through the $\cQ_c$-spin symmetry. Furthermore, we have demonstrated that the entangled hyperqubit-spinor structure enables us to extend the hyperspin symmetry to inhomogeneous hyperspin symmetry. In particular, we have provided a reliable analysis and explanation on the longstanding open questions why there exist leptons and quarks beyond one family in nature and why our observed universe is only four-dimensional spacetime. 

In light of self-conjugated chiral qubit-spinor structure, any entangled qubit-spinor field has been shown to possess the entanglement-correlated translation-like $\cW_e$-spin Abelian-type group symmetry W$^{1,D_h-1}$, which is in analogous to the translational group symmetry P$^{1,D_h-1}$ of coordinates in Minkowski spacetime. As such an entanglement-correlated symmetry W$^{1,D_h-1}$ is associated with a sign flip in chirality with respect to the left-handed and right-handed ($D_h=4$) or so-called westward and eastward ($D_h>4$) entangled qubit-spinor fields in Hilbert space, which is referred to as $\cW_e$-spin symmetry. Therefore, we are able to extend $\cM_c$-spin symmetry SP(1,$D_h$-1) to inhomogeneous $\cM_c$-spin symmetry WS(1,$D_h$-1)=SP(1,$D_h$-1)$\rtimes$ W$^{1,D_h-1}$. For illustration, we have explicitly examined how Majorana and Dirac fermions are regarded as entangled triqubit-spinor and tetraqubit-spinor fields. In a more general case, we have demonstrated in detail four entangled hyperqubit-spinor fields as unified qubit-spinor fields of lepton-quark states in correspondence to hyper-spacetime dimensions $D_h=10, 11, 12, 14$, which are referred to as entangled pentaqubit-spinor, hexaqubit-spinor, heptaqubit-spinor and octoqubit-spinor fields with respective to categoric $\cQ_c$-spin charges $\CQc^{(q_c,k)} = q_c = 0,1,2,3$ with $k=1$. The action of entangled hyperqubit-spinor field in $D_h$-dimensional hyper-spacetime is found to possess an emergent associated symmetry which is characterized by the inhomogeneous hyperspin symmetry WS(1, $D_h$-1) and $\cQ_c$-spin symmetry SP($q_c$) in association with inhomogeneous Lorentz-type/Poincar\'e-type group symmetry PO(1, $D_h$-1) and global scaling symmetry, i.e.:
\be
 G_S = SC(1) \ltimes PO(1, D_h-1) \adjoin WS(1, D_h-1) \rtimes SG(1) \times SP(q_c) , \nn
\ee
where the symbol ``$\adjoin$" is adopted to indicate the associated symmetry in which the transformation of hyperspin symmetry SP(1,$D_h$-1) must be coincidental to that of Lorentz-type group symmetry SO(1,$D_h$-1). 

By considering inhomogeneous hyperspin symmetry WS(1,$D_h$-1) of entangled hyperqubit-spinor field as basic symmetry of lepton-quark states in hyper-spacetime and presenting a systematical analysis on four entangled hyperqubit-spinor fields with respective to hyper-spacetime dimensions $D_h=10, 11, 12, 14$, we have proved that only up to the entangled octoqubit-spinor field in 14-dimensional hyper-spacetime, the basic symmetry of SM can consistently be reproduced as a subgroup symmetry of hyperspin symmetry SP(1,13). It is interesting to notice that two families of leptons and quarks in SM have to appear simultaneously, namely, they must be unified into a single entangled hyperqubit-spinor field as ultra-grand unified qubit-spinor field. The inhomogeneous hyperspin symmetry WS(1,13)=SP(1,13)$\rtimes$W$^{1,13}$ becomes an ultra-grand unified symmetry. As the hyperspin symmetry SP(1,13)$\cong$SO(1,13) contains a maximal subgroup symmetry SP(1,3)$\times$SP(10)$\cong$SO(1,3)$\times$SO(10) with SO(10) as grand unified symmetry, we are able to realize the subgroup symmetry SO(1,3)$\times$SU$_C$(3)$\times$SU$_L$(2)$\times$U$_Y$(1) as the basic symmetry of chiral type leptons and quarks in SM. Meanwhile, the internal $\cQ_c$-spin symmetry SP(3) is shown to be the family-spin symmetry SU$_F$(2) which characterizes two families of chiral type lepton-quark states in SM with including right-handed neutrinos. It is such an intrinsic feature that leads to the prediction that there must exist at least two families of chiral type leptons and quarks in nature, which brings about a natural comprehension on the basic issue why there are leptons and quarks beyond one family in nature although our observed universe is composed of only one family of leptons and quarks. Furthermore, we have verified that the entangled octoqubit-spinor field as ultra-grand unified qubit-spinor field in 14-dimensional hyper-spacetime can be expressed as a superposition of two twisting entangled hyperqubit-spinor fields. Each twisting entangled hyperqubit-spinor field is found to be a grand unified qubit-spinor field, which is exactly composed of one family of leptons and quarks and keeps a free motion only in four-dimensional spacetime. Therefore, when considering the observable fact that the atomic matter in our living universe mainly consists of the up quark and down quark together with the electron, we arrive at a natural explanation on the longstanding open issue why the observed universe with atomic matter is only four-dimensional spacetime. which just belong to one family of leptons and quarks (so-called the first family in SM) and have a free motion only in four-dimensional spacetime,

As the intrinsic $\cQ_c$-spin symmetry appears as direct product group symmetry of the action, the least $\cQ_c$-spin postulate has been proposed to preserve the inhomogeneous hyperspin symmetry as single maximal symmetry of entangle hyperqubit-spinor field. We then arrive at the maximum locally entangled-qubits motion principle as guiding principle, which combines the local coherent-qubits motion postulate and locally entangled-qubits motion postulate together with the least $\cQ_c$-spin postulate. Taking the maximum locally entangled-qubits motion principle and maximum coherence motion principle as two guiding principles, we have demonstrated that the hyperunified qubit-spinor field is an entangled hyperqubit-spinor field which possesses the inhomogeneous hyperspin symmetry as hyperunified symmetry. Namely, the fundamental symmetry of nature should be characterized by a single inhomogeneous hyperspin symmetry in association with inhomogeneous Lorentz-type/Poincar\'e-type group symmetry in $D_h$-dimensional hyper-spacetime, i.e., $PO(1, D_h-1) \adjoin WS(1, D_h-1)$. We have presented a detailed investigation on the locally entangled state of nine qubits as entangled enneaqubit-spinor field with zero $\cQ_c$-spin charge $q_c=0$ and the locally entangled state of ten qubits as entangled decaqubit-spinor field with $\cQ_c$-spin charge $q_c=1$.

It has been shown that the motion-irrelevant intrinsic $\cQ_c$-spin symmetry SP(3) as family-spin symmetry SU$_F$(2) of entangled octoqubit-spinor field in 14-dimensional hyper-spacetime does transmute into the motion-correlation inhomogeneous hyperspin symmetry WS(1,17) of entangled enneaqubit-spinor field in 18-dimensional hyper-spacetime. So that WS(1,17) provides a minimal hyperunified symmetry when taking the entangled enneaqubit-spinor field as minimum hyperunified qubit-spinor field in 18-dimensional hyper-spacetime. Such a minimum hyperunified qubit-spinor field contains two families of chiral type lepton-quark states in SM together with two families of mirror lepton-quark states beyond SM. We have proved that each family of chiral type lepton-quark states in SM is described by a twisting entangled hyperqubit-spinor field which is regarded as grand unified qubit-spinor field with negative U-parity, and each family of mirror lepton-quark states is characterized by a mirror twisting entangled hyperqubit-spinor field which is viewed as mirror grand unified qubit-spinor field with positive U-parity. Therefore, the minimum hyperunified qubit-spinor field is defined to have a negative H-parity and brings about the prediction on mirror lepton-quark states.

We have corroborated that the entangled decaqubit-spinor field provides the minimal entangled hyperqubit-spinor field which unifies all known lepton-quark states into a single hyperunified qubit-spinor field with $\cM_c$-spin charge $\cC_{\cM_c}=19$ and $\cQ_c$-spin charge $\CQc=1$. The hermitian action has been verified to possess an associated symmetry which is characterized by the inhomogeneous hyperspin symmetry WS(1,18) and scalaing symmetry SG(1) in association with inhomogeneous Lorentz-type/Poincar\'e-type group symmetry PO(1,18) and scaling symmetry SC(1) in 19-dimensional Minkowski hyper-spacetime, i.e.:
\be
G_S & = & SC(1)\ltimes PO(1,18) \adjoin WS(1,18) \rtimes SG(1)  \nn \\
& = & SC(1)\ltimes P^{1,18}\ltimes SO(1,18) \adjoin  SP(1,18) \rtimes W^{1,18}\rtimes SG(1) , \nn
\ee
which indicates that the hyperunified field theory based on such a hyperunified symmetry has a conserved H-parity and U-parity as well as the ordinary P-parity. As the leptons and quarks in SM are chiral fermions due to electroweak interaction and have negative U-parity in hyperunified qubit-spinor field, the conserved H-parity and U-parity as well as P-parity and family-spin symmetry have to be  broken down appropriately so as to realize the standard model with three families of chiral leptons and quarks, which is beyond the scope of present paper as part I and will be discussed in another paper as part II of the foundation of the hyperunified field theory\cite{FHUFT-II}.

Before ending this paper, we would like to address that the establishment of hyperunified field theory presented in our previous paper\cite{HUFT} has taken for granted the spinor field in high dimensional hyper-spacetime as hyperunified spinor field and the Poincar\'e-type symmetry in Minkowski hyper-spacetime as fundamental symmetry. In this paper as part I of the foundation of the hyperunified field theory, instead of taking directly such assumptions, we have tried to understand from a more fundamental notion why and how the spinor field emerges as basic constituent of nature and spacetime appears as Minkowski spacetime with only single temporal dimension. Meanwhile, we have also attempted to comprehend how and why the inhomogeneous hyperspin symmetry in Hilbert space together with inhomogeneous Lorentz-type or Poincar\'e-type symmetry in coordinate spacetime appear as an association symmetry. For that, our start point is from the motion nature of differential column vector field by proposing the maximum coherence motion principle when following along the action principle in the path integral formalism. It is such a maximum coherence motion principle that does lead to the automatic presence of high dimensional spinor field and emergence of Minkowski hyper-spacetime with only one temporal dimension as well as appearance of hyperspin symmetry associated with Poincar\'e-type symmetry. When applying such a maximum coherence motion principle to local coherent state of qubits and locally entangled state of qubits, we are allowed to determine explicitly the dimension of hyper-spacetime and symmetry of hyperunified qubit-spinor field for a given qubit number, which motivates us to further make the maximum locally entangled-qubits motion principle. Such a motion principle is shown to extend the hyperspin symmetry to inhomogeneous hyperspin symmetry as fundamental symmetry. Therefore, the maximum locally entangled-qubits motion principle is considered to be a more fundamental principle, which enables us to derive naturally the essential canonical anticommutation relation and Pauli exclusion principle and understand deeply what is the fundamental building block of nature and what is the fundamental symmetry of nature. Meanwhile, it allows us to comprehend why the time is different from space and why there exist more than one family lepton and quark states in nature as well as why our observed universe is four-dimensional spacetime.

Furthermore, we would like to make the following remarks. Firstly, as it has been demonstrated in section VII that in light of the self-conjugated chiral-like qubit-spinor structure, a string-like qubit-spinor field can always be constructed from any qubit-spinor field in $D_h$-dimensional spacetime to keep only a longitudinal free motion along two-dimensional spacetime, we then arrive at the observation that for such a kind of string-like qubit-spinor field, the general $\cM_c$-spin/hyperspin symmetry SP(1,$D_h$-1) of qubit-spinor field is split into a direct product group symmetry SP(1,1)$\times$SP($D_h$-2). 

Secondly, it has been shown that any qubit-spinor field in $D_h$-dimensional spacetime can be expressed as a superposition of two string-like qubit-spinor fields with opposite chirality, so that the relevant transverse free motion in $(D_h-2)$-dimensional space occurs only between two string-like qubit-spinor fields with opposite chirality. Thus we come to the observation that only the qubit-spinor field with nonzero qubit number has a transverse free motion as a zeroqubit-spinor field or bit-spinor field possesses a free motion only in two-dimensional spacetime. We has verified that any hyper-spacetime dimension of qubit-spinor field can be classified as a mod 8 integer number, $D_h^{(q_c,k)} = D_{q_c} + 8k$, with $D_{q_c}$ representing four basic spacetime dimensions $D_{q_c}=3,4,6,10$ with respective to four $\cQ_c$-spin charges $\CQc=q_c=1,2,3,0$ in correspondence to four basic qubit numbers $Q_{N}=1,2,3,4$. Therefore, we arrive at the conclusion that the transverse motions of basic qubit-spinor fields with respective to four basic spacetime dimensions $D_{q_c}=3,4,6,10$ occur just in four basic spatial dimensions $D_{q_c}^{\perp}\equiv D_{q_c}-2=1,2,4,8$. 

Thirdly, we want to point out that the entanglement-correlated translation-like $\cW_e$-spin Abelian-type group symmetry $W^{1,D_h-1}$ as semi-direct product group symmetry of inhomogeneous hyperspin symmetry WS(1,$D_h$-1)  =SP(1,$D_h$-1)$\rtimes$ W$^{1,D_h-1}$ arises originally from the locally entangled-qubits motion postulate, which becomes essential as fundamental symmetry for governing gravitational interaction and understanding the nature of gravity. A systematic analysis and detailed investigation will be presented in another paper as part II of the foundation of the hyperunified field theory\cite{FHUFT-II}. 

Finally, we would like to emphasize that the present paper as the part I of the foundation of the hyperunified field theory has mainly paid to the theoretical analysis and investigation on the existence of fundamental building block and emergence of fundamental symmetry as well as appearance of Minkowski hyper-spacetime. Our main observations and conclusions in this paper are achieved from an entangled qubit-spinor field theory description by following along the maximum coherence motion principle and maximum locally entangled-qubits motion principle as guiding principles. Therefore, such two guiding principles lay the foundation of the hyperunified field theory discussed in the present paper. In another paper as the part II of the foundation of the hyperunified field theory\cite{FHUFT-II}, we will present a detailed and systematic study on the fundamental interaction and evolving universe, which includes how the fundamental interactions are governed by inhomogeneous hyperspin gauge symmetry and scaling gauge symmetry as well as how the inflationary and dark universe can be comprehended from the nature of spacetime and the scaling property of hyperunified field theory. 


\centerline{{\bf Acknowledgement}}

This work was supported in part by the National Key Research and Development Program of China under Grant No.2020YFC2201501, the National Science Foundation of China (NSFC) under Grants  No.~11851302, No.~11851303, No.~11690022, No.~11747601 and special fund for theoretical physics, the Strategic Priority Research Program of the Chinese Academy of Sciences under Grant No. XDB23030100. It is grateful to the organizing committee of SUSY2021 for inviting me to deliver a plenary talk on "Foundation of Unification Theory and Space-based Gravitational Waves Detection".  I would like to dedicate this paper to Professor Zhou Guang-Zhao (K.C. Chou) and Professor Peng Huan-Wu for their great encouragements and stimulating discussions when I began to explore unified field theory.


\begin{thebibliography}{99}

\bibitem{HUFT} Y. L. Wu, ``Hyperunified field theory and gravitational gauge–geometry duality", 
Eur.Phys.J.C 78 (2018) 1, 28; arXiv:1712.04537.
\bibitem{HUFTSB} Y. L. Wu, ``Unified field theory of basic forces and elementary particles with gravitational origin of gauge symmetry". Sci. Bull. {\bf 62}, 1109 (2017); arXiv:1705.06365.
\bibitem{HUFTTK} Y. L. Wu, invited talk at the conference on Particles and Cosmology, ``Hyperunified field theory and Taiji program in space for GWD", Int. J. Mod. Phys. A 33 (2018) 31, 1844014; arXiv:1805.10119.
\bibitem{GQFT} Y. L. Wu, ``Quantum field theory of gravity with spin and scaling gauge invariance and spacetime dynamics with quantum inflation", Phys. Rev. {\bf D 93}, 024012 (2016); arXiv: 1506.01807.
\bibitem{GQFTTK} Y. L. Wu, invited talk presented at the International Conference on Gravitation and Cosmology/the fourth Galileo-Xu Guangqi Meeting, ``Theory of Quantum Gravity Beyond Einstein and Space-time Dynamics with Quantum Inflation", Int. J. Mod. Phys. A 30 (2015) 28, 1545002; arXiv:1510.04720.

\bibitem{GR} A. Einstein, Sitz. Konigl. Preuss. Akad. Wiss., {\bf 25}, 844 (1915).
\bibitem{FGR} A. Einstein, Ann. Phys. (ser. 4), {\bf 49}, 769 (1916).
\bibitem{SR} A. Einstein, Ann. Phys. {\bf 17}, 891(1905).
\bibitem{YM} C.N. Yang and R.L. Mills, Phys. Rev. {\bf 96}, 191 (1954).

\bibitem{SM1} S. Tomonaga, Prog. Theor. Phys. {\bf 1}, 27 (1946).
\bibitem{SM2a}  J. Schwinger, Phys. Rev. {\bf 73}, 416 (1948); 
\bibitem{SM2b} J. Schwinger, Phys. Rev. {\bf 74}, 1439 (1948).
\bibitem{SM3a}  R. P. Feynman, Phys. Rev. {\bf 76}, 769 (1949); 
\bibitem{SM3b} R. P. Feynman, Phys. Rev. {\bf 76}, 749 (1949); 
\bibitem{SM3c} R. P. Feynman, Phys. Rev. {\bf 80}, 440 (1950).
\bibitem{SM4}  S. L. Glashow, Nucl. Phys. {\bf 22}, 579 (1961).
\bibitem{SM5} S. Weinberg, Phys. Rev. Lett. {\bf 19}, 1264 (1967).
\bibitem{SM6} A. Salam, in Proceedings of the Eight Nobel Symposium, Stochholm, Sweden, 1968,
edited by N. Svartholm (Almqvist and Wikell, Stockholm, 1968).
\bibitem{SM7} D. J. Gross and F. Wilczek, Phys. Rev. Lett. {\bf 30}, 1343 (1973).
 \bibitem{SM8}  H. D. Politzer,
  Phys. Rev. Lett. {\bf 30}, 1346 (1973).
 
\bibitem{QFT1a}  F. Dyson, Phys. Rev. {\bf 75}, 486-502 (1949); 
\bibitem{QFT1b}F. Dyson, Phys. Rev. {\bf 75}, 1736 (1949).
\bibitem{QFT2} G. 't Hooft, M. Veltman, 
Nucl. Phys. {\bf B 44}, 189 (1972)

\bibitem{DF} P. A. Dirac,  Proc. R. Soc.  A: Math., Phys. and Eng. Sci. {\bf 117} (778): 610 (1928). 
\bibitem{GGFT6D} Y. L. Wu, ``Maximal symmetry and mass generation of Dirac fermions and gravitational gauge field theory in six-dimensional spacetime", Chin. Phys. {\bf C41}, 103106 (2017).
arXiv:1703.05436, 2017.

\bibitem{QK1} M. Gell-Mann, 
Phys. Lett. {\bf 8}, 214 (1964).
\bibitem{QK2}  G. Zweig,
CERN Report No.8182/TH.401 (1964).

\bibitem{UM} Y. L. Wu and K. C. Chou, Sci. China {\bf A41}, 324 (1998); arXiv:hep-ph/9708206.
\bibitem{GUT1} H. Georgi, in Particles and Fields 1974, ed. by C. Carlson (Amer. Inst. of Physics, New York, 1975).
\bibitem{GUT2} H. Fritzsch and P. Minkowski,  Ann. Phys. {\bf 93}, 193 (1975).


\bibitem{ST1} see, e.g., M.B. Green, J. H. Schwarz and E. Witten, ``Superstring Theory", Cambridge University Press (ISBN: 978052135753), 1988; 
\bibitem{ST2} J. Polchinski, ``String Theory", Cambridge University Press (ISBN: 9780521633031), 1998.

\bibitem{SU5} H. Georgi and S.L. Glashow,  Phys. Rev. Lett. {\bf 32}, 438 (1974).
\bibitem{PS} J. Pati and A. Salam,  Phys. Rev. D {\bf 10},  275 (1974).

\bibitem{PV1} T. D. Lee and C. N. Yang, Phys. Rev. 104, 254 (1956).
\bibitem{PV2} E. Sudarshan and R. Marshak, Phys. Rev. 109, 1860 (1958).
\bibitem{PV3} R. P. Feynman and M. Gell-Mann, Phys. Rev. 109, 193 (1958).

\bibitem{KM} M. Kobayashi and T. Maskawa, 
Prog. Theor. Phys. {\bf 49}, 652 (1973).

\bibitem{SCPV1} T. D. Lee, Phys. Rev. D {\bf 8}, 1226 (1973);  Phys. Rep. {\bf 9}, 143 (1974).
\bibitem{SCPV2} Y. L. Wu and L. Wolfenstein, Phys. Rev. Lett. {\bf 73}, 1762 (1994).
 
\bibitem{FHUFT-II} Y. L. Wu, ``The foundation of the hyperunified field theory II - fundamental interaction and evolving universe", arXiv:2104.11078.

\end{thebibliography}
\end{document}